# GRAVITATIONAL WAVE DETECTION AND LOW NOISE SAPPHIRE OSCILLATORS

Michael Edmund Tobar, B.Sc., B.E. (Hons.)

This thesis is presented for the degree of
Doctor of Philosophy
of
The University of Western Australia,
Department of Physics,
1993.

## ABSTRACT

This thesis describes the development of an ultra-low noise sapphire resonator oscillator that is tunable over X-band. While undertaking this task the author has explained some interesting and very useful phenomena in regards to the design and understanding of multi-mode resonant cavities and oscillators. The oscillator was constructed to operate as the pump oscillator in the superconducting parametric transducer system, attached to a 1.5 tonne niobium resonant bar gravitational wave detector. The effects of incorporating the pump oscillator with the parametric transducer and resonant bar system are analysed to enable prediction of the detector sensitivity, and determine if the measured oscillator specifications are sufficient.

The first part of this thesis describes the development of a Tunable Sapphire Loaded Superconducting Cavity (T-SLOSC), and its use as a high Q filter in an ultra-low noise X-band loop oscillator. Sapphire resonators exploit the ultra-low loss tangent of sapphire which allows Q values of $>10^9$ at 4 K, $>10^7$ at 77 K, and $>10^5$ at 300 K. The tuning element is a mechanically driven sapphire disc that perturbs the resonant frequencies of the dominating cylindrical sapphire crystal situated inside a niobium cavity. Due to the relatively low permittivity of sapphire ($\varepsilon \sim 10$), the operational modes in the sapphire crystals must be "whispering gallery" to enable a sufficient confinement in the sapphire so radiation losses do not limit the Q factor.

To understand the mode structures in a cylindrical sapphire dielectric resonator, Maxwell's equations were solved for the modes inside a uniaxial anisotropic dielectric with the crystal c-axis aligned with the cylindrical z-axis, and in the surrounding free space. Matching fields on the dielectric and free space interface gives a set of transcendental equations that are solved numerically for the resonant frequencies. The theory developed is confirmed by room temperature measurements in three sapphire crystals of different aspect ratios. For the mode families with axial and radial electromagnetic field variations of less than two, the resonant frequency are predicted to within .1% of the experimental values. The sensitivity of mode frequencies to dimensional and permittivity perturbations are also analysed. This theory has for the first time enabled complete identification of the field structure of most observable higher order modes in a cylindrical anisotropic dielectric resonator.

The T-SLOSC operates as an overmoded cavity, with niobium cavity modes and many other dielectric modes other than the main whispering gallery mode families in the X-band spectrum. This thesis analyses for the first time the complex interactions that occur in a multimode cavity, by developing an equivalent circuit from a Lagrangian technique. Coupling between modal fields is shown to be predominantly reactive in the cavity, and can degrade a mode with $Q > 10^8$ by a few orders of magnitude when it is tuned close to a low Q mode. Interactions between co-axial cable resonances and T-SLOSC modes are also observed and seen to be predominantly resistive through the niobium probes. Cross coupling coefficients between some interacting modes have been determined and reflection coefficients modelled with excellent agreement between theory and experiment.

To create a tunable oscillator the T-SLOSC resonator is configured with either room temperature or cryogenic GaAs varactor phase shifter and FET amplifiers to set the phase and gain conditions. Alternatively a fixed frequency Sapphire Loaded Superconducting Cavity (SLOSC) oscillator is mixed with a commercial synthesiser. Phase noise of the two oscillators relative to eachother are measured by phase locking them together in quadrature. The T-SLOSC oscillator phase noise was limited by vibrations of the mechanically tuned element. Experimentally it has been determined that the tuning coefficient of the resonant frequency must be less than $10^6$ Hz/mm to avoid vibrational effects in the noise spectrum. This severely limits the useable tuning range of a T-SLOSC mode to less than 5 MHz. The SLOSC oscillator phase noise was determined to be -140 dBc/Hz at 1 kHz offset frequency. To the author's knowledge this is the best value of phase noise ever measured in an X-band oscillator at this frequency offset.

The gravitational wave detector system at the University of Western Australia consists of a 1.5 tonne niobium resonant bar and a 0.5 kg niobium bending flap that has its resonant frequency tuned to the fundemental resonant frequency of the niobium bar. The bending flap is configured with a 10 GHz re-entrant cavity parametric transducer, which enables very small bending flap displacements to cause a significant modulation of the ultra-low noise pump oscillator incident on the cavity. The cavity transducer is represented in scattering parameter form. The scattering matrix is augmented to include amplitude and phase components. The off diagonal terms can be used directly calculate the amplitude to phase and phase to amplitude conversions that the pump oscillator experiences due to the transducer. The incident pump oscillator causes changes in the resonant frequencies and Q values of the antenna normal modes. Simulations presented have succeeded in modelling these effects, as well as determining parameters such as the electro-mechanical coupling, bending flap effective mass and bonding effects of the secondary mass to the bar. These effects are incorporated into the transfer function of the antenna-transducer system. This work explains some previous unexplained aspects of this type of transducer, as well as simplifying previous work to create a more cohesive understanding.

A noise analysis program was developed utilising the optimal filter technique, allowing the sensitivity of a resonant bar detector with a parametric transducer to be predicted. Utilising the harmonic oscillator model for a resonant-mass antenna transfer function and optimal filter theory was implemented to define the spectral strain sensitivity akin to laser interferometer free-mass detectors. The spectral strain sensitivity is the natural way to compare gravitational wave detectors due to its independence on the gravitational signal. Simple formulae are derived that calculate the optimal detectable strain sensitivity and effective bandwidth in terms of the spectral strain sensitivity for a multi-mode resonant-mass gravitational wave antenna.

For the UWA detector past results show the antenna was limited by vibrations and operated with a spectral strain sensitivity $1.8 \cdot 10^{-20}$ per root Hertz with a 0.8 Hz effective bandwidth. This corresponds to a 1 ms gravitational burst strain sensitivity of $7.1 \cdot 10^{-18}$. If vibrations were eliminated the electronic and thermal noise would have limited the detector to operate with a spectral strain sensitivity $1.5 \cdot 10^{-21}$ per root Hertz with a 1 Hz effective bandwidth. This corresponds to a 1 ms gravitational burst strain sensitivity of $5 \cdot 10^{-19}$. With improvements in the cryogenic demodulation system and secondary mass design, results suggest that the system will operate with a spectral strain sensitivity of less than $10^{-21}$ per root Hertz with about a 18 Hz effective bandwidth. This corresponds to a 1 ms gravitational burst strain sensitivity of less than $10^{-19}$. Further simulations suggest that approaching the thermal limit of $3.7 \cdot 10^{-20}$ strain sensitivity for a 1 ms gravitational burst may be possible.

This thesis shows that the constructed tunable low noise oscillator is adequate for it's required inclusion into The University of Western Australia's gravitational wave detector. This should allow the detector to operate at a comparable sensitivity to the cryogenic resonant bar detectors with SQUID transducers at Louisiana State University and Rome.

## ACKNOWLEDGMENTS

I would like to acknowledge many people who have assisted me in producing this thesis. Firstly I would like to thank my supervisor David Blair for making the project possible, his stimulating method of supervision, and providing excellent research facilities. For the initiation into the laboratory techniques I would like to especially thank Peter Veitch, who answered my many queries in the first six months of my project, and Darryl Ramm for helping me with the "TSLOSC". I would also especially like to thank Adrian Giles for all those stimulating late night discussions about physics over many cups of beverages. Many thanks also go to all fellow roof top goers who have provided the ultimate night of bludging. No person can sanely work forever even if they are a Ph.D. student. Especially I would like to thank Andre Luiten, Tony Mann, Marco Costa, Steve Jones, John Ferrierinho, Ralph James. All of whom have contributed towards my thesis and have provided good company on a Friday. Also I would like to thank the more occasional and distinguished goers, whom I greatly respect and value their knowledge, namely Michael Buckingham, Cyril Edwards and Frank van Kann. Fellow resonant bar and transducer researchers Eugene Ivanov, Peter Turner, Nick Linthorne and Peng Hong are also thanked for their input to my last three chapters of my thesis. Also I would like to thank Jesse Searls of Poseiden Scientific Instruments for taking an interest in the low noise sapphire oscillators.

All other members of the gravity group are thanked including the interferometer dudes, miscellaneous astronomers, cosmologists and extra terrestrial life seekers for making the group meetings interesting and extend beyond the narrow confines of high Qs and no noise. Also I would like to thank ALL work shop personnel for excellent work, and all physics department secretarial staff over the last four years.

Thanks also to the members of the LSU and Stanford groups who I met at various conferences over the last few years. Discussions with them helped me clarify the optimum filter method of analysing a gravitational resonant bar detector.

Thanks to Penny for providing an interesting life beyond physics. Thanks to students and staff at Kingswood and St. Columba college who have provided an interesting abode. Also thanks to my brothers for being close friends and introducing me to the life of a "Blue Meany". Two years of creating colourful tie dye clothes provided a great escape into the real world.

For financial support I would like to thank The University of Western Australia for providing me with a studentship. Also I would like to thank the faculty of engineering at Monash University for supplying me with the funds to travel to many conferences and institutes in 1991. This work was supported by the Australian Research Council.

# CONTENTS

APPENDIX E

# PREFACE

This dissertation is a summary of four years intensive work that has made a number of significant scientific contributions in two areas of engineering and physics. When I first commenced this project the aim was to build a low noise tunable X-band oscillator to be utilised as a pump oscillator in a parametric transducer for a resonant bar gravitational wave detector. The first four chapters of this thesis summarise the research and development under taken on high-Q resonators and low noise oscillators, which contributes to the area of frequency and time standards. The last three chapters summarise the research and development under taken on gravitational wave resonant bar detectors, which contribute to the area of gravitational physics.

My first task was to develop a high-Q tunable X-band resonator based on the Sapphire Loaded Superconducting Cavity (SLOSC) invented at the University of Western Australia. The SLOSC resonator exploits the low loss tangent of sapphire dielectric resonators and the high confinement factor of "whispering gallery" modes inside cylindrical resonators. To obtain extremely high Q factors the resonator operates at 4.2 K, and hence this task involved learning numerous cryogenic techniques.

The first chapter describes the construction and characterisation of the high-Q tunable resonator. The resonator is made from a low loss sapphire crystal with a mechanical tuning element. The density of resonant modes inside this type of device is very large compared to conventional dielectric resonators. When tuning the resonator I noticed that complex mode interactions would occur as they were tuned closely. After reviewing the literature I realised that very little work had been done on interacting resonances because this phenomena did not occur in a conventional resonator. Thus using a Lagrangian technique I developed for the first time an equivalent circuit model of interacting resonant modes. I submitted this work for publication in August 1990 and it was published in September 1991 (Tobar and Blair, Sept. 1991). This work has been summarised in chapter 3. In 1992 I extended this work to calculate quantitatively the effects of a spurious mode on an operational mode in a resonator (Tobar, 1993). The last part of chapter 3 contains some of this work.

During 1990 I became interested in understanding the field structures of the resonant modes inside a cylindrical piece of mono-crystalline sapphire. As far as we knew no one had yet succeeded in calculating accurately the mode structures and frequencies. Attempts by Jet Propulsion Laboratories had large discrepancies between their calculations and results (Dick and Saunders, 1990). It turned out that this was because they assumed the dielectric was isotropic. Sapphire is anisotropic and as soon as I included anisotropy in Maxwell's field equations I was able to determine the mode structures and calculate the mode frequencies to within 0.1 %. This work is presented in chapter 2 of this thesis. Also I presented this work at the 1991 IEEE International Microwave Symposium (Tobar and Mann, June 1991). I received an honourable mention for the presentation of this paper. An expanded version of this paper was also published in the transactions in December 1991 (Tobar and Mann, 1991). Later I discovered that the Russians had done some similar work in 1987, taking into account anisotropy (Bun'kov et al, 1987), however their calculations were more approximate. This work was important for characterising the mode families in my tunable oscillator. Moreover it was invaluable information for the SLOSC frequency standard being developed at the University of Western Australia, where the knowledge about mode structures was paramount in determining the effects of paramagnetic impurities on individual modes (Mann et al, 1992; Giles, 1993).

In parallel to the research presented in chapters 2 and 3 I was trying to build and characterise low noise oscillators based on the SLOSC resonators. This work was done between 1989 and 1992 and is summarised in chapter 4. The performance of an oscillator is dependent on the noise added by the individual components. Besides oscillator noise measurements this chapter presents noise measurements of the amplifiers, phase shifters and SLOSC resonators that comprise the oscillator. For the tunable SLOSC oscillator it was necessary to include a two-stage vibration isolator as the tuning mechanism was sensitive to vibrations. The first round of phase noise measurements were presented at the 45th Annual Symposium on Frequency Control (Tobar and Blair, May 1991). In this paper it was shown that the vibration isolation system behaved as expected and that the oscillator was dominated by excess electronic flicker noise. The measured oscillator noise was -120 dBc/Hz at 1 kHz from the carrier. To investigate ways of improving the oscillator noise I then embarked on an intensive investigation of the component noise. This work was presented at the 1992 IEEE International Microwave Symposium (Tobar and Blair, June 1992). I found by understanding the component noise I was able to reduce the phase noise by 20 dB in a fixed frequency oscillator to -140 dBc/Hz at 1 kHz offset (Tobar and Blair, Feb. 1994). To our knowledge this is the best ever phase noise directly measured in an X-band oscillator. This result implied that a phase noise of -175 dBc/Hz at 1 kHz offset may be possible but not directly measurable with our current techniques.

The mechanically tunable resonator exhibited too much vibrationally induced residual noise to be useful in a low noise pump oscillator that requires a 1% tuning range, even with a two stage vibration isolation system. However I derived an important formula relating vibrational excitation to phase noise (Tobar and Blair, Feb. 1994) which has been useful for other research groups (Santiago and Dick, 1993) who are trying to fine tune their resonators mechanically. In contrast a fixed frequency sapphire dielectric resonator has been shown to have very little residual noise (Tobar et al, 1993), for this reason we were able to obtain the low phase noise (-140 dBc/Hz @ 1 kHz offset) in the fixed frequency oscillator but not the tunable oscillator. An important conclusion from chapter 4 is that to obtain a versatile X-band oscillator that has a greater than 0.2% tuning range and low phase noise, it is better to mix a fixed frequency SLOSC oscillator with a low noise synthesiser rather than build a mechanically tunable oscillator.

While I was investigating sapphire resonators and oscillators I was also continually regarding their operation as a pump oscillator in the UWA microwave parametric transducer, which is attached to the UWA resonant bar gravity wave detector. The contributions that I have made are presented in chapters 5 to 7. The work compliments and extends previous research undertaken at UWA. The main task was to determine the potential detection sensitivity, and calculate the effect of potential improvements.

The first experiment I undertook on the UWA gravity wave detector was to investigate the level of series noise sensed by the parametric transducer system. Experimentally I determined that the observed level of noise ($2 \cdot 10^{-16}$ m/$\sqrt{\text{Hz}}$) was due to background vibration, which revealed that our vibration isolation system was inadequate. This result plus the improvements that we decided to make are outlined in chapter 5. The improvements includes a newly designed vibration isolation system (Blair et al, 1992) and a new non-contacting transducer, the details were published in Rev. of Sci. Intrum. July 1993 (Ivanov and Tobar et al, 1993).

To calculate the detector sensitivity all the noise components must be known as well as the transfer function of the gravitational wave detector. To calculate the transfer function of a resonant mass detector with a parametric transducer, the effects of the parametric transducer must be understood. In chapter 6 I have meticulously calculated the transfer function of the combined detector system and verified the calculation experimentally. This enabled me to optimise the detector transfer function by; calculating the effect of fine tuning the resonant masses and; operating the transducer in a configuration that is insensitive to seismic noise. This work has been published in (Tobar et al, 1991) and (Tobar and Blair, 1993).

In chapter 7 a very detailed noise simulation of the UWA detector is presented using optimal filtering theory. Here I have theoretically determined how to calculate the spectral strain sensitivity per root Hertz and bandwidth for a resonant-mass detector, akin to the laser interferometer free-mass detectors. Also the strain sensitivity for a 1 ms gravitational burst, effective noise temperature and integrated signal to noise ratio were calculated for the UWA detector in various configurations, to determine the limiting noise sources and potential detector sensitivity. I have shown that our detector is capable of operating with a spectral strain sensitivity of less than $10^{-21}$ per root Hertz with about a 18 Hz bandwidth. This is equivalent to a 1 ms burst strain sensitivity of less than $10^{-19}$, an effective noise temperature of less than 20 $\mu$K, and an integrated signal to noise ratio per mK of incident gravitational energy of about 60. This work has now been submitted for publication (Tobar and Blair, 1994), with a summary of the results published in (Tobar and Blair, Sept 1993).

The operation of the UWA gravitational wave detector is a very difficult task. We now have had the detector operating for several months up to September 1993. During this period we have shown that the vibration isolation works very well by recording resonant-mass mode temperatures of close to 4 Kelvin. Work is now under way to have the parametric transducer operating as predicted. It should not be long untill the detector is operating at the theoretically predicted sensitivities I have calculated.

# INTRODUCTION

## I.1 GRAVITATIONAL WAVE DETECTION

Gravitational radiation was first predicted by Albert Einstein (Einstein, 1916) as a consequence of his General Theory of Relativity. Gravitational waves are essentially the propagation of a wave of spacetime curvature, and are generated by perturbations in massive systems. The lowest multipole of this type of radiation is the quadrupole. Even though astrophysical events are expected to emit massive energy fluxes in the form of gravitational radiation, they are yet to be directly detected. This is because gravity waves interact very weakly with matter. However for the last couple of decades experimentalists have been pushing the limits of technology in trying to succeed in this task. Currently there are two types of detectors in operation. They are resonant mass detectors and free mass detectors.

Gravitational wave detectors rely on ultra-sensitive transducers. These transducers detect the displacement change of the system and convert it to an electronic signal. Since the first detector was built (Weber, 1960) technology has been improving rapidly. Numerous detectors use an external frequency source to monitor the displacement of the system. For these types of transducers the sensitivity can be improved by a direct improvement in the probing frequency source. Research groups working on the laser interferometer free mass detectors are developing methods to reduce noise in optical frequency sources. Research groups working on space borne microwave interferometers (Blair et al, 1993) and parametric transducers for resonant mass detectors are developing methods to reduce noise in microwave frequency sources.

If the form of the gravitational radiation and the transfer function of the detector are known, the detection sensitivity of the detector may be improved by the use of an optimal filter. An optimal filter essentially filters through the frequency range in which the gravity wave detector most strongly responds. Sources of gravitational waves may be divided into three classes; bursts, which last only a few cycles or are short compared with the observation time; periodic waves, which are the superposition of gravitational waves of a constant frequency and last a long time compared to the observation time; stochastic waves, stochastically fluctuating gravity waves which last a long time compared to the observation time.

The first gravity waves that are expected to be detected are burst events from the sudden gravitational collapse of dying stars to form neutron stars or black holes. In the electromagnetic spectrum these events are observed as supernovas, and are thought to produce a pulse of gravitational radiation on collapse, with typical pulse widths of the order of milliseconds. These bursts will thus produce a power spectrum that peak at about a kHz. Thus resonant mass detectors are designed to be resonant around this frequency with as large a bandwidth as possible. The energy of a gravitational wave is proportional to the strain on space it exerts. Gravitational wave detectors are compared by quoting the effective strain sensitivity (h) they can detect with a signal to noise ratio of 1. The magnitude of the gravitational wave strain for supernovas at 10 kiloparsecs from our solar system are thought to lie between $h \sim 10^{-18}$ to $10^{-22}$ (Thorne, 1987).

### I.1.1 Resonant Mass Detectors

Joseph Weber (Weber, 1960 & 1969) was the first to build a gravitational wave detector. He developed the theory of a resonant mass detector and showed that a mass quadrupole harmonic oscillator will be weakly excited by a gravitational wave. Resonant mass detectors are usually large cylindrical bars made from low acoustic loss materials. The second generation of such detectors have a smaller resonant mass attached to one end. This secondary mass acts as a transformer, amplifying the primary mass vibrations and thereby improving the impedance match to a displacement sensitive transducer. The combined system forms a two-mode coupled harmonic oscillator.

Because gravitational radiation causes only minute motions in resonant mass detectors, very sensitive transducers are needed. The transducers are electro-mechanical devices which transform the displacement of the secondary mass to an electrical signal. This signal is amplified by a low noise cryogenic amplifier for later analysis and filtering. The most successful transducer has been the RF superconducting quantum interference device (SQUID) transducer which have achieved strain sensitivities of less than $10^{-18}$ (Solomonson et al, 1992; Boughn et al, 1990; Amaldi et al, 1989).

A parametric transducer is an alternative transducer which is also under intense investigation. Parametric transducers involve a mechanically modulated resonant cavity with a low noise pump oscillator incident on the transducer at its resonant frequency. These transducers are somewhat more complicated than the RF SQUID transducers currently in use on the most sensitive resonant mass detectors. The development of a low noise parametric transducer involves developing, high Q displacement sensitive cryogenic resonators (Linthorne and Blair, 1992), low noise cryogenic amplifiers (Mann L.D. et al, 1986; Tobar and Blair, June 1992), low noise cryogenic carrier suppression systems (Ivanov, 1993) and low noise pump oscillators (Tobar and Blair, 1993), all at the same frequency as the transducer. At the University of Western Australia (UWA) these components are all operating well enough to have a transducer operating with a greater sensitivity than the current cryogenic RF SQUID transducers.

### I.1.2 Laser Interferometer "Free Mass" Detectors

The most common free mass detector is the laser interferometer, prototype laser interferometers have been developed with arm lengths in the order of 10 metres (Maischberger et al, 1988; Ward et al, 1988). This is an alternative technique to the resonant mass detector. It involves the monitoring of free test masses at large separation rather than relying on the energy exchange of a gravity wave to a resonant mass. The two main types of gravity wave interferometers are the delay line (Winkler, 1991) and the Fabry-Perot (Drever, 1991). Although these prototypes are less sensitive than the current resonant mass detectors, projects are now just commencing to build these detectors with base lines of the order of a kilometre. These detectors are predicted to have strain sensitivities of the order of $10^{-22}$.

It may seem that the resonant mass detector is doomed to be over taken by this new emerging technology. This may not be true as the Louisiana group has shown that the new super sphere resonant mass detector may be capable of achieving strain sensitivities in the order of $10^{-23}$ (Johnson et al, 1993), and a multi-mode transducer can increase the bandwidth to about 40 Hz (Solomonson et al, 1992). It will be very interesting to see when where and with what detector the first gravity wave will be detected, and which type of detector will ultimately attain the best sensitivity.

## I.2 FREQUENCY AND TIME STANDARDS

In modern communication networks low-noise, ultra-stable oscillators control the frequency and timing. Stability is necessary to maintain synchronisation, and low phase noise is required for the detection of phase/frequency modulated signals that pass from transmitter to receiver.

Frequency standards may be divided into two major types; resonator and atomic oscillators. Resonator oscillators have their frequency determined by a resonator such as a quartz crystal, dielectric resonator or cavity resonator. Atomic oscillators have their frequency determined by the properties of a simple atomic system, they are usually referred to by the type of atom involved. eg. hydrogen, rubidium or caesium.

### I.2.1 Resonator Oscillators

A simplified circuit diagram of a resonator loop oscillator is shown in figure I.1. The amplifier section contains at least one active device and may also include other elements for gain control and impedance matching. The feedback network contains the resonator and may include other elements such as a varactor for frequency tuning, this will allow tuning within the bandwidth of the resonator. Also, the resonator itself may include a mechanical or electronic tuning element, so that tuning greater than the bandwidth of the resonator may be achieved.

The phase condition for oscillation is that the closed loop phase shift is equal to an integral number of half wave lengths. When the oscillator is initially activated there is only noise in the circuit. The components of noise that satisfy the phase condition and have a frequency within the bandwidth of the resonator are propagated around the loop with increasing amplitude. The amplitude continues to increase until the amplifier gain is reduced by the self-limiting non-linearities in the active elements or by an external control. At steady state the closed loop gain is 1.

The noise and stability of these devices depend on the quality factor (Q) of the resonator and the noise properties of the active device. Usually when an amplifier is self-limiting it is highly saturated and non-linear upconversions will degrade the phase noise spectrum. The higher the Q of the resonator in the feedback path the smaller the perturbations in frequency. If a phase perturbation $\Delta\phi$ occurs, the frequency of oscillation must shift by $\Delta f$ to maintain the phase condition, this is given by;

$$\Delta f/f = -\Delta\phi/2Q_L \qquad \text{(I.1)},$$

where $Q_L$ is the quality factor of the loaded resonant cavity. This simple equation tells us that improving the resonator Q factor is very important when trying to reduce the frequency instabilities and phase noise in a resonator oscillator. Thus the frequency and time standard community are forever trying to improve resonator Q values.

### I.2.2 Atomic Oscillators

The properties of atomic systems must be understood in terms of quantum mechanics. Bound energy states in an atomic system are discrete, and when an atom changes from an excited to a lower energy state it will emit a photon. The photon frequency is determined by the energy difference between the two states and is given by Planck's law;

$$\nu = (E_2 - E_1)/h \qquad \text{(I.2)}.$$

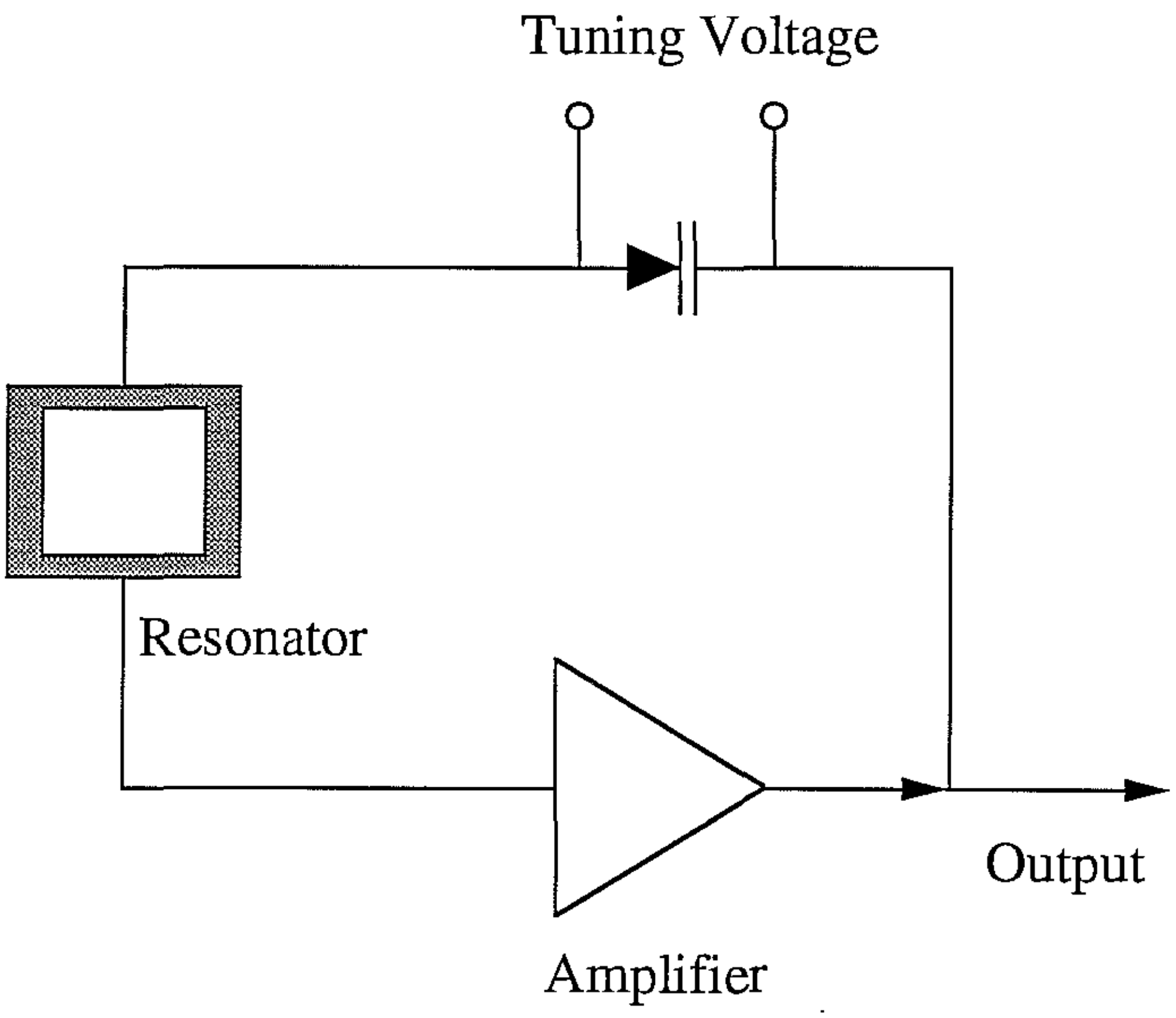

**Figure I.1**

Simplified circuit diagram of a resonator oscillator

Here $E_2$ is the energy of the upper state, $E_1$ the energy of the lower state and h Planck's constant. The photon frequencies are intrinsic to the atom, whereas a resonator oscillator is dependent on the bulk properties of its resonator. For this reason atomic oscillator frequencies are easily reproduced.

Atomic oscillators may be used in a passive or active state. In active atomic oscillators the output signal is derived from the radiation emitted by the atom. In a passive state the atoms are employed as a discriminator to control the frequency of an active device or oscillator, in a similar way to a resonator oscillator.

**I.2.3 Noise in Frequency Standards**

Noise in frequency standards limits timing accuracy, synchronisation, dynamic range, signal to noise ratio, and can cause loss of lock in a phase locked system. In order to compare oscillator instabilities and noise it is necessary to characterise them. There are two methods for characterising the stochastic variations in phase or frequency in an oscillator (Barnes et al, 1971). The first is the spectral density description, used to determine the Fourier frequency ($f_m$) components of the noise. The second is the Allan variance description used to describe the time keeping ability of the oscillator.

The signal from an oscillator may be written as;

$$V(t) = [V_0 + \varepsilon(t)] \sin[2\pi(f_0 t + \phi(t)/2\pi)] \qquad (I.3),$$

where $V_0$ is the nominal amplitude, $\varepsilon(t)$ the amplitude fluctuations, $f_0$ the nominal frequency, and $\phi(t)$ the phase fluctuations. The amplitude noise is usually smaller than the phase noise and will not be discussed further.

The double sideband spectral density of the phase fluctuation, $S_\phi(f_m)$, is defined as the mean square value of phase fluctuations squared per Hz as a function of offset frequency from the carrier. This is a very important parameter because it is directly related to the performance of oscillators in RF signal processing applications. Another common representation is $L(f_m)$, which is one half of $S_\phi(f_m)$, and is defined as the single sideband (SSB) phase noise to carrier power ratio. Sometimes the oscillator noise is characterised in terms of frequency rather than phase noise. The instantaneous fractional frequency is related to the derivative of the total phase in (I.3), by;

$$f(t) = f_0 + (d\phi/dt)/2\pi \qquad (I.4)$$

The frequency noise is described in terms of the instantaneous frequency fluctuations, y, which is defined as;

$$y = (f(t)-f_0)/f_0 = (d\phi/dt)/(2\pi f_0) \tag{I.5}$$

Relating (I.5) to spectral densities gives;

$$S_y(f_m) = (f_m/f_0)^2 \, S_\phi(f_m) \tag{I.6}$$

A simple model is used to describe the noise in precision oscillators. The spectral density can be described as a finite sum of terms proportional to the Fourier frequency raised to the power of a positive or negative integer;

$$S_y(f_m) = \sum_\alpha h_\alpha (f_m)^\alpha \tag{I.7}$$

The values of $\alpha$ for which these processes are defined are; $\alpha=2$, white phase modulation; $\alpha=1$, flicker phase modulation; $\alpha=0$, white frequency modulation; $\alpha=-1$, flicker frequency modulation; $\alpha=-2$, random walk frequency modulation; $\alpha=-4$, random walk frequency aging. However in practice this power is not necessarily an integer.

The spectral method is a powerful way to characterise an oscillator, however it is not useful for characterising its time keeping ability. The Allan variance is used for this and is defined as;

$$\sigma_y(\tau) = \frac{1}{2} E\left\{\left(\frac{\phi(t+2\tau)-2\phi(t+\tau)+\phi(t)}{2\,\pi\, f_0\, \tau}\right)^2\right\} \tag{I.8}$$

where E{ } refers to the expectation value over the ensemble of possible observations. The Allan variance describes the variation in frequency from one interval to the next. The rms time error of a clock after an interval $\tau$ is approximately $\tau\sigma_y(\tau)$, and thus provides a method of characterising the time keeping ability of a clock. The Allan variance is related to the spectral density via a complex integral transform (Barnes et al, 1971).

The atomic standards have very good long term stability. The best atomic standard is the Hydrogen Maser, which has a square root Allan variance of less than $10^{-15}$ for integration times greater than $10^3$ seconds. For smaller integration times the Hydrogen Maser degrades rapidly to about $10^{-13}$ at 1 second. To obtain good stability in this regime a high-Q resonator oscillator can be used. The best short term stability ever measured was in the cryogenic superconducting cavity oscillators at Stanford (Stein and Turneaure, 1978). They measured $3 \cdot 10^{-16}$ for integration times of 30 to 300 seconds, however the long term stability above 300 seconds degrades rapidly. The superconducting resonators in these oscillators had loaded Q factors of greater than $10^{10}$ at 1.6 K. These high Q factors were highly dependent on the surface impurities on the cavity walls, and therefore very dependent on the superconductor cleaning technique. These results are only reproducible if elaborate cleaning facilities are available. Also the frequency stability is dependent on dimensional changes in the walls of the superconductor. These changes can be induced by temperature and power dependent variations in penetration depth, radiation pressure, and coupled vibrations.

To reduce the Q value and frequency dependence on the superconducting walls, a low loss dielectric may be loaded inside a superconducting cavity (Jones 1988). The field inside the resonator will mainly exist within the dielectric, outside the dielectric the field will decay evanescently reducing the effect of the superconductor by a few orders of magnitude. For this type of cavity the Q factor and frequency are determined mainly by the losses and resonant modes in the dielectric respectively. The Sapphire Loaded Superconducting Cavities (SLOSC) developed at UWA are cavities of this type. They have obtained ultra-high Q values approaching $10^{10}$ at 2 K (Luiten et al, 1993), that are very reproducible with minimal cleaning. This is comparable to the superconducting cavities developed at Stanford. Thus the performance of the SLOSC resonator oscillator has the potential to surpass the superconducting cavity oscillators as they are less dependent on wall perturbations, which was the limiting effect in the Stanford oscillators (Stein, 1974). The SLOSC oscillators at UWA (Giles et al, 1989) have already achieved an Allan variance of $9 \cdot 10^{-15}$ for integration times from 1 to 300 seconds with loaded resonator Q values of greater than $10^{8}$.

When considering the accuracy of RF signal processing the short term stability is very important. The sapphire resonator oscillators are already in use for VLBI astronomy with the potential for superior results (Costa et al, 1992). However these oscillators are not primary standards and suffer greatly due to long term effects, ie. aging. By locking a sapphire oscillator to a primary standard a combined clock that provides a broader range of timing capabilities than either alone can be obtained.

# CHAPTER 1

# THE TUNABLE SAPPHIRE LOADED SUPERCONDUCTING CAVITY RESONATOR

## 1.1 INTRODUCTION

This chapter describes the construction and characterisation of a tunable sapphire loaded superconducting cavity (T-SLOSC) resonator. The principle aim of its construction was to utilise it as a high Q frequency discriminating element in a loop oscillator circuit, thus creating an ultra-low noise oscillator tunable over X-band. It was then to be determined if it was good enough to operate as a low noise pump oscillator in a parametric transducer system, for a resonant bar gravitational wave antenna at the University of Western Australia (Veitch et al, 1987). Its operation as an oscillator and its inclusion in the gravitational wave detector system is discussed in the later chapters of this thesis, while this chapter and chapters 2 and 3 are concerned only with analysing the properties of sapphire cavity resonators.

Fixed frequency (Jones, 1988), (Giles et al, 1989) and tunable (Blair and Sanson, 1989), (Tobar and Blair, Sept 1991) sapphire resonators have been described previously. These resonators exploit the low loss tangent of sapphire, with achievable Q values of greater than $10^9$ at 4.2 K, greater than $10^7$ at 77 K, and greater than $10^5$ at 290 K. However the permittivity of sapphire is only about ten, so to obtain confinement of about 95% in the sapphire part of the loaded cavity, a high circumferential mode number is required. Thus the cavity operates in an overmoded environment with many dielectric and cavity resonant modes in the vicinity of X-band. The overmoded environment means that as an operational mode is tuned a spurious mode may tune close and interact. This phenomena is analysed in detail in chapter 3. In this chapter the T-SLOSC is considered to be in a tuning state that contains a set of non-interacting normal modes so the interaction between closely tuned modes can be ignored.

In this chapter the design and construction of the T-SLOSC resonator is presented. Following this, an equivalent circuit model of one of the loop probes coupled to a cavity resonance is developed. Important parameters of the resonator are defined in conjunction with this model. The structure of the resonant modes are then identified and the modal parameters of a selected mode are measured with respect to the equivalent circuit model.

## 1.2 T-SLOSC CONSTRUCTION

A schematic of the T-SLOSC under investigation is illustrated in figure 1.1. The heart of this resonator is a cylindrical sapphire mushroom and a tuning disc inside a cylindrical niobium cavity. The sapphire mushroom's height by diameter dimension is 27.5 × 30.1 mm, while the sapphire tuning disc is 3 × 30.1 mm. Both are inside a 50 × 50 mm cylindrical niobium cavity. The sapphire tuning disc affects the evanescent field outside the sapphire, and perturbs the resonant frequency. Tuning is achieved by adjusting the axial position of the tuning disc using the tuning stepper motor. Typically a tuning range of the order of tens of MHz is achieved for a mode with an unloaded Q of a few hundred million. Spurious movement of the tuning disk will create an unwanted perturbation in the cavity's resonant frequency. Hence a decoupling mechanism which isolates the tuning plunger from the tuning mechanism has been incorporated into the design to help reduce this effect. Three niobium probes couple to and extract electromagnetic energy from within the cavity. Stepper motors attached to the probes vary the coupling to particular modes.

For operation at cryogenic temperatures the T-SLOSC cavity was situated in a vacuum can inside a cryogenic dewar capable of holding liquid helium. It was held to the top of the vacuum can by a two-stage vibration isolation system which is discussed further in chapter 4. Microwave power was coupled into the resonator via stainless steel coaxial cables from the top of the dewar to the vacuum can. Then from the top of the vacuum can to the movable loop probes, very thin coaxial cables were used to inhibit vibrational short circuits, and to allow the probes to move up and down freely.

## 1.3 LOOP COUPLED EQUIVALENT CIRCUIT

A single resonant mode coupled inductively by an external loop probe may be represented by the series LCR circuit shown in figure 1.2a. This circuit can then be transformed to a parallel LCR circuit in series with an inductance as shown in figure 1.2b (Montgomery, Dicke and Purcell, 1948). These circuit representations of the T-SLOSC resonator are derived in chapter 3 for the general case when two normal modes are interacting.

The four individual mode circuit elements of figure 1.2b may be expressed in terms of the four cavity - coupling parameters; $Q_1$, the unloaded cavity Q factor; $\omega_1$, the unloaded cavity resonant frequency; $L_0$ the loop probe inductance; $\kappa_{01} = L_{01}/(L_0L_1)^{1/2}$, the transformer coupling between the loop probe and cavity;

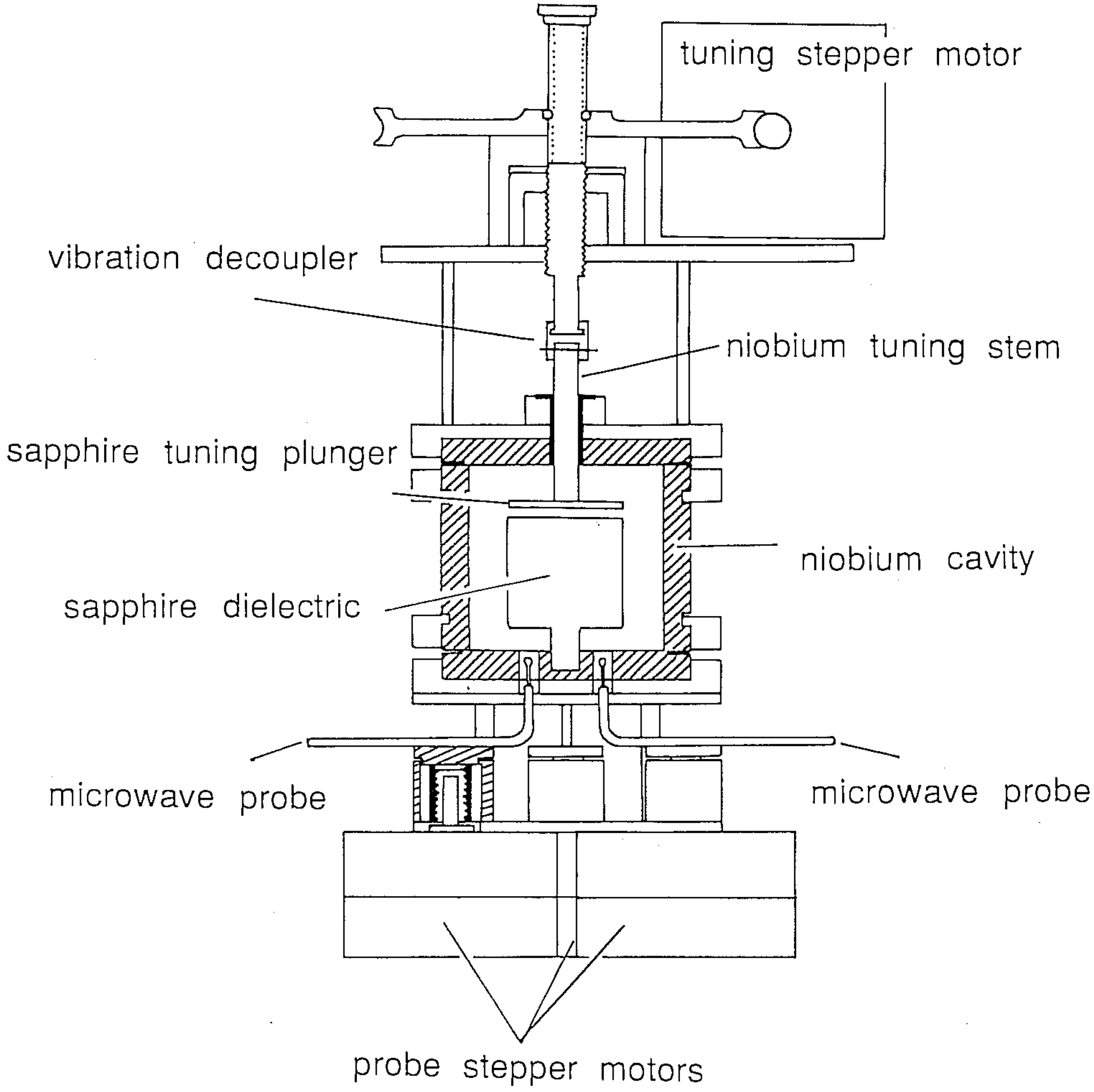

# Figure 1.1

Scale diagram of the construction of the T-SLOSC, where the internal dimension of the niobium cavity is 50 by 50 mm,

$$l_0 = L_0(1-\kappa_{01}^2) \tag{1.1a}$$
$$r_1 = \kappa_{01}^2 L_0 \omega_1 Q_1 \tag{1.1b}$$
$$l_1 = \kappa_{01}^2 L_0 \tag{1.1c}$$
$$c_1 = \frac{1}{\kappa_{01}^2 L_0 \omega_1^2} \tag{1.1d}$$

The input impedance of figure 1.2b can be written in normalised form as;

$$\frac{Z_{in}}{Z_0} = \frac{\beta_1}{1 + 2j Q_1 y_1} + j\alpha \tag{1.2}$$

where $\beta_1 = \frac{r_1}{Z_0}$, $\alpha = \frac{l_0 \omega_1}{Z_0}$, $y_1 = \frac{\omega - \omega_1}{\omega_1}$, $Q_1 = \frac{\omega_1}{\Delta\omega_1}$ ($\Delta\omega_1$ is the unloaded cavity bandwidth and $Z_0$ is the co-axial cable characteristic impedance). From the reflection coefficient defined by (1.3) the loaded resonant frequency can be derived (1.4) by finding the frequency of minimum reflection, and the loaded Q can be derived (1.5) by finding the loaded cavity bandwidth from the half power points of the reflected signal.

$$\rho = \frac{\frac{Z_{in}}{Z_0} - 1}{\frac{Z_{in}}{Z_0} + 1} \tag{1.3}$$

$$\omega_{1L} = \omega_1\left(1 + \frac{\alpha \kappa_1}{2 Q_1}\right) \tag{1.4}$$

$$Q_{1L} = \frac{Q_1}{1 + \kappa_1} \tag{1.5}.$$

Here $\kappa_1$ is the definition of the coupling to a cavity with a series reactance (Ginzton, 1957), (Kajfez, 1984), (Kajfez and Hwan, 1984) and is related to the normalised cavity parameters by;

$$\kappa_1 = \frac{\beta_1}{\alpha^2+1} \tag{1.6}$$

From (1.4) the frequency pulling due to the probe reactance can be calculated to be;

$$\omega_{1L}-\omega_1 = \frac{\alpha \kappa_1 \Delta\omega_1}{2} \tag{1.7}.$$

Thus the frequency pulling is proportional to the coupling coefficient and series inductance. If the inductance was zero there would be no frequency pulling and the coupling would be resistive. For electric field probes the series reactance is capacitive giving a negative value of $\alpha$.

The series inductance given by (1.1a) is less than the probe inductance. Defining $\alpha' = \frac{L_0 \omega_1}{Z_0}$ and combining (1.1a&b) in normalised form, it can be shown that;

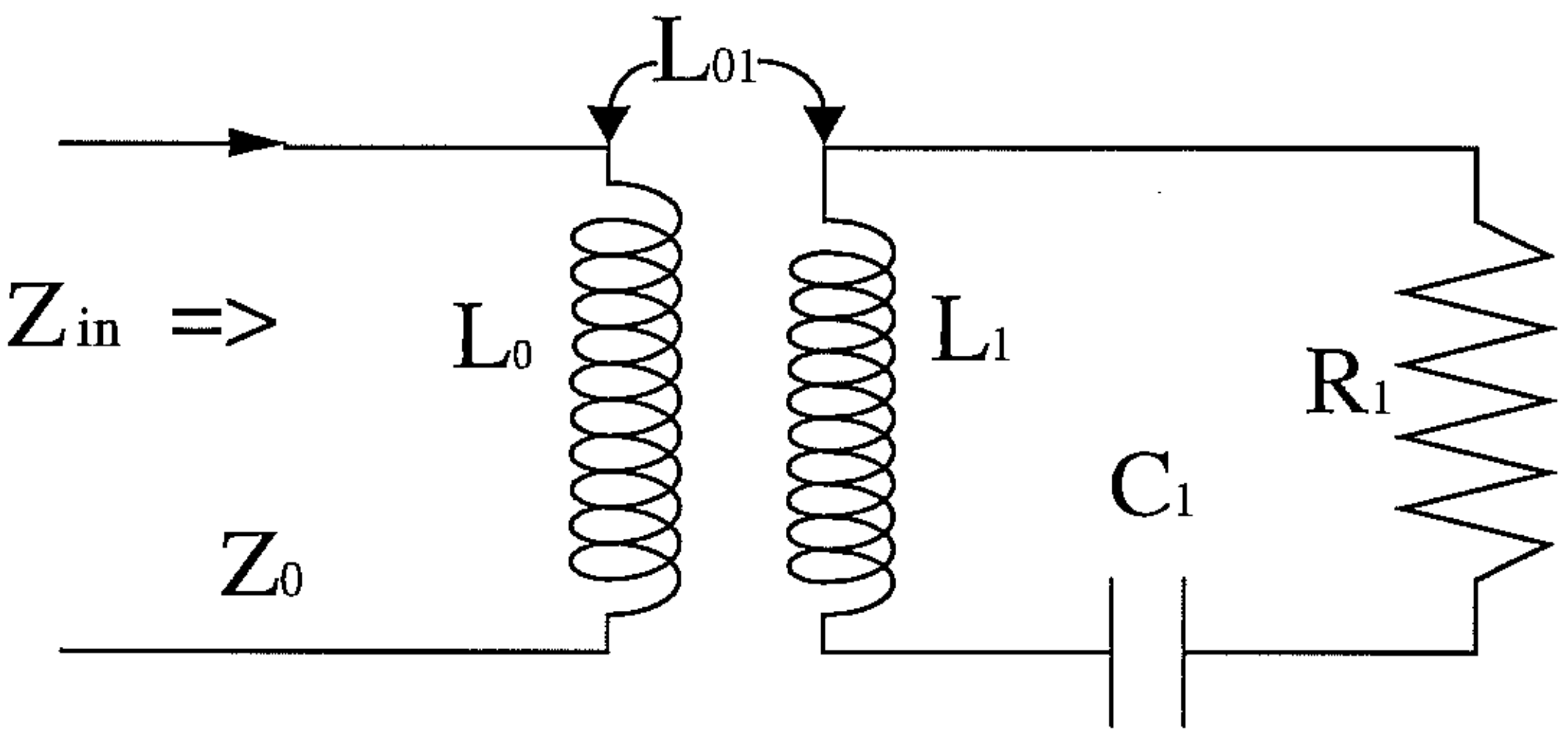

Figure 1.2a

A single cavity normal mode coupled inductively by an external loop probe can be represented by the above series LCR circuit.

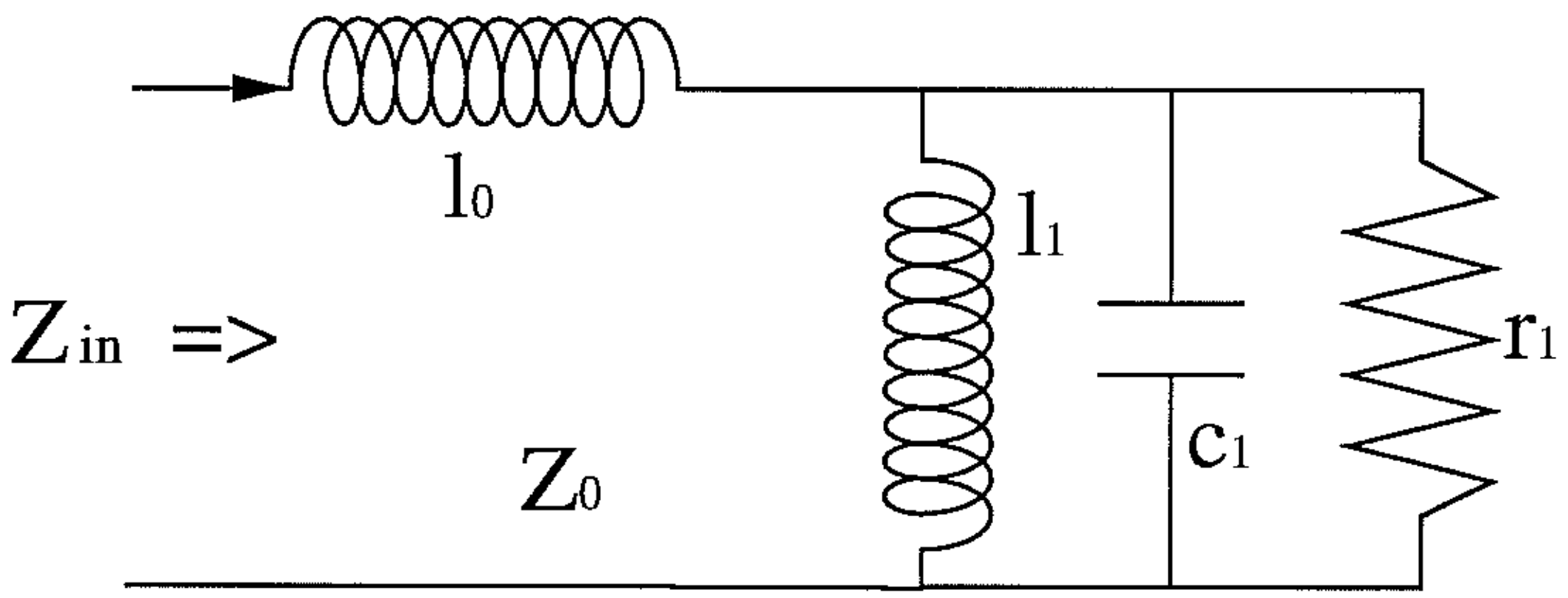

Figure 1.2b

The above circuit shown in figure 1.2a can be mathematically transformed to a parallel LCR circuit in series with a finite inductance.

$$\alpha = \alpha' - \frac{\beta_1}{Q_1} \qquad (1.8)$$

Thus for any moderately high Q cavity the series reactance ($\alpha$) will be equivalent to the probe reactance ($\alpha'$), ie the transformer coupling $\kappa_{01}$ in (1.1a) can be ignored.

## 1.4 T-SLOSC PARAMETERS

### 1.4.1 Mode analysis

The higher order modes in a dielectric resonator are hybrid modes, but can be classified as quasi $TE_{mnp+\delta}$ (Transverse Electric, $E_z$~0) or quasi $TM_{mnp+\delta}$ (Transverse Magnetic, $H_z$~0) depending whether the dominant field in the resonant structure is either $H_z$ or $E_z$. Here m is the azimuthal mode number, n the radial mode number and p the axial mode number. Details of how to identify and calculate the field components and frequencies are presented in chapter 2.

Tuning ranges may be predicted by calculating the resonant frequency of the cylindrical resonator with height h and h+t. For the quasi $TE_{6\,1\,1+\delta}$ at 4.2 K a tuning range of 94 MHz is predicted, this is consistent with the measurement presented in figure 1.3. From table 1.1 we can see that modes with greater axial mode number p have larger tuning ranges.

Frequency shifts with temperature are predicted by the change in dielectric constant, which is the dominant effect. TM modes have a larger frequency shift with temperature than TE modes, because the permittivity parallel to the c-axis changes more than the permittivity perpendicular to the c-axis. From table 1.1 the frequency shift of the $TE_{6\,1\,1+\delta}$ and $TM_{8\,1\,1+\delta}$ from 290 K to 4.2 K is calculated to be approximately 72 and 110 MHz respectively, which is to within 5% of what is actually measured.

### 1.4.2 Q factor and coupling

To measure the T-SLOSC parameters the circuit illustrated in figure 1.4a was constructed. A high stability fixed frequency SLOSC was mixed with a HP-8662A synthesiser to create a highly stable tunable sweep oscillator.

For a slow sweep rate the steady state characteristic of $|\rho|^2$ was traced out. The resonant frequency was measured from the point of minimum reflection, the loaded Q factor from the bandwidth given by the half power points, and the coupling from (1.9);

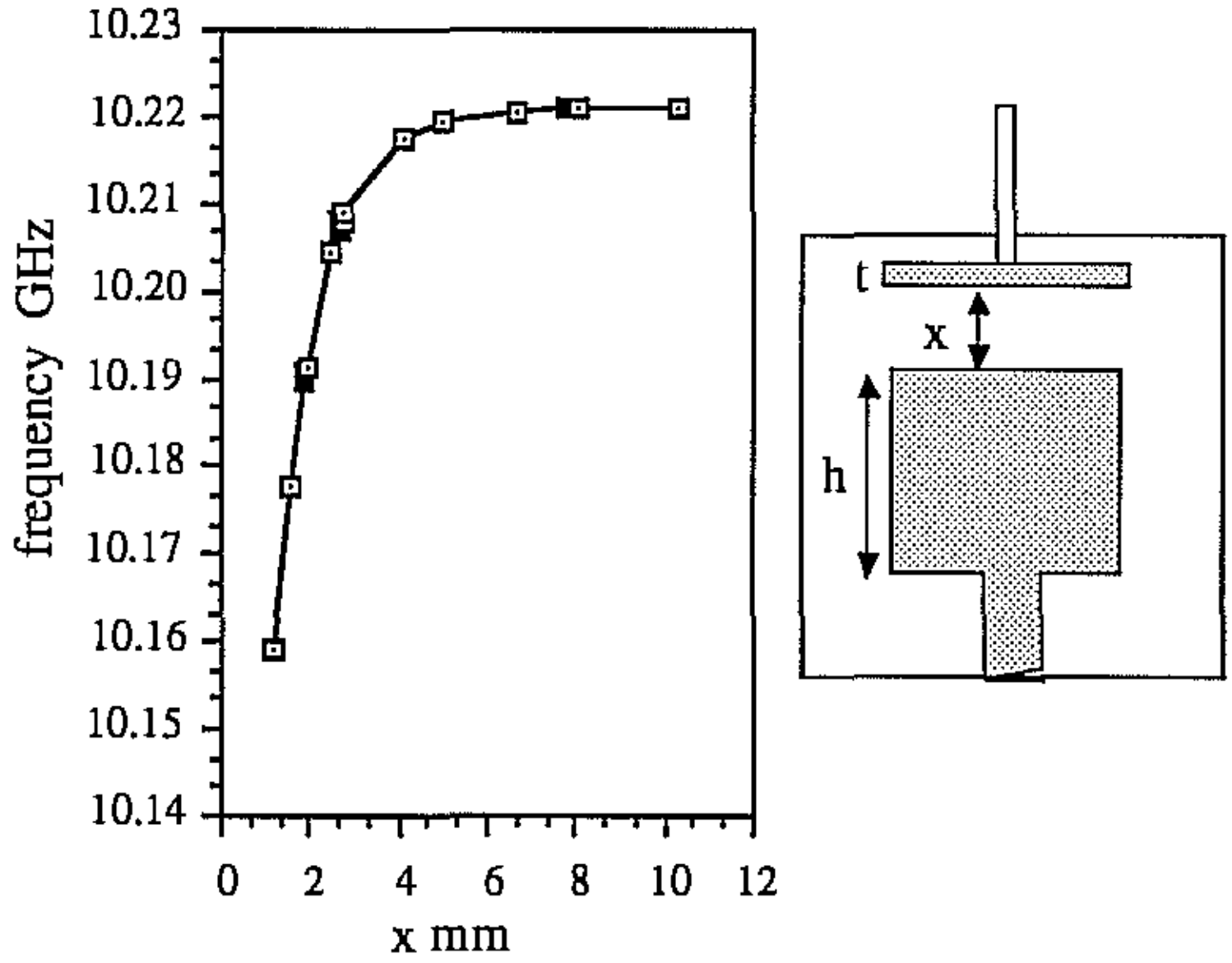

Figure 1.3

Tuning curve for the quasi $TE_{6\ 1\ 1+\delta}$ mode, with a schematic of the T-SLOSC, with a sapphire mushroom of height h, tuning disk of thickness t, and tuning parameter x defined as the distance from the top of the mushroom to the disk.

| TE Modes m,n,p+ δ | Predicted Frequency @ 290 K (GHz) | Experimental Frequency @ 290 K (GHz) | Predicted Frequency Shift @ 4K (MHz) | Predicted Tuning Range (MHz) |
|---|---|---|---|---|
| 8, 1, 0 | 11.921 | 11.911 | 95.9 | 20.6 |
| 7, 1, 0 | 10.828 | 10.809 | 74.3 | 22.4 |
| 6, 1, 0 | 9.712 | 9.688 | 66.5 | 25.1 |
| 5, 1, 0 | 8.586 | 8.553 | 58.9 | 28.5 |
| 7, 1, 1 | 11.223 | 11.220 | 79.0 | 85.6 |
| 6, 1, 1 | 10.154 | 10.151 | 71.8 | 94.3 |
| 5, 1, 1 | 9.087 | 9.090 | 64.5 | 104.8 |
| TM Modes m,n,p+ δ | | | | |
| 10, 1, 0 | 11.890 | 11.888 | 131.9 | 24.8 |
| 9, 1, 0 | 10.909 | 10.882 | 114.2 | 21.1 |
| 8, 1, 0 | 9.917 | 9.888 | 107.7 | 28.0 |
| 7, 1, 0 | 8.920 | 8.887 | 96.9 | 30.0 |
| 9, 1, 1 | 11.289 | 11.320 | 121.8 | 90.6 |
| 8, 1, 1 | 10.336 | 10.368 | 110.3 | 98.6 |
| 7, 1, 1 | 9.384 | 9.415 | 99.9 | 108.2 |
| 6, 1, 1 | 8.436 | 8.463 | 88.9 | 120.7 |

Table 1.1

Theoretical and experimental comparisons of frequency for the four lowest order in p mode families. Also included are the predicted frequency shifts from 290 K to 4.2 K and the expected tuning range for each mode.

$$\kappa = (1-\sqrt{\Gamma_0/\Gamma_S})/(1+\sqrt{\Gamma_0/\Gamma_S}) \qquad (1.9)$$

Here $\Gamma_0$ is defined as the reflected power on resonance and $\Gamma_S$ as the reflected power off resonance. An example of $|\rho|^2$ for the quasi $TM_{8\ 1\ 1+\delta}$ mode at 4.2 K is shown in figure 1.4b. Thus the loaded Q and coupling were calculated to be $2.7\cdot10^8$ and 0.1 respectively at a frequency of 10.44 GHz.

If the sweep rate is increased, at a certain rate the reflected response started to ring (Schmitt and Zimmer, 1966). As long as the coupling to the mode is small, the Q factor will be proportional to the ring down time. Finite coupling causes an asymmetry in the ring down. An example of the fast sweep technique is shown in figure 1.4c. The unloaded Q factor can be calculated from the half amplitude decay time ($t_{1/2}$) by;

$$Q = \frac{f_1\, t_{1/2}\, \pi}{\ln[2]} \qquad (1.10)$$

The unloaded Q factor of the quasi $TM_{8\ 1\ 1+\delta}$ mode at 4.2 K using this method was calculated to be $2.5\cdot10^8$, which is to within 8% of the slow sweep method.

### 1.4.2 Probe reactance

The T-SLOSC is configured with movable probes so the coupling to the required operational mode may be varied. By increasing the coupling gradually and observing the frequency pulling, from (1.7) the probe reactance may be calculated. These measurements were done on the quasi $TE_{6\ 1\ 1+\delta}$ at 4.2 K, and are plotted in figure 1.5. The normalised probe reactance is given by the slope of figure 1.5, thus $\alpha \approx 1.2$. From this result the inductance of the probe can be inferred to be of the order of $10^{-9}$ henrys at 10 GHz, with 50 ohm characteristic impedance coaxial cables.

The inductance of a single loop may also be calculated using the definition $L = \Lambda/i$, where $\Lambda$ is the flux linkage and i is the current in the loop. Using the Biot Savart law the flux linkage of a single coil may be estimated, and it can be shown that $L \sim 10^{-6}$ r, where r is the radius of the coupling loop in meters. Thus to obtain an inductance of $10^{-9}$ henrys the radius of the loop must be in the order of 1mm, which is approximately the radius of the T-SLOSC loop probes (shown in figure 1.1). This calculation confirms the model of the cavity - coupling system presented in section 1.3.

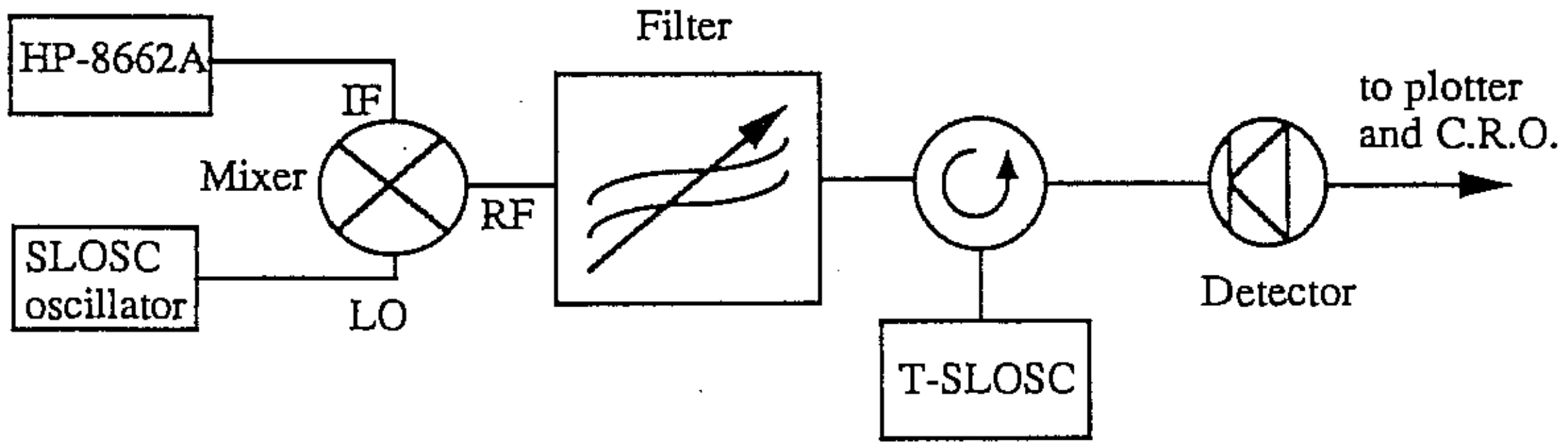

Figure 1.4a

A tunable sweep oscillator was created by mixing a fixed frequency SLOSC oscillator with a HP-8662A synthesizer. This set up to was used to measure the T-SLOSC parameters, including the resonant frequencies, Q factors and couplings.

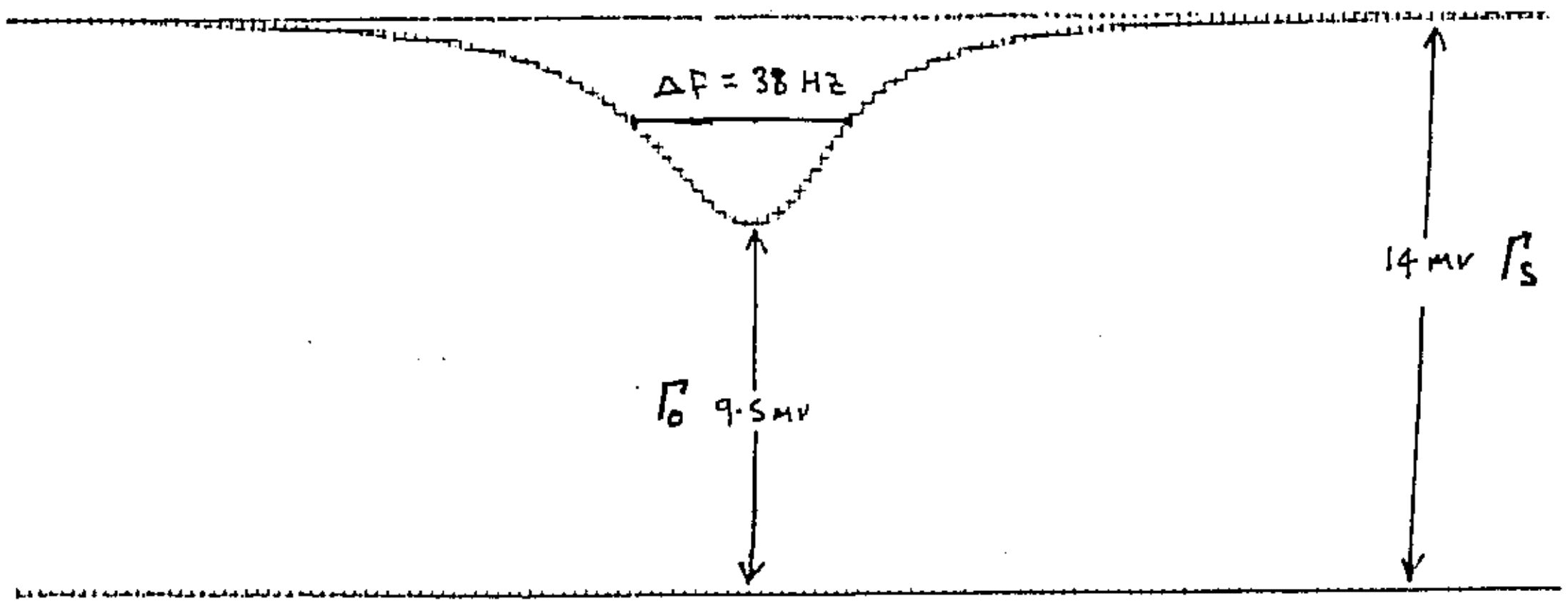

Figure 1.4b

$|\rho|^2$ as measured by the detector in figure 1.4a when the probing oscillator was swept slowly across the quasi $TM_{8\,1\,1+\delta}$ mode at 10.44 GHz, at a rate of 100 seconds per frequency span. From this plot the resonant frequency, coupling and loaded Q factor can be measured.

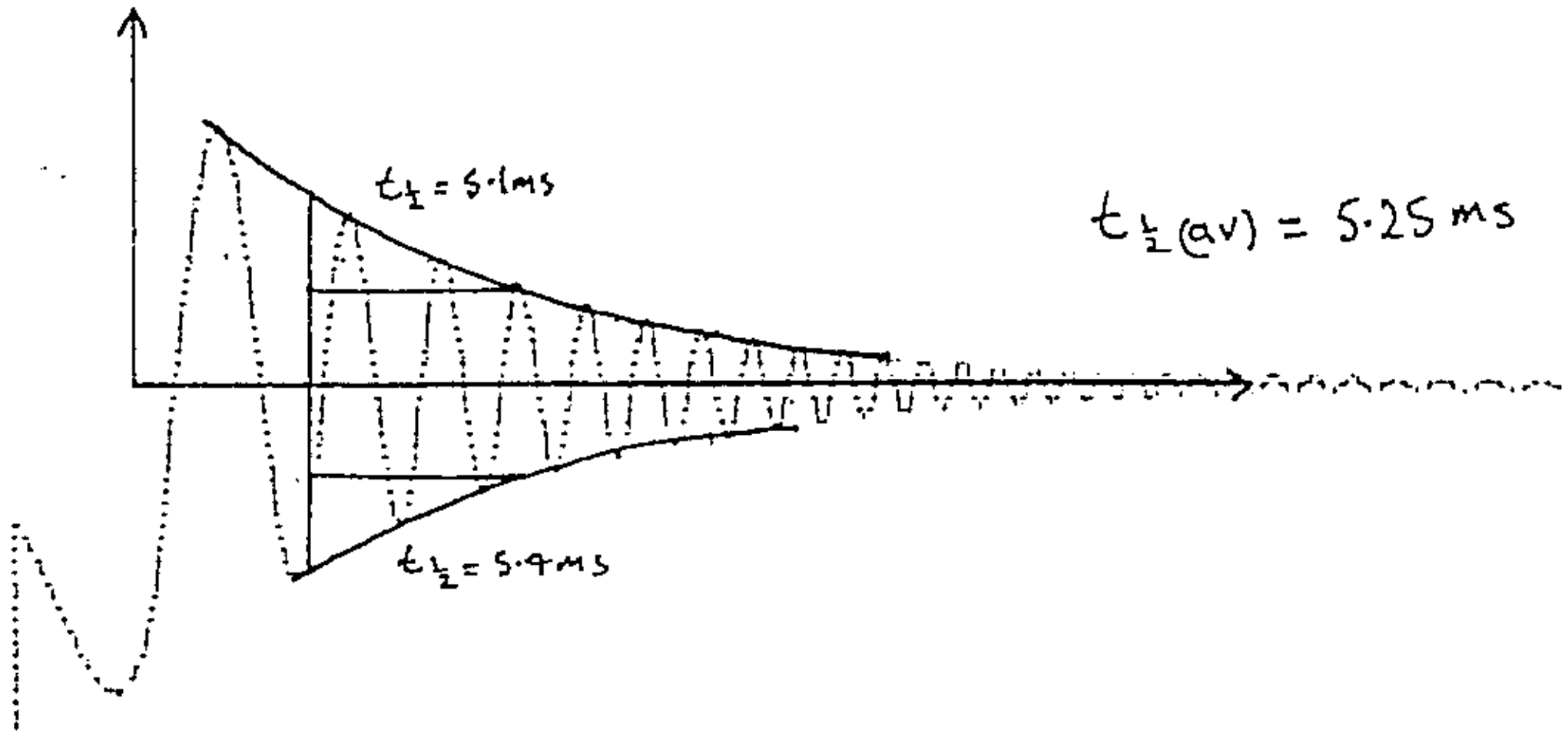

Figure 1.4c

With the sweep rate increased to 50 milliseconds per frequency span the above ringing is observed. The Q factor can be measured from the half amplitude decay time ($t_{1/2}$), using (1.9). The ringing is slightly asymmetric due to the finite coupling.

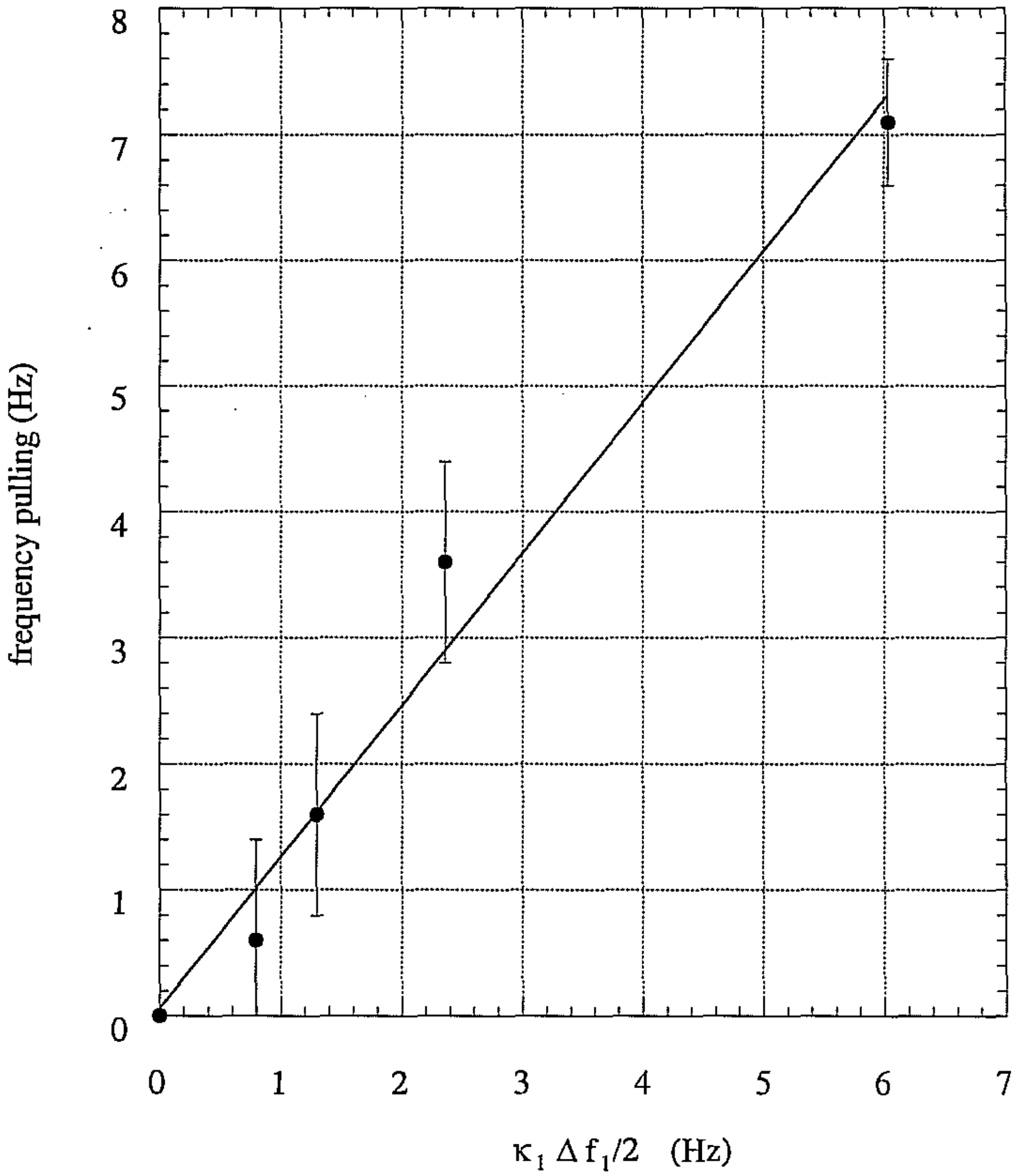

Figure 1.5

Frequency pulling versus the product of the halfbandwidth and coupling. From (1.7) the slope gives the value of the probe normalized reactance $\alpha$.

# CHAPTER 2

# RESONANT FREQUENCIES OF HIGHER ORDER MODES IN CYLINDRICAL ANISOTROPIC DIELECTRIC RESONATORS

## 2.1 INTRODUCTION

The higher order modes in cylindrical sapphire dielectric resonators at cryogenic temperatures (Braginsky et al 1987) can exhibit extremely high Q factors (>$10^7$) and can be used to construct ultra-stable low noise microwave oscillators (Dick and Saunders, 1990), (Giles et al, 1989 & 1990), (Tobar and Blair, May 1991 & June 1992). Other anisotropic dielectric crystals such as rutile (Okaya, 1960), (Okaya and Barash, 1962) and lithium niobate (Bourreau, Guillon and Chartard-Moulin, 1986) offer higher permittivities but also much higher losses.

In this chapter an improved method is developed which allows the determination of mode frequencies to high accuracy in cylindrical anisotropic dielectric resonators (Tobar and Mann, 1991). This is an extension of Garault and Guillon's method (Garault and Guillon, 1976) from isotropic to anisotropic dielectrics, applied to four different classes of field patterns. Previous equations are shown only to be valid for quasi TE modes with even mode number in the axial direction. Four different axial match equations are derived depending whether the modes are quasi TE or quasi TM, and have an odd or even axial mode number. A general radial match equation is derived. Combining it with the relevant axial equation forms a set of two coupled transcendental equations which can be solved numerically. This is a general treatment of higher order modes in an anisotropic medium, although previously whispering gallery mode approximations for anisotopic crystals have been used successfully (Bun'kov et al, 1987), (Ivanov and Kalinichev, 1988).

Theory is confirmed by room temperature measurements in two sapphire crystals of different aspect ratios. Very good agreement is found even though the permittivity is only about ten. The anisotropy forces the TM mode families to be lower in frequency than the TE, which explains the discrepancy between theory and experiment of previous work (Dick and Saunders, 1990). Successful predictions of frequency shifts from 300 K to 4.2 K are made using the known change in permittivity. Also the tuning range of a tunable sapphire resonator has been calculated due to the effective change in height (presented in chapter 1). This work has lead to a very good understanding of electromagnetic resonances in sapphire crystals, with potential applications to design.

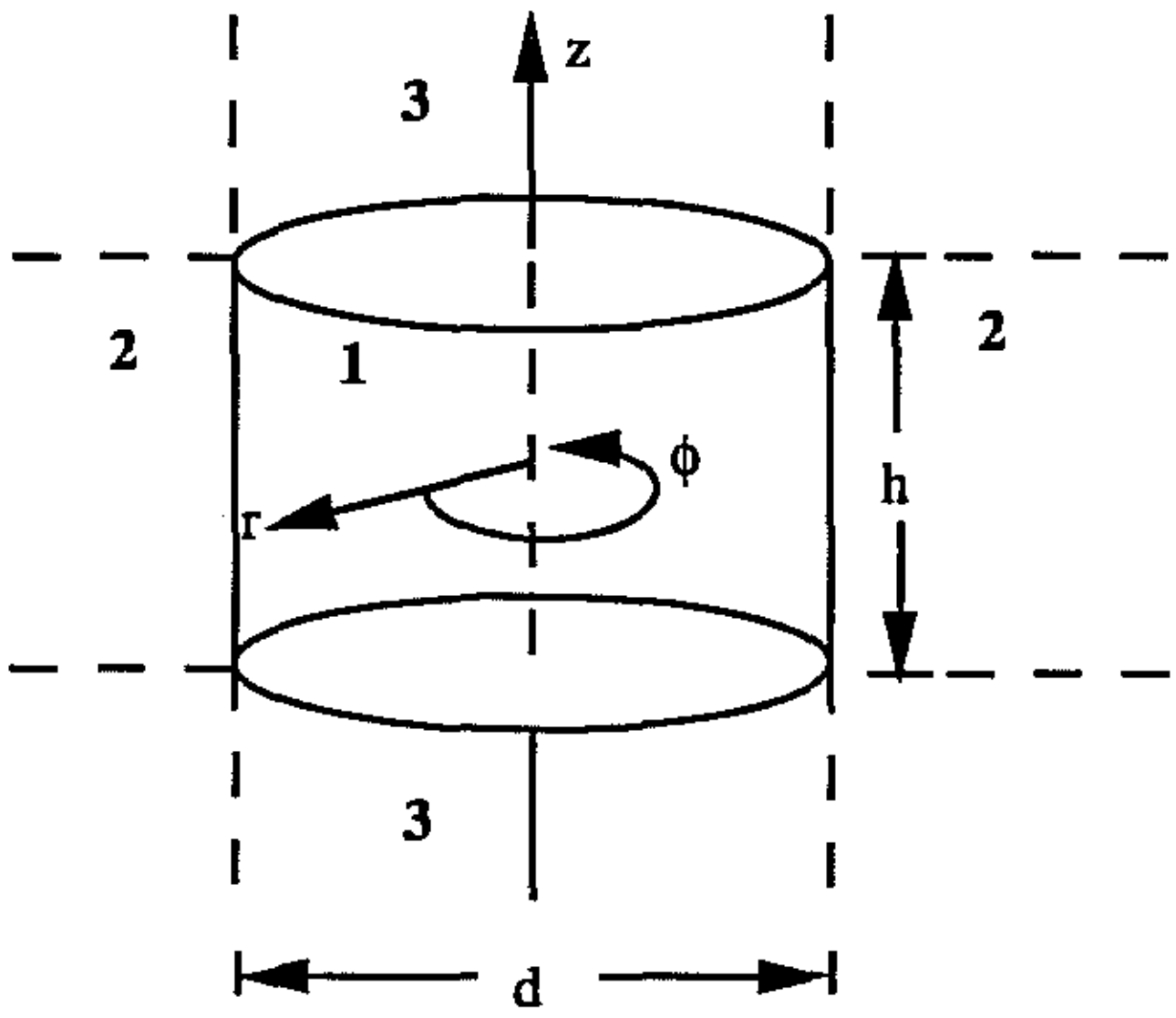

# Figure 2.1

The open dielectric crystal is analyzed in cylindrical coordinates {r,φ,z}. Resonant frequencies are solved by matching tangential fields between regions 1 and 2, and regions 1 and 3.

## 2.2 THEORY

Cylindrical anisotropic crystals in free space are analysed relative to the coordinate system defined in figure 2.1, with the c-axis of the crystal parallel to the z axis. The permittivity parallel and perpendicular to the c-axis is defined as $\varepsilon_z$ and $\varepsilon_r$ respectively. Thus $\varepsilon_\phi = \varepsilon_r$ and we assume no off diagonal terms in the permittivity tensor.

The problem is solved using Maxwell's equations for anisotropic media, a complete derivation of the field components is given in appendix A. Applying separation of variables on the z component of the electromagnetic field (see appendix A), it follows that;

$$E_{z1} = A\, J_m(k_E r) \begin{matrix} \mathrm{Cos}(m\phi) \\ \mathrm{Sin}(m\phi) \end{matrix} \left(P_1 \exp(-j\beta z) + P_2 \exp(+j\beta z)\right) \quad (2.1a)$$

$$E_{z2} = C\, K_m(k_{out} r) \begin{matrix} \mathrm{Cos}(m\phi) \\ \mathrm{Sin}(m\phi) \end{matrix} \left(P_1 \exp(-j\beta z) + P_2 \exp(+j\beta z)\right) \quad (2.1b)$$

$$E_{z3} = E\, J_m(k_E r) \begin{matrix} \mathrm{Cos}(m\phi) \\ \mathrm{Sin}(m\phi) \end{matrix} \exp(-\alpha z) \quad (2.1c)$$

$$H_{z1} = B\, J_m(k_H r) \begin{matrix} \mathrm{Sin}(m\phi) \\ \mathrm{Cos}(m\phi) \end{matrix} \left(P_1 \exp(-j\beta z) + P_2 \exp(+j\beta z)\right) \quad (2.1d)$$

$$H_{z2} = D\, K_m(k_{out} r) \begin{matrix} \mathrm{Sin}(m\phi) \\ \mathrm{Cos}(m\phi) \end{matrix} \left(P_1 \exp(-j\beta z) + P_2 \exp(+j\beta z)\right) \quad (2.1e)$$

$$H_{z3} = F\, J_m(k_H r) \begin{matrix} \mathrm{Sin}(m\phi) \\ \mathrm{Cos}(m\phi) \end{matrix} \exp(-\alpha z) \quad (2.1f)$$

where; $k_E^2 = \varepsilon_z k_0^2 - \beta^2$, $k_H^2 = \varepsilon_r k_0^2 - \beta^2$ and $k_{out}^2 = \beta^2 - k_0^2$ (2.1g)

Here m is the azimuthal mode number, $\beta$ the axial propagation constant inside the dielectric, $k_0$ the free space wave number, $k_{out}$ the radial propagation constant outside the dielectric, and $k_E$ and $k_H$ the radial propagation constants inside the dielectric for TM and TE modes respectively. From the z component, Maxwell's equations can then be used to obtain all other electromagnetic field components (Auda and Kajfez, 1986).

### 2.2.1 Radial Match

By matching tangential components of the $\underline{H}$ and $\underline{E}$ fields between regions 1 and 2, the following transcendental equation is obtained;

$$\left(\frac{\varepsilon_r J'_m(x_E)\, x_E}{x_H^2 J_m(x_E)} + \frac{K'_m(y)}{y\, K_m(y)}\right)\left(\frac{J'_m(x_H)}{x_H J_m(x_H)} + \frac{K'_m(y)}{y\, K_m(y)}\right) = m^2 \frac{\left(x_H^2+\varepsilon_r y^2\right)\left(x_H^2+y^2\right)}{x_H^4\, y^4} \tag{2.2}$$

where $x_E = k_E\, d/2$ , $x_H = k_H\, d/2$ and $y = k_{out}\, d/2$ . For a fixed diameter this equation is a function of two variables, $k_0$ and $\beta$. To find the solutions graphically (2.2) is expressed in terms of $x_H^2$ and $y^2$ using the propagation constant relations in (2.1g) to eliminate $x_E$.

In general y becomes imaginary for the lower order whispering gallery mode families. In this case the Evanescent Bessel Function $K_m(y)$ becomes a Hankel Function of the second kind (Wait, 1967). Hence we employ algorithms which allow complex arguments.

### 2.2.2 Axial Match

A second transcendental equation may be derived by matching H and E fields between regions 1 and 3. The field components of $E_z$ and $H_z$ inside a resonator must be orthogonal in space and hence can not coexist with the same dependence on z. However assuming the same dependence simplifies proceedings by allowing the axial match to be calculated independently of the radial match. Therefore for the z dependence of (2.1a) - (2.1f) to be consistant, quasi TE modes ($E_z \approx 0$) or quasi TM modes ($H_z \approx 0$) must be assumed. Four different transcendental equations are derived by matching tangential fields between regions 1 and 3, and are given by:

$TE_{m\,n\,p+\delta}$ p even; $$k_0^2 = \beta^2 \frac{1+\mathrm{Tan}^2(\beta\, h/2)}{\varepsilon_r - 1} \tag{2.3a}$$

p odd; $$k_0^2 = \beta^2 \frac{1+\mathrm{Cot}^2(\beta\, h/2)}{\varepsilon_r - 1} \tag{2.3b}$$

$TM_{m\,n\,p+\delta}$ p even; $$k_0^2 = \beta^2 \frac{1+\left(\mathrm{Tan}(\beta\, h/2)/\varepsilon_r\right)^2}{\varepsilon_z - 1} \tag{2.3c}$$

p odd; $$k_0^2 = \beta^2 \frac{1+\left(\mathrm{Cot}(\beta\, h/2)/\varepsilon_r\right)^2}{\varepsilon_z - 1} \tag{2.3d}$$

Here the radial and axial mode numbers are n and p respectively. Equation (2.3a) is the same as derived by Garault and Guillon, which was solved with the isotropic version of equation (2.2). As before, to find the soloutions graphically (2.3) is expressed in terms of $x_H^2$ and $y^2$ using the propagation constant relations in (2.1g) to eliminate $x_E$.

### 2.2.3 Solving the coupled equations

Equations (2.2) and (2.3) were expressed in terms of $x_H^2$ and $y^2$ then solved using Mathematica (Wolfram, 1988). Solutions were initially found graphically on a $x_H{}^2$ versus $y^2$ graph, then more accurately using a Newton-Raphson technique. For a given value of m, (2.2) gives an infinite set of solutions in $\{x_H{}^2, y^2\}$ space, which are almost perpendicular to the $x_H{}^2$ axis. This distinguishes the radial mode number n, and whether they are TE or TM. Equation (2.3) gives a infinite set of solutions nearly parallel to the $x_H{}^2$ axis. Mode frequencies are solved from the intersection of these two solution sets. Care must be taken to avoid spurious solutions due to the restriction of p being either even or odd, or modes being TE or TM. Figure 2.2 illustrates how the solution of the $TE_{6\,1\,\delta}$ in a sapphire crystal of height by diameter of 0.03 by 0.03 metres is found graphically.

## 2.3 EXPERIMENTAL VERIFICATION

### 2.3.1 Cylindrical orientated sapphire

Figure 2.3 shows how quasi TE and TM modes are distinguished. To excite a TM or TE mode an $E_z$ or $H_z$ field is excited respectively. Analysis of azimuthal and axial mode numbers is done by observing the $H_\phi$ or $E_\phi$ field respectively. In reality all modes are hybrid, and experimentally one can excite higher order axial mode number families with either a TM or TE probe. Figure 2.4 illustrates the power density of the quasi $TE_{61\delta}$ mode's $H_z$ field given by (2.1d) inside a cylindrical sapphire crystal of height by diameter of 0.03 by 0.03 metres.

An experimental study of resonant modes was conducted from 8 to 12 GHz for two cylindrical pieces of sapphire with different aspect ratios. Figures 2.5 and 2.6 compare experimental and theoretical mode frequencies, and observed open resonator Q values. There is better agreement for the whispering gallery type families of low axial mode number p, as they are more TM or TE like. The difference between the calculated and measured mode frequencies generally increases with p and decreases with m. Percentage difference between calculated and measured frequencies are presented in tables 2.1 and 2.2 for the 50.0 mm diameter sapphire crystal. For modes with p=0 errors are less than .1% which is smaller than estimated uncertainties in permittivities and dimensions. Above p=2 errors can be of the order 1% in TE modes and 2% in TM modes (Tobar and Mann, June 1991). Tables 2.3 and 2.4 show the percentage uncertainties for the 31.8 mm diameter sapphire crystal, in general they are less accurate due to the smaller azimuthal mode numbers present in this crystal at X-band, ie. they are less whispering gallery like.

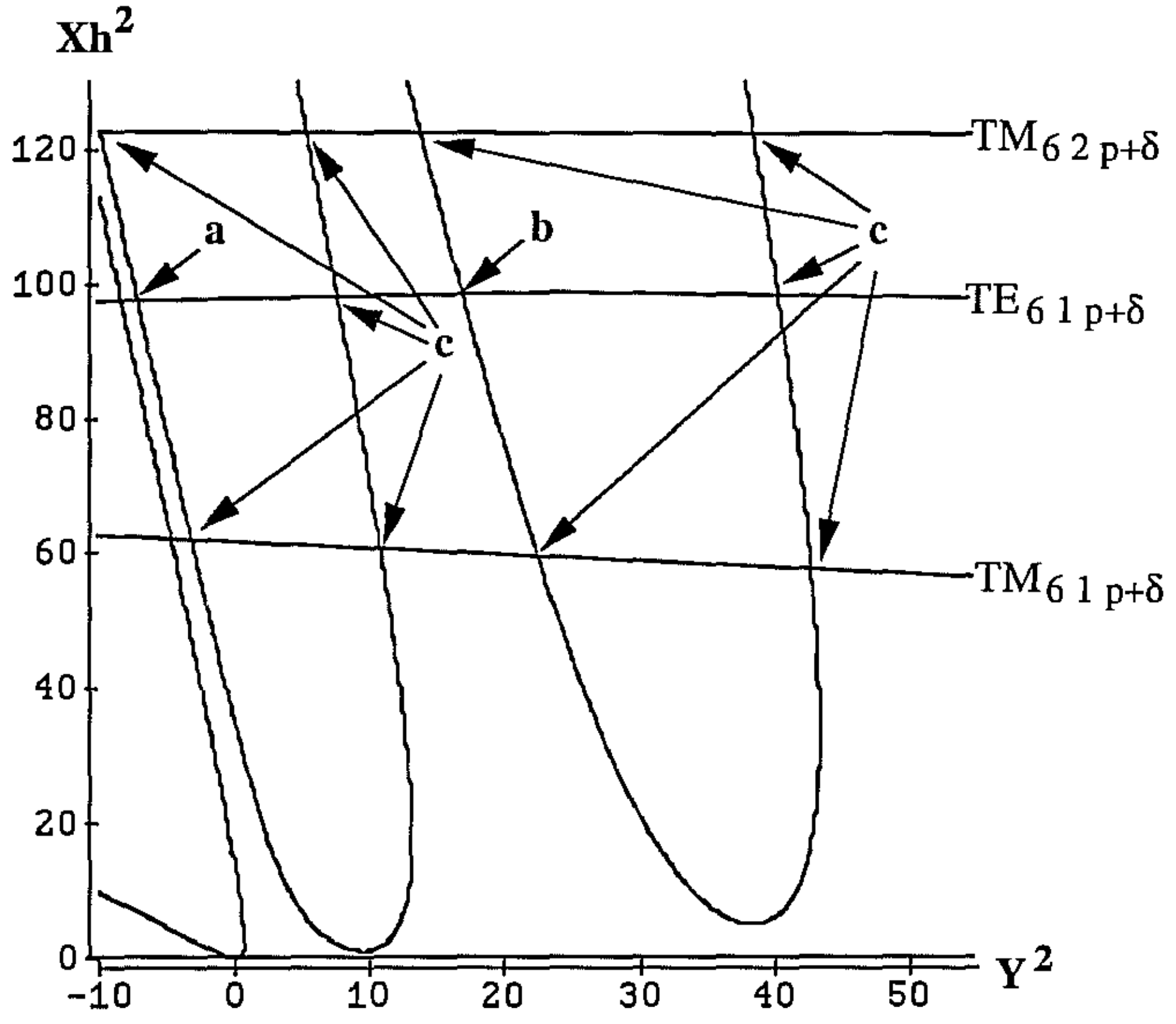

# Figure 2.2

Intersections of (2.2) and (2.3a) are illustrated in $\{x_h^2, y^2\}$ space.
(a) $TE_{6\,1\,\delta}$ mode. (b) $TE_{6\,1\,2+\delta}$ mode. (c) Spurious solutions.

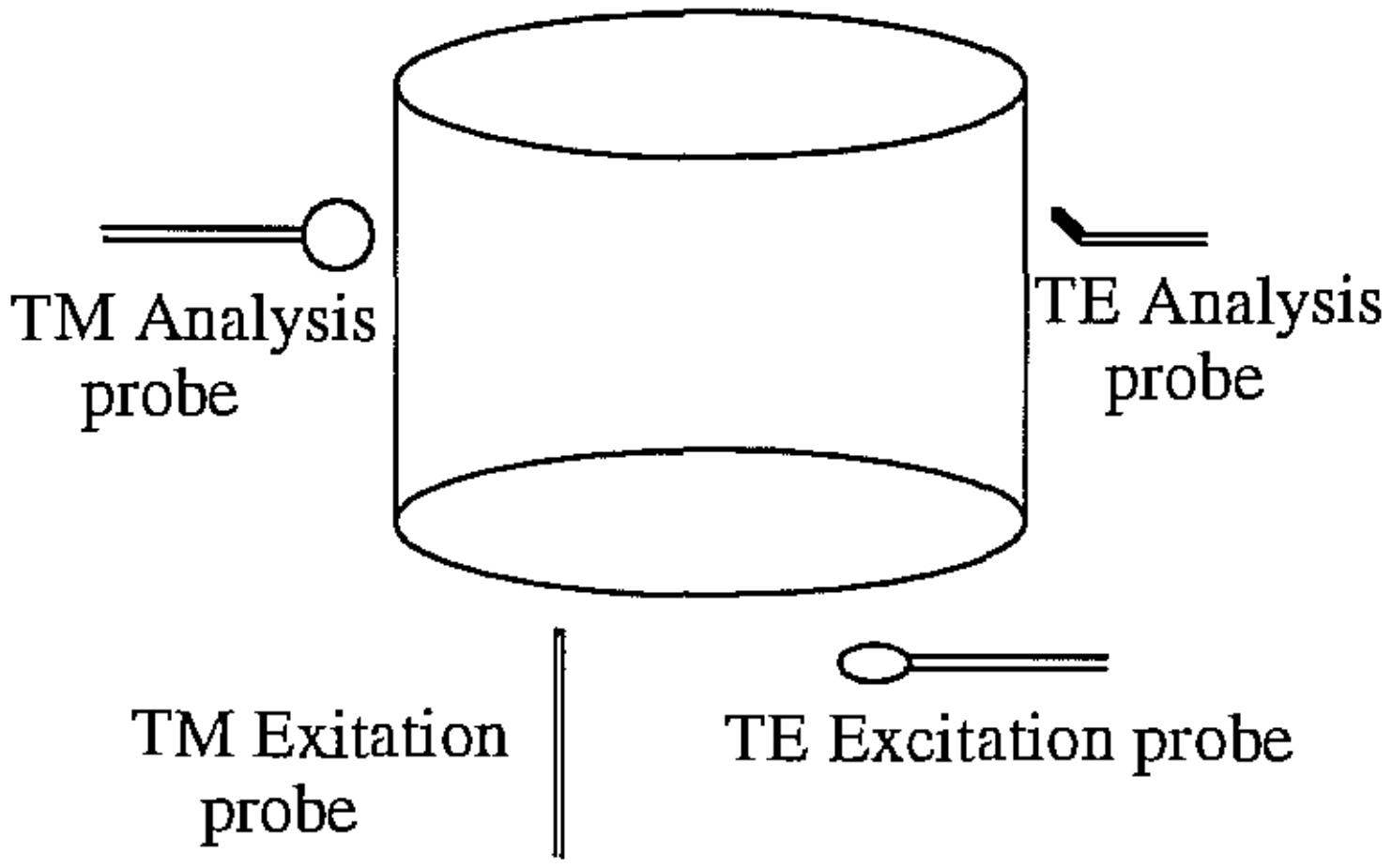

# Figure 2.3

TM modes are excited by creating an $E_z$ field, while TE modes are excited by creating a $H_z$ field. TM modes are analysed by coupling to the $H_\phi$ field, while TE modes are analysed by coupling to the $E_\phi$ field.

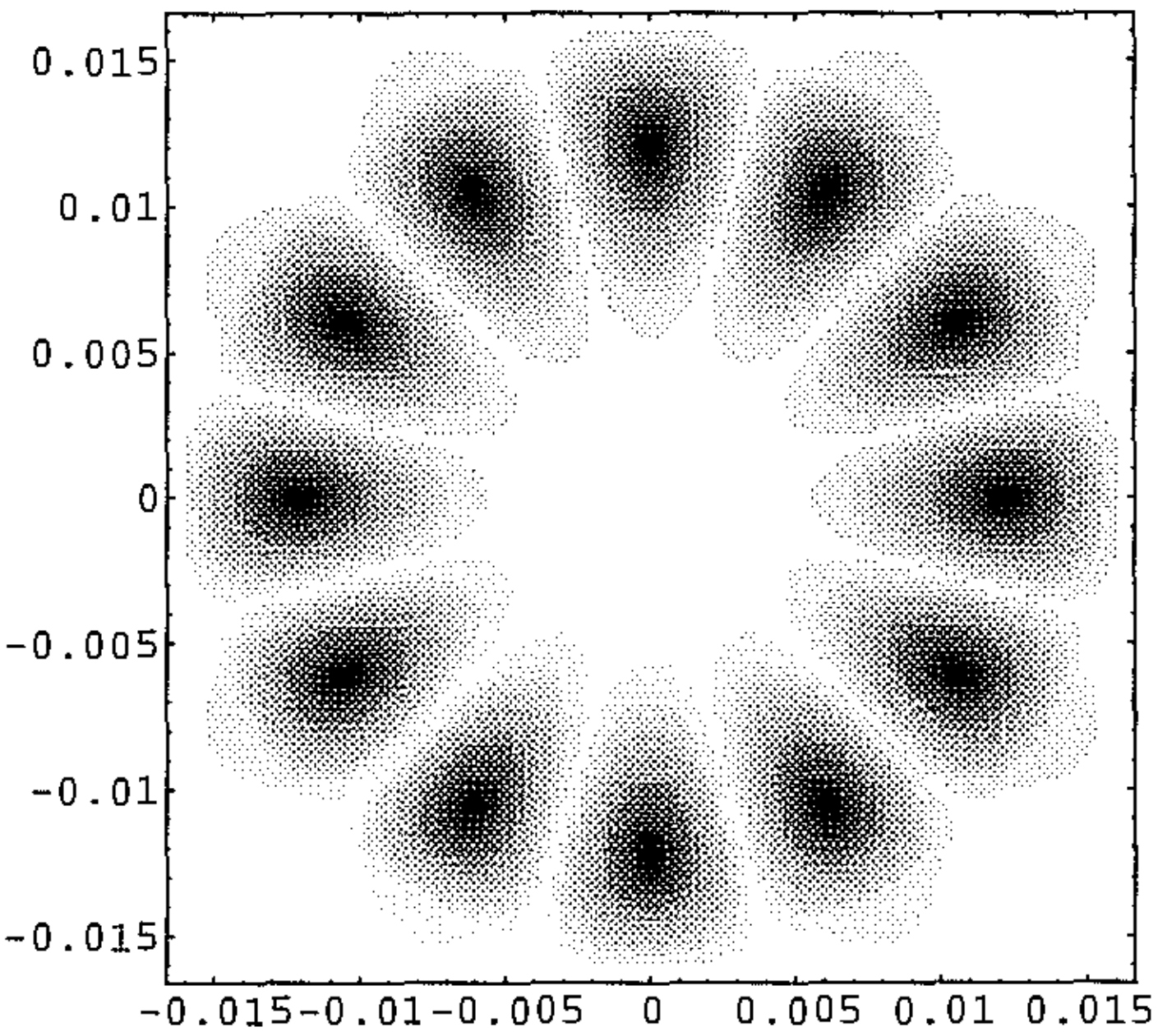

**(a)**

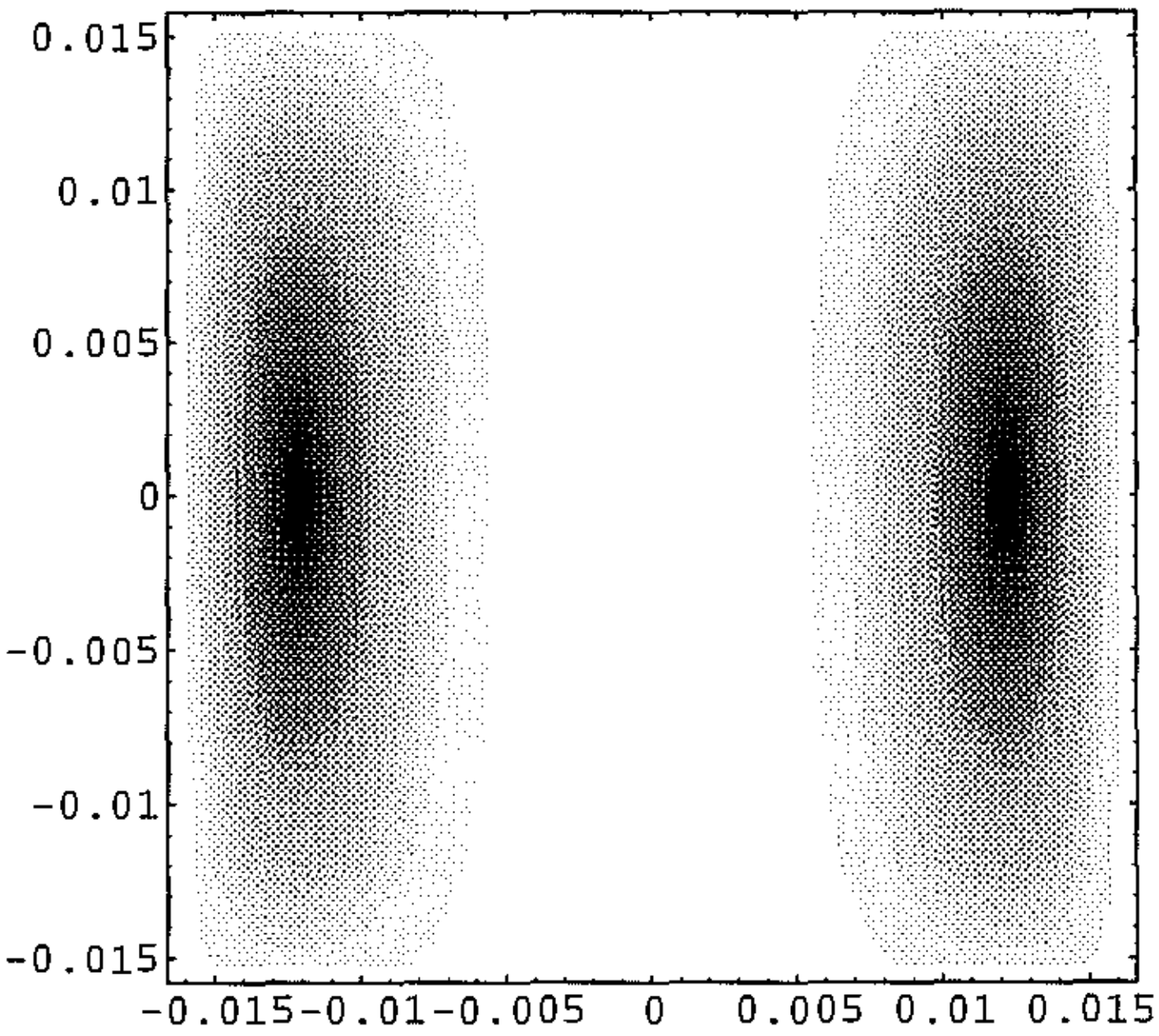

**(b)**

# Figure 2.4

Density plot of $|H_z|^2$ for the $TE_{6\ 1\ \delta}$ mode in a sapphire crystal with a diameter and height of approximately 30 mm. (a) Shows the density in a $\{r,\phi\}$ plane through the centre of the crystal, highlighting the mode numbers m = 6 and n =1. (b) Shows the density in a {r,z} plane through the centre of the crystal, highlighting the mode numbers n =1 and p = 0.

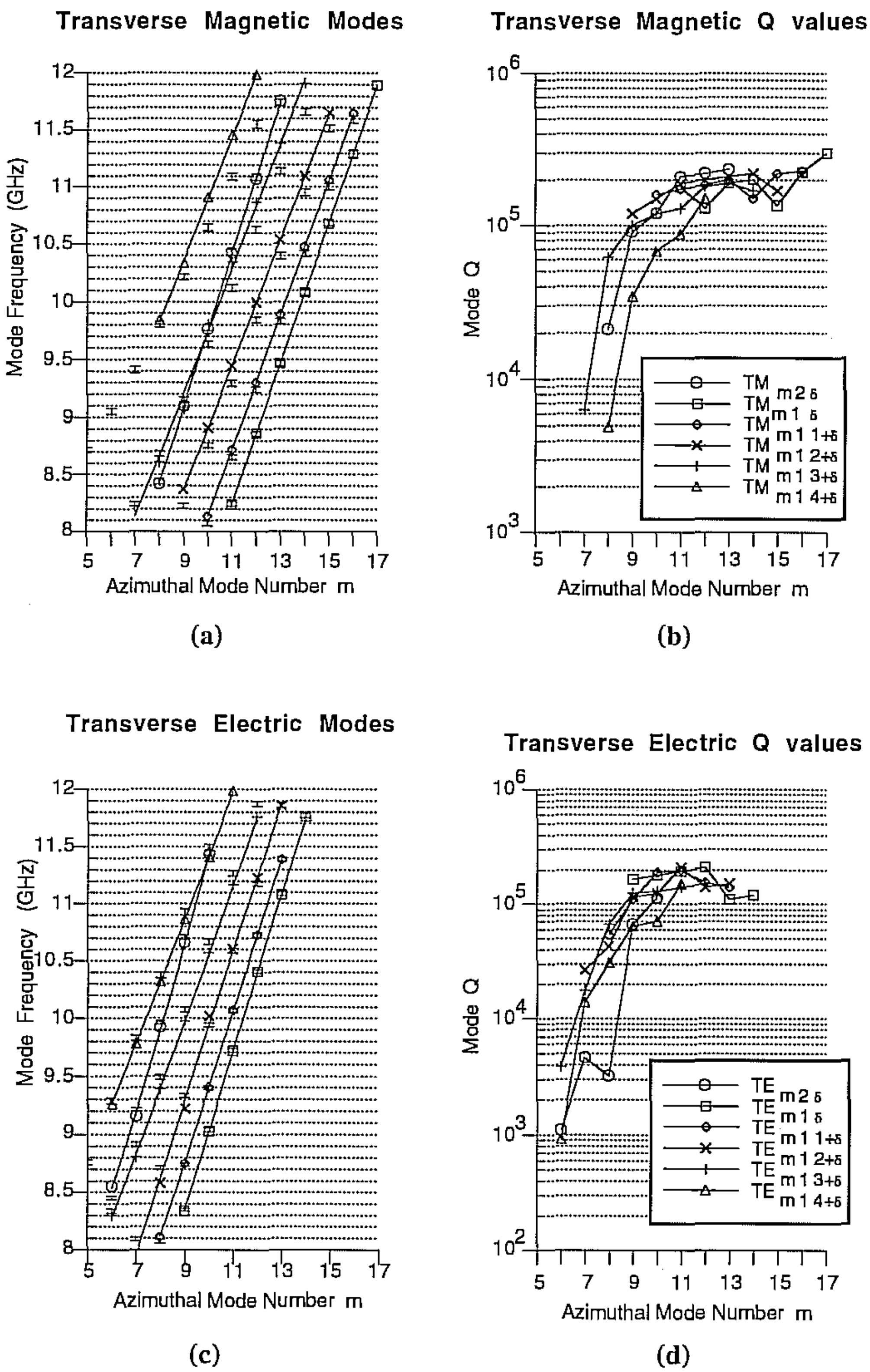

# Figure 2.5

Mode frequencies and Q values as a function of azimuthal mode number, for a 50.0 mm diameter and 30.0 mm high sapphire crystal. Theoretical points are plotted as error bars due to uncertainties in dimensions and permittivities, while experimental points are joined by lines.

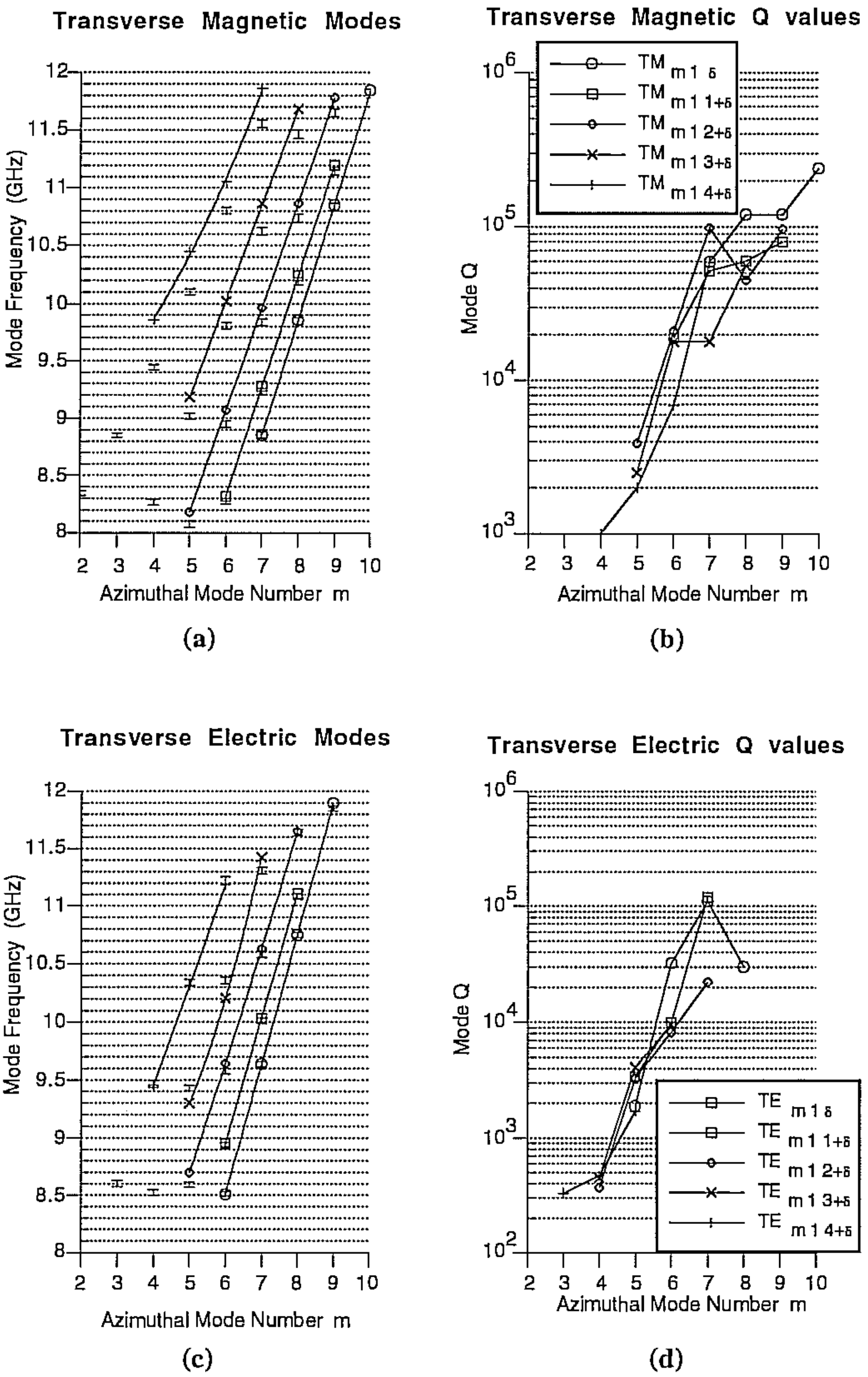

# Figure 2.6

Mode frequencies and Q values as a function of azimuthal mode number, for a 31.8 mm diameter and 30.2 mm high sapphire crystal. Theoretical points are plotted as error bars due to uncertainties in dimensions and permittivities, while experimental points are joined by lines.

| m | n=2 ; p=0 | n=1 ; p=0 | n=1 ; p=1 | n=1 ; p=2 | n=1 ; p=3 | n=1 ; p=4 |
|---|---|---|---|---|---|---|
| 7 | -0.209 | | | | -0.219 | |
| 8 | -0.135 | | | | -0.910 | 0.415 |
| 9 | -0.123 | | | 1.73 | -0.962 | 1.15 |
| 10 | -0.0786 | | 0.800 | 1.70 | 1.78 | 2.46 |
| 11 | -0.115 | -0.0239 | 0.710 | 1.61 | 2.19 | 3.27 |
| 12 | 0.130 | 0.000678 | 0.656 | 1.50 | 2.24 | 3.67 |
| 13 | | -0.00306 | 0.601 | 1.40 | 2.18 | |
| 14 | | -0.000994 | 0.547 | 1.30 | 2.12 | |
| 15 | | -0.0627 | 0.505 | 1.20 | | |
| 16 | | 0.0319 | 0.454 | | | |
| 17 | | -0.0555 | | | | |

Table 2.1

Percentage difference between theoretical and experimental TM mode frequencies for the 50.0 mm diameter sapphire.

| m | n=2 ; p=0 | n=1 ; p=0 | n=1 ; p=1 | n=1 ; p=2 | n=1 ; p=3 | n=1 ; p=4 |
|---|---|---|---|---|---|---|
| 6 | 1.19 | | | | -0.411 | -0.352 |
| 7 | -0.600 | | | -1.33 | -1.21 | -0.495 |
| 8 | -0.416 | | 0.429 | -1.46 | -1.06 | -0.535 |
| 9 | -0.430 | -0.0920 | 0.152 | -1.20 | -0.976 | -0.558 |
| 10 | -0.0926 | -0.0820 | 0.0657 | 0.802 | -0.905 | -0.694 |
| 11 | | 0.0718 | 0.0308 | 0.144 | -0.884 | -0.607 |
| 12 | | -0.0557 | 0.00746 | 0.428 | -0.938 | |
| 13 | | -0.0550 | 0.000871 | -0.0632 | | |
| 14 | | -0.0689 | | | | |

Table 2.2

Percentage difference between theoretical and experimental TE mode frequencies for the 50.0 mm diameter sapphire.

| m | n=1 ; p=0 | n=1 ; p=1 | n=1 ; p=2 | n=1 ; p=3 | n=1 ; p=4 |
|---|---|---|---|---|---|
| 4 | | | | | 4.34 |
| 5 | | | 1.33 | 1.86 | 3.40 |
| 6 | | 0.458 | 1.33 | 2.14 | 2.28 |
| 7 | 0.0165 | 0.459 | 1.27 | 2.26 | 2.63 |
| 8 | -0.0974 | 0.456 | 1.19 | 1.85 | |
| 9 | 0.173 | 0.448 | 1.10 | | |
| 10 | -0.203 | | | | |

Table 2.3

Percentage difference between theoretical and experimental TM mode frequencies for the 31.8 mm diameter sapphire.

| m | n=1 ; p=0 | n=1 ; p=1 | n=1 ; p=2 | n=1 ; p=3 | n=1 ; p=4 |
|---|---|---|---|---|---|
| 4 | | | | | -0.127 |
| 5 | | | 1.27 | -1.38 | -0.242 |
| 6 | -0.0247 | 0.114 | 0.656 | -1.48 | -0.330 |
| 7 | 0.00 | 0.252 | 0.471 | 1.04 | |
| 8 | -0.0279 | 0.261 | 0.0515 | | |
| 9 | 0.396 | | | | |

Table 2.4

Percentage difference between theoretical and experimental TE mode frequencies for the 31.8 mm diameter sapphire.

The permittivity of sapphire above and below X-band has been measured previously with slightly conflicting results (Loewenstein, Smith and Morgan, 1973), (Shelby and Fontanella, 1980). To be consistent with both reports we can be confident that $\varepsilon_z$ is 11.6245±.0355 and 11.355±.015 at 300 K and 4 K respectively, and $\varepsilon_r$ is 9.407±.012 and 9.2895±.0255 at 300 K and 4 K respectively.

The power radiated from a given mode of the open dielectric resonator could be calculated from the field components (2.1) by integrating the Poynting vector over the resonator surface, but is beyond the scope of this thesis. Within each mode family the power radiated from the open resonator decreases as the azimuthal mode number increases (Bun'kov et al, 1987). The observed open resonator Q thus increases monotonically with m to a limit set by the dielectric loss tangent. At room temperature and 10 GHz the limit to the Q value is about $2\cdot10^5$ and scales inversely with frequency (Braginsky et al, 1987). Some modes do not follow this pattern as they may be reactively coupled to a nearby mode. This will cause the Q factor of the lower Q mode to increase at the expense of the other degrading (Tobar and Blair, Sept. 1991). For the 50 mm diameter sapphire the density of modes in the 11 GHz region is large and interacting modes are common. In this region some mode Q values may actually decrease for an increase in azimuthal mode number.

### 2.3.2 Frequency Sensitivity

The sensitivities of mode frequency to perturbations ( $\Delta\varepsilon_r$ , $\Delta\varepsilon_z$ , $\Delta d$ and $\Delta h$ ) have been calculated and are shown in tables 2.5 to 2.8, for the large and small sapphires. To within the precision of the calculation (0.2 to 2%) the sums of the sensitivities obey the following relations;

$$\frac{\partial f}{\partial \varepsilon_r}\frac{\varepsilon_r}{f} + \frac{\partial f}{\partial \varepsilon_z}\frac{\varepsilon_z}{f} = -\frac{1}{2} \tag{2.4}$$

$$\frac{\partial f}{\partial d}\frac{d}{f} + \frac{\partial f}{\partial h}\frac{h}{f} = -1 \tag{2.5}$$

Equation (2.4) is true if $f \propto (u\,\varepsilon_r^2 + v\,\varepsilon_z^2)^{-1/4}$, as one might expect, where u and v are constants. Equation (2.5) can be justified by considering the electromagnetic path length perpendicular and parallel to the c-axis, $\varepsilon_r^{1/2}\,d$ and $\varepsilon_z^{1/2}\,h$ respectively. Frequency is inversely proportional to the path length, therefore a small fractional perturbation in the height or diameter will have twice the effect as the same fractional perturbation in $\varepsilon_z$ or $\varepsilon_r$ respectively.

General trends are observed in the sensitivities as shown in tables 2.5 to 2.8. Within a mode family, as azimuthal mode number increases the dependence of mode frequency on diameter increases. For both T.E. and T.M. modes with high azimuthal and low axial mode number the frequency depends mainly (~98%) on the diameter. As the axial mode number increases the frequency dependence on the height increases, and becomes comparable with the diameter when $p > 3$ for TE modes and $p > 2$ for TM modes. Quasi T.E. modes have most of their electric field perpendicular to the z-axis and are seen to mainly depend on $\varepsilon_r$, while quasi T.M. modes have most of their electric field parallel to the z-axis and mainly depend on $\varepsilon_z$. Within a mode family this dependence on the respective dielectric constant increases with azimuthal mode number. For the higher order axial mode families ($p > 3$ for the TE modes and $p > 2$ for the TM modes), solutions are well within the hybrid regime of (2.2) and the dependence on $\varepsilon_r$ and $\varepsilon_z$ are about equal. This is why solutions are less accurate for these families as (2.3) assumes pure T.E. and T.M. modes while (2.2) does not.

### 2.3.3 Tilting of the c-axis

If the c-axis is tilted with respect to the cylinder z-axis then mode field patterns will distort. This is a more complicated situation to analyse as the permittivity perpendicular to the z-axis is no longer constant, but becomes an elliptical function of the azimuthal angle. Whispering gallery like modes, which have the highest Q values, are highly dependent on the cylindrical symmetry inhibiting radiation losses. If the c-axis is tilted, the symmetry is destroyed and the fields will radiate more easily out of the edges of the dielectric cylinder; for example, a 3 cm diameter cylindrical sapphire resonator, with c-axis tilted at 60 degrees, does not exhibit any Q values exceeding $10^5$ over X-band (Peng Hong private communication, 1991).

### 2.3.4 Applications to Sapphire Loaded Superconducting Cavities (SLOSC)

Details of a high stability 9.73 GHz SLOSC oscillator have been presented previously (A.J. Giles et al, 1989 & 1990). The operational sapphire resonance can now be identified as quasi $TE_{6\,1\,\delta}$. Theory predicts a 66 MHz shift in frequency when cooled due to the change in dielectric constant, which is equivalent to the measured value to within 5%.

Details of a tunable SLOSC are discussed in detail in chapters 1 and 3. It consists of a 3 cm diameter cylindrical sapphire crystal, and an axially driven tuning disc .3 cm thick. Tuning ranges can be predicted with presented theory, and a detailed mode map with tuning and cooling dependencies are presented in chapter 1. Modes analysed previously (Tobar and Blair, Sept. 1991) at 4.2 K can now be identified as $TE_{6\,1\,\delta+1}$ at 10.221

| TM Modes m, n, p+δ | $-\frac{\Delta f}{f}\frac{\varepsilon_r}{\Delta\varepsilon_r}$ | $-\frac{\Delta f}{f}\frac{\varepsilon_z}{\Delta\varepsilon_z}$ | $-\frac{\Delta f}{f}\frac{d}{\Delta d}$ | $-\frac{\Delta f}{f}\frac{h}{\Delta h}$ | $-\frac{\Delta f}{f}\frac{\varepsilon_r}{\Delta\varepsilon_r} - \frac{\Delta f}{f}\frac{\varepsilon_z}{\Delta\varepsilon_z}$ | $-\frac{\Delta f}{f}\frac{d}{\Delta d} - \frac{\Delta f}{f}\frac{h}{\Delta h}$ |
|---|---|---|---|---|---|---|
| 11, 1, 0 | .005 | .502 | .966 | .039 | .507 | 1.005 |
| 12, 1, 0 | .020 | .498 | .985 | .031 | .518 | 1.016 |
| 13, 1, 0 | -.007 | .512 | .994 | .032 | .505 | 1.026 |
| 14, 1, 0 | .001 | .510 | .994 | .024 | .511 | 1.018 |
| 15, 1, 0 | -.021 | .508 | .986 | .026 | .487 | 1.012 |
| 16, 1, 0 | .001 | .505 | .993 | .022 | .506 | 1.015 |
| 17, 1, 0 | .032 | .481 | .969 | .008 | .513 | 0.977 |
| 10, 1, 1 | .019 | .475 | .860 | .143 | .494 | 1.003 |
| 11, 1, 1 | .015 | .482 | .882 | .126 | .497 | 1.008 |
| 12, 1, 1 | .015 | .483 | .895 | .111 | .498 | 1.006 |
| 13, 1, 1 | .008 | .486 | .907 | .097 | .494 | 1.004 |
| 14, 1, 1 | .016 | .486 | .918 | .085 | .502 | 1.003 |
| 15, 1, 1 | .014 | .491 | .926 | .078 | .505 | 1.004 |
| 16, 1, 1 | .009 | .493 | .932 | .070 | .502 | 1.002 |
| 9, 1, 2 | .054 | .441 | .703 | .308 | .495 | 1.011 |
| 10, 1, 2 | .046 | .451 | .737 | .271 | .497 | 1.008 |
| 11, 1, 2 | .034 | .460 | .767 | .244 | .494 | 1.011 |
| 12, 1, 2 | .031 | .467 | .793 | .219 | .498 | 1.012 |
| 13, 1, 2 | .028 | .471 | .814 | .189 | .499 | 1.003 |
| 14, 1, 2 | .024 | .477 | .835 | .179 | .501 | 1.014 |
| 15, 1, 2 | .018 | .479 | .849 | .157 | .497 | 1.006 |
| 7, 1, 3 | .119 | .371 | .476 | .541 | .490 | 1.017 |
| 8, 1, 3 | .095 | .393 | .527 | .488 | .488 | 1.015 |
| 9, 1, 3 | .081 | .413 | .573 | .440 | .494 | 1.013 |
| 10, 1, 3 | .064 | .430 | .612 | .398 | .494 | 1.010 |
| 11, 1, 3 | .048 | .445 | .652 | .362 | .493 | 1.014 |
| 12, 1, 3 | .047 | .447 | .682 | .330 | .494 | 1.012 |
| 13, 1, 3 | .049 | .451 | .711 | .297 | .500 | 1.008 |
| 14, 1, 3 | .042 | .460 | .740 | .269 | .502 | 1.009 |
| 5, 1, 4 | .214 | .268 | .260 | .748 | .482 | 1.008 |
| 6, 1, 4 | .175 | .308 | .315 | .694 | .483 | 1.009 |
| 7, 1, 4 | .143 | .342 | .368 | .642 | .485 | 1.010 |
| 8, 1, 4 | .118 | .369 | .419 | .594 | .487 | 1.013 |
| 9, 1, 4 | .099 | .390 | .466 | .548 | .489 | 1.014 |
| 10, 1, 4 | .084 | .409 | .490 | .505 | .493 | 0.995 |
| 11, 1, 4 | .076 | .421 | .550 | .467 | .497 | 1.017 |
| 12, 1, 4 | .061 | .430 | .580 | .428 | .491 | 1.008 |
| 7, 2, 0 | .008 | .491 | .957 | .037 | .499 | 0.994 |
| 8, 2, 0 | .012 | .485 | .973 | .039 | .497 | 1.012 |
| 9, 2, 0 | -.001 | .497 | .974 | .034 | .496 | 1.008 |
| 10, 2, 0 | -.004 | .493 | .963 | .018 | .489 | 0.981 |
| 11, 2, 0 | .030 | .489 | .985 | .020 | .519 | 1.005 |
| 12, 2, 0 | .019 | .485 | .975 | .021 | .504 | 0.996 |
| 13, 2, 0 | .004 | .482 | .982 | .019 | .486 | 1.001 |

## Table 2.5

Fractional frequency dependencies due to small perturbations in the dimensions and permittivities for TM modes in the 50.0 mm diameter sapphire crystal, with; $\Delta d/d = 1.0\cdot10^{-3}$, $\Delta h/h = 1.7\cdot10^{-3}$, $\Delta\varepsilon_r/\varepsilon_r = 1.3\cdot10^{-3}$, $\Delta\varepsilon_z/\varepsilon_z = 3.0\cdot10^{-3}$.

| TE Modes m, n, p+δ | $-\frac{\Delta f}{f}\frac{\varepsilon_r}{\Delta\varepsilon_r}$ | $-\frac{\Delta f}{f}\frac{\varepsilon_z}{\Delta\varepsilon_z}$ | $-\frac{\Delta f}{f}\frac{d}{\Delta d}$ | $-\frac{\Delta f}{f}\frac{h}{\Delta h}$ | $-\frac{\Delta f}{f}\frac{\varepsilon_r}{\Delta\varepsilon_r} - \frac{\Delta f}{f}\frac{\varepsilon_z}{\Delta\varepsilon_z}$ | $-\frac{\Delta f}{f}\frac{d}{\Delta d} - \frac{\Delta f}{f}\frac{h}{\Delta h}$ |
|---|---|---|---|---|---|---|
| 9, 1, 0 | .480 | .009 | .972 | .030 | .489 | 1.002 |
| 10, 1, 0 | .482 | .008 | .976 | .026 | .490 | 1.002 |
| 11, 1, 0 | .484 | .007 | .979 | .022 | .491 | 1.001 |
| 12, 1, 0 | .485 | .006 | .981 | .020 | .491 | 1.001 |
| 13, 1, 0 | .486 | .006 | .983 | .017 | .492 | 1.000 |
| 14, 1, 0 | .487 | .005 | .985 | .015 | .492 | 1.000 |
| 8, 1, 1 | .447 | .043 | .884 | .122 | .490 | 1.006 |
| 9, 1, 1 | .454 | .037 | .900 | .106 | .491 | 1.006 |
| 10, 1, 1 | .460 | .033 | .912 | .093 | .493 | 1.005 |
| 11, 1, 1 | .463 | .029 | .922 | .082 | .492 | 1.004 |
| 12, 1, 1 | .467 | .026 | .931 | .073 | .493 | 1.004 |
| 13, 1, 1 | .470 | .025 | .938 | .065 | .495 | 1.003 |
| 14, 1, 1 | .472 | .023 | .944 | .059 | .495 | 1.003 |
| 7, 1, 2 | .368 | .126 | .757 | .255 | .494 | 1.012 |
| 8, 1, 2 | .389 | .106 | .785 | .227 | .495 | 1.012 |
| 9, 1, 2 | .404 | .092 | .808 | .203 | .496 | 1.011 |
| 10, 1, 2 | .415 | .081 | .828 | .182 | .496 | 1.010 |
| 11, 1, 2 | .425 | .072 | .846 | .163 | .497 | 1.009 |
| 12, 1, 2 | .432 | .066 | .861 | .146 | .498 | 1.007 |
| 13, 1, 2 | .437 | .062 | .875 | .134 | .499 | 1.009 |
| 6, 1, 3 | .239 | .260 | .625 | .391 | .499 | 1.016 |
| 7, 1, 3 | .277 | .224 | .659 | .358 | .501 | 1.017 |
| 8, 1, 3 | .306 | .195 | .689 | .326 | .501 | 1.015 |
| 9, 1, 3 | .330 | .172 | .718 | .297 | .502 | 1.015 |
| 10, 1, 3 | .350 | .154 | .744 | .275 | .504 | 1.019 |
| 11, 1, 3 | .364 | .139 | .765 | .249 | .503 | 1.014 |
| 12, 1, 3 | .373 | .130 | .787 | .227 | .503 | 1.014 |
| 4, 1, 4 | .082 | .408 | .445 | .568 | .490 | 1.013 |
| 5, 1, 4 | .125 | .369 | .489 | .525 | .494 | 1.014 |
| 6, 1, 4 | .164 | .333 | .530 | .486 | .497 | 1.016 |
| 7, 1, 4 | .198 | .302 | .567 | .449 | .500 | 1.016 |
| 8, 1, 4 | .227 | .276 | .602 | .415 | .503 | 1.017 |
| 9, 1, 4 | .251 | .252 | .633 | .385 | .503 | 1.018 |
| 10, 1, 4 | .271 | .233 | .661 | .356 | .504 | 1.017 |
| 11, 1, 4 | .290 | .217 | .687 | .329 | .507 | 1.016 |
| 6, 2, 0 | .467 | .018 | .971 | .031 | .485 | 1.002 |
| 7, 2, 0 | .470 | .017 | .976 | .025 | .487 | 1.001 |
| 8, 2, 0 | .473 | .017 | .980 | .022 | .490 | 1.002 |
| 9, 2, 0 | .475 | .017 | .983 | .019 | .492 | 1.002 |
| 10, 2, 0 | .474 | .019 | .985 | .016 | .493 | 1.001 |

# Table 2.6

Fractional frequency dependencies due to small perturbations in the dimensions and permittivities for TE modes in the 50.0 mm diameter sapphire crystal, with; $\Delta d/d = 1.0 \cdot 10^{-3}$, $\Delta h/h = 1.7 \cdot 10^{-3}$, $\Delta\varepsilon_r/\varepsilon_r = 1.3 \cdot 10^{-3}$, $\Delta\varepsilon_z/\varepsilon_z = 3.0 \cdot 10^{-3}$.

| TM Modes $m, n, p+\delta$ | $-\frac{\Delta f}{f}\frac{\varepsilon_r}{\Delta\varepsilon_r}$ | $-\frac{\Delta f}{f}\frac{\varepsilon_z}{\Delta\varepsilon_z}$ | $-\frac{\Delta f}{f}\frac{d}{\Delta d}$ | $-\frac{\Delta f}{f}\frac{h}{\Delta h}$ | $-\frac{\Delta f}{f}\frac{\varepsilon_r}{\Delta\varepsilon_r} - \frac{\Delta f}{f}\frac{\varepsilon_z}{\Delta\varepsilon_z}$ | $-\frac{\Delta f}{f}\frac{d}{\Delta d} - \frac{\Delta f}{f}\frac{h}{\Delta h}$ |
|---|---|---|---|---|---|---|
| 7, 1, 0 | .009 | .494 | .958 | .034 | .503 | 0.992 |
| 8, 1, 0 | -.005 | .502 | .968 | .028 | .497 | 0.996 |
| 9, 1, 0 | .001 | .481 | .978 | .019 | .482 | 0.997 |
| 10, 1, 0 | -.002 | .512 | .970 | .021 | .510 | 0.991 |
| 6, 1, 1 | .035 | .461 | .847 | .144 | .496 | 0.991 |
| 7, 1, 1 | .028 | .472 | .878 | .116 | .500 | 0.994 |
| 8, 1, 1 | .019 | .479 | .898 | .096 | .498 | 0.994 |
| 9, 1, 1 | .017 | .486 | .915 | .081 | .503 | 0.996 |
| 5, 1, 2 | .096 | .405 | .653 | .334 | .501 | 0.987 |
| 6, 1, 2 | .063 | .429 | .716 | .272 | .492 | 0.998 |
| 7, 1, 2 | .043 | .453 | .753 | .228 | .496 | 0.981 |
| 8, 1, 2 | .027 | .466 | .795 | .192 | .493 | 0.987 |
| 9, 1, 2 | .025 | .470 | .829 | .164 | .495 | 0.993 |
| 4, 1, 3 | .177 | .308 | .433 | .552 | .485 | 0.985 |
| 5, 1, 3 | .133 | .358 | .517 | .468 | .491 | 0.985 |
| 6, 1, 3 | .096 | .380 | .589 | .398 | .476 | 0.987 |
| 7, 1, 3 | .078 | .418 | .634 | .340 | .496 | 0.974 |
| 8, 1, 3 | .062 | .434 | .690 | .294 | .496 | 0.984 |
| 2, 1, 4 | .388 | .082 | .149 | .843 | .470 | 0.992 |
| 3, 1, 4 | .296 | .184 | .236 | .748 | .480 | 0.984 |
| 4, 1, 4 | .221 | .263 | .329 | .658 | .484 | 0.987 |
| 5, 1, 4 | .171 | .317 | .408 | .577 | .488 | 0.985 |
| 6, 1, 4 | .130 | .358 | .479 | .506 | .488 | 0.985 |
| 7, 1, 4 | .103 | .387 | .541 | .445 | .490 | 0.986 |

# Table 2.7

Fractional frequency dependencies due to small perturbations in the dimensions and permittivities for TM modes in the 31.8 mm diameter sapphire crystal, with; $\Delta d/d = 1.6\cdot10^{-3}$, $\Delta h/h = 1.7\cdot10^{-3}$, $\Delta\varepsilon_r/\varepsilon_r = 1.3\cdot10^{-3}$, $\Delta\varepsilon_z/\varepsilon_z = 3.0\cdot10^{-3}$.

| TE Modes m, n, p+δ | $-\frac{\Delta f}{f}\frac{\varepsilon_r}{\Delta\varepsilon_r}$ | $-\frac{\Delta f}{f}\frac{\varepsilon_z}{\Delta\varepsilon_z}$ | $-\frac{\Delta f}{f}\frac{d}{\Delta d}$ | $-\frac{\Delta f}{f}\frac{h}{\Delta h}$ | $-\frac{\Delta f}{f}\frac{\varepsilon_r}{\Delta\varepsilon_r} - \frac{\Delta f}{f}\frac{\varepsilon_z}{\Delta\varepsilon_z}$ | $-\frac{\Delta f}{f}\frac{d}{\Delta d} - \frac{\Delta f}{f}\frac{h}{\Delta h}$ |
|---|---|---|---|---|---|---|
| 5, 1, 0 | .470 | .011 | .965 | .033 | .481 | 0.998 |
| 6, 1, 0 | .487 | .008 | .972 | .026 | .495 | 0.998 |
| 7, 1, 0 | .477 | .006 | .978 | .021 | .483 | 0.999 |
| 8, 1, 0 | .482 | .007 | .981 | .017 | .489 | 0.998 |
| 5, 1, 1 | .438 | .043 | .878 | .117 | .481 | 0.995 |
| 6, 1, 1 | .450 | .034 | .902 | .094 | .484 | 0.996 |
| 7, 1, 1 | .459 | .027 | .920 | .077 | .486 | 0.997 |
| 4, 1, 2 | .344 | .138 | .734 | .258 | .482 | 0.992 |
| 5, 1, 2 | .380 | .103 | .778 | .215 | .483 | 0.993 |
| 6, 1, 2 | .404 | .083 | .813 | .180 | .487 | 0.993 |
| 7, 1, 2 | .420 | .066 | .841 | .152 | .486 | 0.993 |
| 3, 1, 3 | .166 | .320 | .600 | .392 | .486 | 0.992 |
| 4, 1, 3 | .241 | .245 | .645 | .347 | .486 | 0.992 |
| 5, 1, 3 | .295 | .191 | .687 | .304 | .486 | 0.991 |
| 6, 1, 3 | .335 | .153 | .725 | .265 | .488 | 0.990 |
| 2, 1, 4 | .130 | .476 | .465 | .525 | .506 | 0.990 |
| 3, 1, 4 | .093 | .397 | .518 | .472 | .490 | 0.990 |
| 4, 1, 4 | .159 | .331 | .566 | .424 | .490 | 0.990 |
| 5, 1, 4 | .213 | .276 | .609 | .381 | .489 | 0.990 |

Table 2.8

Fractional frequency dependencies due to small perturbations in the dimensions and permittivities for TE modes in the 31.8 mm diameter sapphire crystal, with; $\Delta d/d = 1.6 \cdot 10^{-3}$, $\Delta h/h = 1.7 \cdot 10^{-3}$, $\Delta\varepsilon_r/\varepsilon_r = 1.3 \cdot 10^{-3}$, $\Delta\varepsilon_z/\varepsilon_z = 3.0 \cdot 10^{-3}$.

GHz and $TM_{8\,1\,\delta+1}$ at 10.44 GHz. Tuning ranges of these modes are predicted to be 94 and 99 MHz respectively, which agrees favorably with experiment. These T-SLOSC modes are analysed in the following chapter in regards to interactions with low Q spurious modes as they tune close by.

## 2.4 CONCLUSION

Improved theory for multimode analysis of anisotropic dielectric resonators was presented. This has lead us to a very good understanding of electromagnetic resonances in sapphire crystals, with potential application to design.

# CHAPTER 3

# MODAL INTERACTIONS BETWEEN ELECTROMAGNETIC RESONANCES IN THE T-SLOSC

## 3.1 INTRODUCTION

For the high-Q higher order modes in the T-SLOSC resonator only approximately half a percent tuning of eigen-frequencies is experienced. Therefore eigenfunction field patterns only experience small field perturbations. One might expect the quality factor to thus remain high throughout the tuning range. This is desirable as the phase noise obtainable from the constructed loop oscillator is dependent on this high quality factor. Experimental results reveal that while tuning, modes interact. If a high-Q mode interacts with a low-Q mode, the high-Q mode will experience a severe drop in Q.

Complete field solutions of coupled cavities are very difficult. However a cavity coupling system may be replaced by impedance or admittance elements of known magnitude and frequency dependence. For a right cylindrical resonator an equivalent circuit based on a Lagrangian technique (Montgomery et al, 1948) has been used to derive equivalent circuit parameters, from known field patterns and cavity properties. In a fixed right cylinder there are no interactions between modes, ie. the associated eigenfunction field patterns are orthogonal. In this case cross couplings between modes are nonexistent. Previous work (Banos, 1944) has shown that a mutual resistive interaction was apparent in a Tunable Echo Box, between degenerate modes. This also occurs in the T-SLOSC when modes interact with co-axial cable resonances (Tobar and Blair, Sept. 1991). However for internal modal interactions a mutual reactive effect dominates due to the reactive sapphire dielectric (Tobar and Blair, Sept. 1991). This type of effect has also been observed between two similar sapphire disk resonators (Vzyatyshev, 1987).

All internal elements add possibilities for mutual interactions. In this chapter previous theory has been generalized to include all possible mutual interacting terms (Crout, 1944; Banos, 1944; Montgomery, 1948). This keeps Lagranges equations and the equivalent circuit representation general. These interacting modes have been modelled using the comercial computer package, Mathematica (Wolfram, 1988). Results show the Q degradation as observed for reactive coupling between modes. Resistive interaction modeling reveals that a higher Q mode distorts as it tunes across the bandwidth of a lower Q mode. In our experiments this effect was apparent when tuning a mode across co-axial cable resonances.

This chapter describes high precision measurements of mode interactions, made possible by the use of an ultra-stable sapphire-loaded superconducting cavity (SLOSC) oscillator, within a new theoretical framework for understanding the observed behaviour. The newly developed coupled mode model enables the derivation of the reflection coefficient for two interacting modes. Taking a very precise look at some reactively and resistively coupled interactions experimentally verifies the model.

## 3.2 EQUIVALENT CIRCUIT MODEL- THEORETICAL DEVELOPMENT

In this section general equivalent circuits for a resonant cavity and an isotropic dielectric resonator, coupled to by a magnetic probe are calculated (Tobar and Blair, Sept. 1991). Separating the sapphire dielectric and niobium cavity parts, the loop coupled Lagrangian equations for a given mode n, are presented. Electric charge $q_n$, or Magnetic flux $\phi_n$, are chosen as normal coordinates, corresponding to a resonant wave number $k_n$. Duality exists between these electric and magnetic quantities. It turns out that the electromagnetic field in a cavity or a dielectric resonator can be expanded in terms of the normal coordinates $\phi_n$ , $\dot{\phi}_n$ or $q_n$ , $\dot{q}_n$ respectively, with eigenvalues $k_n$ and dimensionless vector eigenfunctions $\xi_n$ , $\eta_n$. Thus;

$$\mathbf{H} = \sum_n k_n \dot{q}_n \eta_n \tag{3.1a}$$

$$\mathbf{D} = \sum_n k_n^2 q_n \xi_n \tag{3.1b}$$

in a cavity resonator, or

$$\mathbf{B} = \sum_n k_n^2 \phi_n \eta_n \tag{3.2a}$$

$$\mathbf{E} = \sum_n k_n \dot{\phi}_n \xi_n \tag{3.2b}$$

in a dielectric resonator.

The dimensionless vector eigenfunctions obey the wave equations:

$$\nabla^2 \xi_n + k_n^2 \xi_n = 0 \qquad \nabla \cdot \xi_n = 0 \tag{3.3a}$$

$$\nabla^2 \eta_n + k_n^2 \eta_n = 0 \qquad \nabla \cdot \eta_n = 0 \tag{3.3b}$$

and relevant boundary conditions on the sapphire dielectric or superconducting cavity walls. If no coupling between modes occurs the following orthogonality conditions hold true for a dielectric resonator or cavity resonator respectively;

$$\int_{\text{dielectric volume}} \xi_n \cdot \xi_m \, dt = \delta_{nm} \upsilon_d \tag{3.4a}$$

$$\int_{\text{dielectric volume}} \eta_n \cdot \eta_m \, dt = \delta_{nm} \upsilon_d \tag{3.4b}$$

$$\int_{\text{cavity volume}} \xi_n \cdot \xi_m \, dt = \delta_{nm} \upsilon_c \tag{3.5a}$$

$$\int_{\text{cavity volume}} \eta_n \cdot \eta_m \, dt = \delta_{nm} \upsilon_c \quad . \tag{3.5b}$$

Here $\upsilon_d$ and $\upsilon_c$ are the dielectric and cavity volume respectively and $\delta_{nm}$ the Kronecker delta. Also $\xi_n$ and $\eta_n$ are related by;

$$k_n \xi_n = \nabla \times \eta_n \qquad -k_n \eta_n = \nabla \times \xi_n \quad . \tag{3.6}$$

Using the electric charge coordinates $q_n$ and $\dot{q}_n$, a series circuit naturally evolves from the theory. Modelling the cavity, the series loss represents the surface resistance losses of the niobium superconductor. Using the magnetic flux coordinates $\phi_n$ and $\dot{\phi}_n$, a parallel circuit naturally evolves from the theory. Modelling the sapphire, the dissipative term represents a shunt dielectric loss. The series circuit can be transformed to a parallel circuit for a general mode with Q >> 1 (Montgomery et al, 1948), and vice versa. This dual representation can be taken further when regarding interactions between modes by introducing cross coupling terms. When attempting to isolate these parameters to either superconducting niobium or sapphire dielectric, both representations are needed to understand meaningful equivalent circuit parameters in terms of the modal fields.

### 3.2.1 Series Circuit Representation of the niobium cavity

In this section we shall derive the series equivalent circuit of figure 3.1, with respect to the electric charge coordinates $q_n$ and $q_m$. Assuming only two excited modes, n and m, and using (3.1a) and (3.1b), the magnetic and electric field energies in the cavity can be represented by the respective Lagrangian variables;

$$T_{nm} = \frac{\mu}{2} \int_{\text{cavity volume}} \mathbf{H}_n \cdot \mathbf{H}_m \, d\tau = \frac{\mu \, k_n \, k_m \, \dot{q}_n \dot{q}_m}{2} \int_{\text{cavity volume}} \eta_n \cdot \eta_m \, d\tau \tag{3.7}$$

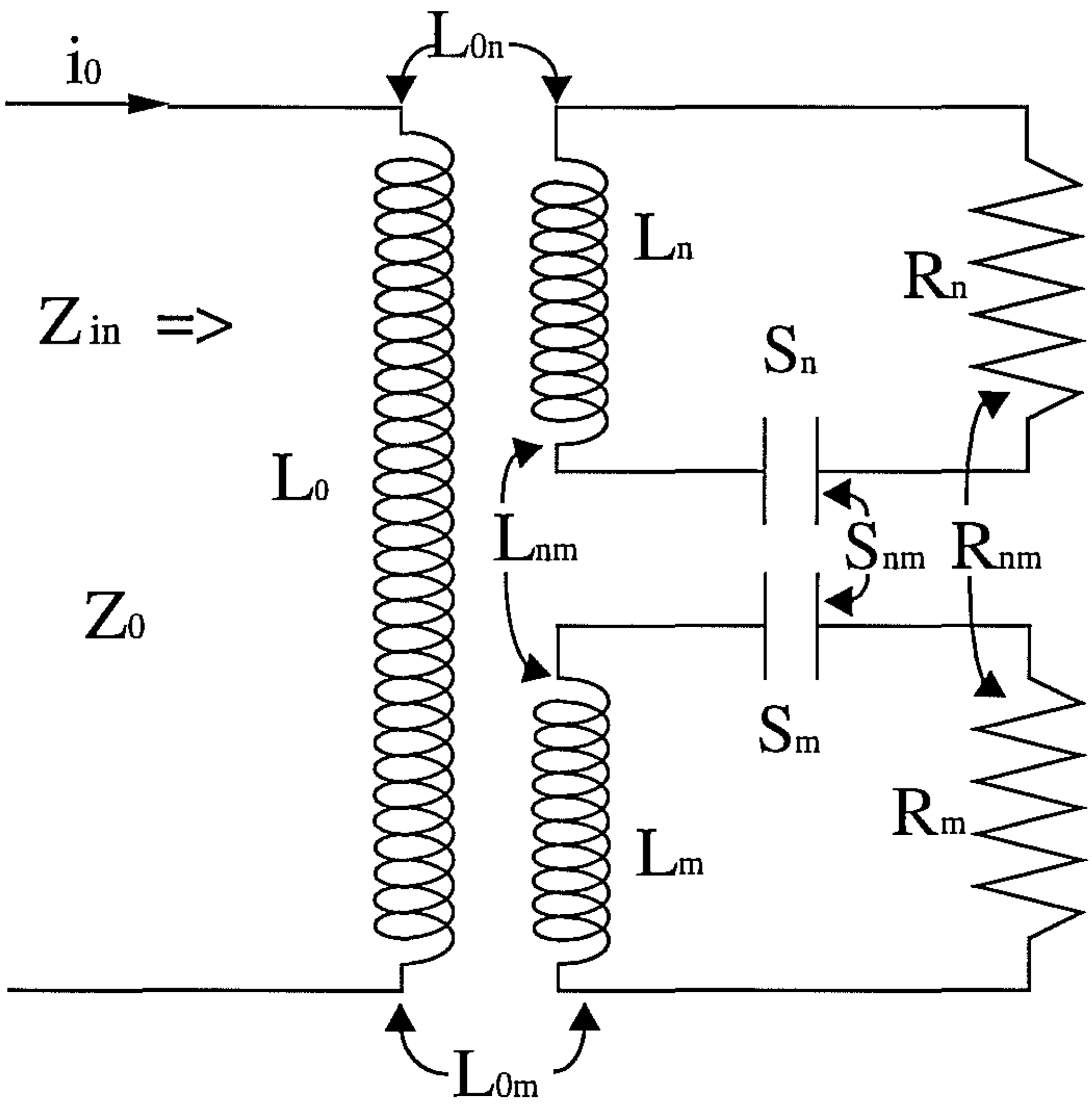

# Figure 3.1

General equivalent circuit of two excited modes in a loop-coupled cavity, as derived from normal coordinates $q_n$ and $q_m$.

$$V_{nm} = \frac{1}{2\varepsilon} \int_{\text{cavity volume}} \mathbf{D}_n \cdot \mathbf{D}_m \, d\tau = \frac{k_n^2 \, k_m^2 \, q_n \, q_m}{2\varepsilon} \int_{\text{cavity volume}} \xi_n \cdot \xi_m \, d\tau \qquad (3.8)$$

Here $\mu$ and $\varepsilon$ are the permeability and permittivity within the cavity respectively. In general they are functions of position internal to the cavity.

In the case n = m, these field energies have the form of inductive or capacitive stored energies, and;

$$L_n = \mu k_n^2 \int_{\text{cavity volume}} |\eta_n|^2 \, d\tau \qquad (3.9a)$$

$$S_n = \frac{k_n^4}{\varepsilon} \int_{\text{cavity volume}} |\xi_n|^2 \, d\tau \quad , \qquad (3.9b)$$

where $L_n$ is the equivalent series inductance and $S_n$ the equivalent series elastance. Combining (3.9) with the relation between wave number and resonant frequency; $k_n = \omega_n \sqrt{\varepsilon\mu}$, it can be shown that ;

$$\omega_n = \sqrt{\frac{S_n}{L_n}} \quad . \qquad (3.10)$$

In the case n ≠ m, cross coupling terms are defined;

$$L_{nm} = \mu k_n k_m \int_{\text{cavity volume}} \eta_n \cdot \eta_m \, d\tau \qquad (3.11a)$$

$$S_{nm} = \frac{k_n^2 k_m^2}{\varepsilon} \int_{\text{cavity volume}} \xi_n \cdot \xi_m \, d\tau \qquad (3.11b)$$

These are energy storage terms which couple energy between two given modes, and in general may include surface reactance effects of the niobium.

Series losses in the superconducting niobium surfaces are represented by the dissipation function F (Goldstein, 1959), given by;

$$F = \frac{1}{2} \int_{\text{niobium surface}} R_s \, |H|^2 \, ds \quad . \qquad (3.12)$$

Here $R_s$ is defined as the surface resistance, and in general is a function of temperature according to the BCS theory (Bardeen et al 1957). Combining (3.12) and (3.1) ;

$$F_{nm} = \frac{1}{2} \dot{q}_n \dot{q}_m R_{nm} \qquad (3.13)$$

where $$R_{nm} = k_n k_m \sqrt{R_{s\,n} R_{s\,m}} \int_{\text{niobium surface}} \eta_n \cdot \eta_m \, ds \tag{3.14}$$

For a normal conductor the relation $R = \omega L Q^{-1}$ can be generalized to;

$$R_{nm} = \frac{\sqrt{\omega_n \omega_m L_n L_m}}{Q_{nm}} \tag{3.15}$$

where;

$$\frac{1}{Q_{nm}} = \frac{\sqrt{\delta_n \delta_m}}{2\upsilon} \int_{\text{conductor surface}} \eta_n \cdot \eta_m \, ds \quad . \tag{3.16}$$

Here $\delta$ is the skin depth if the cavity is a normal conductor, or the analogous effective skin depth if the cavity is a superconductor. If n = m, (3.16) defines the quality factor associated with niobium losses for an individual mode. However when $n \neq m$ (3.16) defines a quality factor due to the two modes interacting or aligning in the niobium surface. This mutual dissipative term describes an energy coupling path between the two modes.

Using $T_{nm} - V_{nm}$ as the Lagrangian, the generalized coupled Lagrange equations with respect to $q_n$ and $\dot{q}_m$ become;

$$\begin{bmatrix} L_n & L_{nm} \\ L_{nm} & L_m \end{bmatrix} \begin{bmatrix} \ddot{q}_n \\ \ddot{q}_m \end{bmatrix} + \begin{bmatrix} R_n & R_{nm} \\ R_{nm} & R_m \end{bmatrix} \begin{bmatrix} \dot{q}_n \\ \dot{q}_m \end{bmatrix} + \begin{bmatrix} S_n & S_{nm} \\ S_{nm} & S_m \end{bmatrix} \begin{bmatrix} q_n \\ q_m \end{bmatrix} = \begin{bmatrix} \Theta_n \\ \Theta_m \end{bmatrix} \tag{3.17}$$

where $\Theta_n$ and $\Theta_m$ are the emfs induced in the normal mode meshes by currents in the coupling loop. Assuming the loop is small enough that the current distribution along the loop is uniform, then for a loop current of $i_0$;

$$\Theta_n = L_{0n} \frac{di_0}{dt} \quad \text{and} \quad \Theta_m = L_{0m} \frac{di_0}{dt} \quad , \tag{3.18}$$

where $L_{0n}$ is the mutual inductance between the loop and normal mode mesh n. Defining the mesh currents $i_n = \dot{q}_n$ and $i_m = \dot{q}_m$ and taking the Fourier transform of (3.17) one obtains;

$$j\omega I_0 \begin{bmatrix} L_{0n} \\ L_{0n} \end{bmatrix} = \begin{bmatrix} j\omega L_n + R_n + \frac{S_n}{j\omega} & j\omega L_{nm} + R_{nm} + \frac{S_{nm}}{j\omega} \\ j\omega L_{nm} + R_{nm} + \frac{S_{nm}}{j\omega} & j\omega L_m + R_m + \frac{S_m}{j\omega} \end{bmatrix} \begin{bmatrix} I_n \\ I_m \end{bmatrix} \; . \tag{3.19}$$

The equivalent circuit relating to (3.19) is illustrated in figure 3.1, where $L_0$ is the loop inductance, Zin the input impedance and $Z_0$ the characteristic line impedance. A standard transformer coupling coefficient due to the probe coupling to the cavity may then be defined;

$$\kappa_n = \frac{L_{0n}}{\sqrt{L_0 L_n}} \quad . \tag{3.20}$$

It is also possible to define the following cross coupling coefficients;

(1) Magnetic field cross coupling: $\Delta_L = \dfrac{L_{nm}}{\sqrt{L_n L_m}}$

$$= \frac{\int_{\text{cavity volume}} \eta_n \cdot \eta_m \, d\tau}{\left( \int_{\text{cavity volume}} \eta_n \cdot \eta_n \, d\tau \cdot \int_{\text{cavity volume}} \eta_m \cdot \eta_m \, d\tau \right)^{\frac{1}{2}}} \qquad -1 \le \Delta_L \le 1 \tag{3.21a}$$

(2) Electric field cross coupling: $\Delta_C = \dfrac{S_{nm}}{\sqrt{S_n S_m}}$

$$= \frac{\int_{\text{cavity volume}} \xi_n \cdot \xi_m \, d\tau}{\left( \int_{\text{cavity volume}} \xi_n \cdot \xi_n \, d\tau \cdot \int_{\text{cavity volume}} \xi_m \cdot \xi_m \, d\tau \right)^{\frac{1}{2}}} \qquad -1 \le \Delta_C \le 1 \tag{3.21b}$$

(3) Dissipative cross coupling: $\Delta_R = \dfrac{R_{nm}}{\sqrt{R_n R_m}} = \dfrac{\sqrt{Q_n Q_m}}{Q_{nm}}$

$$= \frac{\int_{\text{niobium surface}} \eta_n \cdot \eta_m \, ds}{\left( \int_{\text{niobium surface}} \eta_n \cdot \eta_n \, ds \cdot \int_{\text{niobium surface}} \eta_m \cdot \eta_m \, ds \right)^{\frac{1}{2}}} \qquad -1 \le \Delta_R \le 1 \, . \tag{3.21c}$$

These coupling coefficients are normalized mutual parameters, which are invariant, regardless of equivalent circuit representation. Thus when we write the equivalent circuit impedance in terms of normalized parameters, we shall write the mutual component in terms of these coupling coefficients.

### 3.2.2 Parallel Circuit Representation of the sapphire dielectric

Applying the formalism of section 3.2.1, the shunt equivalent circuit of figure 3.2 shall be derived, with respect to the magnetic flux coordinates $\phi_n$ and $\phi_m$. Using (3.2a) and (3.2b) the electric and magnetic field energies of two modes, n and m, can be represented by the respective Lagrangian variables;

$$T_{nm} = \frac{\varepsilon k_n k_m \dot{\phi}_n \dot{\phi}_m}{2} \int_{\text{dielectric volume}} \xi_n \cdot \xi_m \, d\tau \tag{3.22}$$

$$V_{nm} = \frac{k_n^2 k_m^2 \phi_n \phi_m}{2\mu} \int_{\text{dielectric volume}} \eta_n \cdot \eta_m \, d\tau \tag{3.23}$$

These again have the form of energy storage elements as an inductor and a capacitor. In the case n = m;

$$C_n = \varepsilon k_n^2 \int_{\text{dielectric volume}} |\xi_n|^2 \, d\tau \tag{3.24a}$$

$$\Gamma_n = \frac{k_n^4}{\mu} \int_{\text{dielectric volume}} |\eta_n|^2 \, d\tau \quad , \tag{3.24b}$$

where $\Gamma_n$ is the equivalent parallel reciprocal inductance and $C_n$ the equivalent parallel capacitance. It follows from (3.24) and $k_n = \omega_n \sqrt{\varepsilon\mu}$;

$$\omega_n = \sqrt{\frac{\Gamma_n}{C_n}} \quad . \tag{3.25}$$

In the case n ≠ m, cross coupling terms are defined;

$$C_{nm} = \varepsilon k_n k_m \int_{\text{dielectric volume}} \xi_n \cdot \xi_m \, d\tau \tag{3.26a}$$

$$\Gamma_{nm} = \frac{k_n^2 k_m^2}{\mu} \int_{\text{dielectric volume}} \eta_n \cdot \eta_m \, d\tau \quad . \tag{3.26b}$$

Shunt Losses in the cavity dielectric, are represented by the dissipation function F, given by;

$$F = \frac{1}{2} \int_{\text{dielectric volume}} \sigma_d \, |E|^2 \, d\tau \quad , \tag{3.27}$$

where $\sigma_d$ is the dielectric conductivity. The paramagnetic loss associated with a chromium impurities, has been shown to be negligible (Jones, 1988) and has been omitted. Substituting (3.2b) into (3.27) yields:

$$F_{nm} = \frac{1}{2} \dot{\phi}_n \dot{\phi}_m G_{nm} \tag{3.28}$$

where

$$G_{nm} = k_n k_m \sqrt{\sigma d_n\, \sigma d_m} \int_{\text{dielectric volume}} \xi_n \cdot \xi_m \, d\tau = \frac{\sqrt{\omega_n\, \omega_m\, C_n\, C_m}}{Q_{nm}} \quad , (3.29)$$

and

$$\frac{1}{Q_{nm}} = \frac{\sqrt{\frac{\varepsilon_n''}{\varepsilon} \frac{\varepsilon_m''}{\varepsilon}}}{\upsilon_d} \int_{\text{dielectric volume}} \xi_n \cdot \xi_m \, d\tau \quad . \quad (3.30)$$

Here $\frac{\varepsilon_n''}{\varepsilon} = \frac{\sigma d_n}{\varepsilon\, \omega_n}$ is the $n^{th}$ mode loss tangent or ratio of imaginary to real permittivity, in the sapphire dielectric.

Using $T_{nm} - V_{nm}$ as the Lagrangian, and defining $\Im_n$ and $\Im_m$ as the mmf induced in the normal mode meshes by currents in the coupling loop, the generalized coupled Lagrange equations with respect to $\phi_n$ and $\dot{\phi}_m$ become;

$$\begin{bmatrix} C_n & C_{nm} \\ C_{nm} & C_m \end{bmatrix} \begin{bmatrix} \ddot{\phi}_n \\ \ddot{\phi}_m \end{bmatrix} + \begin{bmatrix} G_n & G_{nm} \\ G_{nm} & G_m \end{bmatrix} \begin{bmatrix} \dot{\phi}_n \\ \dot{\phi}_m \end{bmatrix} + \begin{bmatrix} \Gamma_n & \Gamma_{nm} \\ \Gamma_{nm} & \Gamma_m \end{bmatrix} \begin{bmatrix} \phi_n \\ \phi_m \end{bmatrix} = \begin{bmatrix} \Im_n \\ \Im_m \end{bmatrix}$$

(3.31)

Conservation of mmf in each mesh implies that, $\frac{N_0}{N_n} \Im_n = \frac{N_0}{N_m} \Im_m = i_0$ . The effective turns ratio $\frac{N_0}{N_n}$ , is calculated from the ratio of the primary and secondary magnetizing inductances, $\kappa_n^{\,2} L_0$ and $\Gamma_n^{\,-1}$ respectively, and given by $\frac{N_0}{N_n} = \sqrt{\kappa_n^{\,2} L_0 \Gamma_n}$ . Subtracting the magnetizing inductances from the loop inductance, the overall leakage inductance referred to the primary can be written as $L_0(1-\kappa_n^{\,2}-\kappa_m^{\,2})$. Expressing (3.31) in terms of node voltages $V_n = \dot{\phi}_n$ and $V_m = \dot{\phi}_m$, (3.31) Fourier transforms to;

$$I_0 \begin{bmatrix} \frac{N_0}{N_n} \\ \frac{N_0}{N_m} \end{bmatrix} = \begin{bmatrix} j\omega C_n + G_n + \frac{\Gamma_n}{j\omega} & j\omega C_{nm} + G_{nm} + \frac{\Gamma_{nm}}{j\omega} \\ j\omega C_{nm} + G_{nm} + \frac{\Gamma_{nm}}{j\omega} & j\omega C_m + G_m + \frac{\Gamma_n}{j\omega} \end{bmatrix} \begin{bmatrix} V_n \\ V_m \end{bmatrix} \quad (3.32)$$

This describes the circuit illustrated in figure 3.2, after adding the series leakage inductance.

The parallel cross coupling coefficients are evaluated as;

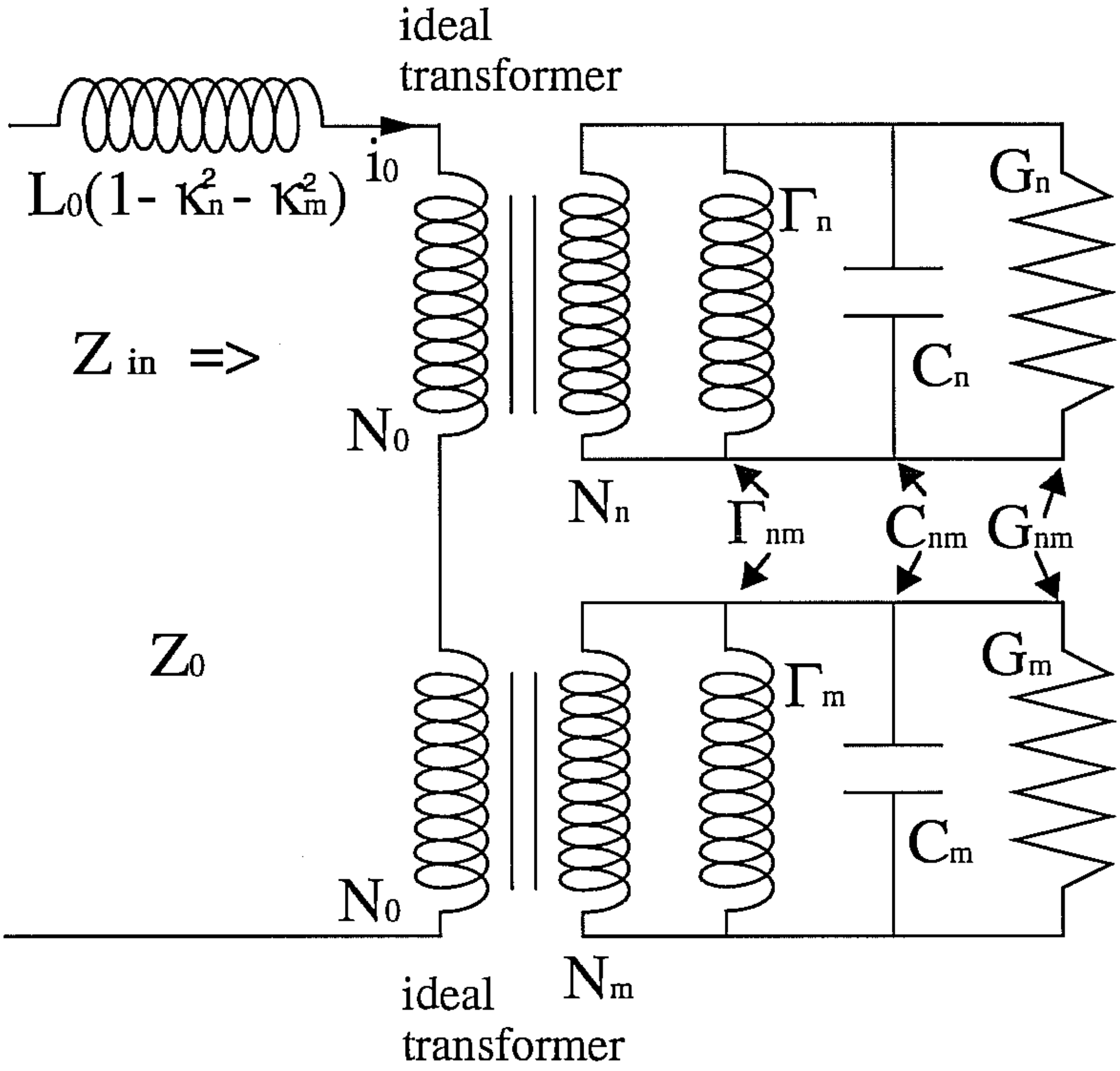

# Figure 3.2

General equivalent circuit of two excited modes in a loop-coupled dielectric resonator, as derived from normal coordinates $\phi_n$ and $\phi_m$.

(1) Magnetic field cross coupling: $$\Delta_L = \frac{\Gamma_{nm}}{\sqrt{\Gamma_n \Gamma_m}}$$

$$= \frac{\int_{\text{dielectric volume}} \eta_n \cdot \eta_m \, d\tau}{\left( \int_{\text{dielectric volume}} \eta_n \cdot \eta_n \, d\tau \cdot \int_{\text{dielectric volume}} \eta_m \cdot \eta_m \, d\tau \right)^{\frac{1}{2}}} \qquad -1 \leq \Delta_L \leq 1 \quad (3.33a)$$

(2) Electric field cross coupling: $$\Delta_C = \frac{C_{nm}}{\sqrt{C_n C_m}}$$

$$= \frac{\int_{\text{dielectric volume}} \xi_n \cdot \xi_m \, d\tau}{\left( \int_{\text{dielectric volume}} \xi_n \cdot \xi_n \, d\tau \cdot \int_{\text{dielectric volume}} \xi_m \cdot \xi_m \, d\tau \right)^{\frac{1}{2}}} \qquad -1 \leq \Delta_C \leq 1 \quad (3.33b)$$

(3) Dissipative cross coupling: $$\Delta_R = \frac{G_{nm}}{\sqrt{G_n G_m}} = \frac{\sqrt{Q_n Q_m}}{Q_{nm}}$$

$$= \frac{\int_{\text{dielectric volume}} \xi_n \cdot \xi_m \, d\tau}{\left( \int_{\text{dielectric volume}} \xi_n \cdot \xi_n \, d\tau \cdot \int_{\text{dielectric volume}} \xi_m \cdot \xi_m \, d\tau \right)^{\frac{1}{2}}} \qquad -1 \leq \Delta_R \leq 1 \ . \quad (3.33c)$$

These cross coupling coefficients are invariant when referred to the primary or secondary coils of the ideal transformers in figure 3.2. The mutual capacitance and reciprocal inductance are not invariant. This fact highlights why the former are useful when considering a normalized impedance of the cavity for theoretical purposes.

### 3.2.3 Significance to the Overall T-SLOSC Equivalent Circuit

Deriving the input impedances of figure 3.1 and figure 3.2, gives the same input impedance (figure 3.3), illustrating the duality of the circuits. However the parameters relating to the series circuit involve the niobium superconductor, while the parallel circuit involves the sapphire dielectric. A general circuit shown in figure 3.3 is considered, where each parameter may have more than one contribution due to the sapphire and niobium parts internal to the cavity.

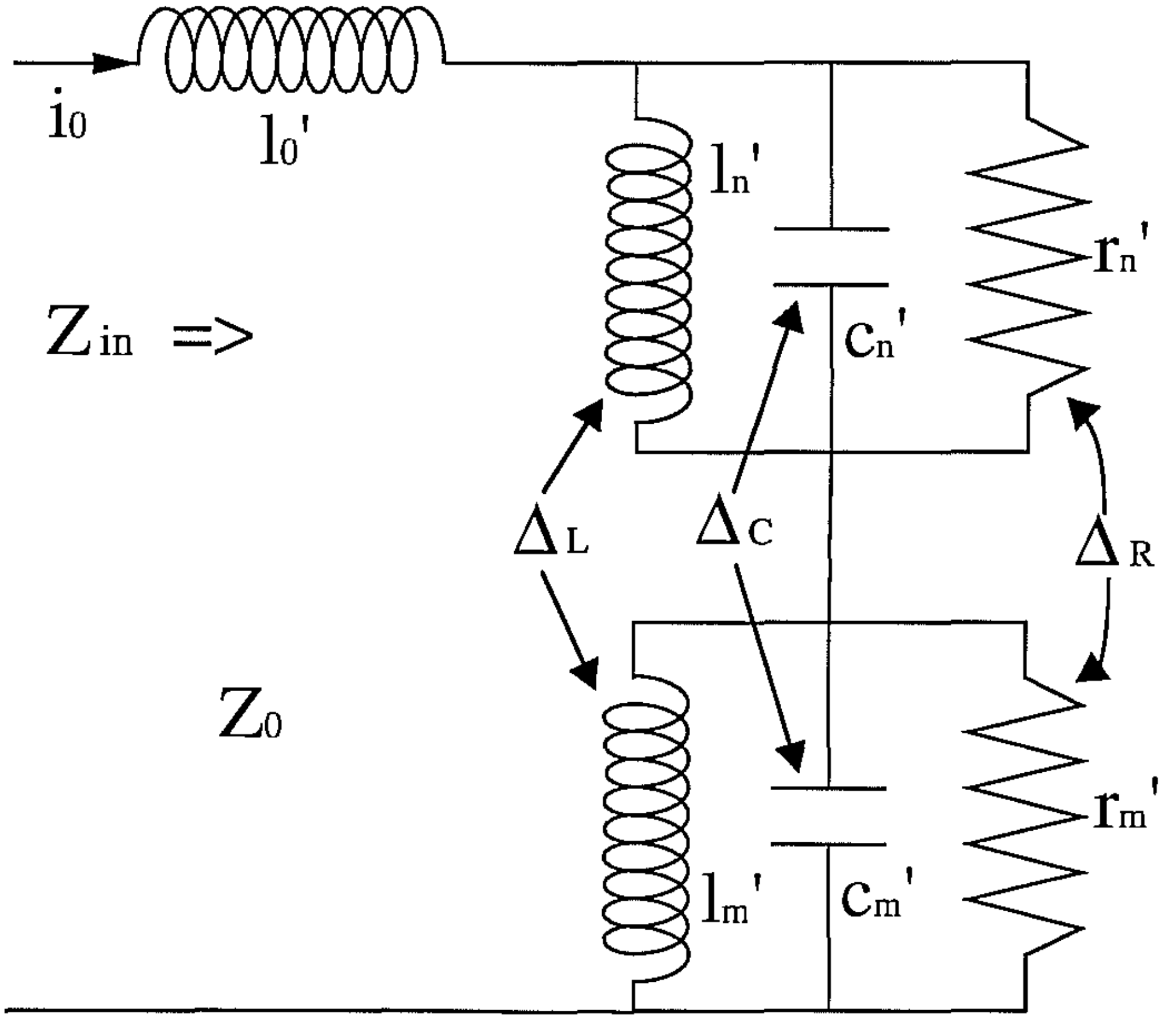

Figure 3.3

Both figure 3.1 and figure 3.2 can be represented as a series combination of parallel LCR circuits, with a defined coupling coefficient between like elements.

The 4 individual mode circuit elements may be expressed in terms of the 4 cavity - coupling parameters, $Q_n$ , $L_0$ , $\omega_i$ , $\kappa_n$ ;

$$l_0' = L_0(1-\kappa_n^2-\kappa_m^2) \quad (3.34a)$$

$$r_n' = \kappa_n^2 L_0 \omega_n Q_n \quad (3.34b)$$

$$l_n' = \kappa_n^2 L_0 \quad (3.34c)$$

$$c_n' = \frac{1}{\kappa_n^2 L_0 \omega_n^2} \quad . \quad (3.34d)$$

$L_0$ and $\kappa_n$ depend on the probe properties and its placement in the cavity. If the transformer coupling is zero then the probe does not couple to any modes, and the input impedance is equal to the loop reactance. As the transformer coupling increases, the series reactance diminishes and coupling to the mode increases. Probe stepper motors that work cryogenically, are used to adjust penetration into the cavity to vary the coupling. Both the niobium and sapphire contribute to $Q_n$ by; $Q_n^{-1} = Q_{Sapp\,n}^{-1} + Q_{Nb\,n}^{-1}$. For high Q modes $\omega_n$ is dominated by the sapphire dielectric while the niobium cavity only perturbs it very slightly.

For modelling purposes it is convenient to express the input impedance of the circuit in figure 3.3 in terms of the normalized parameters: $\beta_n = \frac{r_n'}{Z_0}$ , $\alpha = \frac{l_0' \omega_n}{Z_0}$ , $y_n = \frac{\omega - \omega_n}{\omega_n}$ , $Q_n$ and cross coupling coefficients from either (3.21) or (3.33). Thus it can be shown that;

$$\frac{Zin}{Z_0} = \frac{\beta_n \Lambda_n + \beta_m \Lambda_m - 2\sqrt{\beta_n \beta_m}\, \Lambda_n \Lambda_m D_{nm}}{1 - \Lambda_n \Lambda_m D_{nm}^2} + j\alpha \quad (3.35)$$

where

$$\Lambda_n = \frac{1}{1+2jQ_n y_n}$$ is a Lorentzian of unity coupling and

$$D_{nm} = \Delta_R + j\sqrt{Q_n Q_m}\left(\Delta_C \left((y_n+1)(y_m+1)\right)^{\frac{1}{2}} - \Delta_L \left((y_n+1)(y_m+1)\right)^{\frac{-1}{2}}\right)$$

is a mutual cross term due to derived cross coupling coefficients.

The reflection coefficient is defined by;

$$\rho = \frac{\frac{Zin}{Z_0} - 1}{\frac{Zin}{Z_0} + 1} \quad (3.36)$$

$$= \frac{\beta_n \Lambda_n + \beta_m \Lambda_m - 2\sqrt{\beta_n \beta_m}\, \Lambda_n \Lambda_m D_{nm} + (j\alpha - 1)(1 - \Lambda_n \Lambda_m D_{nm}^2)}{\beta_n \Lambda_n + \beta_m \Lambda_m - 2\sqrt{\beta_n \beta_m}\, \Lambda_n \Lambda_m D_{nm} + (j\alpha + 1)(1 - \Lambda_n \Lambda_m D_{nm}^2)}$$

By expressing both mode frequencies in terms of a tuning parameter, and graphing the reflection coefficient at various tunings, the behaviour of closely tuned modes can be investigated. Mathematica allows the algebra to be handled with ease.

## 3.3 OBSERVATION OF COUPLED MODES

The model describing coupled mode behavior was tested using the apparatus shown schematically in figure 3.4. The sapphire resonator has two unperturbed high-Q modes at 10.439 (quasi $TM_{8,1,1+\delta}$) and 10.22 GHz (quasi $TE_{6,1,1+\delta}$), both occur in doublet pairs. Experimental tuning curves for one of each pair of doublets are shown in figure 3.5. A decrease in the observed Q of each mode occurs when a low Q mode is tuned close to the high Q mode under investigation. The tuning variable x is measured in millimeters from the top of the sapphire resonator to the tuning disc. It has been determined after the publication of the tuning data (Tobar and Blair, Sept. 1991) that there was approximately -2mm systematic error in the absolute measurements of x. This was caused by movement when cooling, hence all values of x presented here after should have 2 mm subtracted. However relative values of tuning are still very accurate.

Q degradation for the 10.439 GHz modes was observed for x between 7 and 8 mm. This effect was seen on three experimental runs and has never been observed for the 10.22 GHz mode. Either the niobium tuning stem links a significant fraction of the field creating additional dissipation, or there is a reactive coupling to some distant mode. This would however require larger cross coupling coefficients than those measured in this section.

The 10.439 GHz mode in figure 3.5 was plotted using a room temperature circulator rather than the cryogenic circulator shown in figure 3.4. The 1.5 meters of coaxial cable leading out of the cryogenic dewar greatly enhanced line resonances. Section 3.5 shows that co-axial cable line resonances affect cavity resonances through a resistive interaction with the probes. To measure cavity properties, a line resonance can be detuned away from a cavity resonances by varying a line stretcher between the room temperature circulator and the T-SLOSC. After the addition of the cryogenic circulator, the interaction around x = 10.2 mm was observed with greater accuracy, and is shown in figure 3.6.

## 3.4 REACTIVE COUPLING

For reactive coupling (ignoring damping), the method of Goldstein calculates the resonant frequencies from, det $\left(V_{ik} - \omega^2 T_{ik}\right) = 0$. Using circuit elements of figure 3.3, it follows that;

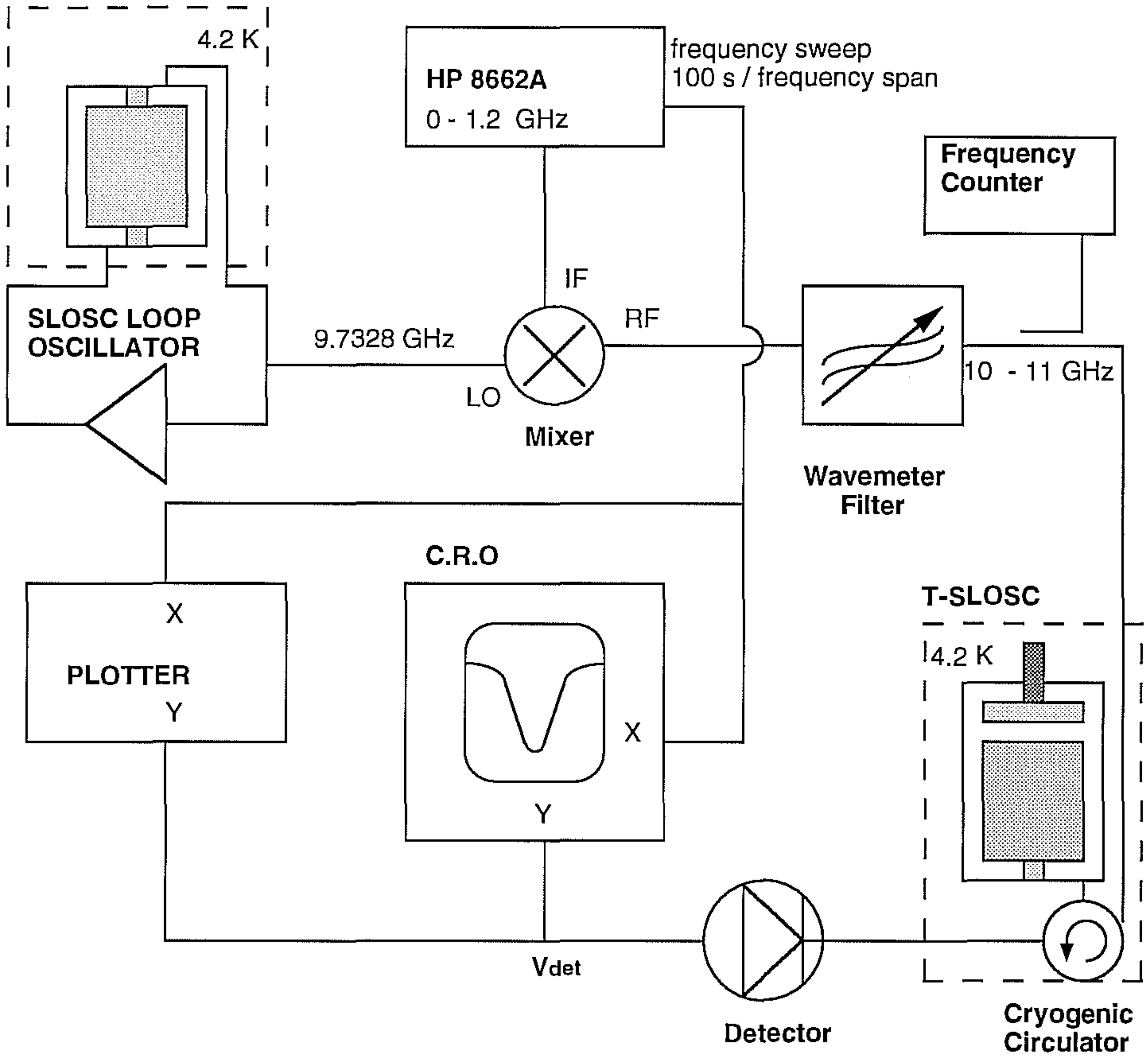

# Figure 3.4

An ultra-stable swept microwave frequency source is obtained by mixing a fixed-frequency SLOSC oscillator with an HP 8662A synthesizer locked to a frequency standard. The signal is swept slowly over the T-SLOSC in reflection, at 100 sec per frequency span. The voltage Vdet is measured from the detector in the range where it is proportional to $|\rho|^2$. From this plot Q and coupling can be measured.

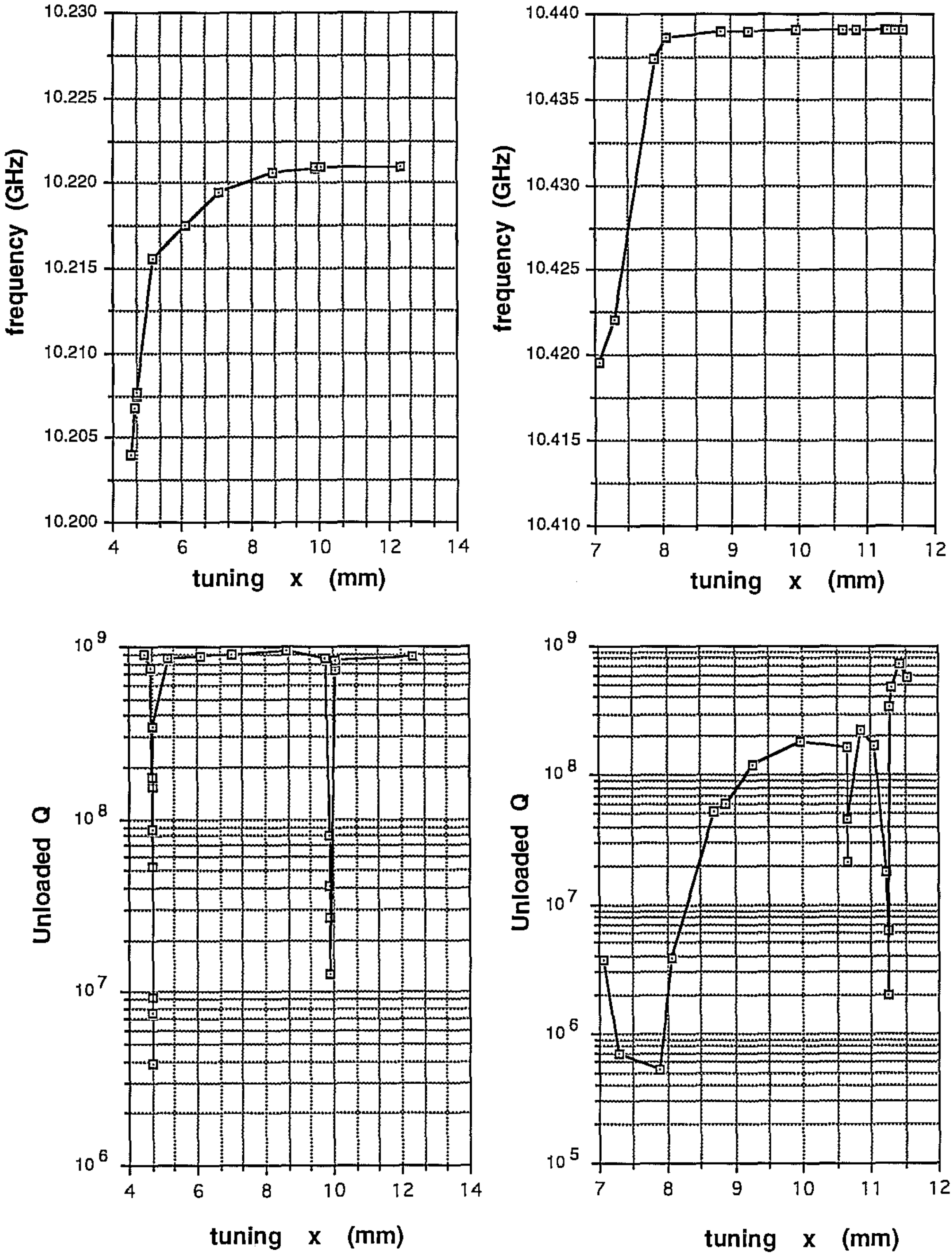

Mode frequency and unloaded Q curves for one of the 10.22 GHz (quasi $TE_{6,1,1+\delta}$) and 10.439 GHz (quasi $TM_{8,1,1+\delta}$) high Q doublet pairs. Unloaded Q curves are beneath the corresponding frequency curves.

Figure 3.5

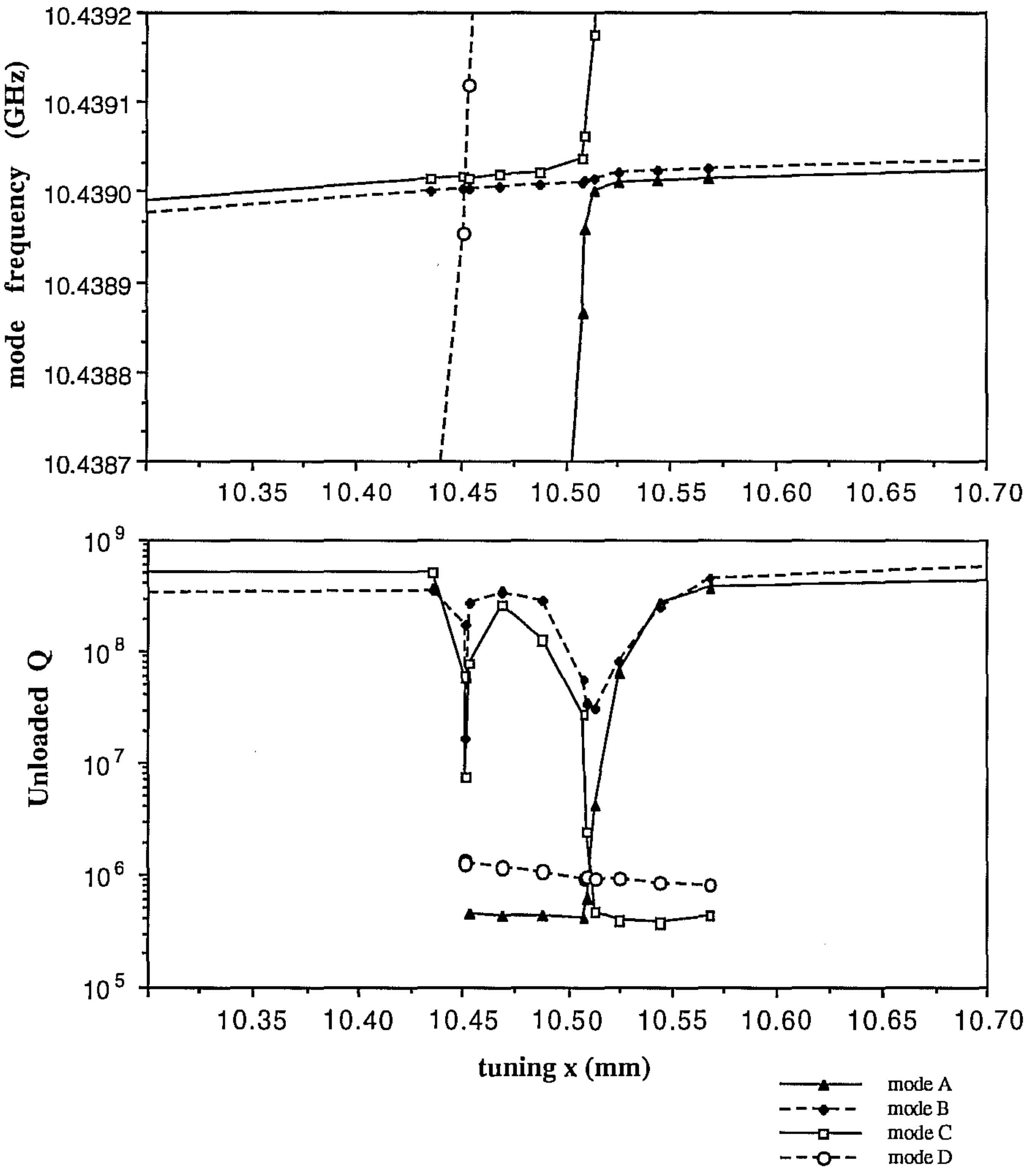

Frequency and unloaded Q versus tuning curves, of the interaction of a low Q doublet pair with the 10.439 GHz high Q doublet pair

Figure 3.6

$$\det\begin{pmatrix} \dfrac{1}{l_n'} - \omega^2 c_n' & \dfrac{1}{l_{nm}'} - \omega^2 c_{nm}' \\ \dfrac{1}{l_{nm}'} - \omega^2 c_{nm}' & \dfrac{1}{l_m'} - \omega^2 c_m' \end{pmatrix} = 0 \quad , \qquad (3.37)$$

or

$$\omega^4 (1-\Delta_C) + \omega^2 (2\,\Delta_C\,\Delta_L\,\omega_n\,\omega_m - \omega_n^{\,2} - \omega_m^{\,2}) + (1-\Delta_L)\,\omega_n^{\,2}\,\omega_m^{\,2} = 0 \ .$$

Assuming the interacting modes have linear tuning coefficients, so $\omega_n = \omega_0\left(1 + \frac{\Delta x'}{2}\right)$ and $\omega_m = \omega_0\left(1 - \frac{\Delta x'}{2}\right)$, where $\Delta x'$ is an arbitrary dimensionless tuning unit. The solution to first order in $\Delta x'$ and $\Delta_L - \Delta_C$ is;

$$\omega_{1,2} = \omega_0 \left(1 \pm \frac{\sqrt{\Delta x'^{\,2} + (\Delta_L - \Delta_C)^2}}{2}\right) \ . \qquad (3.38)$$

This shows that the two modes always have a minimum frequency separation when $\Delta x' = 0$, due to the reactive coupling between the two modes.

Experimentally it is possible to measure the net reactive coupling by;

$$|\Delta_T| = |\Delta_L - \Delta_C| = \left.\frac{\omega_1 - \omega_2}{\omega_0}\right|_{\Delta x' = 0} \ . \qquad (3.39)$$

Equations (3.21) and (3.33) allow couplings of either sign. Negative inductive coupling has the same effect as positive capacitive coupling, thus it is impossible to determine experimentally whether the coupling is through the magnetic or electric field. However, the reactive coupling can be measured and substituted into the Mathematica program for simulation. The interaction is not symmetrical, so the coupling coefficient sign can be determined by comparing the simulation with experiment.

A general interaction of two doublet pairs requires four mode equivalent circuits, with cross coupling elements between each mode circuit. For the higher frequency interaction illustrated in figure 3.6 and figure 3.7, the dominant coupling occurs between mode A and mode C. Mode D is too far away in frequency, while Mode B only couples very weakly in this interaction. This is highlighted by $Q_B$ dropping an order of magnitude less than $Q_A$ , and mode B's tuning curve only deviating slightly, while mode A and mode C swap characteristics. Thus the model derived for two interacting modes should be applicable.

The linear tuning curves for mode A and C after the interaction, are measured to be (from figure 3.6);

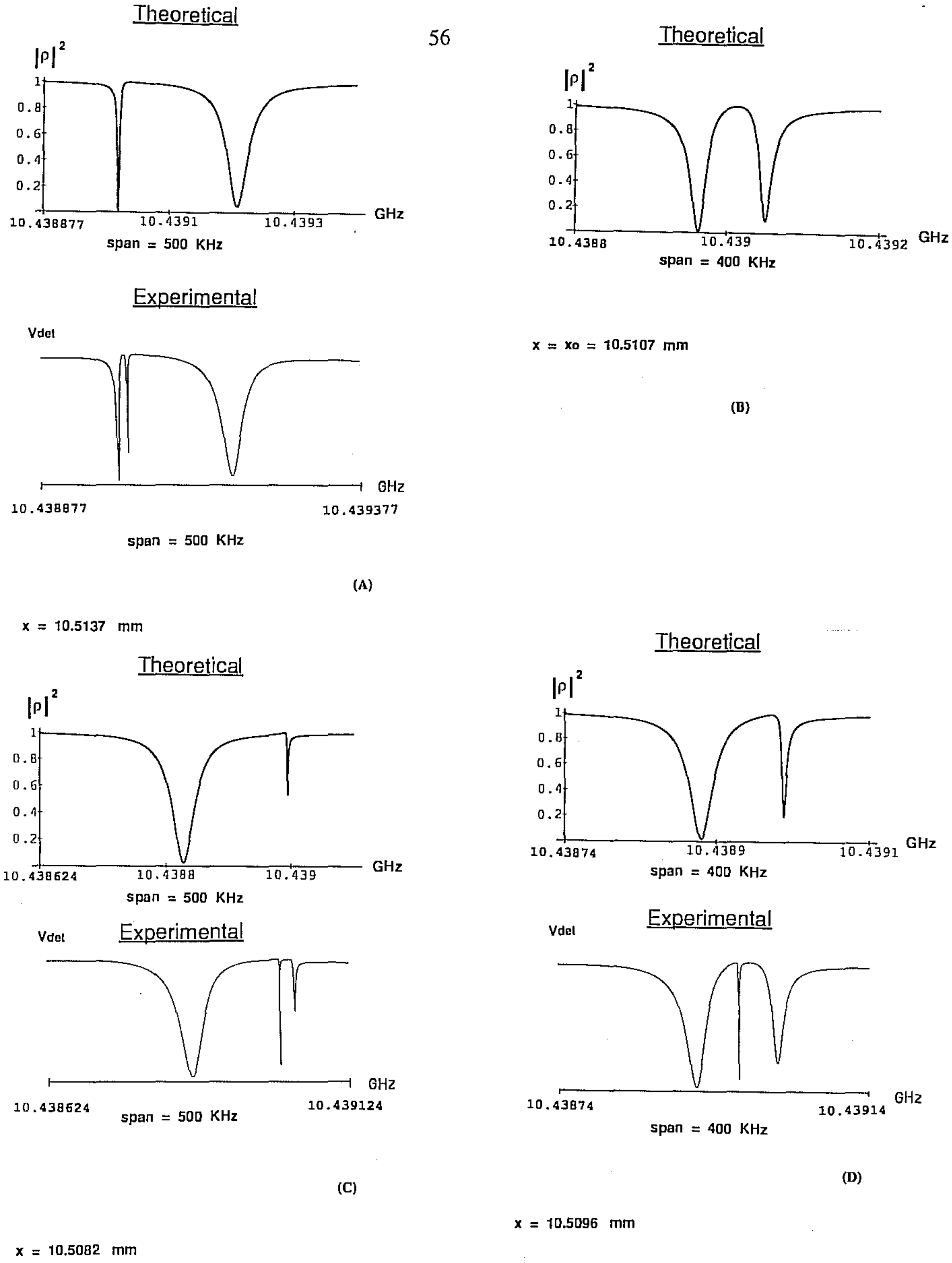

Figure 3.7 A close look at the higher frequency interaction from fig. 3.6, with the Experimentally observed interaction compared with theoretical simulation.

# Figure 3.7

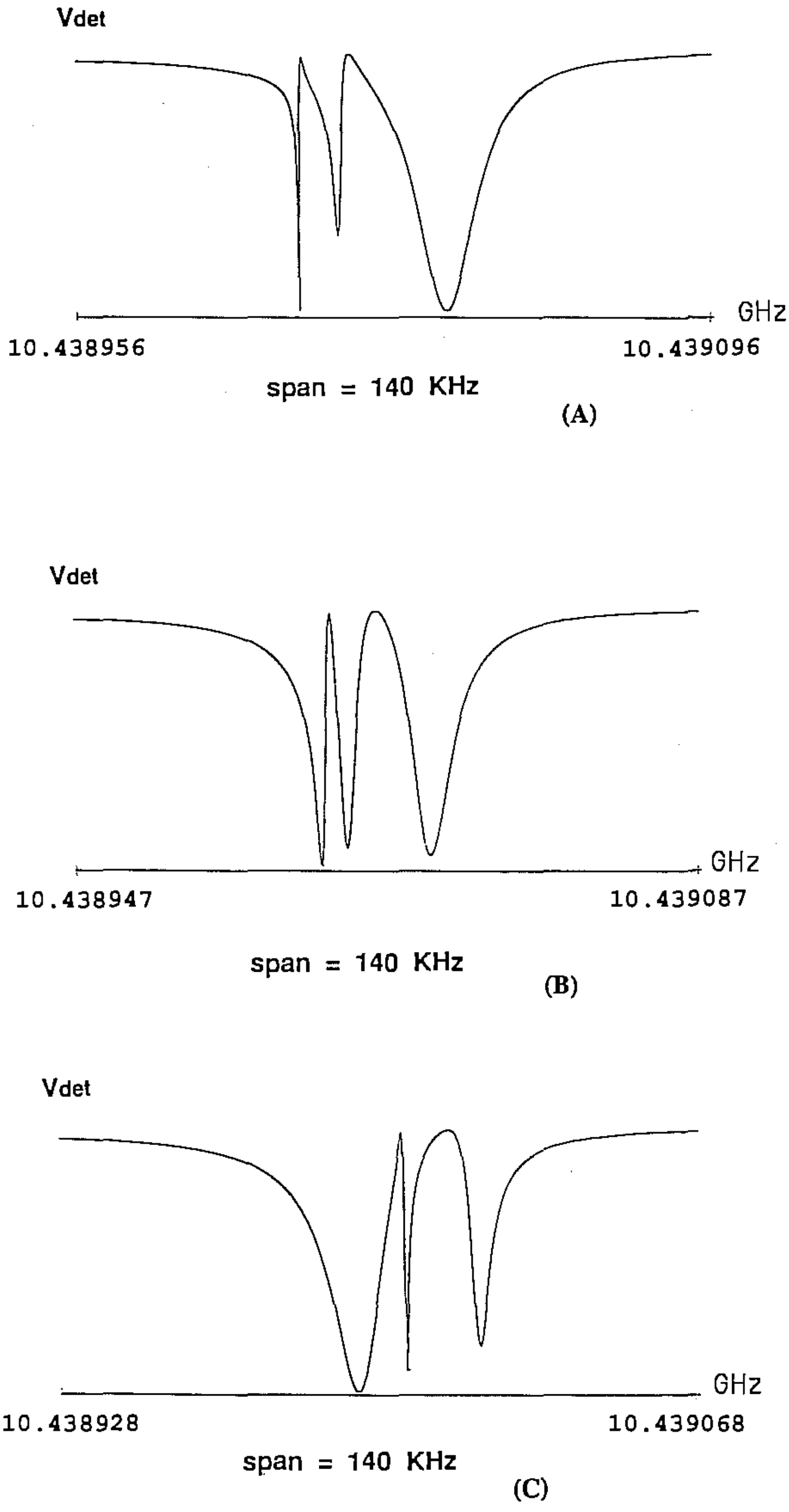

Figure 3.8 A close look at the lower frequency interaction from fig. 3.6.

# Figure 3.8

$$f_A = f_0[1.0281\ 10^{-5}\,\Delta x\ +1] \qquad f_C = f_0[5.4276\ 10^{-3}\,\Delta x\ +1]\,. \quad (3.40)$$

The frequency of intersection between the two lines in (3.40) is given by $f_0 = 10.43900$ GHz, and $\Delta x$ is given by $\Delta x = x - x_0$ , where $x_0 = 10.5107$ mm. High precision in relative values of x was achieved by a stepper motor, although the systematic error in the absolute value is much greater (~ -2mm) due to the possibility of thermal differential movement during cooling.

The normalized series probe inductance ($\alpha$) was calculated to be 1.2 in chapter 1. In the following simulations the series inductance was ignored as it only has small loading effects that will cause frequency perturbations within the bandwidth of the cavity resonance, so simulations including the series reactance hardly differ. Following this assumption the coupling can be assumed to be only resistive. Thus the measured values of coupling are substituted directly into the values of $\beta$. Before the interaction the following parameters were measured (assuming $\alpha = 0$): $Q_A = 10^8$, $Q_C = 4{\cdot}10^5$, $\beta_A = 2$, $\beta_C = 0.7$. Using (3.39) the mutual coupling is estimated from the experimental curve in figure 3.7d, to be; $\Delta_T = -1.0{\cdot}10^{-5}$. This is an over estimate since $x \neq x_0$. After a few trials $\Delta_T = -8.5{\cdot}10^{-6}$ gives a reasonable agreement with experiment. Experimental and theoretical curves are compared in figure 3.7.

The lower frequency interaction in figure 3.6 is more complicated, as modes B and C interact simultaneously with mode D, to about the same order of magnitude. The mutual coupling is estimated from (3.39) to give $\Delta_T \approx -2{\cdot}10^{-6}$. This smaller coupling enables the high Q modes to get closer to the low Q mode (see fig. 3.8). For the 10.22 GHz mode interactions, cross coupling coefficients are of the order of $-5{\cdot}10^{-5}$

The mechanism of this reactive coupling is either via the electric field as one might expect in a dielectric resonator, or via the magnetic field through a chromium electron spin resonance. This chromium resonance has been observed to cause a power dependent coupling between modes in a fixed frequency SLOSC (Giles, 1993), which causes frequency pulling of one mode when power is injected into another. It is possible that this coupling is another manifestation of this effect.

## 3.5 RESISTIVE COUPLING

Resistive couplings were observed between line resonances and most cavity resonances before the cryogenic circulator was added to the circuit in figure 3.4. The line resonances had Q factors of about $5.5{\cdot}10^2$ with varying couplings. After adding the cryogenic

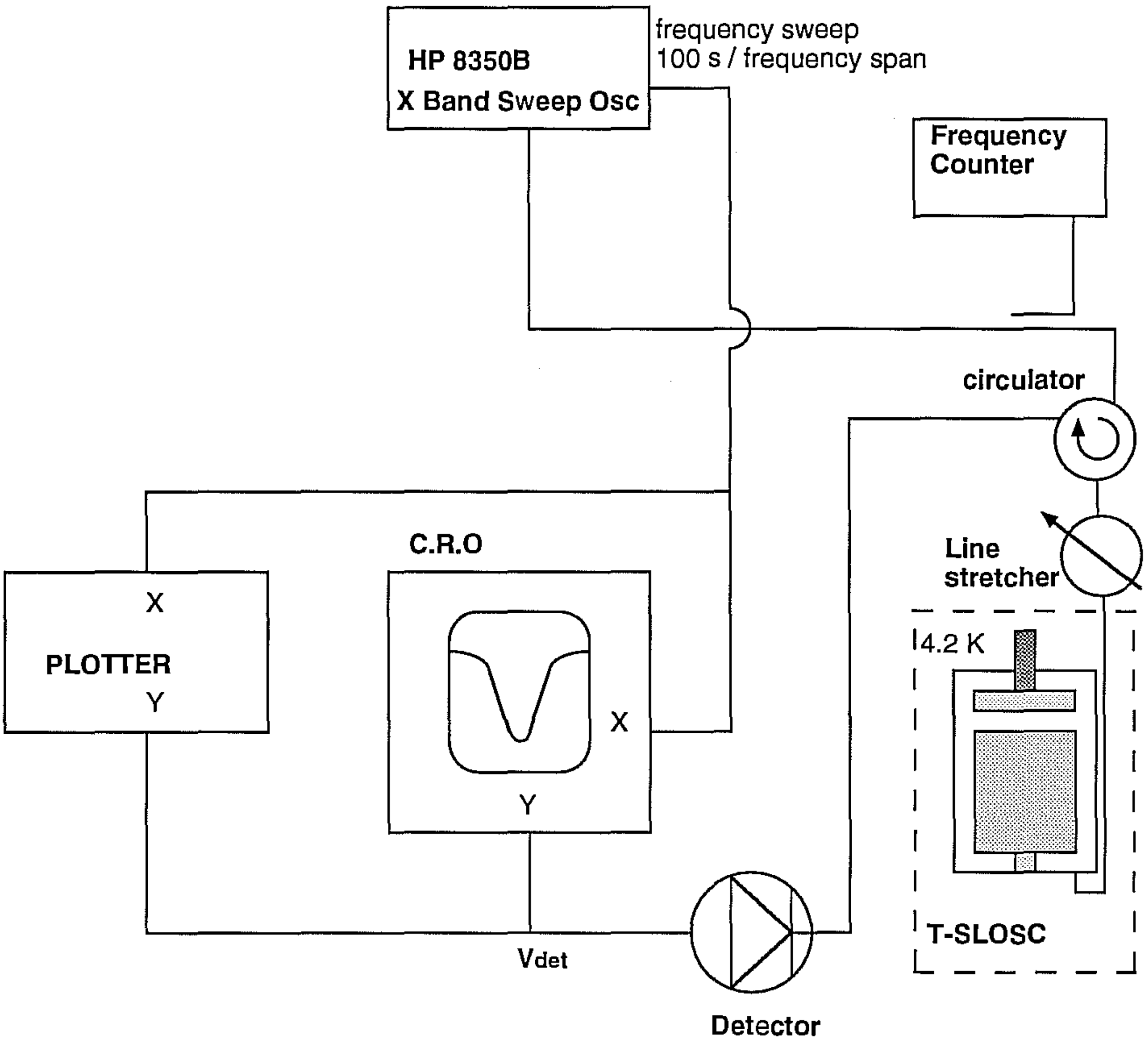

A HP 8350B is used to span over a line resonance so interactions with a T-SLOSC resonance can be observed. The line stretcher is used to frequency pull the line resonance across the cavity resonance.

# Figure 3.9

circulator the effects of line resonances was negligible, due to reducing the line length and the number of connectors from the probe to the circulator.

The circuit in figure 3.4 can not span over the bandwidth of a line resonance due to the wavemeter in the circuit. Thus the circuit in figure 3.9 was used to observe a line resonance interact with a mode of $Q \approx 10^5$ and $\beta \approx 3.5$. An over coupled mode is observed because from (3.36) it is apparent that mutual effects will be enhanced for an over coupled mode. The frequency stability of the microwave sweeper only enabled the study of low Q modes with this circuit, however the same effects on high-Q modes were observed using the circuit of figure 3.4 with a room temperature circulator. The line stretcher in figure 3.9 was used to frequency pull a line resonance across the T-SLOSC mode under investigation. Figure 3.10 shows the results of comparison of experiment with theory.

Line resonances couple to cavity resonances through the niobium loop probes. From (3.21c);

$$\Delta_R = \frac{\displaystyle\int_{\text{niobium probe}} \eta_n \cdot \eta_m \, ds}{\left( \displaystyle\int_{\text{niobium probe}} \eta_n \cdot \eta_n \, ds \int_{\text{niobium probe}} \eta_m \cdot \eta_m \, ds \right)^{\frac{1}{2}}} ,$$

$\Delta_R = 0.9$ was found to give a good agreement between the model and experiment (figure 3.10). The cavity resonance distorts as it tunes across the line resonance.

## 3.6 GENERAL ANALYSIS OF INTERACTIONS BETWEEN COUPLED RESONANT MODES

In this section the coupled mode theory of two interacting coupled resonant modes is simplified and presented in terms of normalized equivalent circuit parameters so it may be easily applied to any resonant interacting system. This allows analytical equations to be derived that explain the effects on the mode frequencies, Q factors and couplings as a function of detuning for two reactively coupled resonant modes. All equations are verified experimentally by presenting the mode interaction data between mode A and mode C (see figures 3.6 and 3.7) in terms of the normalised parameters and comparing them with theory. This work enables the theoretical analysis of the effect of a spurious mode on a operational mode in any resonant cavity (Tobar, 1993). The frequencies, Q factors and couplings can all be affected beyond the bandwidth of the interacting modes resonances, with the coupling being affected the most and the frequency the least.

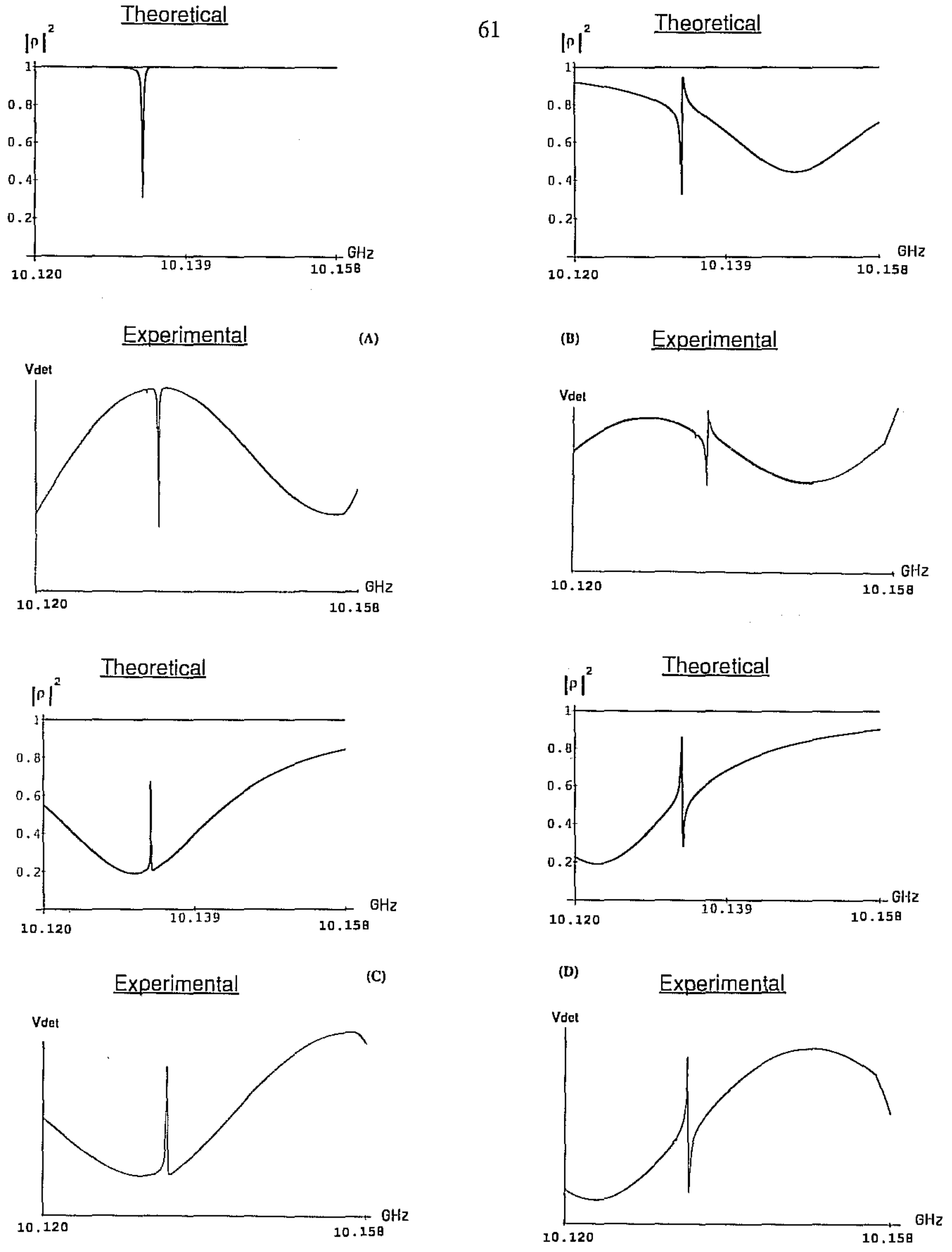

Figure 3.10 A comparison of the experimental effect and theoretical simulation of frequency pulling a line resonance across a T-SLOSC resonant mode.

# Figure 3.10

In the past the effect of a spurious mode interaction has only been analysed in terms of the change in effective series reactance of the probing mechanism (Kajfez and Crnadak, 1985), (Montgomery et al, 1948). This is less significant and is ignored in this section. This section studies the more dramatic interaction effect due to a mutual admittance between resonances in terms of normalized parameters, so that these results can be applied to any interacting resonant system.

### 3.6.1 Theory

Ignoring the series impedance added by the probing mechanism, it has been shown experimentally in sections 3.5 and 3.6 that two coupled cavity resonances, may be represented by the circuit illustrated in figure 3.3. Defining $\delta_n = 2Q_n y_n$, the normalised admittance for the cavity resonance in terms of the normalized parameters is given by;

$$Y_n = \frac{1 + j\delta_n}{\beta_n} \tag{3.41}$$

The following approximation (3.42) of the admittance of mode m, in terms of parameters of mode n, is valid if the entire range of measured frequencies in $\omega$ is small with respect to the resonant frequencies, $\omega_n$ and $\omega_m$. This will generally be true for a moderately high Q resonant system.

$$Y_m \approx \frac{1 + j\left(Q_m/Q_n\right)\left(\delta_n - \delta_{nm}\right)}{\beta_m} \tag{3.42}$$

Here $\delta_{nm}$ is the detuning between mode n and m in terms of half bandwidths of mode n. The mutual admittance $G_{nm} + j\left(\omega C_{nm} - \Gamma_{nm}/\omega\right)$, given in figure 3.2 is non-resonant over $\omega$ and hence a slowly varying function which can be approximated as a constant. Introducing the cross coupling terms derived in section 3.2 and 3.3, the mutual admittance normalised with respect to the characteristic line impedance may be represented as;

$$Y_{nm} \approx \frac{\Delta_R + j\sqrt{Q_m/Q_n}\left(\Delta\delta_n/2\right)}{\sqrt{\beta_n\beta_m}} \tag{3.43}$$

Here $\Delta\delta_n = 2Q_n\Delta_T$ is the normalised detuning when $\delta_{nm} = 0$, due to the inductive coupling between the two modes, and $\Delta_T = \Delta_C - \Delta_L$ as given by (3.39).

The normalised input impedance given by (3.35) can thus be approximated by (3.44), substituting in (3.41), (3.42) and (3.43) gives the normalised input impedance as a function of $Q_m/Q_n$, $\delta_n$, $\delta_{nm}$, $\beta_n$, $\beta_m$, $\Delta_R$, and $\Delta\delta_n$, which are all normalised parameters.

$$\frac{Z_{in}}{Z_0} \approx \frac{Y_n + Y_m - 2Y_{nm}}{Y_n Y_m - Y_{nm}^2} \tag{3.44}$$

This equation is similar to (3.35) except here it is expressed in terms of normalised admittance elements rather than impedance elements and the approximations greatly simplify the expression. This allows an analytical solution of the reactively coupled problem to be determined.

#### 3.6.1.1 Reactively Coupled Resonant Modes

Equating real and imaginary parts of (3.44) with the input impedance of two independent second order modes (normal modes) given by (3.45), the normal mode parameters can be solved in terms of the uncoupled mode parameters (3.46) - (3.48).

$$\frac{Z_{in}}{Z_0} = \frac{\beta_+}{1 + j(Q_+/Q_n)(\delta_n - \delta_{n+})} + \frac{\beta_-}{1 + j(Q_-/Q_n)(\delta_n - \delta_{n-})} \tag{3.45}$$

$$\delta_{n\pm} = \frac{\left(\delta_{nm} \pm \sqrt{\delta_{nm}^2 + \Delta\delta_n^2}\right)}{2} \tag{3.46}$$

$$\frac{Q_\pm}{Q_n} = \frac{\sqrt{\delta_{nm}^2 + \Delta\delta_n^2}}{\pm\,(Q_n/Q_m)\delta_{n\pm} \mp \delta_{n\mp}} \tag{3.47}$$

$$\beta_\pm = \frac{\pm\left(\beta_m(Q_n/Q_m)\delta_{n\pm} - \beta_n\delta_{n\mp} - \sqrt{\beta_n\beta_m(Q_n/Q_m)}\,\Delta\delta_n^2\right)}{\pm\,(Q_n/Q_m)\delta_{n\pm} \mp \delta_{n\mp}} \tag{3.48}$$

Here the + and - subscript refer to the positively and negatively detuned normal modes respectively.

Figures 3.11 - 3.13 show the theoretical effects of tuning two reactively coupled modes together, as various parameters are changed, with the dashed and undashed lines showing the behaviour of the positively (+) and negatively (-) detuned normal modes respectively. The frequency and Q factor plots (fig.3.11-3.12) are symmetrical, and do not depend on the sign of the mutual coupling term $\Delta\delta_n$, however the coupling curves (fig. 3.13) are asymmetric, with the symmetry depending on the sign.

### 3.6.2 Experimental Verification

Equations (3.46-3.48) are verified using the mode interaction data of the high Q quasi $TM_{8\,1\,1+\delta}$ whispering-gallery type mode interacting with a low Q spurious mode as they are tuned close together. Denoting the low Q spurious mode by the subscript 1, and the high Q mode by the subscript 2, the following parameter values were found to give good agreement between experiment and theory; $\beta_1 = 0.65$, $\beta_2 = 1.9$, $Q_1 = 4.1\cdot10^5$, $Q_2 = 6.56\cdot10^8$, $\Delta\delta_1 = 3.5$, with the frequency offset ($\delta_{12}$) of the high Q mode from the low Q spurious mode, normalized in terms of $Q_1$. In section 3.4 the + mode was designated as mode C and the - mode as mode A. Results are illustrated in figures 3.14-3.16.

In figures 3.15 and 3.16 one can see an excursion from the theory at around $\delta_{12} \approx 90$. This is due to the spurious mode occuring as a doublet. Generally throughout this

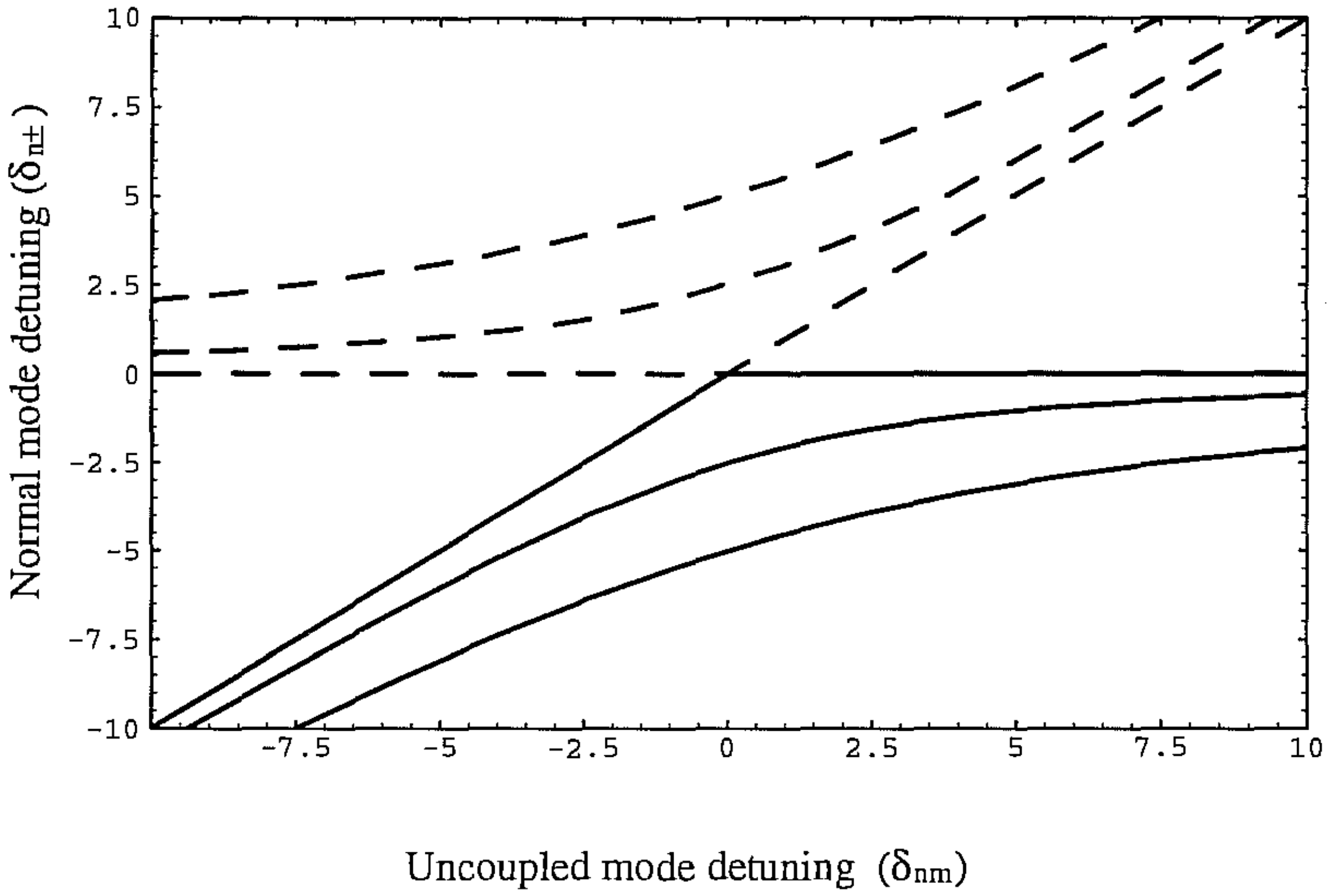

# Figure 3.11

Normal mode detuning ($\delta_{n\pm}$) versus uncoupled mode detuning ($\delta_{nm}$) for varying values of $\Delta\delta_n$. For $|\Delta\delta_n| = 0$ there is no interaction, as $|\Delta\delta_n|$ increases an increasing frequency separation between normal modes occurs as they are tuned together. This figure shows $|\Delta\delta_n| = 0$, 5 and 10, which is equivalent to the separation of the normal modes when $\Delta\delta_{nm} = 0$. The positvely detuned normal mode is represented by the dashed line, while the negatively detuned normal mode is represented by the undashed line.

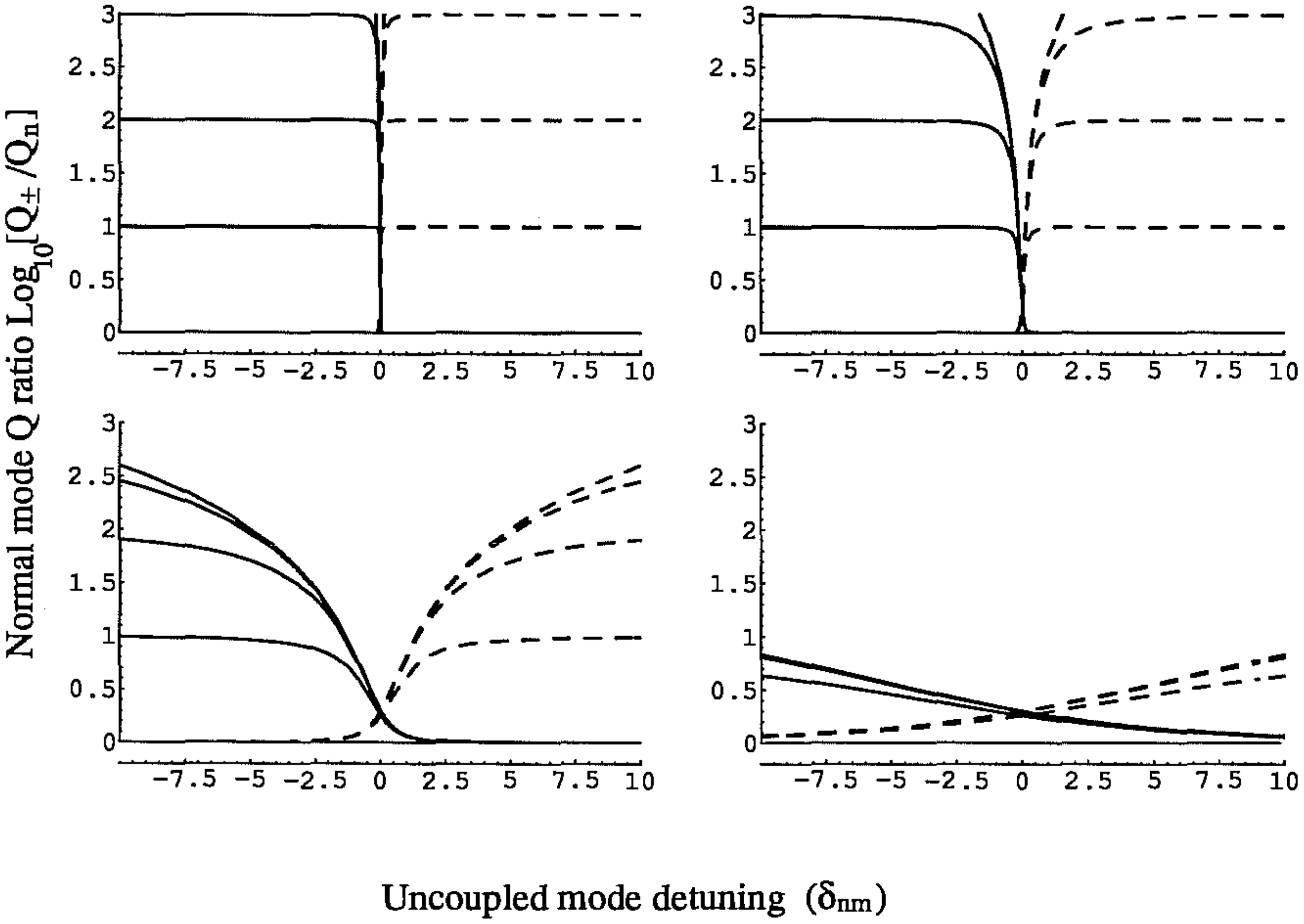

# Figure 3.12

Normal mode Q ratio ($Q_\pm/Q_n$) versus uncoupled mode detuning ($\delta_{nm}$) for varying values of $\Delta\delta_n$ and $Q_m/Q_n$. On each graph $Q_m/Q_n$ = 1,10, 100, 1000 and $\infty$, with $\Delta\delta_n$ varying from top left to bottom right as; $|\Delta\delta_n|$ = .01, 0.1, 1, 10. The positvely detuned normal mode is represented by the dashed line, while the negatively detuned normal mode is represented by the undashed line.

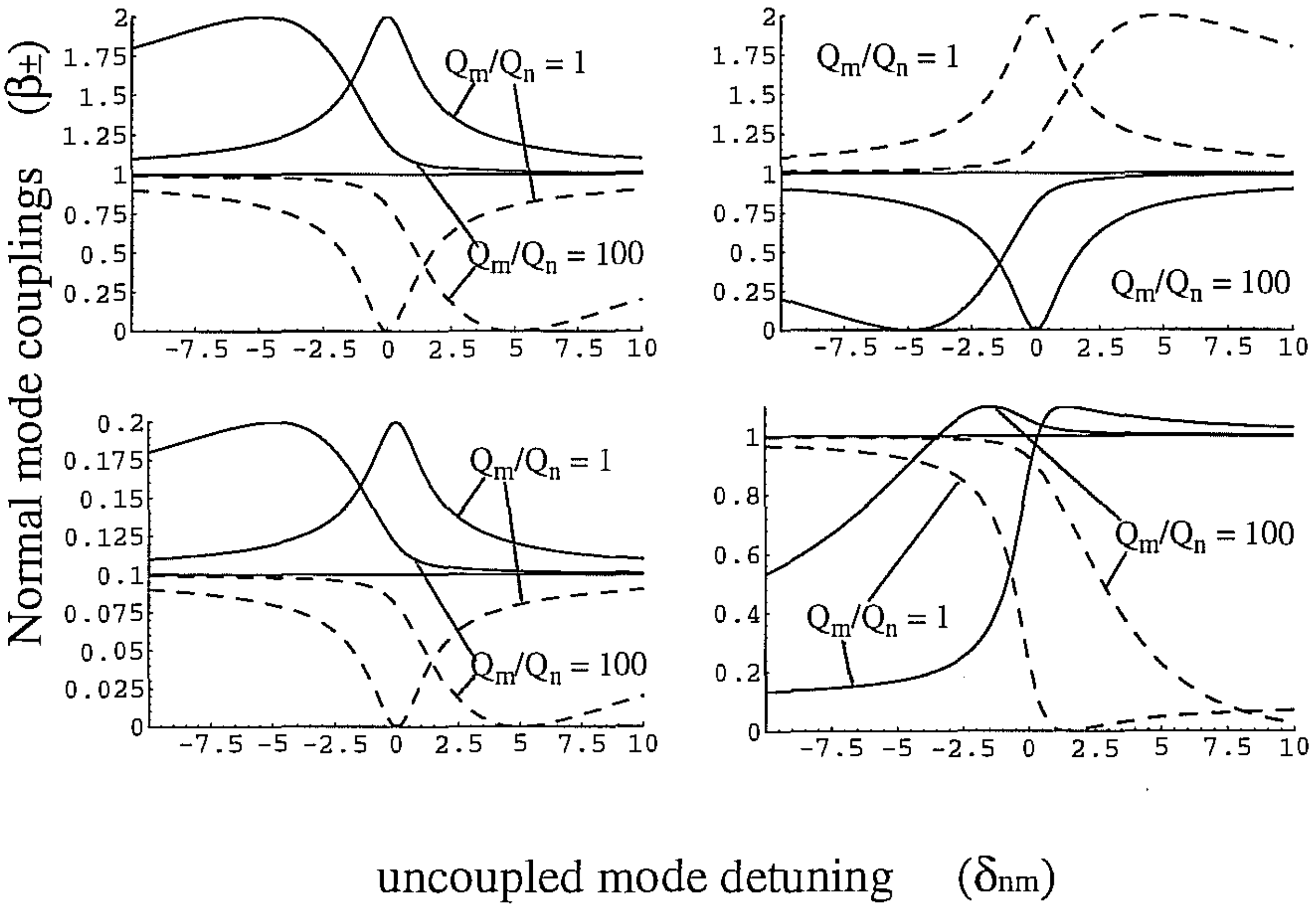

# Figure 3.13

Normal mode coupling ($\beta_{\pm}$) versus uncoupled mode detuning ($\delta_{nm}$) for varying values of $\beta_n$ , $\beta_m$, $\Delta\delta_n$ and $Q_m/Q_n$. On each graph $Q_m/Q_n$ = 1, and 100, with the positvely detuned normal mode represented by the dashed line, and the negatively detuned normal mode represented by the undashed line. From top left to bottom right ; $\beta_n = \beta_m = \Delta\delta_n = 1$; $\beta_n = \beta_m = 1$, $\Delta\delta_n = -1$; $\beta_n = \beta_m = 0.1$, $\Delta\delta_n = 1$; $\beta_n = 1, \beta_m = 0.1$, $\Delta\delta_n = 1$. The interaction is symmetric when the Q ratio =1 and $\beta_n = \beta_m$ , otherwise it is generally asymmetric. From the asymmetry of the interaction the sign of $\Delta\delta_n$ can be determined.

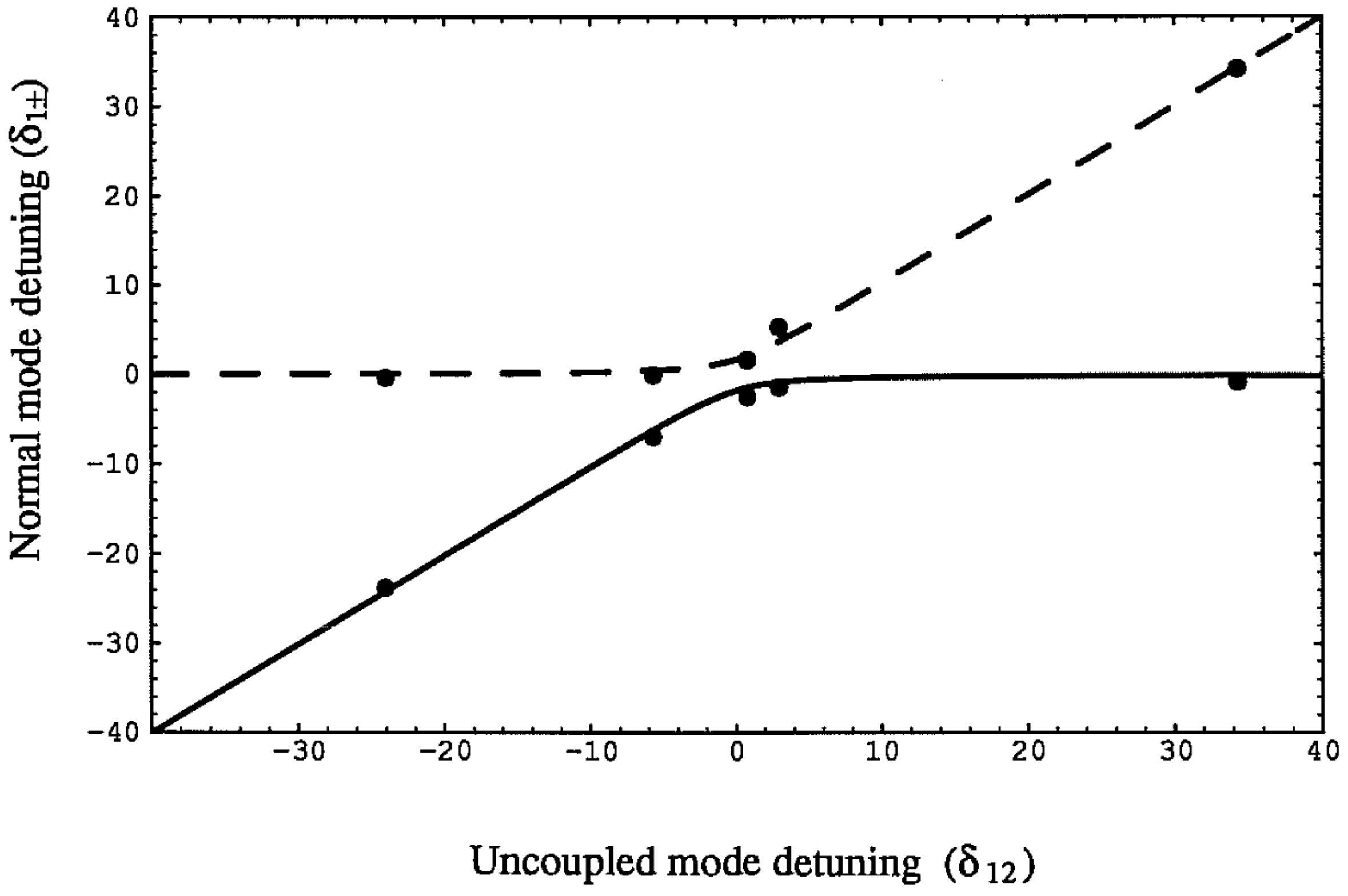

# Figure 3.14

Experimental and theoretical plots of the normal mode detuning ($\delta_{1\pm}$) versus uncoupled mode detuning ($\delta_{12}$), where experimental points are represented by the points, the positively detuned normal mode represented by the dashed line, and the negatively detuned normal mode represented by the undashed line.

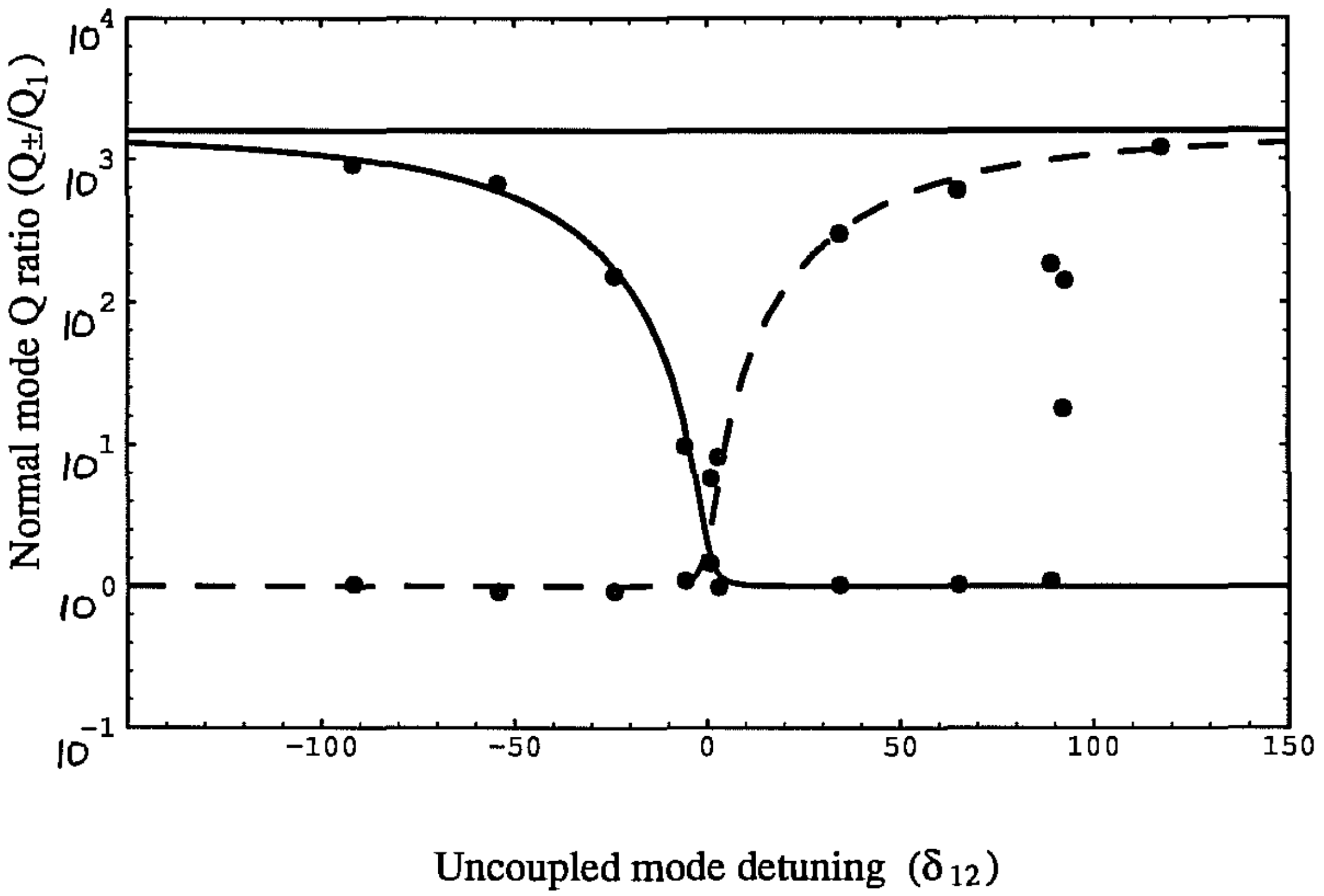

# Figure 3.15

Experimental and theoretical plots of the normal mode Q ratio ($Q_{\pm}/Q_1$), versus uncoupled mode detuning ($\delta_{12}$), where experimental points are represented by the points, the positively detuned normal mode represented by the dashed line, and the negatively detuned normal mode represented by the undashed line.

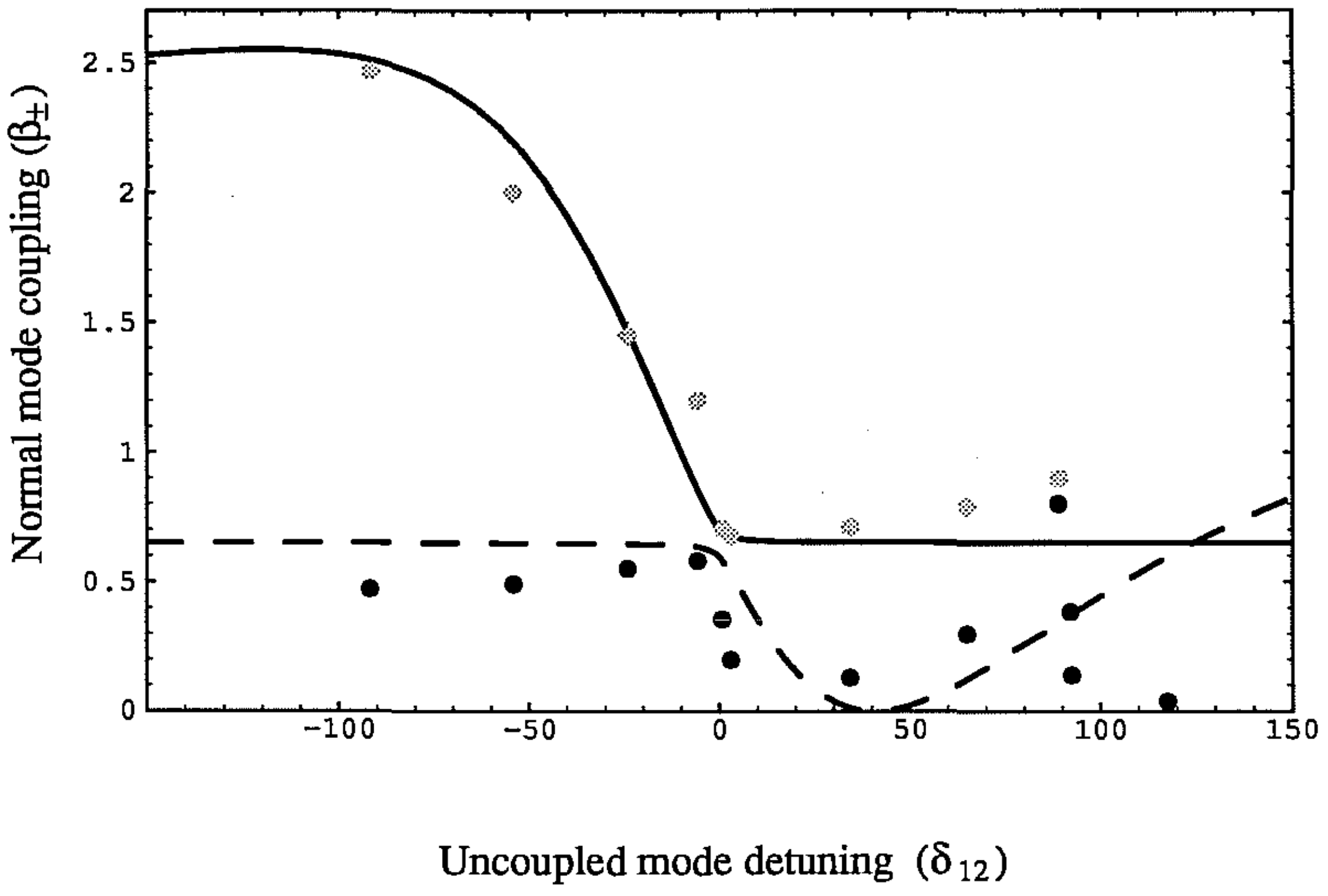

# Figure 3.16

Experimental and theoretical plots of the normal mode coupling ($\beta\pm$), versus uncoupled mode detuning ($\delta_{12}$), where experimental points are represented by the points (faded points for the negatively detuned mode), the positively detuned normal mode represented by the dashed line, and the negatively detuned normal mode represented by the undashed line.

interaction its influence was not significant, except at this point when it tuned past the high Q mode. Because the mutual reactive coupling was an order of magnitude lower during this interaction, it was less dramatic than the one analysed.

These results show decisively that the Q factors and couplings can be affected when modes are far removed from each other. The coupling more so, which is highlighted in figure 3.16, where at $\delta_{12} = \pm 150$ the high Q mode's coupling is far from the limiting detuned value of 1.9 while the Q factor has reached the limiting value (figure 3.15). This means that the spurious mode is still affecting the operational mode, with any perturbations in its frequency causing a change in the operational modes coupling. It is not until $\delta_{12} = \pm 700$ ($\pm$ 18 MHz) that the coupling of the high Q mode settles down close to the limiting value of 1.9. When creating an oscillator using an overmoded resonator, even if the Q factor is high, one must be careful that a nearby spurious mode is still not causing unwanted a.m. perturbations. If the reactance of the probe is significant ($\alpha \neq 0$) these a.m. perturbations are converted to p.m. perturbations.

## 3.7 CONCLUSION

Interactions between various modes in a Tunable Loaded Sapphire Superconducting Cavity have been studied, the main features that emerge are:

1. Reactive coupling causes modes to influence each other beyond their bandwidths, because in (3.35) the reactive factor in $D_{nm}$ is multiplied by the geometric mean of $Q_n$ and $Q_m$ .

2. Reactively coupled modes do not tune across eachother, they become hybrids of each other and eventually change identities. The reactive cross coupling can be calculated from the closest point of tuning between the two modes.

3.A reactively coupled low Q spurious modes can cause perturbations in the operational modes coupling, even when the modes are detuned far enough that the mode Q value remains unaffected.

4. Resistively coupled modes only influence each other if one can be tuned into the bandwidth of another.

5. To measure the T-SLOSC mode properties amongst a line resonance, the line resonance must be tuned away by a line stretcher, so that the line resonance does not affect the cavity resonance through a mutual resistance in the probe.

In terms of the uncoupled parameters, analytical solutions for the normal mode frequencies, Q factors and couplings of two reactively coupled normal modes have been derived. Solutions are expressed using normalized parameters so they may be applied to any interacting system. Equations were verified experimentally, and can be used to analyse the effect of a spurious mode on a resonator's operational mode.

This chapter has successfully modelled some of the complex phenomena that occurs in a tunable multimode cavity. In spite of the complex interactions, Q factors greater than $10^8$ can be achieved, with a useful tuning range in the order of tens of MHz.

# CHAPTER 4

# OSCILLATOR AND COMPONENT PHASE NOISE

## 4.1 INTRODUCTION

Previous chapters have described the T-SLOSC and SLOSC as high Q resonators. This chapter describes how they were combined with X-band amplifiers and other components to create ultra-low noise oscillators.

Ultra-low noise high stability sapphire oscillators are under development by research groups at Jet Propulsion Laboratories (JPL) (Wang and Dick, 1992), Westinghouse (Driscoll and Weinert, 1992), University of Maryland (Zaki and Liang, 1992), University of Western Australia (UWA) (Giles et al, 1989; Tobar and Blair 1993), Dupont (Shen et al, 1992), East-Siberian Research Institute and the Moscow Power Engineering Institute (Bun'kov et al 1990). A sapphire dielectric resonator (SDR) is the key element due to the crystal's low loss tangent at microwave frequencies. The properties of these resonators were discussed in detail in previous chapters. The high Q resonator is either configured in the feedback path of a loop oscillator or used as a stabilisation element for an external voltage controlled oscillator (ie. GUNN or SAW oscillator). For a loop oscillator the frequency is determined by the resonant frequency of the resonator in the feed back path, while the stability and noise performance is determined by its Q factor (Leeson 1966). To stabilise an external voltage controlled oscillator it must be tuned to the frequency of the resonator, with sensing circuitry returning an error signal to its voltage control, thereby locking the frequency of the oscillator to that of the resonator (Stein and Turneaure, 1975). Higher Q resonators give a greater stabilisation sensitivity due to the narrower line width of the stabilisation resonator.

The two ways to characterise the operation of an oscillator are by either measuring the Allan variance as a function of integration time or by measuring the phase noise as a function of offset frequency (Allan, 1966), (Barnes et al., 1971), (also see introduction). The two are related by complicated integral transform equations (Rutman, 1974). The best frequency stability in X-band SDR oscillators have been demonstrated at the University of Western Australia (UWA) (Giles et al, 1989) and the Jet Propulsion Laboratories (JPL) (Dick and Wang, 1991), where both operate with state of the art frequency stability (square root Allan variance) better than $10^{-14}$ for integration times from 1 to 300 seconds. These oscillators have frequency stabilisation circuitry built in, reducing the near carrier phase noise below 10 Hz offset at the expense of adding noise above 10 Hz offset. The lowest phase noise above 10 Hz offset has been measured in the free running SLOSC oscillators at UWA (Tobar and Blair 1993), these results are presented in section 4.5. The central aim of this work was to build a low noise pump oscillator for the re-entrant cavity parametric transducer attached to the resonant bar gravitational wave antenna at UWA. The signal detection is at 710 Hz with a bandwidth of approximately 20 Hz. At this offset frequency a free running loop oscillator was preferable to a stabilised oscillator. It is possible to build a servo system that will cancel phase noise at 710 Hz (Mann and Blair, 1983), however in chapter 7 it will be shown that the phase noise of the free running loop oscillator is sufficient.

It is almost impossible for a SLOSC oscillator and a re-entrant cavity to be built with exactly the same frequency. Therefore the pump oscillator must be designed as a tunable oscillator. Thus the T-SLOSC cavity resonator as discussed in chapter 1 was developed as the high Q element in a free running loop oscillator. Vibrational noise that causes axial movement of the tuning disc with respect to the resonator will add to the residual phase noise of the cavity. In general a mode which exhibits a large tuning range will have greater sensitivity to tuning and hence be more susceptible to vibrations of the tuning mechanism. Therefore this component of residual phase noise can be reduced at the expense of the cavity tuning range. A vibration isolation system was built for the T-SLOSC resonator to help limit the residual noise added by vibrations. In section 4.4 I show that the oscillator can operate with a tuning range of 5 MHz or less before the phase noise becomes seriously degraded.

A tuning range of 5 MHz has not been sufficient for the oscillator's inclusion into the re-entrant cavity transducer. Therefore a fixed frequency SLOSC was mixed with a HP 8662A synthesiser to create a low noise oscillator with a 1 GHz tuning range. The phase noise of this system is presented in section 4.5.

## 4.2 PHASE NOISE MODEL

A phase modulation of an external signal is caused by the right combination of an upper and a lower sideband. A common representation of phase noise is $S_\phi(f)$, which is defined as the double sideband (DSB) phase noise because its level is measured relative to both the upper and lower noise sidebands. It is also common to represent phase noise relative to one sideband as $\mathcal{L}(f)$, which is defined as the single sideband (SSB) phase noise. The single sideband and double sideband noise are related by; $\mathcal{L}(f)=S_\phi(f)-3$ [dBc/Hz]. The basic phase noise model (Leeson, 1966) relevant for a loop oscillator is shown in figure 4.1a. Here $S_{\phi\,a}$ is the phase noise introduced by the active components, $S_{\phi\,c}$ the oscillator phase noise from the non filtered port (port 2), and $S_{\phi\,osc}$ the oscillator phase

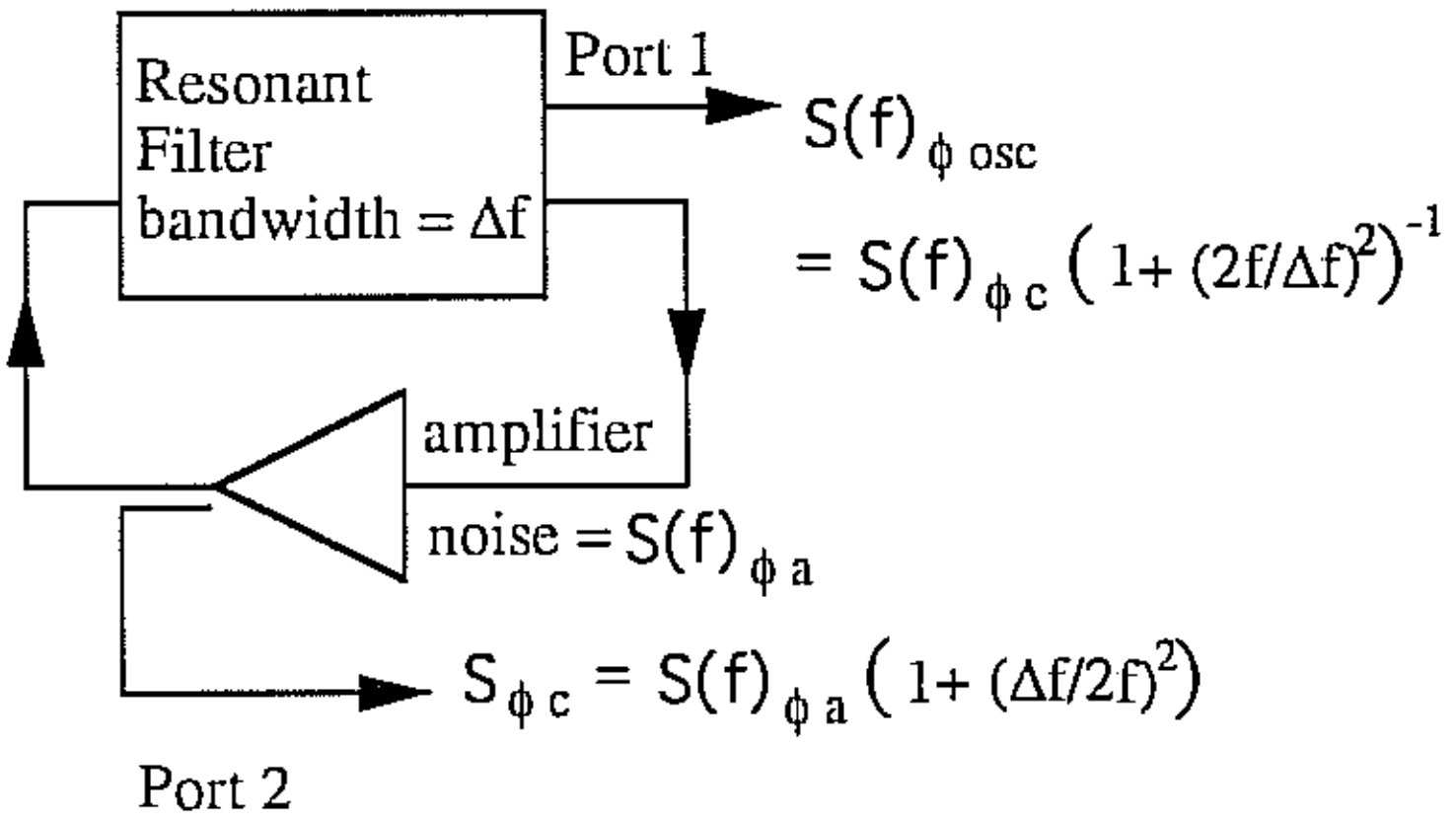

(4.1 a)

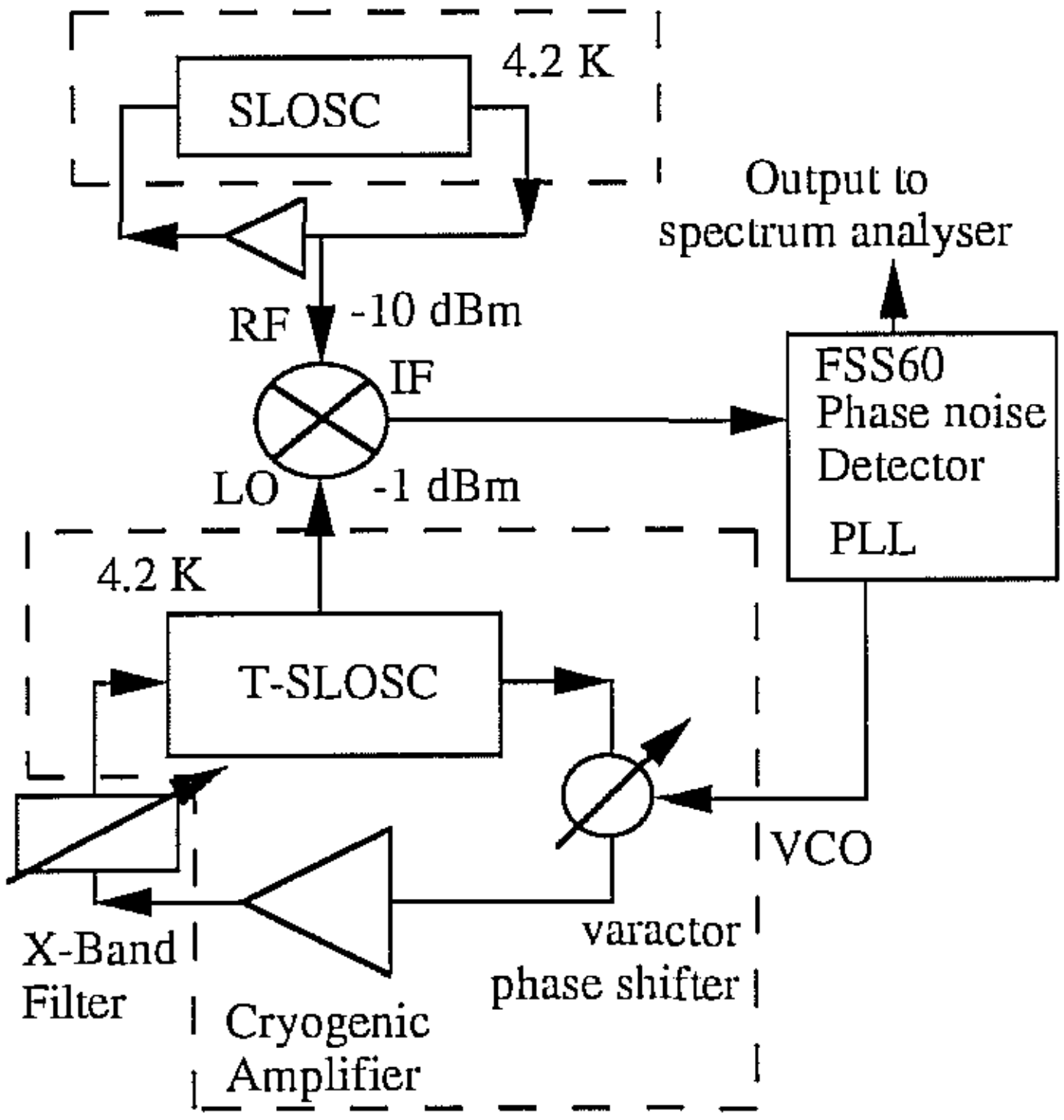

(4.1 b)

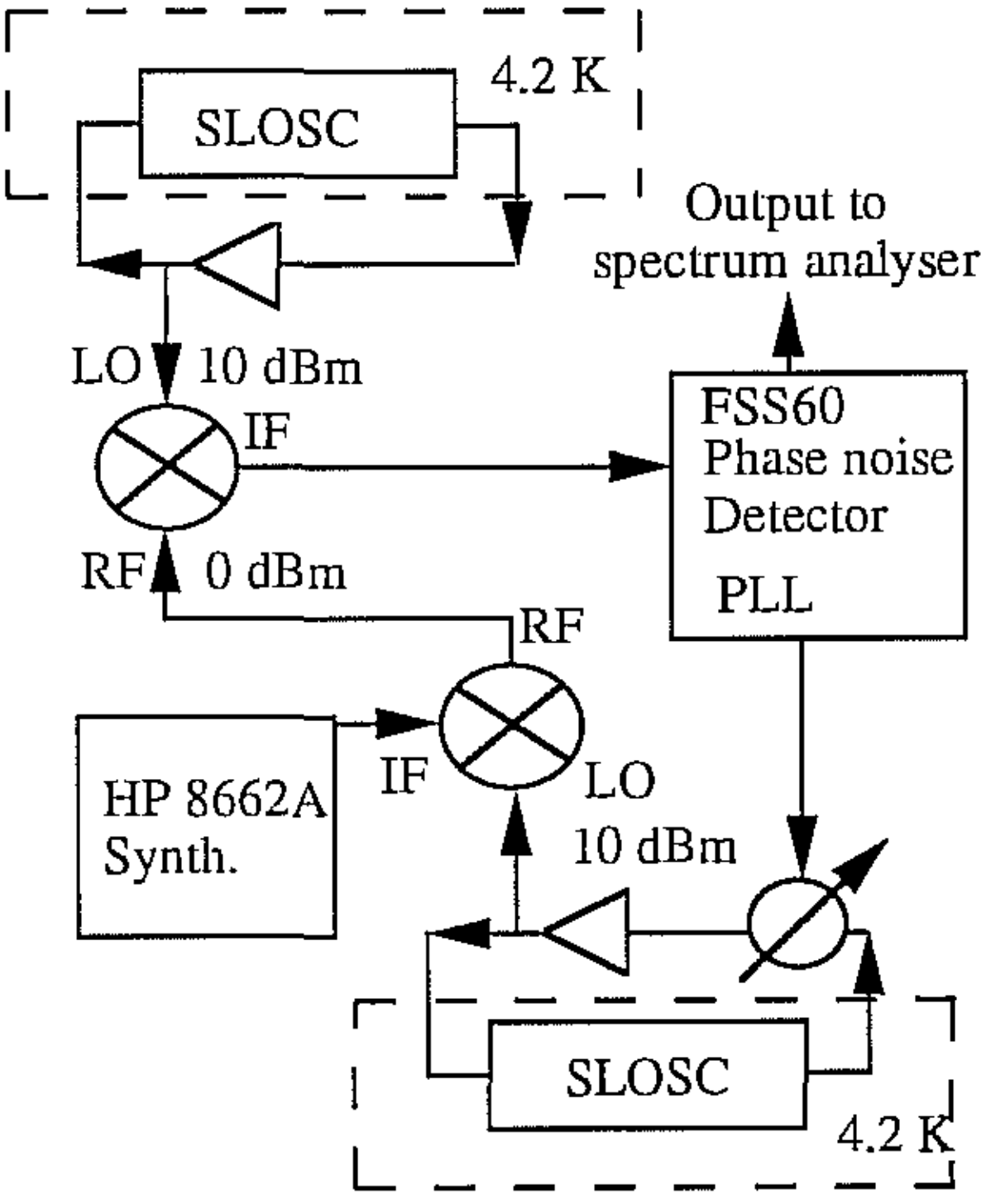

(4.1 c)

# Figure 4.1

4.1a shows a resonator of bandwidth Δf configured in a loop oscillator, along with the phase noise dependency at output ports 1 and 2 as predicted by Leeson's model. 4.1b and c illustrate the schematics of the SLOSC oscillator phase noise measurement circuits that were implemented, 4.1b shows the T-SLOSC oscillator quadrature locked to a fixed frequency SLOSC oscillator, while 4.1c shows two fixed frequency SLOSC oscillators quadrature locked with a frequency synthesiser used to make up the frequency difference.

noise from the filtered port (port 1). To obtain the lowest phase noise the oscillator signal should be extracted after the high Q resonator from port 1. The phase noise from this port is filtered for offset frequencies above the cavity half bandwidth, at the expense of a decrease in the available output power due to attenuation by the resonator. In the cryogenic SLOSC resonators the output power will also be attenuated by long cryogenic coaxial cables (~1.5 meters) running from the cryogenic resonator to the room temperature output. An amplifier cannot be used to increase output power as this will add phase noise which will not be filtered out by the cavity.

At the low phase noise levels achievable using sapphire resonators, extreme care must be taken when measuring phase noise. Two approaches have been used: a) Phase locking a T-SLOSC oscillator in quadrature to a fixed frequency SLOSC oscillator. This requires a very difficult tuning operation (Tobar and Blair, Sept. 1991): b) Phase locking the beat frequency of two SLOSC oscillators to a low noise synthesiser. Schematics of these alternative measurement systems are shown in figure 4.1b and 4.1c. The former is limited by vibrations of the tuning mechanism, while the latter is limited by phase noise in the synthesiser. Both measurements are presented in this chapter.

The oscillator phase noise is dependent on the phase noise of its components. In the following section we describe residual phase noise due to vibrations in the T-SLOSC resonator. This is followed by measurements of the phase noise performance of two GaAs amplifiers and a varactor phase shifter, which were assessed for use in an all-cryogenic SLOSC oscillator.

## 4.3 COMPONENT PHASE NOISE

### 4.3.1 Cavity residual noise

The residual phase noise of a fixed frequency SDR cavity has been measured to be less than -150 dBc/Hz at 1 kHz offset frequency at room temperature (Tobar et al, 1993). The residual phase noise for these types of resonators are insignificant. However this is not true for the T-SLOSC resonator; vibrational noise that causes axial movement of the tuning disk with respect to the resonator will add to the residual phase noise of this type of cavity.

To analyse the effects of vibration on the T-SLOSC resonator the circuit in figure 4.2a was constructed. This bridge circuit had a phase shifter in the local oscillator arm and the T-SLOSC in the RF arm, with the IF port of the mixer read directly by an audio spectrum analyser. The T-SLOSC was placed inside a cryogenic dewar inside a vacuum can submerged in a 4.2 K liquid helium bath. To provide some vibration isolation the T-

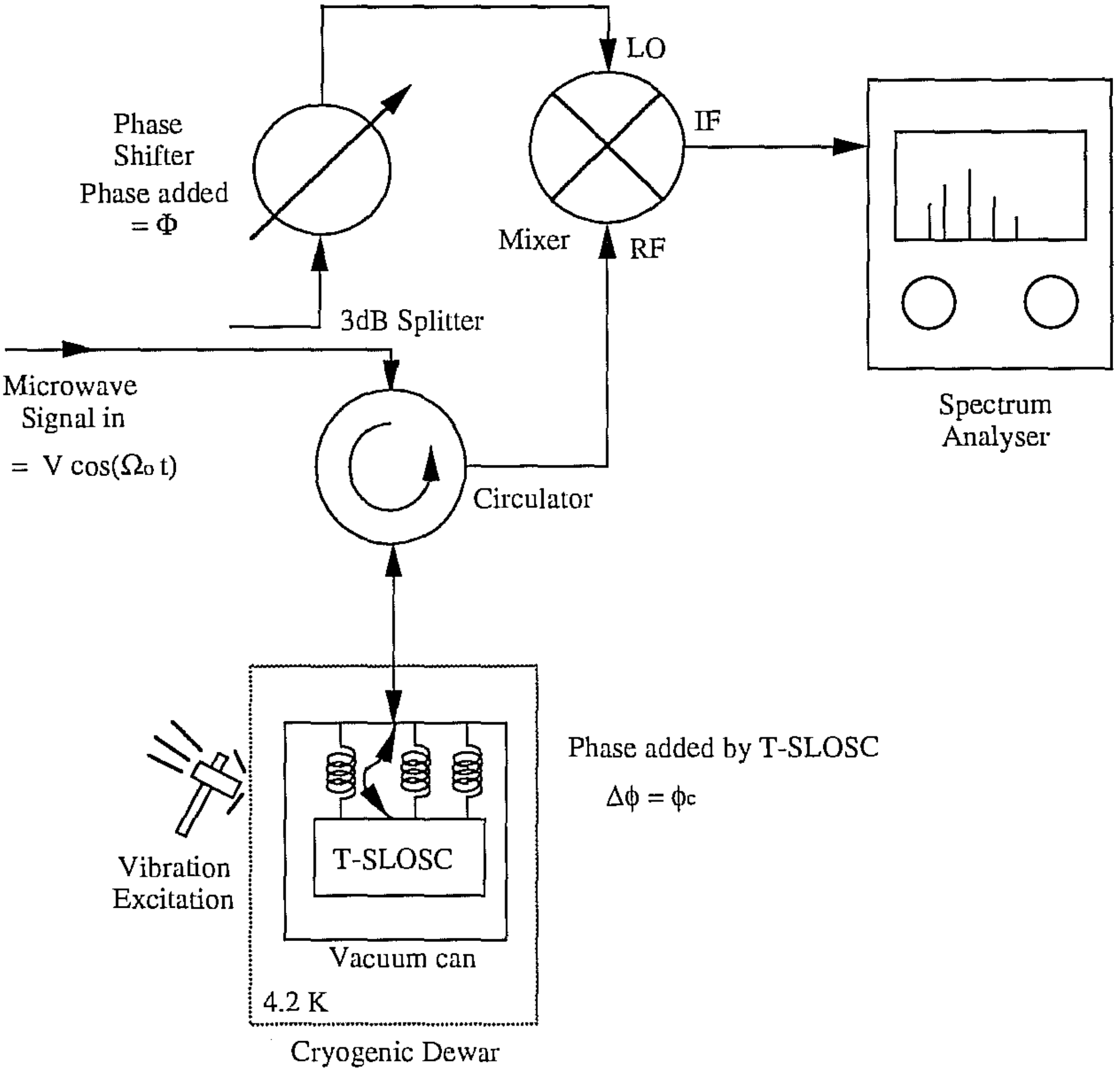

# Figure 4.2a

Schematic of the experimental set up implemented to measure the effect of a vibrational impulse inflicted upon the T-SLOSC. This consisted of a bridge circuit with the T-SLOSC in the RF arm and a phase shifter in the local oscillator arm, with the IF port of the mixer read by an audio spectrum analyser. The T-SLOSC was placed inside a cryogenic dewar inside a vacuum can submerged in a 4.2 K liquid helium bath. To provide some vibration isolation the T-SLOSC was supported by three soft springs from the top of the vacuum can.

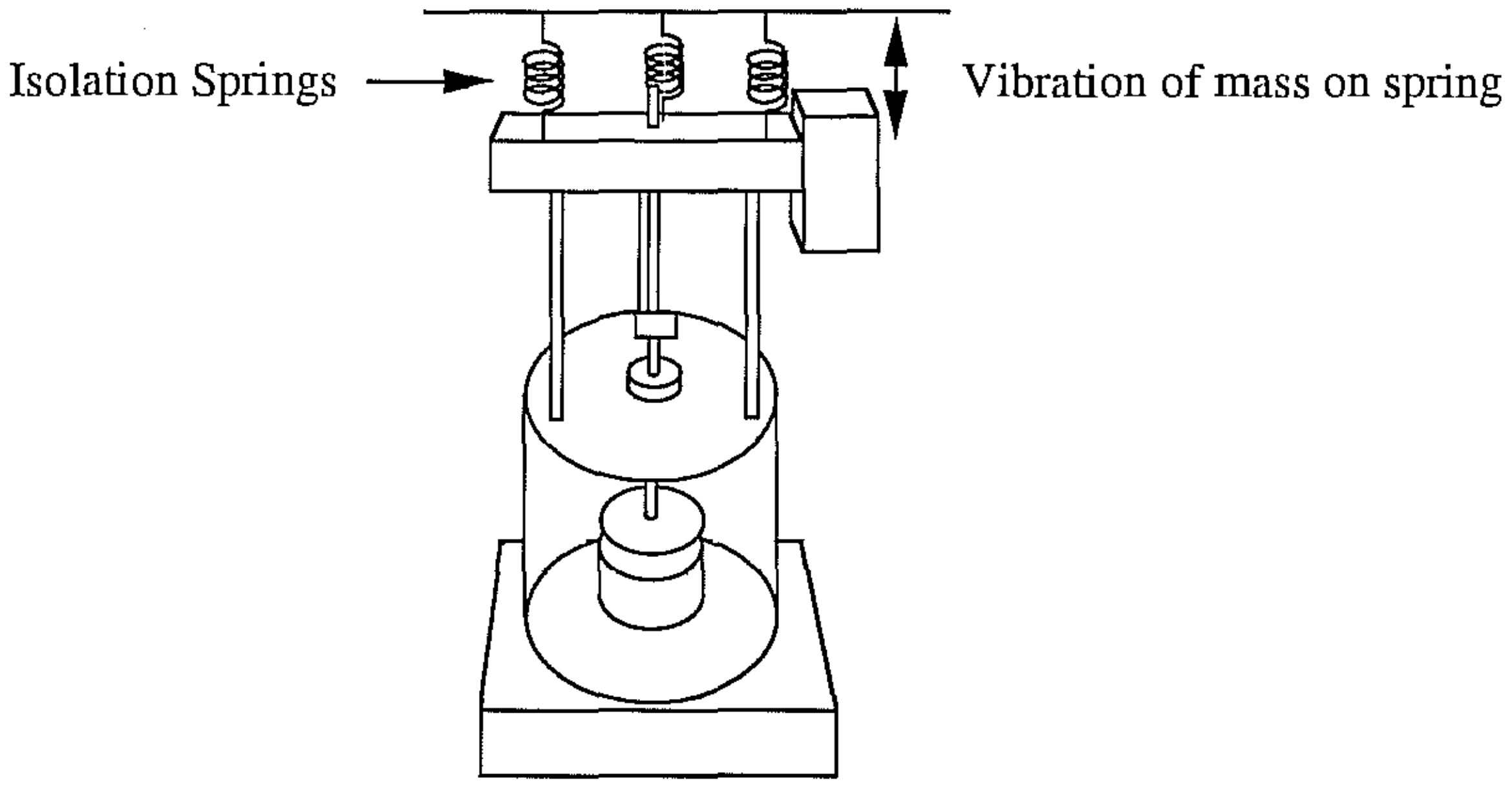

(4.2 b)

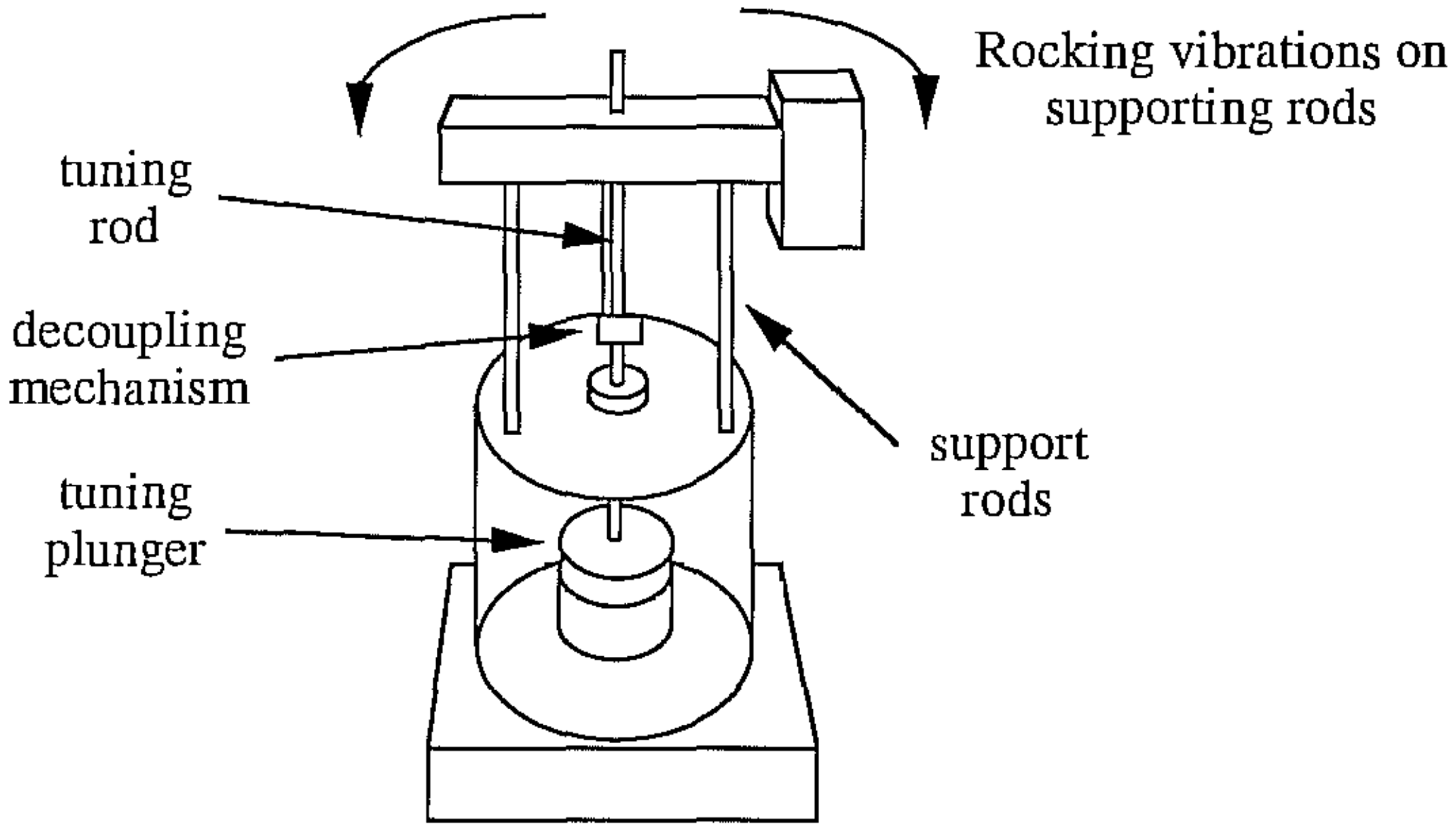

(4.2 c)

# Figure 4.2 b&c

Schematic of the low frequency vibrational modes in the T-SLOSC support structure. Figure 4.2b shows the resonant motion of the T-SLOSC (mass = 6kg) on soft isolation springs at a frequency of 5 Hz. Figure 4.2c shows the rocking resonant motion of the suport rods. When the tuning plunger was free the frequency occurred at 119 Hz at all tuning states. The exception was when the plunger was tuned close to the sapphire crystal and the decoupling mechanism was jammed against the top of the cavity. This caused the tuning rod to act as another support rod increasing the effective spring constant which raised the frequency to 144 Hz.

SLOSC was supported by three soft springs from the top of the vacuum can. The signal at the local oscillator port is proportional to $V\cos(\Omega_0 t + \Phi)$ where $\Phi$ is the phase added by the phase shifter. The signal at the RF port is proportional to $|\Gamma| V\cos(\Omega_0 t + \phi_c)$ where $\Gamma$ is the T-SLOSC reflection coefficient given by (1.3), and $\phi_c = \angle\Gamma$ the phase added by the T-SLOSC. Therefore at the IF port the mixed signal $\Theta_v$ is proportional to;

$$\Theta_v \propto |\Gamma| V^2 \left[\cos(2\Omega_0 t + \Phi + \phi_c) + \cos(\Phi + \phi_c)\right] \qquad (4.1).$$

On the audio spectrum analyser only the D.C. term was observed as the $2\Omega_0$ term has a frequency of about 20 GHz and was filtered by the spectrum analyser.

If the incident microwave signal has a frequency equal to the T-SLOSC resonant frequency then $\phi_c$ is zero. If the T-SLOSC is subject to vibrations that cause a small change in its resonant frequency then $\phi_c$ will vary, causing a modulation of the microwave signal. If the phase shifter is set in quadrature with $\Phi = \pi/2$ the output is proportional to $|\Gamma| V\sin(\phi_c)$ and if $\phi_c$ is small the output is then proportional to $\phi_c$. This measurement is essentially a residual phase noise measurement of the T-SLOSC resonator. If the phase shifter is set in phase with $\Phi = 0$ the output is proportional to $|\Gamma| V\cos(\phi_c)$ and if $\phi_c$ is small the output is then proportional to $|\Gamma|$. This measurement is essentially a residual amplitude noise measurement of the T-SLOSC resonator. To analyse the vibrational effects a large vibrational impulse was applied to the T-SLOSC. In this case it is assumed that $\phi_c$ consists of the sum of many harmonic components such that;

$$\phi_c = \sum_n C_n \cos(\omega_n t + \phi_n) \qquad (4.2).$$

Substituting (4.2) into (4.1) and ignoring the $2\Omega_0$ term, when $\Phi = \pi/2$ (4.1) becomes;

$$\Theta_v \propto \sum_{m=0}^{\infty} (-1)^m \, 2J_{2m+1} \sum_n C_n \cos\left[(2m+1)(\omega_n t + \phi_n)\right] \qquad (4.3),$$

when $\Phi = 0$ (4.1) becomes;

$$\Theta_v \propto \sum_{m=0}^{\infty} (-1)^m \, 2J_{2m} \sum_n C_n \cos\left[(2m)(\omega_n t + \phi_n)\right] \qquad (4.4).$$

Due to the non-linearity of the mixer, (4.3) and (4.4) show that when the bridge circuit is either phase or amplitude sensitive a mechanical vibrational mode will have all its odd or even harmonics produced respectively. Harmonics that are upconverted mechanically will exhibit all harmonics independent of the phase of the bridge circuit detector, as they are real mechanical vibrations and not upconversions due to the non-linear output of the bridge circuit. This determines a method of investigating whether observed harmonics are due to mechanical upconversion or the non-linear output.

The two most sensitive modes to vibrational excitations are shown in figure 4.2b and c, and are in the T-SLOSC support structure. Figure 4.2b shows the T-SLOSC held to the top of the vacuum can by soft springs. The combined spring constant of these springs was ~ 150 N/m and the mass of the T-SLOSC was ~ 6 kg. Therefore this resonance motion occurred at 5 Hz. The tuning plunger was supported by 3 support rods, as shown in figure 4.2. A rocking resonance of the tuning mechanism on these support rods occurred at 119 Hz at all tuning states of the plunger. Figure 4.2c shows the mechanism of this resonant mode. However when the tuning plunger was tuned close to the sapphire crystal the decoupling mechanism would jam against the top of the cavity. Thus the plunger would act as an extra support rod which increased the effective spring constant of this motion and hence the resonant frequency from 119 Hz to 144 Hz. Figure 4.3a and 4.3b show the excited vibrational spectrum of these two situations. When setting the phase shifter in figure 4.2a such that $\Phi = \pi/2$ or 0 the excited spectrum did not change. This implies that the upconversions observed are mechanical in nature. They were probably due to friction between the tuning rod and where it enters the cavity. As the tuning rod rocks from side to side this friction would cause a violin type effect that can cause upconversions. Figure 4.4 shows the steady state amplitude of the mass - spring mode at 5.16 Hz. This amplitude would vary quite a bit depending on the level of background excitation in the laboratory. There were no friction mechanisms to cause mechanical upconversions in this resonant mode. When a forced excitation was applied upconversions were present due to the non-linear electrical output. This is verified by figure 4.5a & b, which shows the excited spectra when the phase added by the phase shifter was $\Phi = \pi/2$ and 0 respectively. As predicted odd harmonics given by (4.3) are present in figure 4.5a and even harmonics given by (4.4) are present in figure 4.5b. At very high excitations the 5 Hz mode would modulate the support modes as shown in figure 4.6.

The soft spring vibration isolation was quite inadequate: when the tuning state of the T-SLOSC was in the high tuning coefficient regime, the high Q electromagnetic resonances were unstable to many times their frequency bandwith. This caused a great number of problems with low frequency jitter and did not allow phase locking during phase noise measurements. The excessive low frequency vibrations were caused by the high Q of the 5 Hz spring-mass resonance. To remove this problem one could either actively damp the resonance or, provided that sufficient isolation could be obtained at the frequency of interest, raise the resonant frequency of the isolator to a frequency at which the lower ambient level of vibrations would not result in a large motion. The latter route was chosen, and a more rigid vibration isolation system was designed.

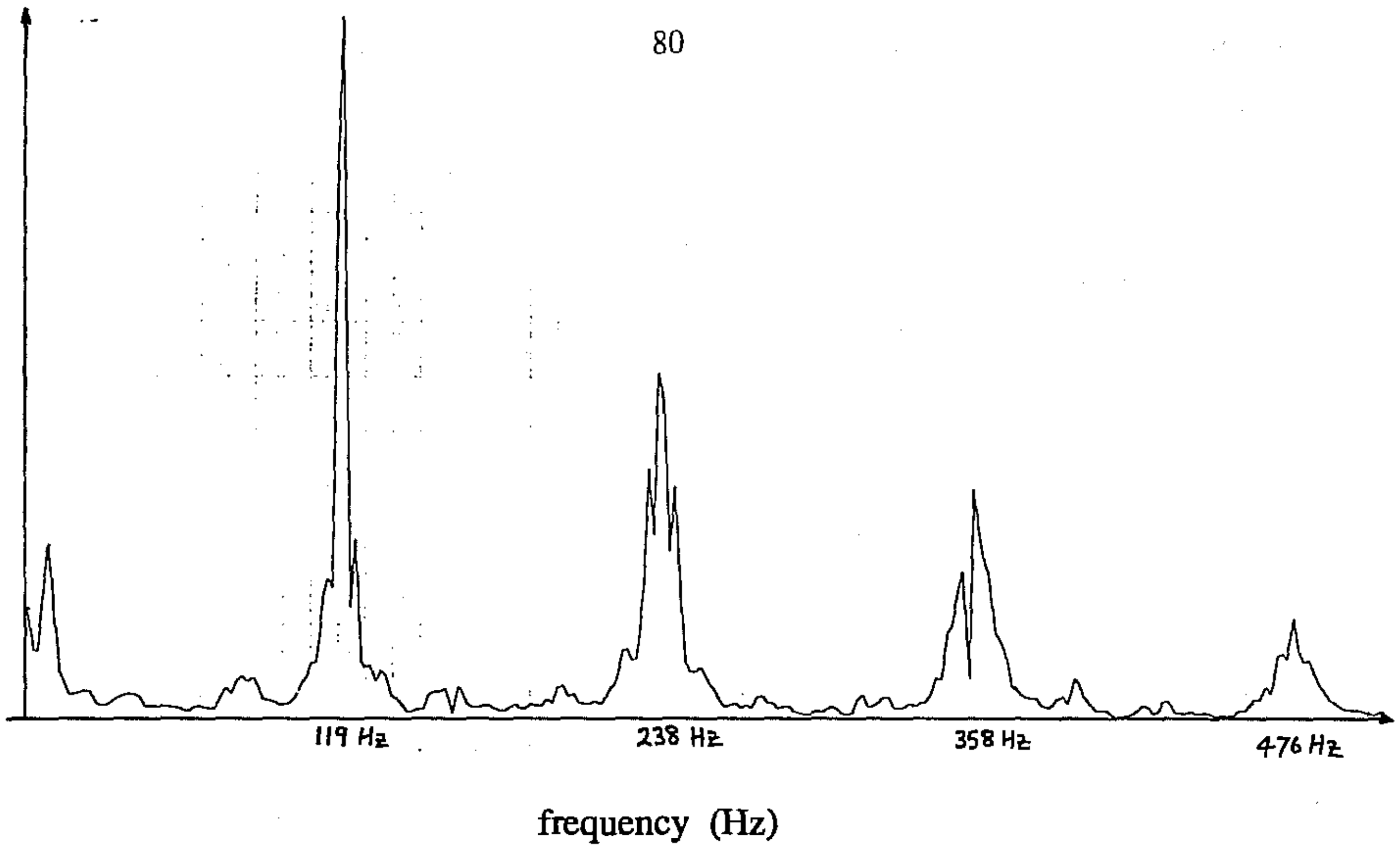

(4.3 a)

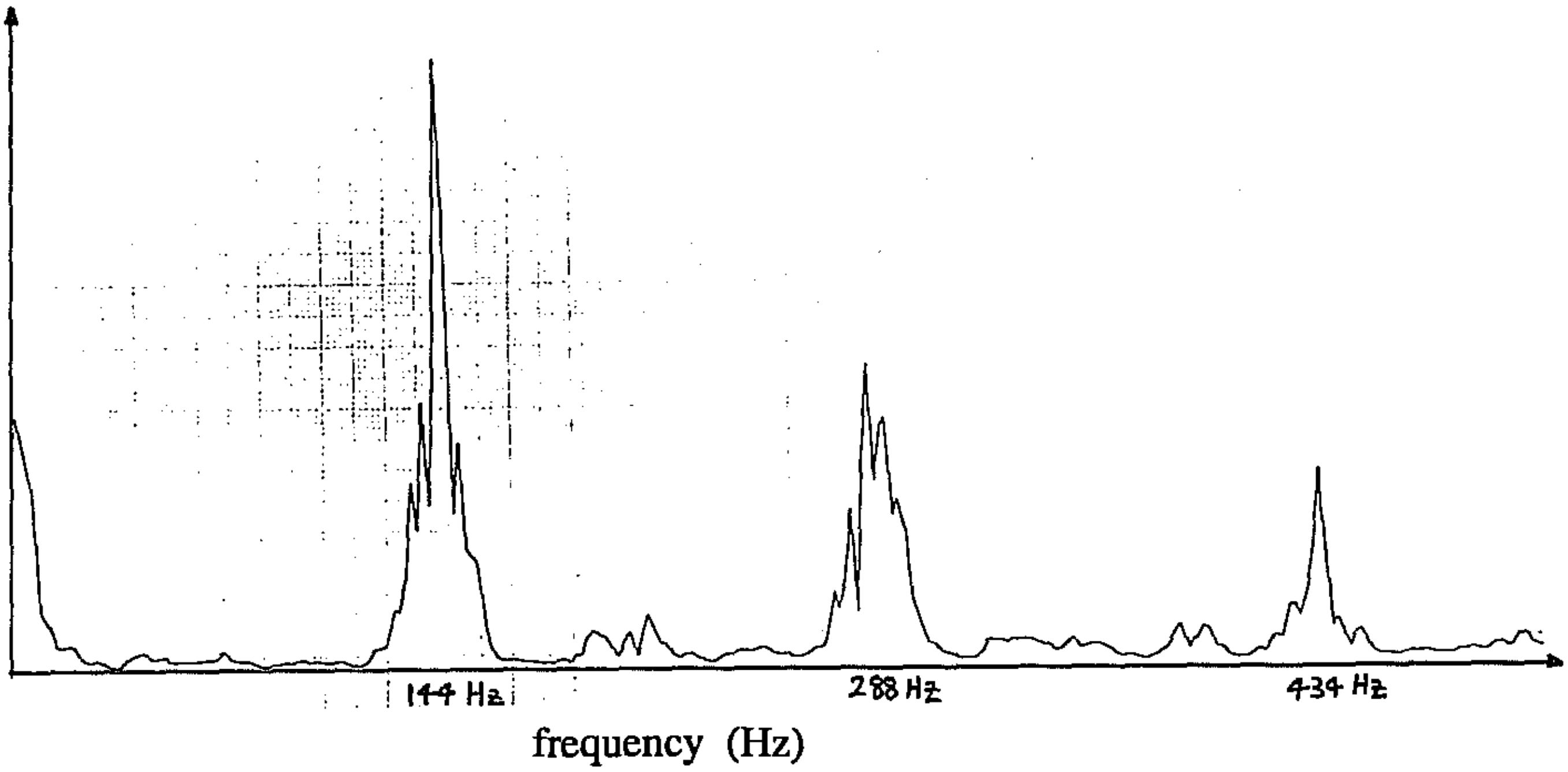

(4.3 b)

# Figure 4.3

Excited spectrum of modes created by a rocking resonance of the support structure shown in figure 4.2b. At high amplitude excitations mechanical upconversion are present due to the friction between the tuning rod and where it enters the cavity.

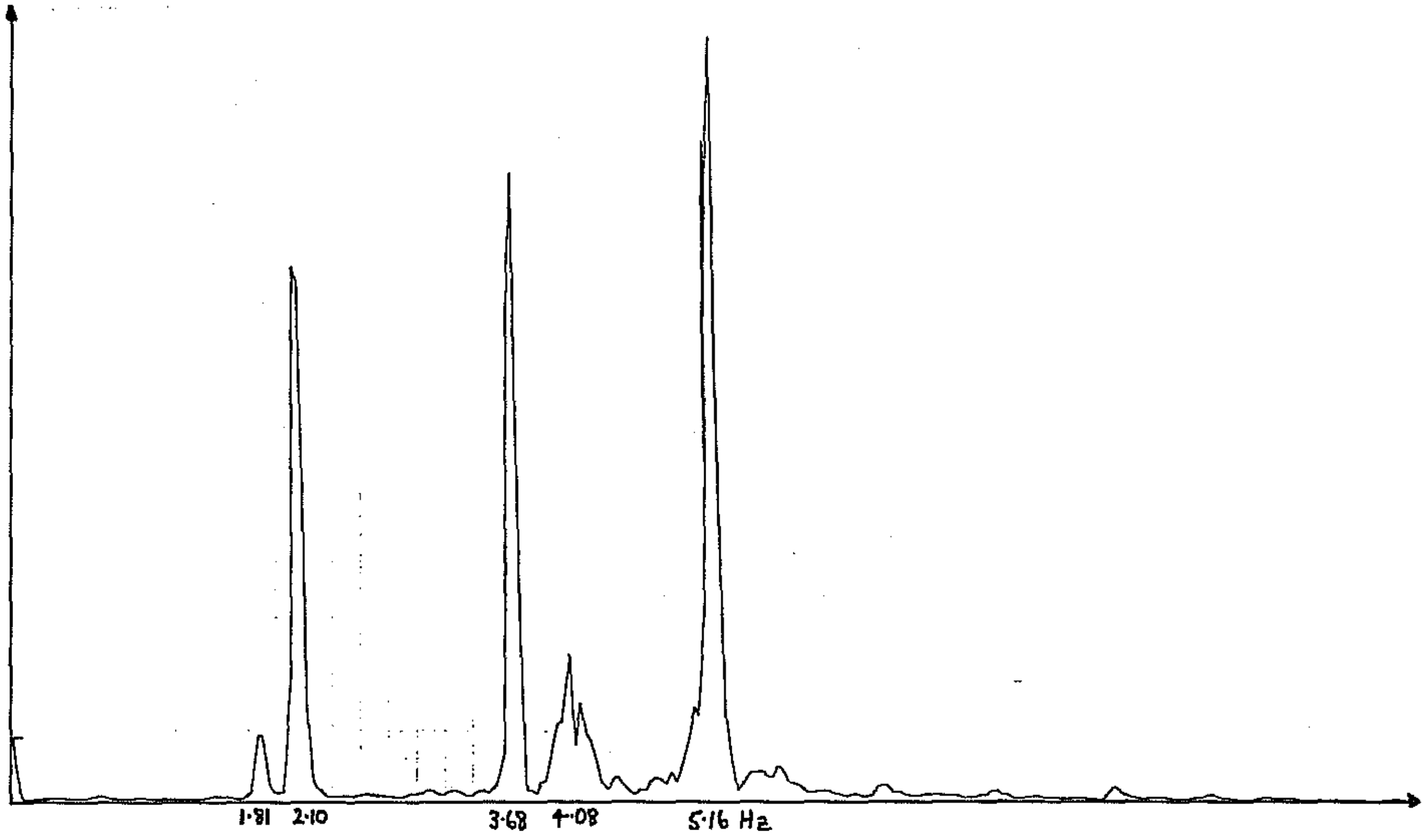

Frequency (Hz)

# Figure 4.4

Steady state spectrum of low frequency T-SLOSC modes between 0 to 10 Hz. The resonant frequency of the T-SLOSC mass on the isolation springs was 5.16 Hz . The amplitude of this mode was quite large causing the resonant frequencies of the T-SLOSC resonator to be unstable at certain tuning states. This effect was enhanced as the tuning coefficient of the T-SLOSC was increased.

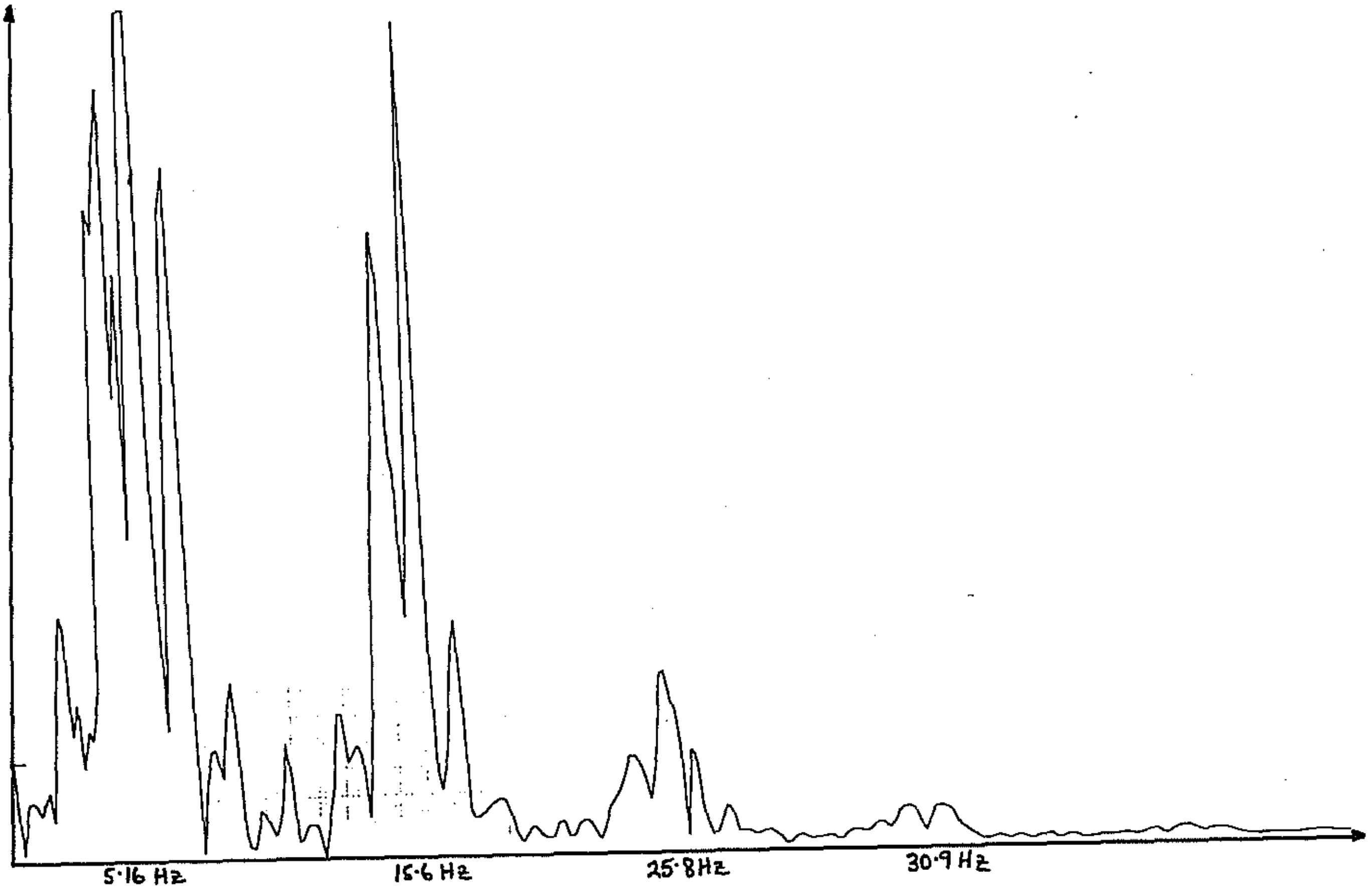

frequency (Hz)

(4.5 a)

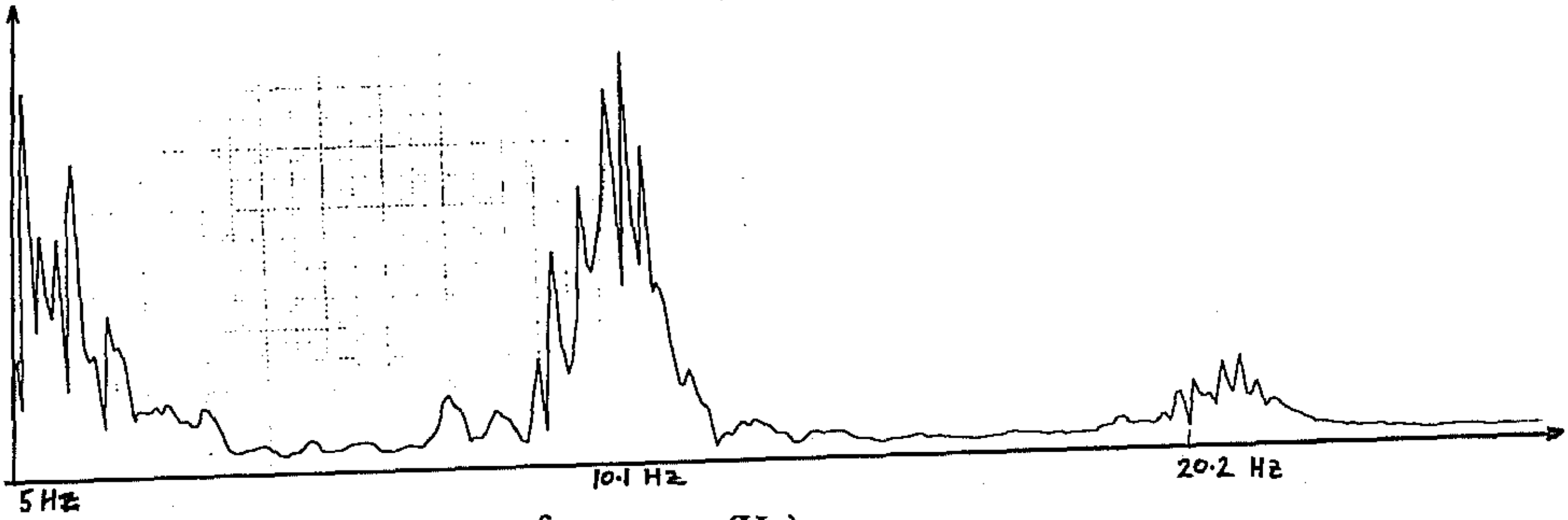

frequency (Hz)

(4.5 b)

# Figure 4.5

Excited spectrum of modes created by the resonance of the T-SLOSC on the isolation springs shown in figure 4.2a. At high amplitude excitations upconversions were present due to the non-linear electrical output, no mechanical upconversions were present. 4.4a shows the spectrum when the phase added by the phase shifter in figure 4.1 was $\Phi = \pi/2$. Odd harmonics given by (4.3) were present. 4.4b shows the spectrum when $\Phi = 0$, Even harmonics given by (4.4) were present.

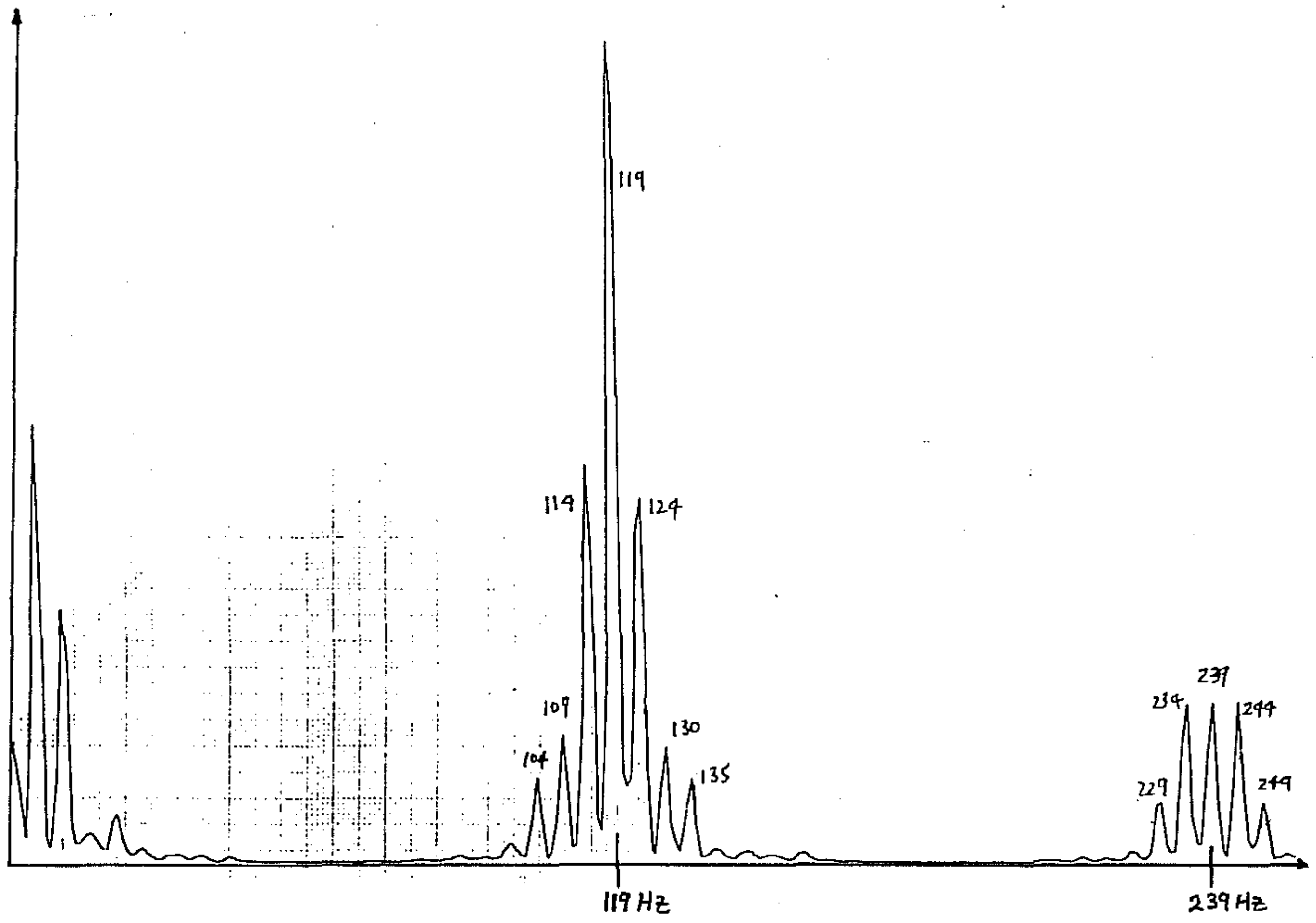

frequency (Hz)

# Figure 4.6

At very high levels of excitations the 119 Hz support resonance was modulated by the 5 Hz spring - mass resonance.

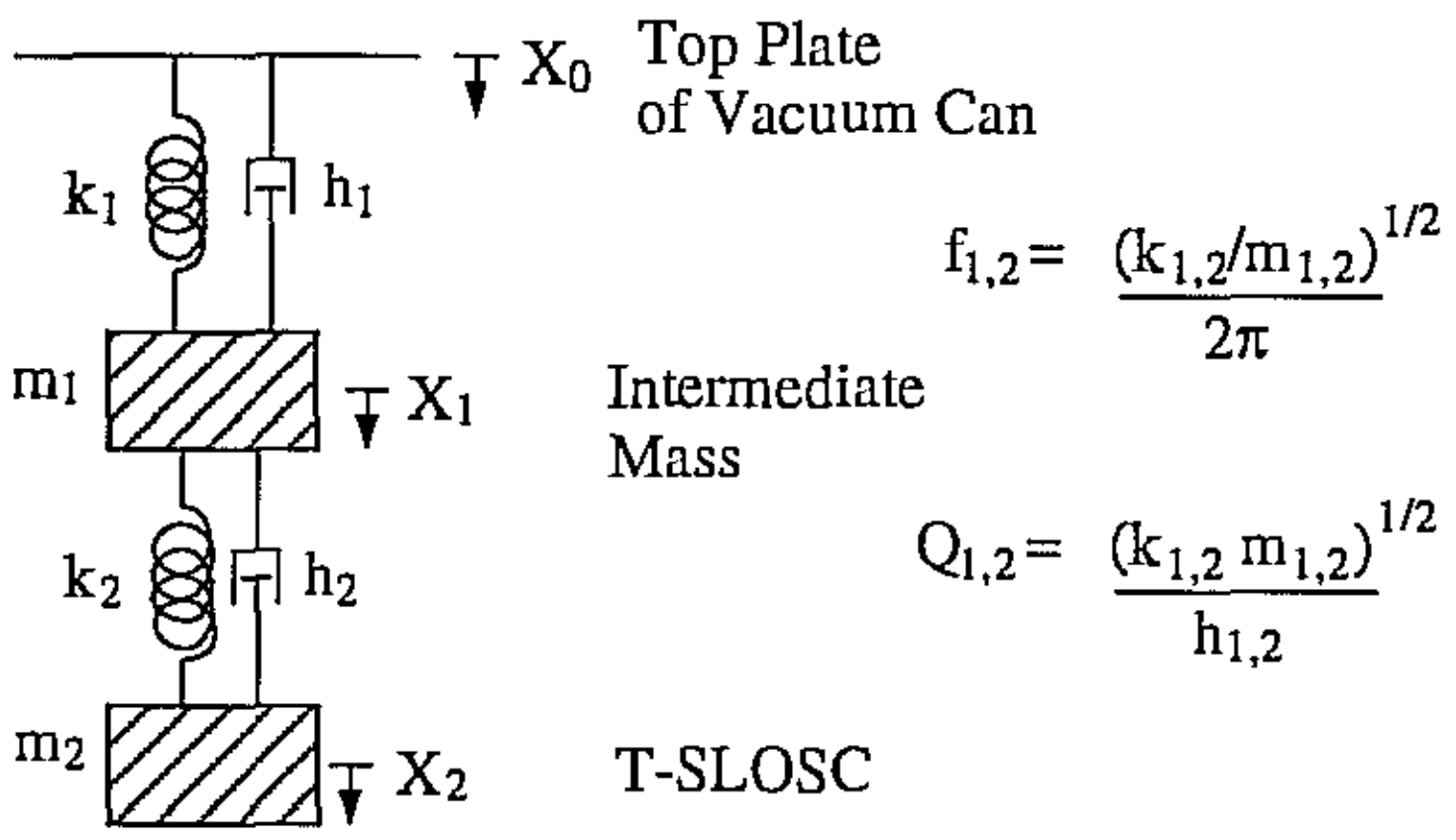

(4.7a)

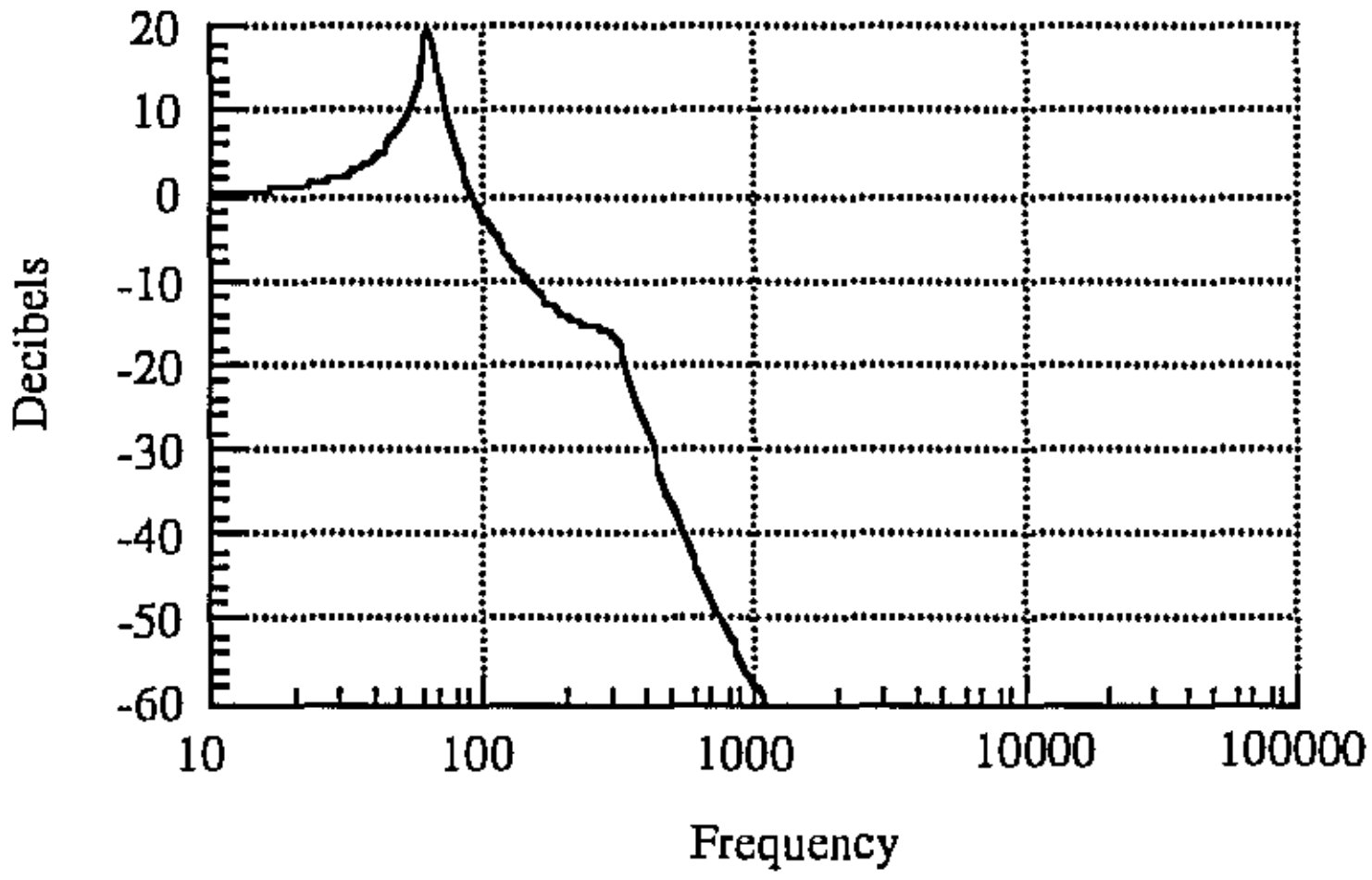

(4.7b)

# Figure 4.7

4.7a shows the one dimensional model of the vibrational isolation system from the top plate of the vacuum can to the T-SLOSC. 4.7b shows the predicted transfer function $X_2/X_0$. Subscript 1 refers to the intermediate mass and 2 to the T-SLOSC, where $f_{1,2}$ are the mechanical resonant frequencies, $Q_{1,2}$ are the mechanical Q values $k_{1,2}$ the mechanical spring constants $h_{1,2}$ the mechanical dissapation constants, $X_{1,2}$ the displacements and $m_{1,2}$ the masses.

A two mode vibration isolation system was designed to eliminate the excessive low frequency vibrations, and still give a significant attenuation at 700 Hz. It comprised of an intermediate mass attached via leaf springs to the top plate of the vacuum can, and the T-SLOSC in the same way attached to the intermediate mass. Vibrations cause movement of the tuning plunger, probes and walls with respect to the sapphire resonator. These vibrations modulate the loop oscillator signal and manifest as phase noise. A one dimensional model of the isolation system designed is shown in figure 4.7a. The transfer function $X_2 / X_0$, of this system has been modelled and is shown in figure 4.7b. The intermediate mass and T-SLOSC mass were measured to be; $m_1 = 2.9$ kg and $m_2 = 5.95$ kg respectively. The spring constant of the leaf spring were designed to give; $f_1 = 117$ Hz and $f_2 = 167$ Hz. The transfer function in figure 4.8 was calculated with $Q_{1,2} = 5$. Thus the vibration isolation system was stiff enough to limit low frequency vibrations while still theoretically achieving 50 dB attenuation at 700 Hz. Experimentally over 30 dB of isolation was measured above a noise floor (see figure 4.21). If there were any internal resonances in the structure within 2 kHz, 50 dB may not be achieved at 700 Hz (Veitch, 1991).

### 4.3.2 Gallium arsenide phase and amplitude components

The single side band phase noise characteristics of an electronic varactor phase shifter and two cryogenic amplifiers, Miteq AMF-8012-CRYO and Miteq AFS3-4K, are presented in this section. Measurements were obtained by mixing in quadrature two arms of a phase bridge at 9.7 GHz. One arm contained the device under test (DUT) mounted in a vacuum sealed chamber inside a cryogenic dewar, the device was heat sunk and mounted on the vibration isolation system designed for the T-SLOSC. The other arm included a phase shifter to set the quadrature as shown in figure 4.8. Phase noise was measured at room temperature, 77 K and 4.2 K.

#### 4.3.2.1 GaAs varactor phase shifter

The GaAs varactor phase shifter was required to create a voltage controlled oscillator (VCO) from a SLOSC loop oscillator, and set the oscillator's phase condition. To operate the phase shifter in a VCO a low noise voltage adder was built with a battery driven offset voltage to set the loop oscillator phase condition. The error voltage from the phase lock loop was then added to the d.c. voltage to keep the oscillator in phase lock with another.

This phase shifter was specified to operate at room temperature with a 0 to 30 volt bias, supplying at least 0 to 400 degrees phase shift across X-band. At 290 K and 77 K the phase sensitivity did not change significantly, however at 4.2 K it slightly increased, the phase versus voltage characteristic is plotted in figure 4.9b. Also at 290 and 77 K the

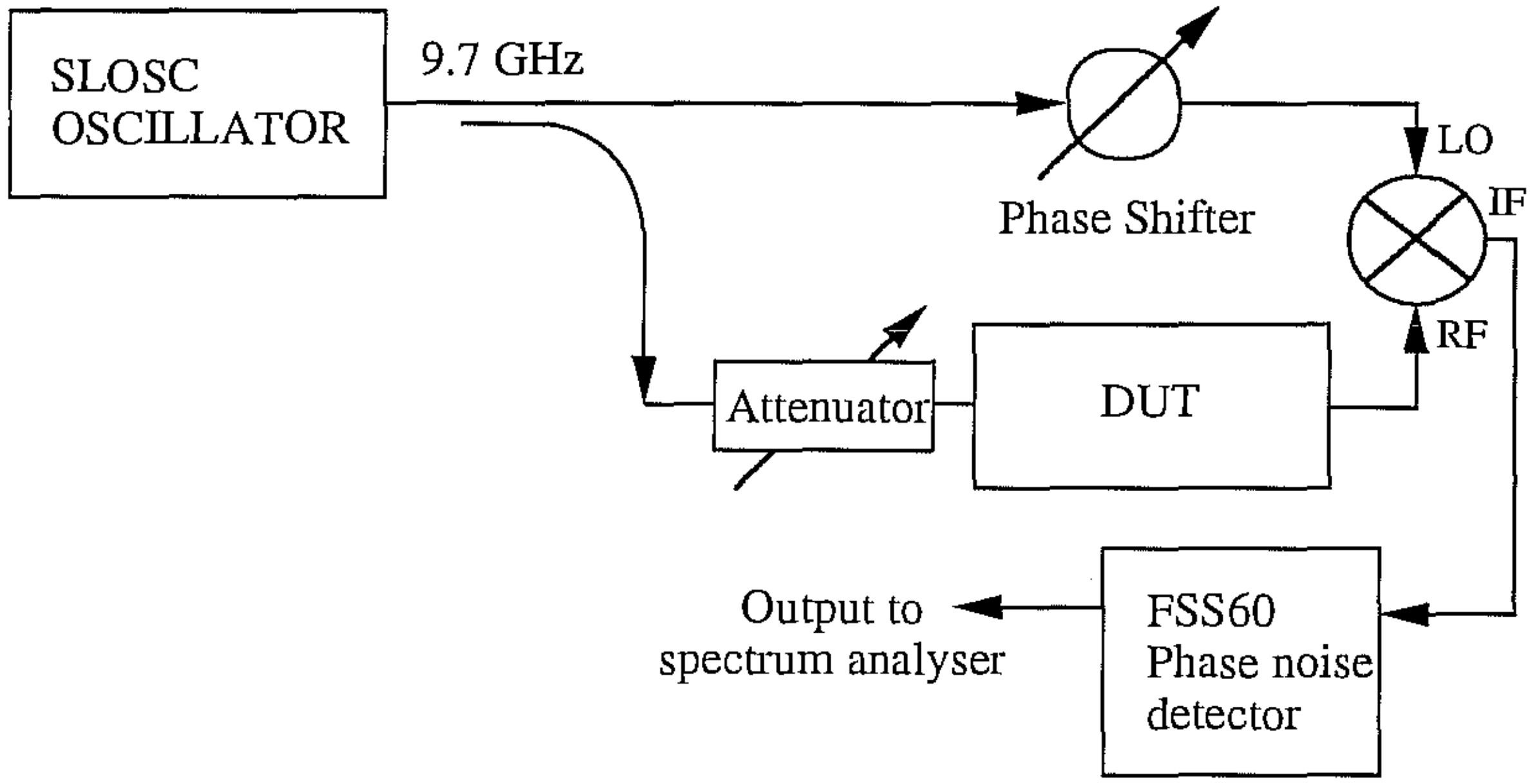

Figure 4.8

Bridge circuit used to measure phase noise. The device under test (DUT) is either the phase shifters or amplifiers under measurement

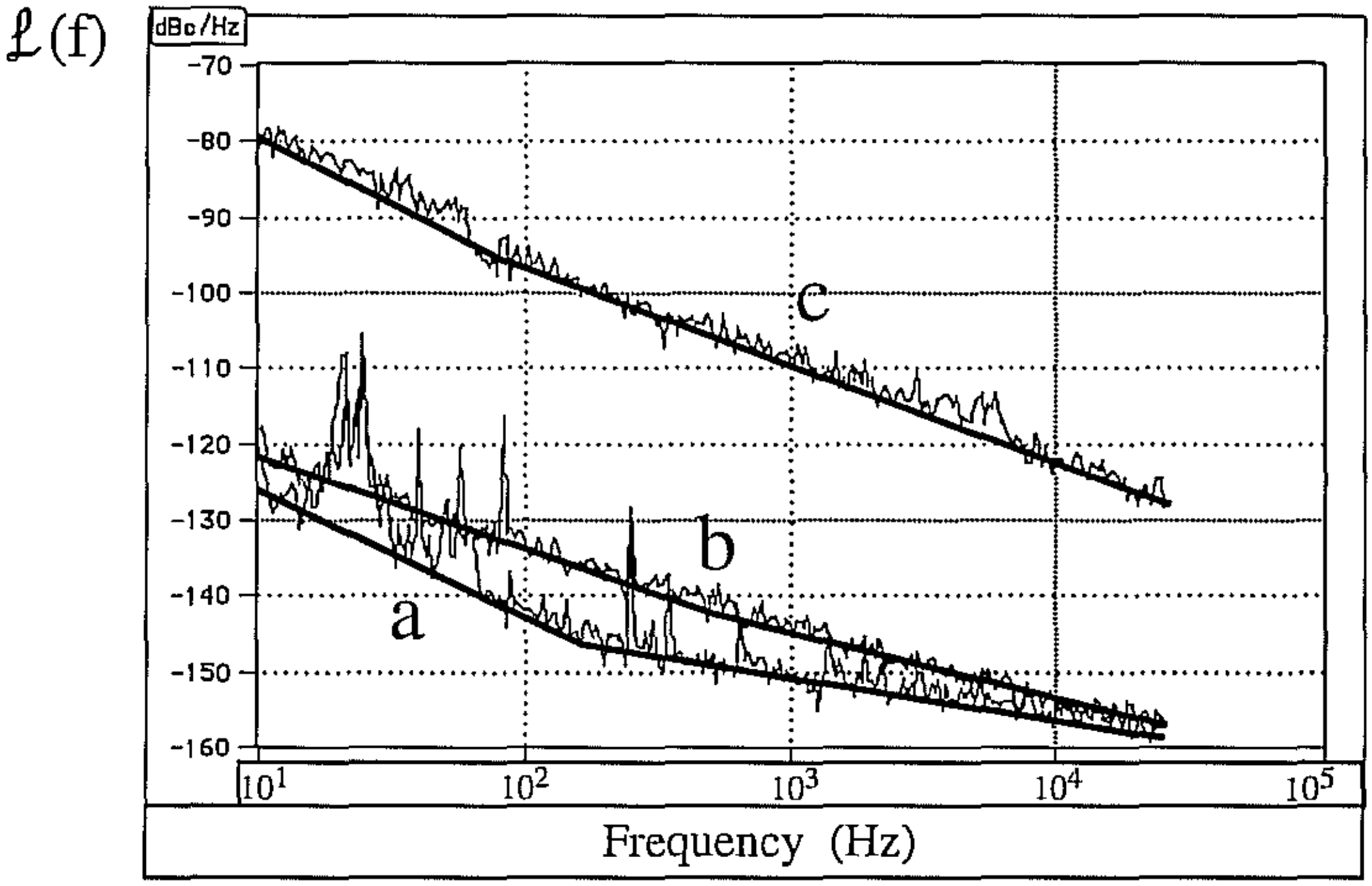

(4.9a)

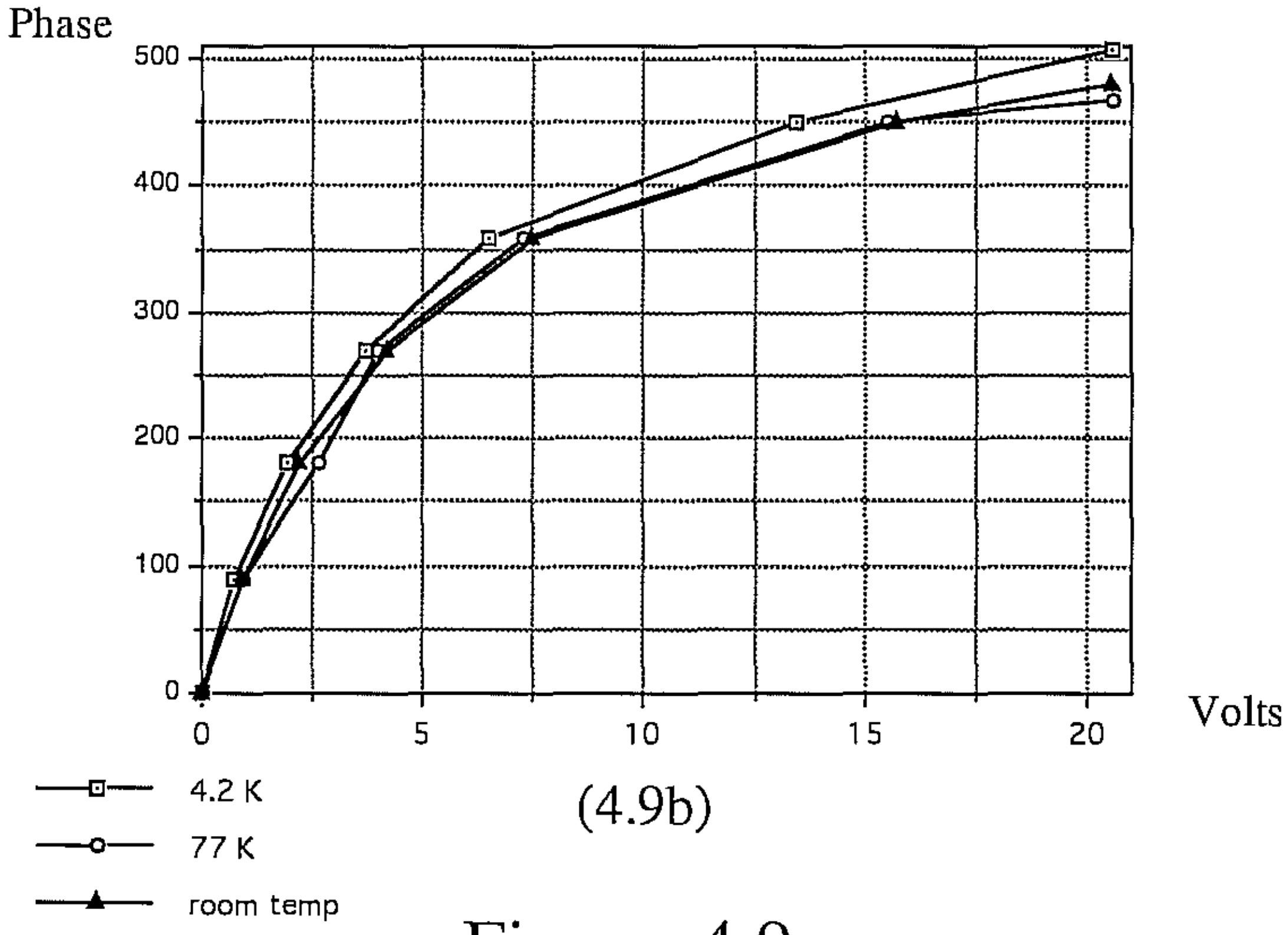

(4.9b)

# Figure 4.9

4.9a shows varactor phase shifter SSB phase noise at 4.2 K; **a**. measurement noise floor; **b**. 13.48 volts bias; **c**. 20.6 volts bias. Below 10 volts bias, the noise can not be measured above the noise floor. However 0 to 10 volts is sufficient to achieve 0 to 400 degrees of low noise phase shift at 4.2 K. At 77 K and room temperature the SSB phase noise could not be measured above the noise floor for all phase states. 4.9b shows the phase versus voltage characteristic at room temperature, 77 K and 4.2 K.

phase noise could not be measured above the noise floor at any phase state. This was not true at 4.2 K for bias voltages above 10 volts, as shown in figure 4.9.

#### 4.3.2.2 GaAs amplifiers

In the past there have been conflicting results concerning the flicker phase noise performance of GaAs FET amplifiers as a function of input power, bias conditions and physical temperature (Lusher and Hardy, 1989), (Hurlimann and Hardy, 1987), (Mann, Blair and Wellington, 1986). Results seem to depend on the characteristics of the individually constructed amplifiers. Flicker phase noise is not well understood and its level can vary substantially between similar amplifiers. Picking out the best transistor in a batch can mean over a 10dB improvement from the average (Celeritek, 1991). The active device noise must be determined if the oscillator noise is to be understood. Both amplifiers analysed had different flicker phase noise characteristics as a function of input power, bias conditions and physical temperature. Most significant was the variation of flicker phase noise with FET bias voltage, and a temperature dependence of the optimum bias conditions (where phase noise was a minimum). In the following tests only one sample unit of each type of amplifier was tested, so it is possible that the same type of amplifier will have differing flicker noise levels. Both amplifiers were designed to operate at cryogenic temperatures which means that on board voltage regulators were omitted as they fail at cryogenic temperatures. The bias voltages quoted in this chapter are the drain - source bias voltages supplied to the transistor.

1) Miteq AFS3-4K Amplifier: This GaAs FET amplifier was designed to operate at 6.0 V bias at room temperature with a gain of 27 dB (Tobar and Blair, June 1992). This bias voltage also gave the minimum phase noise at room temperature. Figures 4.10 to 4.13 show bias, saturation and temperature effects. When submerged in a 4.2 K liquid helium bath the optimum bias voltage became 3.5 V. The amplifier operated with a gain of 28.5 dB and had a physical temperature of 10 K. When submerged in a 2 K superfluid helium bath the physical temperature of the amplifier changed to 8 K, no measurable change in the phase noise and gain characteristics compared to the 10 K values occurred. Figure 4.13 shows the change of amplifier phase noise when the device temperature changed from room temperature to 8 K at a constant bias voltage of 6.0 volts. Comparing measurement **b** in figure 4.12 and measurement **a** in figure 4.12 there was a 17 dB degradation in phase noise at 10 Hz which reduced to no change at 10 kHz, when cooling from room temperature to 8 K. The phase noise of this amplifier also increased significantly when the input power saturated the device (figure 4.11).

2) Miteq AMF-8012-CRYO Amplifier: This GaAs FET amplifier was designed to operate at 3.9 V bias at room temperature with a gain of 28.5 dB (Tobar and Blair, June

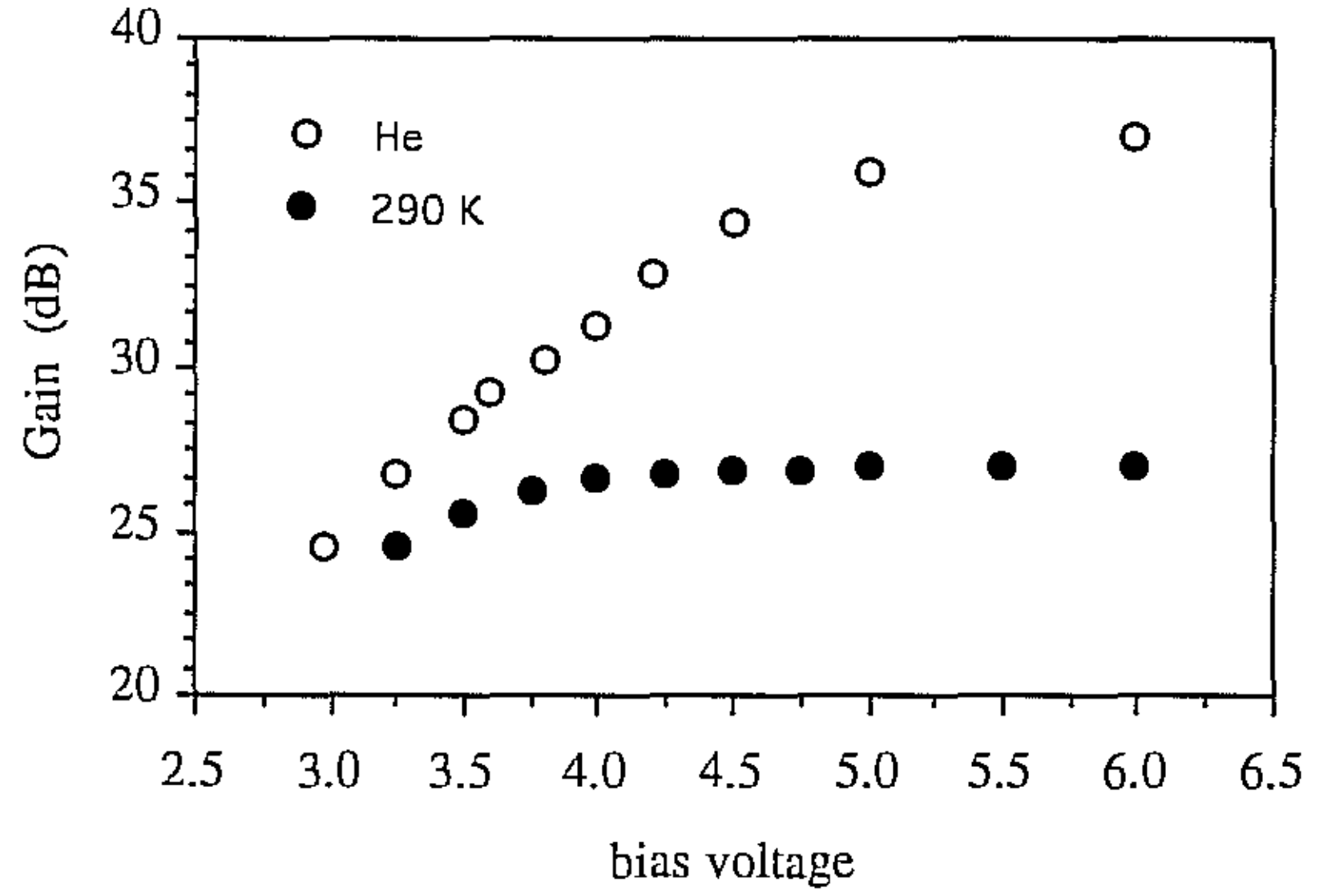

# Figure 4.10

Miteq AFS3-4K gain versus bias voltage for a 1 μw input at liquid helium and room temperature.

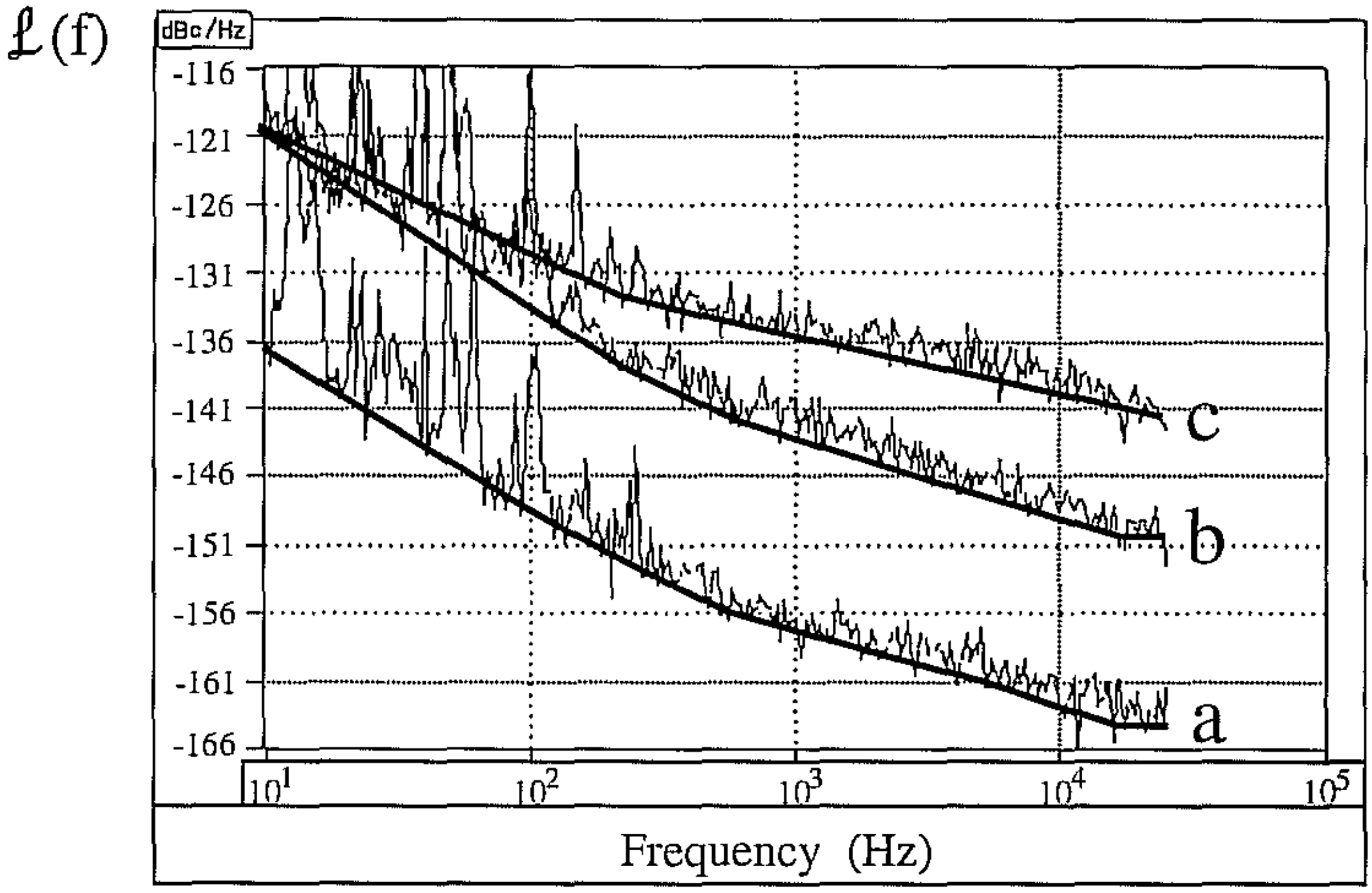

# Figure 4.11

Miteq AFS3-4K saturation effects; **a**. noise floor for all proceeding amplifier measurements; **b**. phase noise with input power of 1.75 μw; **c**. phase noise with input power of 755 μw.

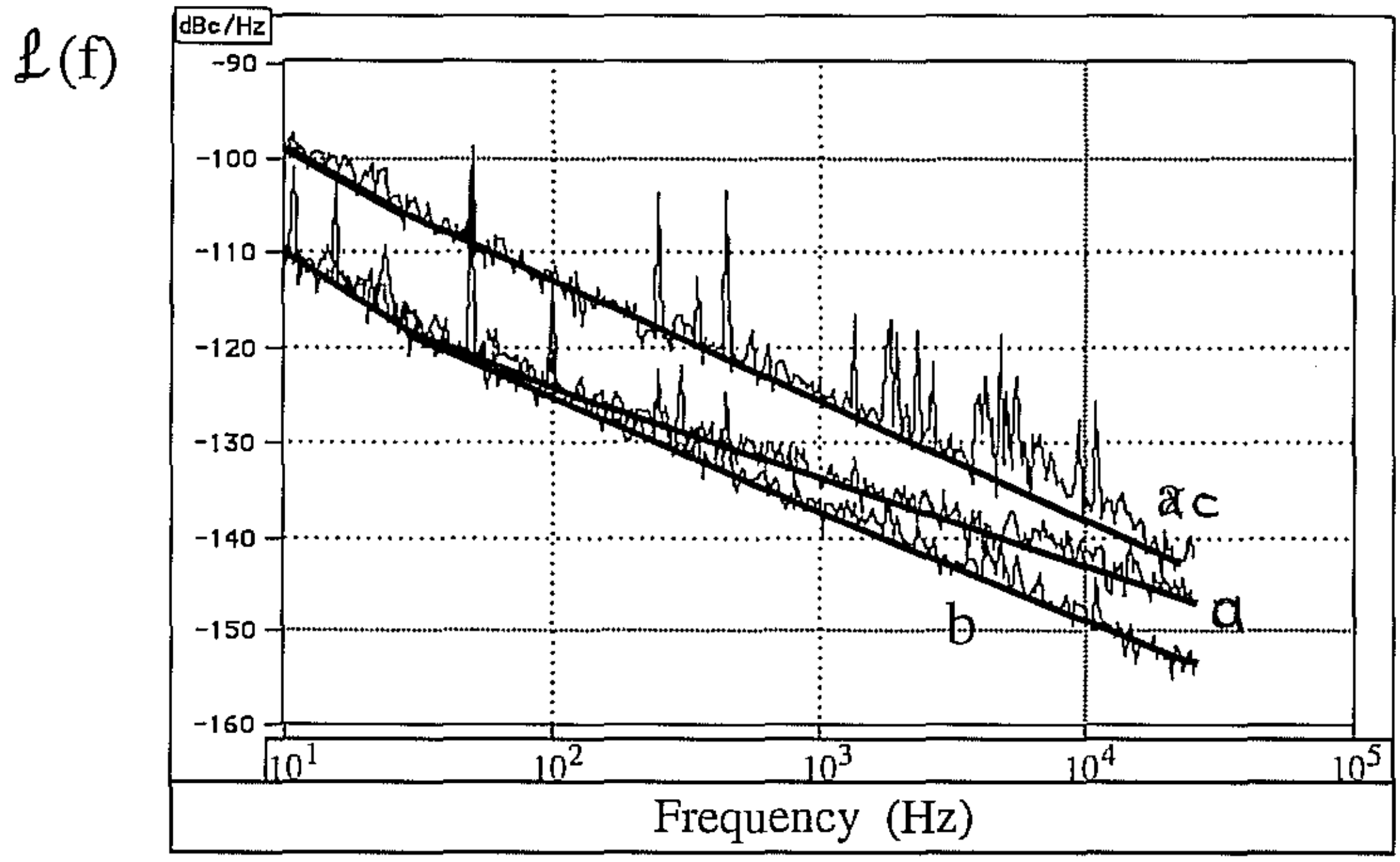

# Figure 4.12

Miteq AFS3-4K amplifier phase noise at liquid helium temperature; **a**. bias = 6.0 V; **b**. bias = 3.5 V; **c**. bias = 3.0 V.

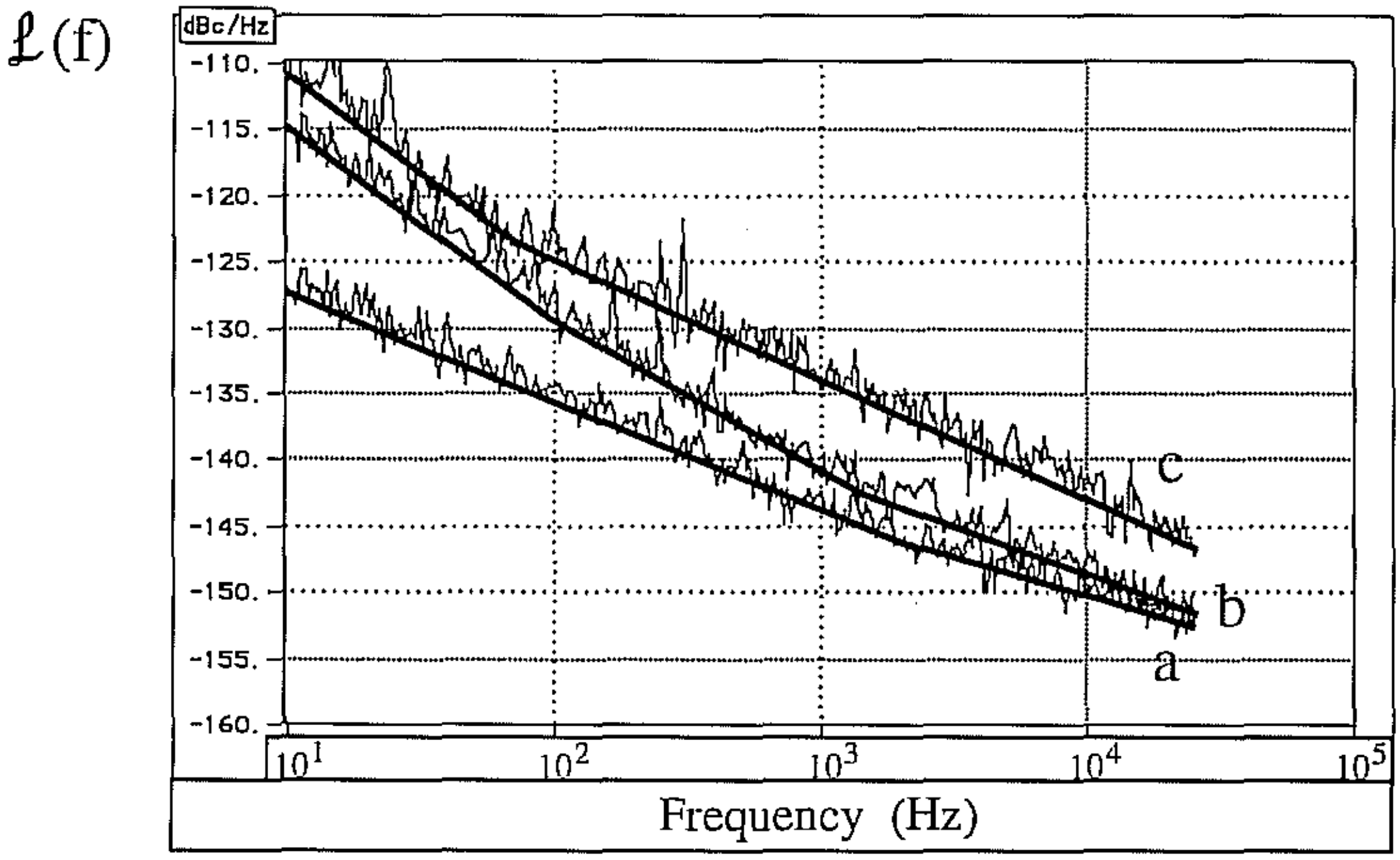

# Figure 4.13

Miteq AFS3-4K amplifier phase noise at a bias of 6.0 V; **a**. room temperature; **b**. liquid nitrogen temperature; **c**. liquid helium temperature.

1992). This bias voltage also gave the minimum phase noise at room temperature. Figures 4.14 to 4.17 show bias, saturation and temperature effects. When submerged in a 4.2 K liquid helium bath the optimum bias voltages ranged between 2.6 V to 2.2 V as shown in figure 4.16. The amplifier operated with a gain of 27 dB when biased at 2.6 V and 15 dB when biased at 2.2 V. The physical temperature of the amplifier was 8 K when immersed in a 4.2 K liquid helium bath and reduced to 7 K when immersed in a 2 K superfluid helium bath, with no measurable change in the phase noise and gain characteristics. Figure 4.17 shows the change in amplifier phase noise when the device temperature changed from room temperature to 8 K at a constant bias voltage of 3.9 volts. Comparing measurement **a** and **c** in figure 4.16, there was 10 dB degradation in phase noise at 10 Hz, which reduced to no change at 500 Hz and a 10 dB improvement at 20 kHz, when cooling from room temperature to 7 K. At room temperature the amplifier phase noise only increases by a few dB when the input power saturated this device (figure 4.15).

At liquid helium temperatures with a bias voltage above 2.8 volts, the amplifier saturated at the very low input power of 4.56 μw, due to a non-linear kink manifesting in the input versus output power characteristic. This effect has been dubbed the kink effect (Tehrani et al, 1989) and is due space-charge-limited current associated with deep traps in the substrate. Recently reports on its effect on white noise levels was reported (Reynoso-Hernandez, 1992). Results presented here show that the flicker phase noise was also enhanced in this kink regime, as shown in figure 4.17.

3) Amplifier amplitude noise results: Figure 4.18 and 4.19 show the a.m. noise of the amplifiers under test. This noise was measured using the phase bridge illustrated in figure 4.8. The phase shifter in the bridge circuit was adjusted so the mixer was sensitive to a.m. perturbations, rather than p.m. perturbations when the phase noise was measured. The a.m. noise was generally a couple of dB lower than the phase noise. This measurement is also sensitive to the a.m. noise in the oscillator. Changing the SLOSC oscillator to a very low amplitude noise GUNN oscillator made no difference to this result. Thus this measurement also determines an upper limit to the amplitude noise in a SLOSC loop oscillator.

## 4.4 T-SLOSC OSCILLATOR PHASE NOISE

Before the above component noise measurements were made, the first round of oscillator noise measurements were performed (Tobar and Blair, May 1991). A tunable reference oscillator was constructed by mixing a fixed frequency SLOSC with a HP 8662A synthesiser (figure 4.1c). The SLOSC was configured with one narrow band high gain

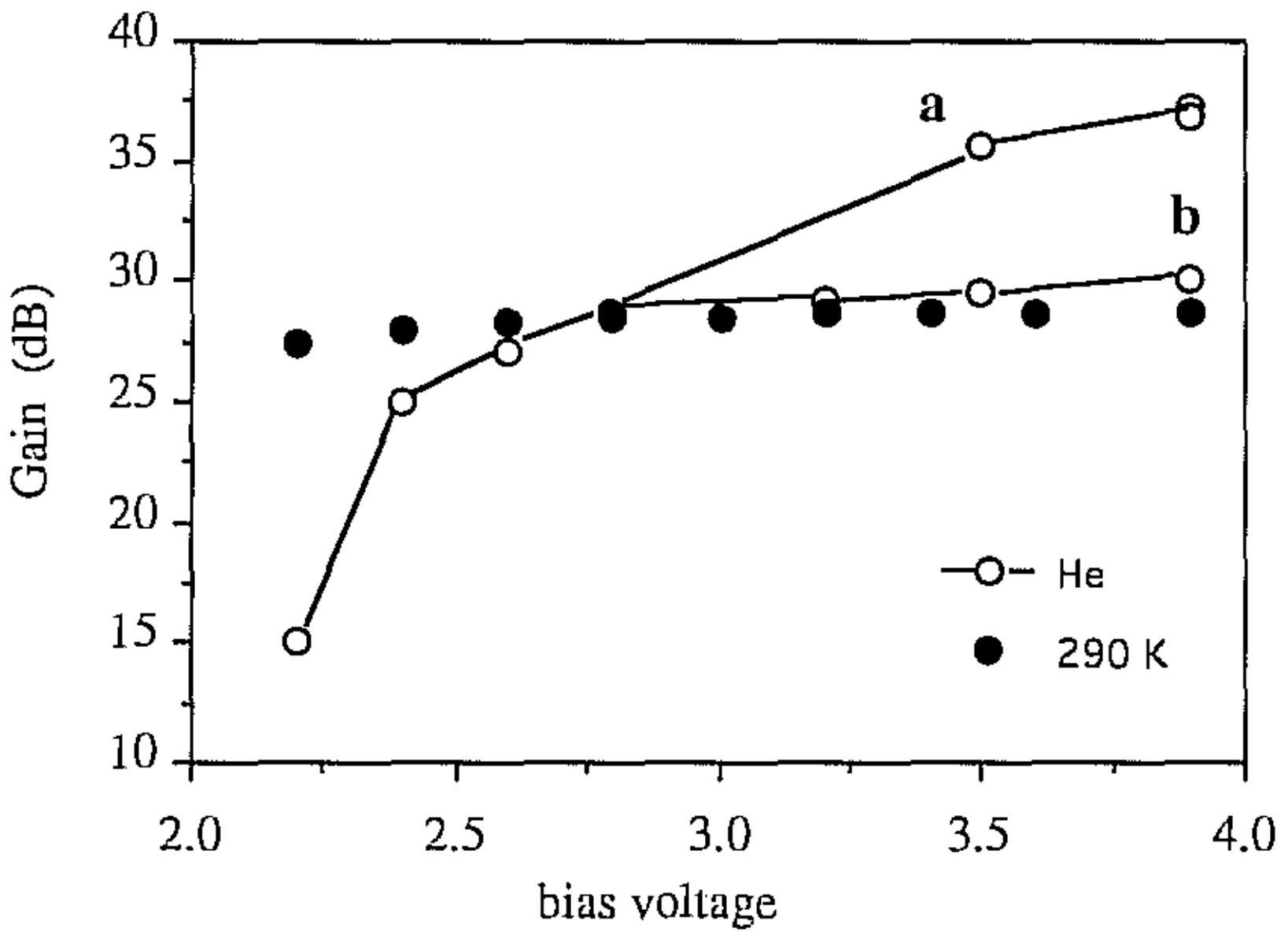

# Figure 4.14

Miteq AMF-8012-CRYO gain versus bias voltage for a 1 μw input at liquid helium and room temperature. When the bias voltages were greater than 2.8 V the amplifier became easy to saturate at low power inputs. This was due to a manifesting non-linear kink in the input to output power realation. **a**. Input power = 1.75 μw, **b**. Input power = 4.56 μw (saturated gain response).

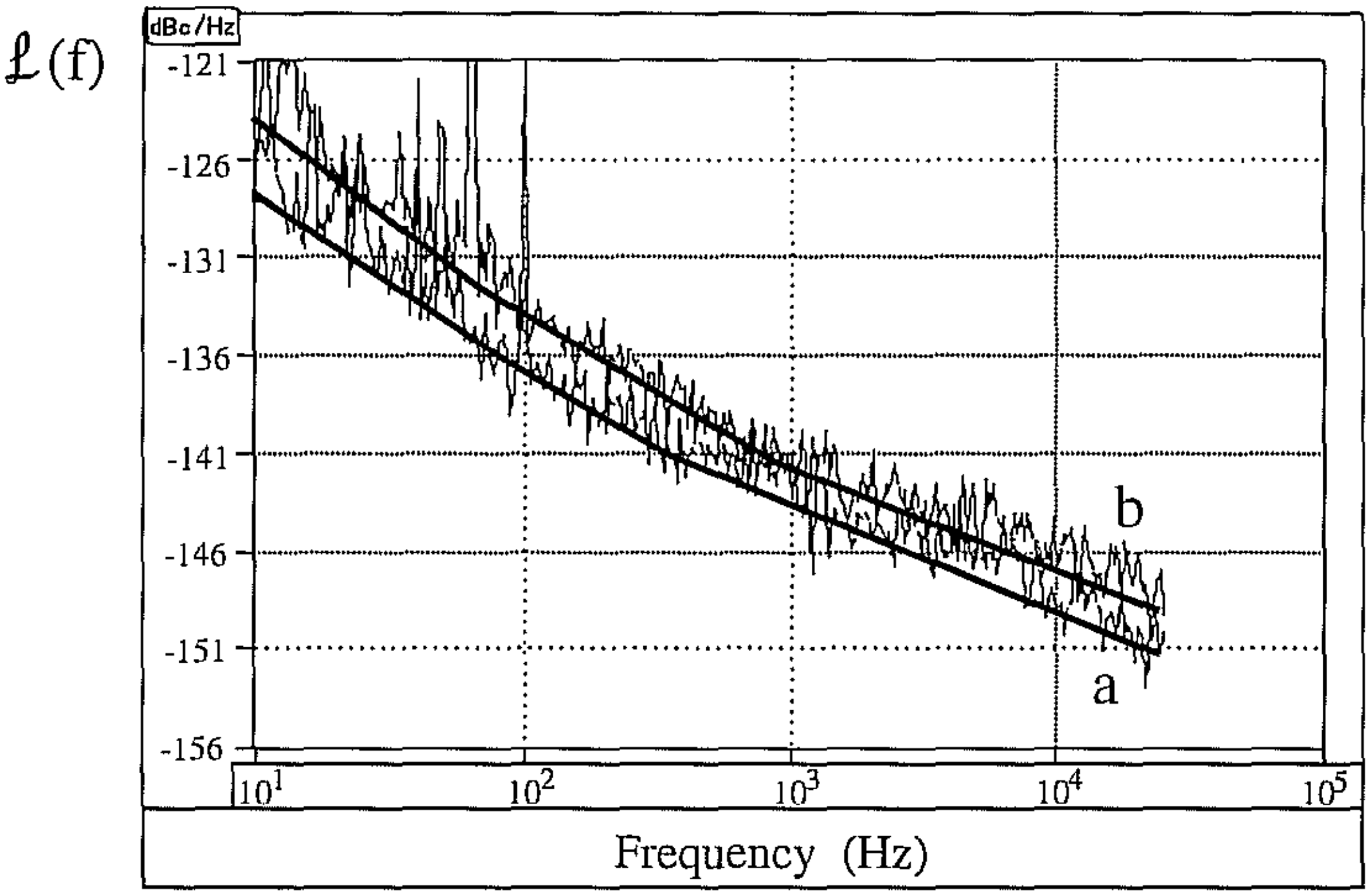

# Figure 4.15

Miteq AMF-8012-CRYO saturation effects; **a**. phase noise with input power of 1 μw; **b**. phase noise with input power of 866 μw.

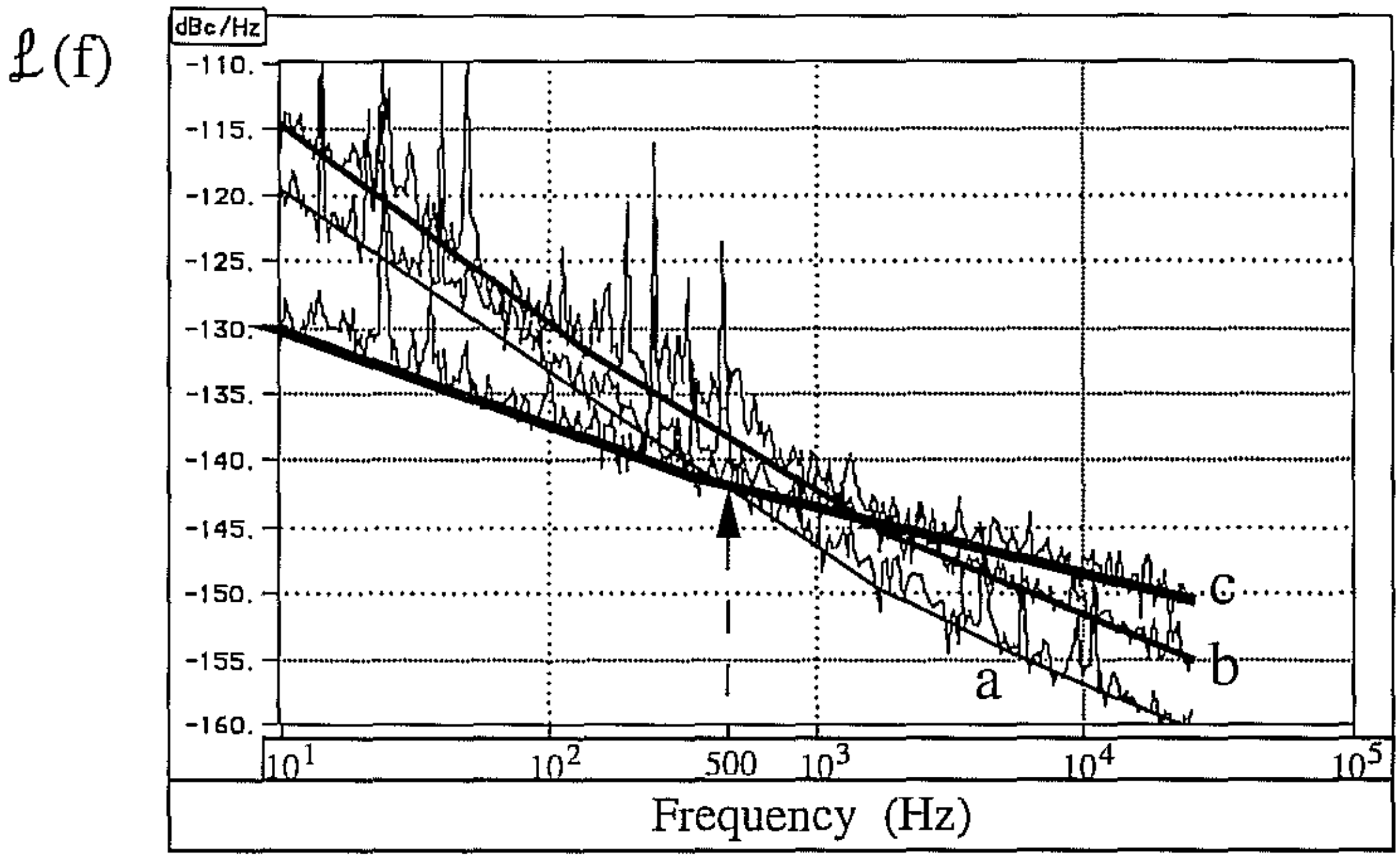

# Figure 4.16

Miteq AMF-8012-CRYO amplifier SSB phase noise at liquid helium temperature; **a**. bias = 2.6 V, the noise measurement between 2.6 to 2.2 V did not change substantially; **b**. bias = 3.9 V, input power = 1.75 μw (non-saturated); **c**. room temperature, bias = 3.9 V.

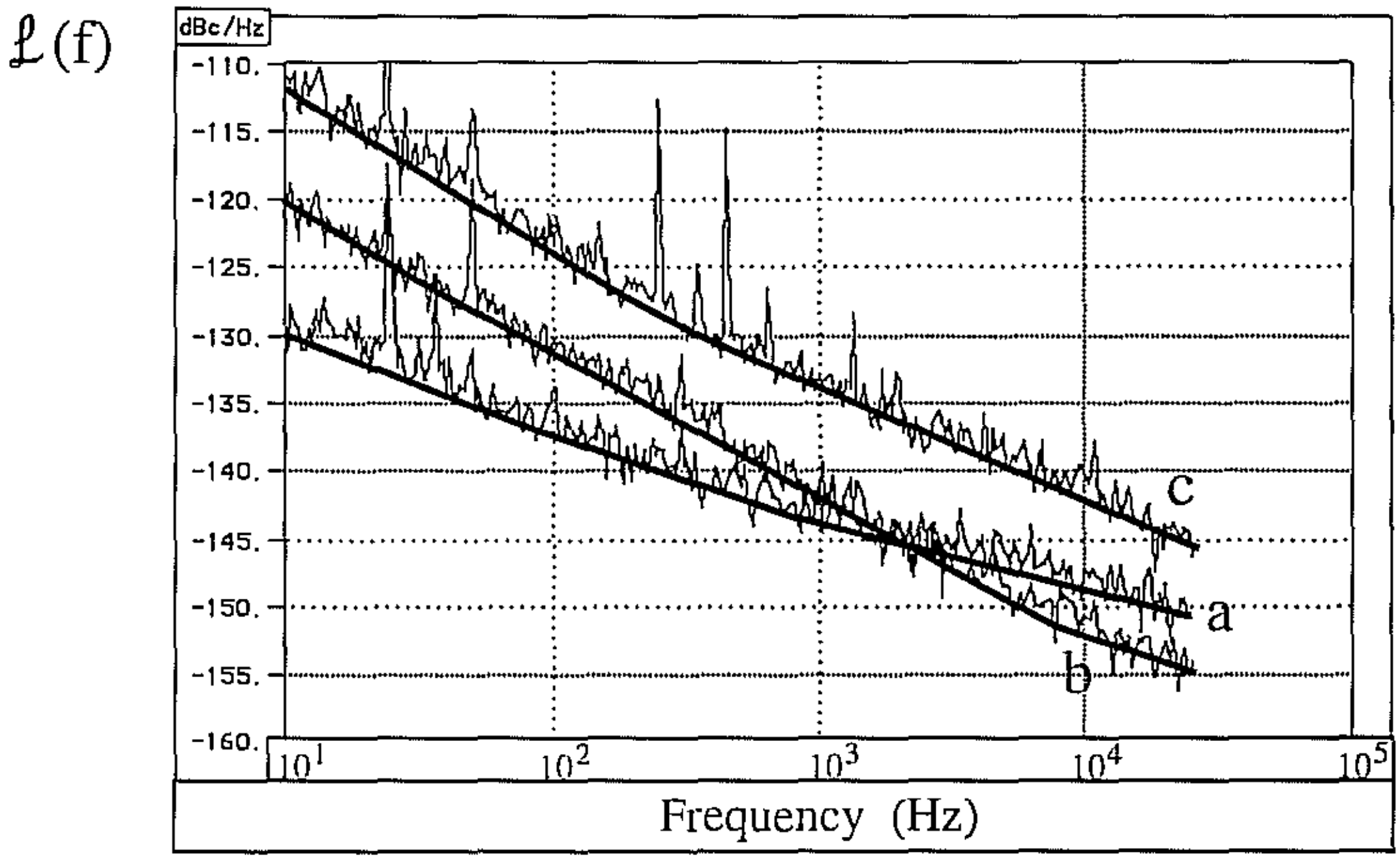

# Figure 4.17

Miteq AMF-8012-CRYO amplifier SSB phase noise at a bias of 3.9 V; **a**. room temperature; **b**. liquid nitrogen temperature; **c**. liquid helium temperature, this level is greater than curve b in figure 4.16, the non-linear kink caused the amplifier to saturate with an input power of only 4.56 μw.

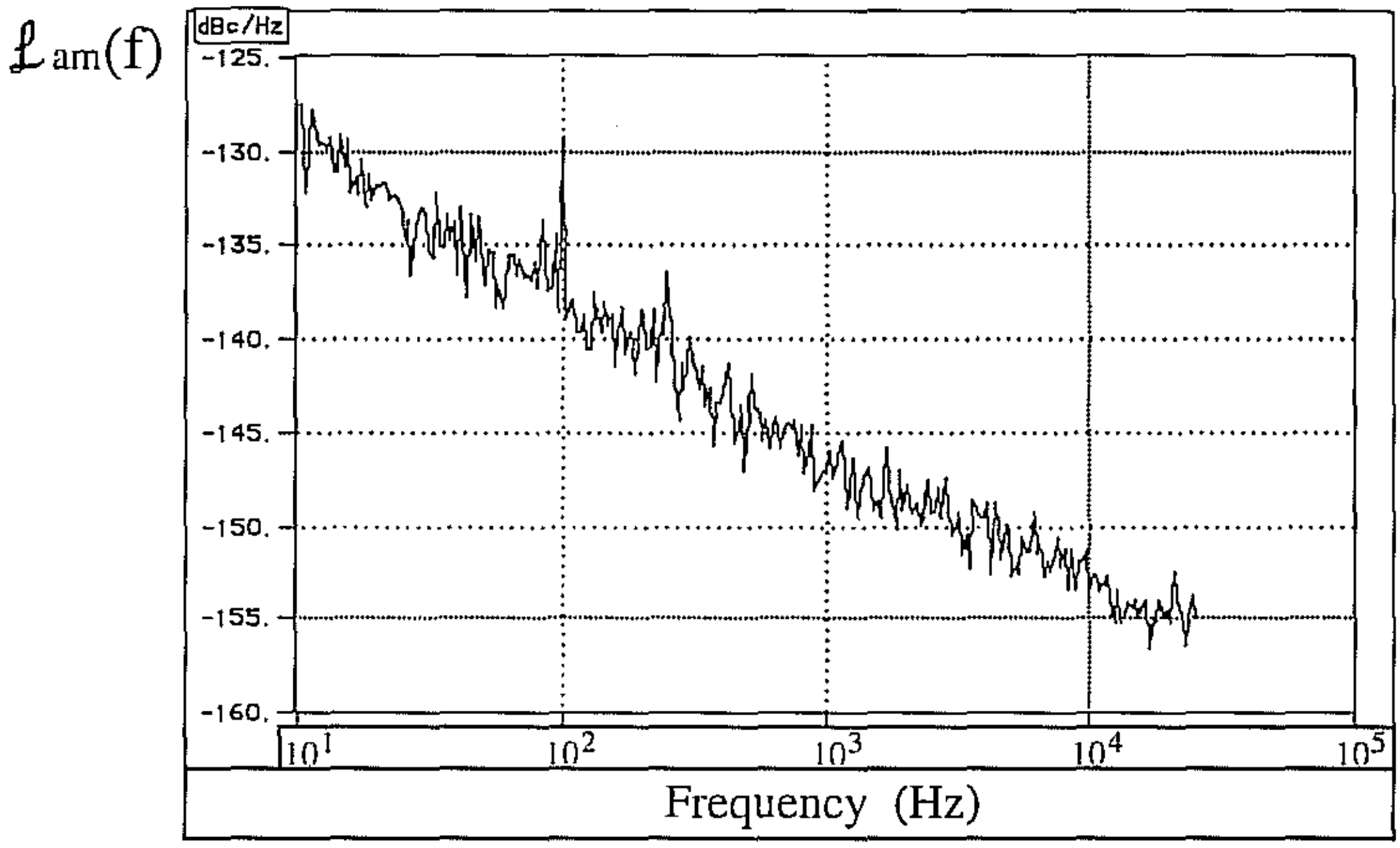

# Figure 4.18

Miteq AMF-8012-CRYO amplifier SSB amplitude noise at room temperature.

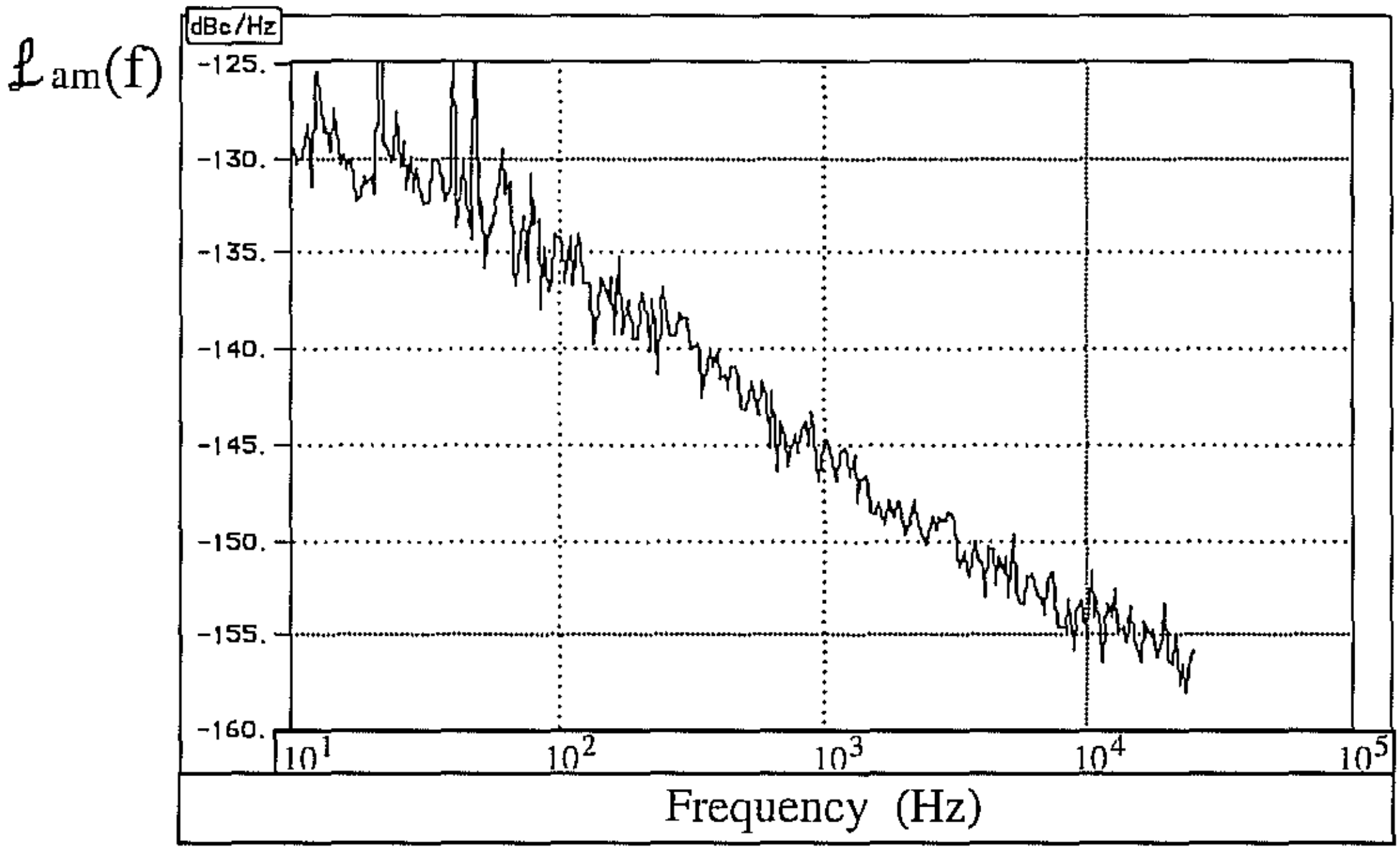

# Figure 4.19

Miteq AFS3-4K amplifier SSB amplitude noise at room temperature.

Systron Donner amplifier, optimised to operate in its unsaturated regime to limit the amplifier flicker noise. The single side band phase noise floor of the synthesiser was measured to be between -140 dBc/Hz and -135 dBc/Hz between 10 hz to 25 kHz offset frequency, by phase locking the synthesiser in quadrature to an ultra low noise fixed frequency quartz oscillator. Forth coming T-SLOSC oscillator measurements reveal that its feedback amplifier phase noise was present above this level. Thus the reference oscillator sufficed for the preliminary measurements.

A series of measurements from room temperature to 4.2 K were implemented with the T-SLOSC oscillating on the $TE_{6\ 1\ 1+\delta}$ mode. The tuning plunger was detuned to eliminate vibrational effects, then the T-SLOSC loop oscillator was phase locked in quadrature to the tunable reference oscillator.

At room temperature the $TE_{6\ 1\ 1+\delta}$ mode had a resonant frequency of 10.15 GHz, with a loaded Q of $1.3 \cdot 10^5$. To supply enough gain for oscillation, the AFS3-4K amplifier was combined with an Avanteq AMT-12034 amplifier, saturating the AFS3-4K. Oscillator phase noise is presented in figure 4.20a. The device noise of the feedback amplifiers can be determined from the equations given in figure 4.1a, to be -85/f dBc/Hz (SSB). This is over 40 dB worse than the noise presented in figure 4.13, probably due to the AFS3-4K amplifier being highly saturated. However this value of device noise is consistant with other phase noise measurements in SLOSC loop ocillators (Giles et al, 1989).

At 77K the $TE_{6\ 1\ 1+\delta}$ mode split into a doublet pair, at 10.21966 and 10.21957 GHz, with loaded Q values of $2.2 \cdot 10^6$ and $7.6 \cdot 10^5$ respectively. The phase noise measurement presented in figure 4.20b was taken while oscillating on the lower Q mode. For this measurement the loop consisted of the AMF-8012-CRYO amplifier, two long coaxial cables from the cryogenic vacuum can to room temperature, a room temperature filter, a cryogenic phase shifter and the AFS3-4K amplifier. The attenuation between one amplifier and the other was large enough to limit some of the saturation of the AFS3-4K amplifier. The device noise of the amplifiers can be determined from the equations given in figure 4.1a, as -96/f dBc/Hz (SSB). This is still greater than 10 dB worse than the noise measured at 77 K for the AFS3-4K amplifier with a bias voltage of 6.0 volts (figure 4.13). However the saturation effects were reduced from 40 dB at room temperature to about 10 dB at 77 K.

As the experiment warmed up from 4.2 K, at about 40 K the phase noise was measured (figure 4.20c). The T-SLOSC was oscillating on the lower Q mode ($Q = 5 \cdot 10^6$) at 10.2209 GHz, in the same configuration when at 77 K. The device noise of the

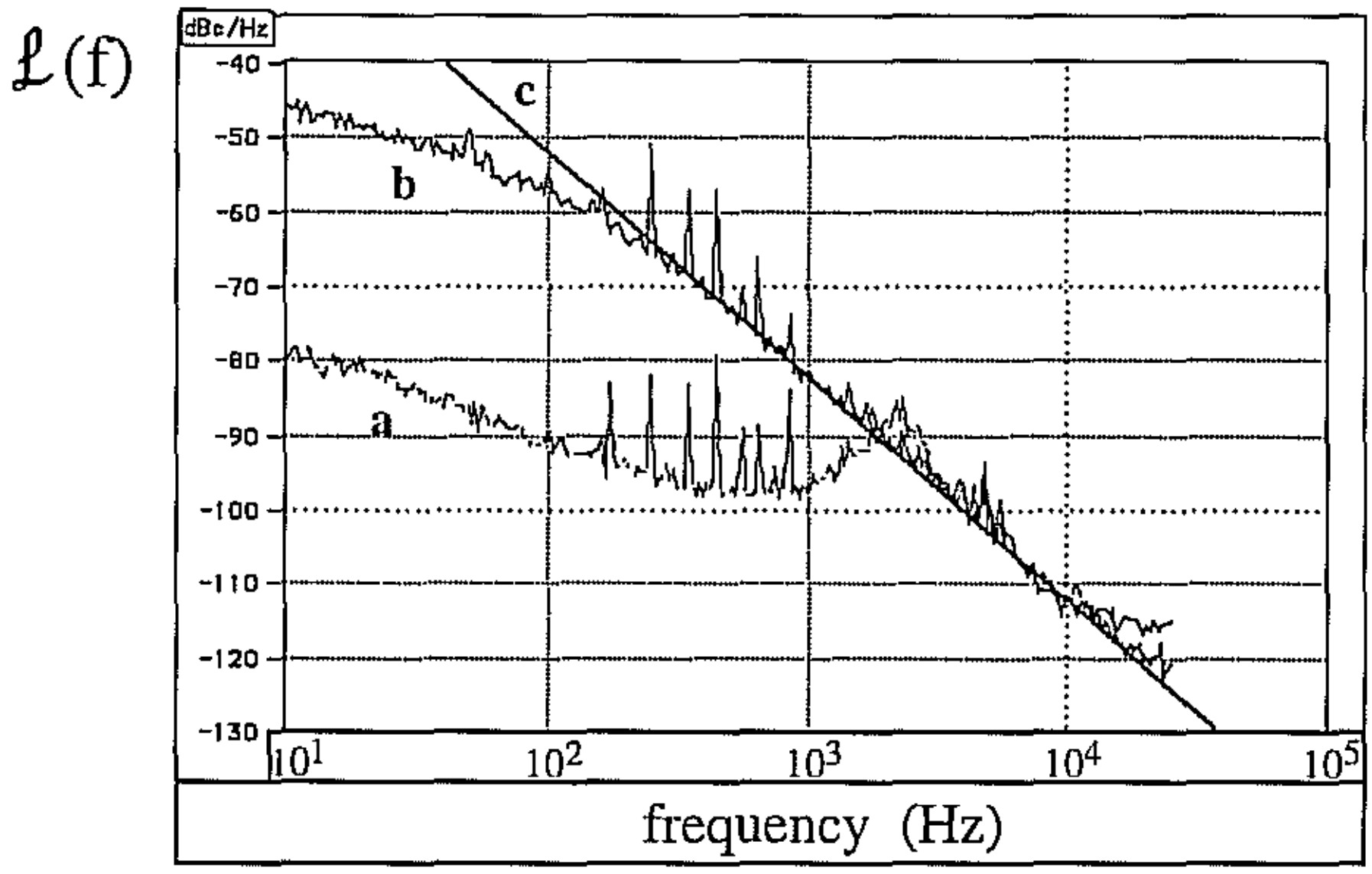

(4.20a)

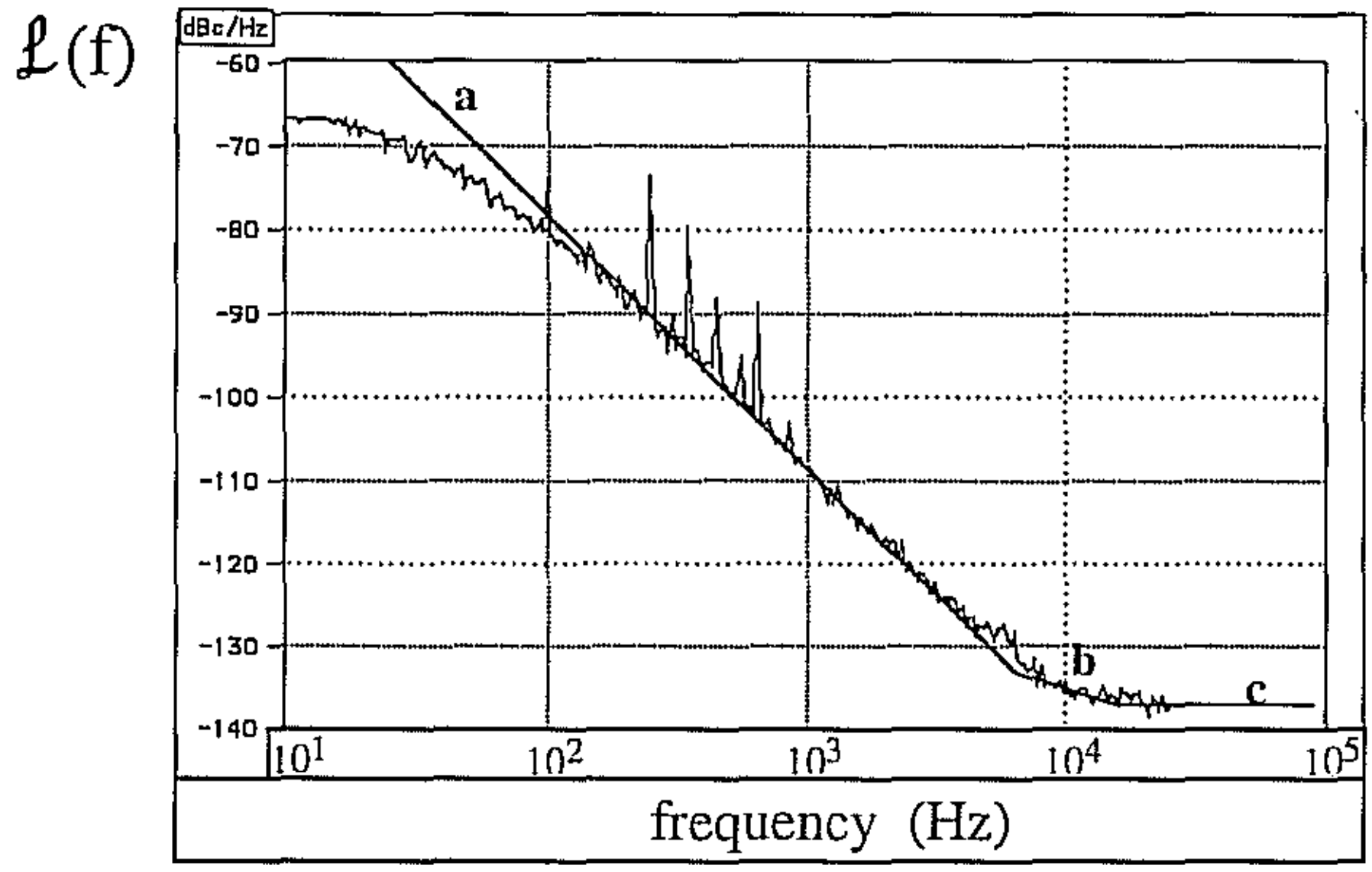

(4.20b)

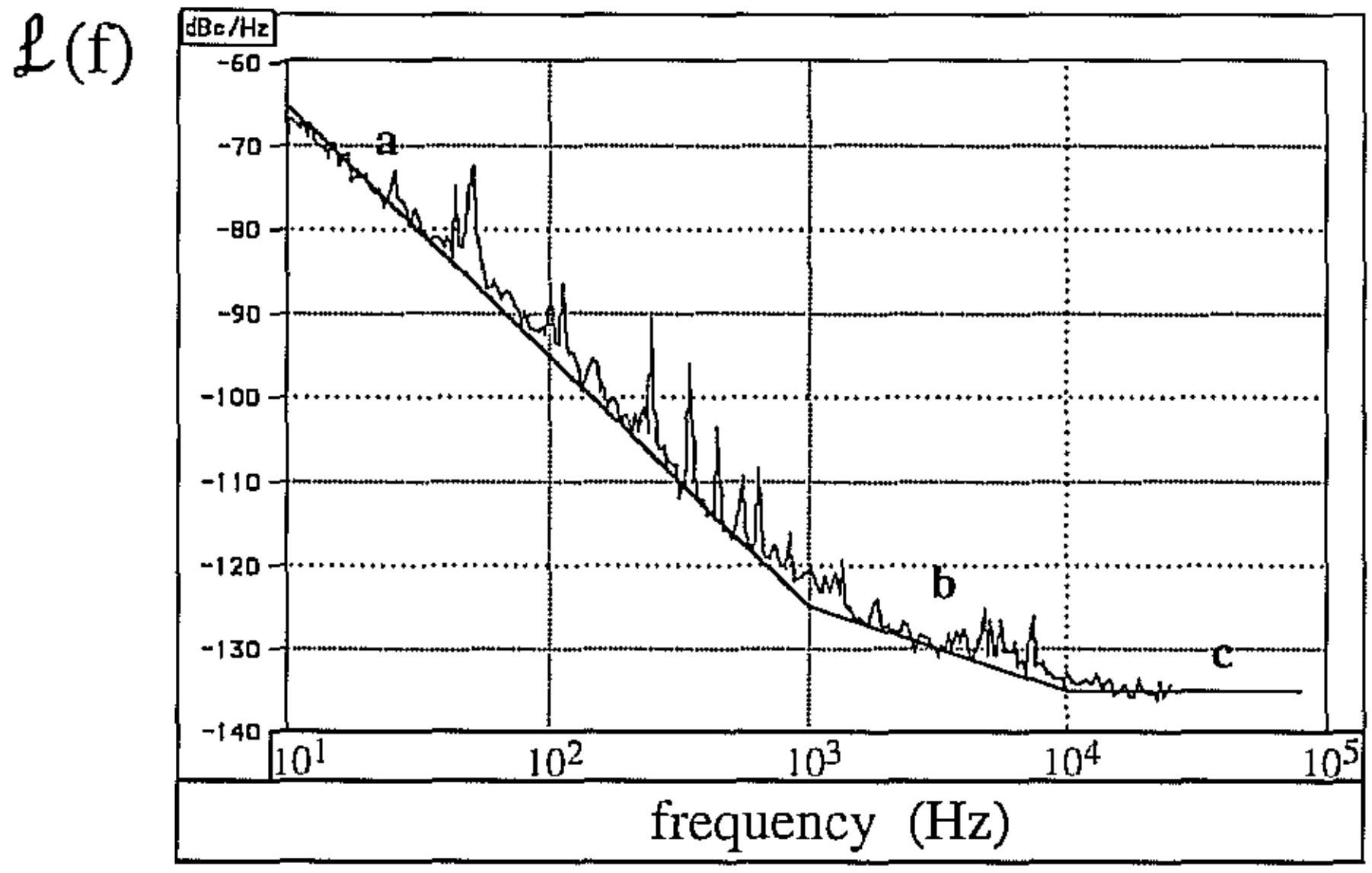

(4.20c)

# Figure 4.20

SSB phase noise of the T-SLOSC oscillator: <u>4.20a Room temperature;</u> **a**. phase noise with a high loop gain in the PLL; **b**. the effect of decreasing the PLL gain; **c**. the extrapolated phase noise disregarding the effects of the PLL: <u>4.20b Liquid Nitrogen temperature & 4.20c 40 Kelvin;</u> **a**. flicker noise within the bandwidth of the resonator with 30 dB per decade roll off; **b**. flicker noise outside the bandwidth of the resonator with 10 dB per decade roll off; **c**. noise floor of the HP 8662A synthesizer.

amplifiers can be determined from the equations given in figure 4.1a, to be -96/f dBc/Hz (SSB), close to the value at 77 K.

At liquid helium temperature the Miteq AMF-8012-CRYO amplifier failed. This port was thus decoupled from the oscillator. A room temperature amplifier was combined with the tunable wavemeter filter to complete the feedback loop with the two remaining ports. A directional coupler was placed before the amplifiers to extract the signal. Oscillation was achieved on the higher Q mode of the doublet, with the results presented in figure 4.21. The loaded Q was measured to be $1.5 \cdot 10^8$ and frequency 10.22107 GHz. The device noise of the amplifiers can be determined from the equations given in figure 4.1a, as -91/f dBc/Hz (SSB). This is 5 dB worse than the estimated device noise at 77 K. This is consistent with the results presented in figure 4.13 as the AFS3-4K amplifier phase noise increased by about 5 dB when cooled from 77 K to 8 K and biased at 6.0 V.

The oscillator noise spectrum was excited by vibrations from a pump that was needed to pump on the vacuum can due to a leak in its indium seal. When the dewar was raised on soft springs the lower frequency spectrum below 20 Hz disappeared. The increase in tension in the rubber tube connecting the pump to the vacuum can enhanced frequencies between 20 and 300 Hz. When the pump was turned off this spectrum disappeared, however the frequency drift due to the leak was enough to inhibit a phase noise measurement with the pump off. It is interesting to compare the structure of the low frequency noise in figure 4.21 with the theoretical transfer function (figure 4.7) of the vibration isolation. This is strong evidence that the vibration isolation works as designed.

The oscillator measurements just discussed were taken before the GaAs phase and amplitude components presented in section 4.3.2 were measured. The oscillator measurements implied that electronic flicker noise was a limiting component. The AMF-8012-CRYO amplifier was sent back to the manufactures and fixed so it could work at liquid helium temperature. Once returned I decided to do extensive component phase noise measurements to see if changing the operating conditions would reduce the phase noise of these components and hence the oscillator's.

After understanding the phase noise of the cryogenic components, it is possible to optimise the oscillator design to minimise phase noise. The all cryogenic tunable oscillator was configured with the cryogenic AMF amplifier, biased between 2.2 to 2.8 volts, in the linear low phase noise regime. We were able to extract -1 dBm from the filtered port of the quasi $TE_{61\delta}$ T-SLOSC mode at 9.7 GHz. To obtain the most sensitive phase noise measurements, the measurement systems illustrated in figure 4.1 need 4-10 dBm to drive the local oscillator port of the mixer. Amplification of the signal from the

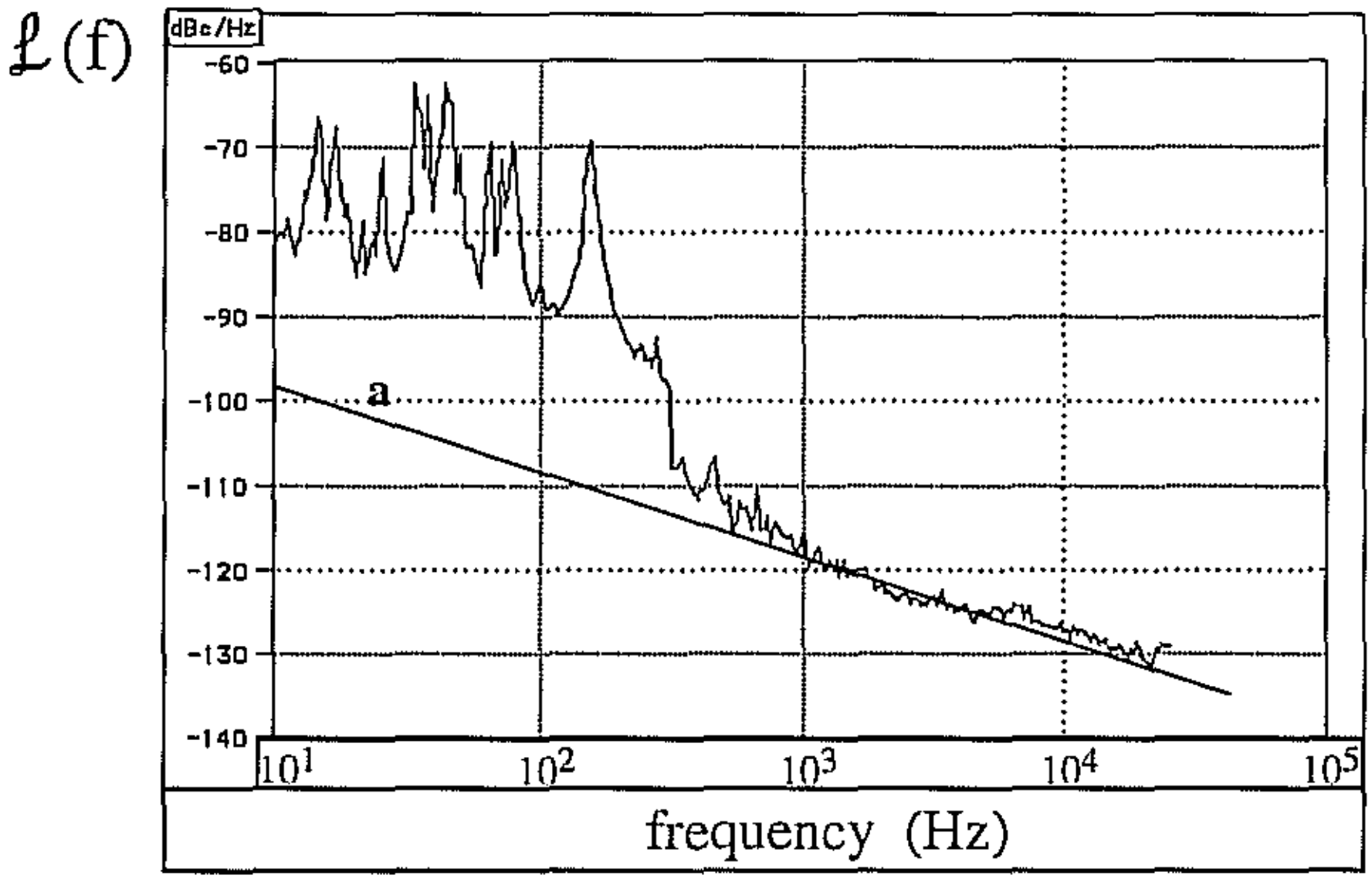

(4.21a)

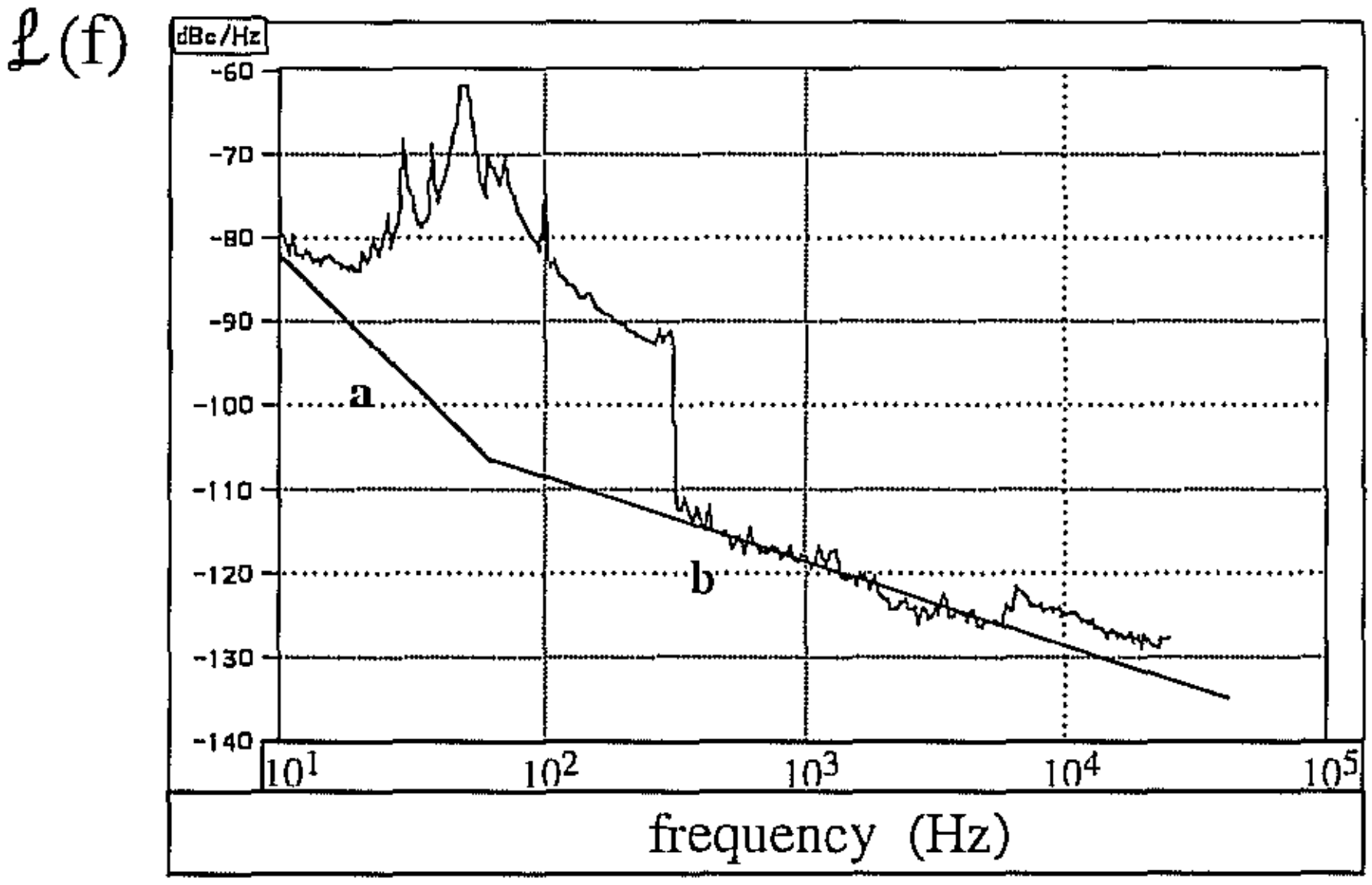

(4.21b)

# Figure 4.21

SSB phase noise of the T-SLOSC oscillator at liquid heluim temperature: 4.21a: dewar on ground; **a**. flicker noise outside the bandwidth of the resonator with 10 dB per decade roll off: 4.21b: dewar on springs; **a**. flicker noise within the bandwidth of the resonator with 30 dB per decade roll off; **b**. flicker noise outside the bandwidth of the resonator with 10 dB per decade roll off.

filtered transmission port must be avoided to prevent amplifier phase noise from dominating. Hence the measurement in figure 4.1b was performed with a power starved local oscillator port limiting the measurement noise floor. However the noise floor was sufficient to measure the vibrational effects of the tuning mechanism, but was too high to measure the oscillator noise. Figure 4.22 shows some results taken in this way.

To reduce the vibrational levels exciting the tunable oscillator at 700 Hz, the resonator was housed in a vibration isolation system with the dewar raised of the ground by a set of soft springs, and the tuning plunger decoupled from the tuning mechanism. Equation (4.5) below is derived in appendix D, and relates the vibration of the tuning mechanism to a phase noise level.

$$S_{\phi}(f) = \left(\frac{df}{dx}\right)^2 \left(\frac{2}{\Delta f}\right)^2 \left(1 + \left(\frac{2f}{\Delta f}\right)^2\right)^{-1} S_x(f) \qquad (4.5)$$

Here $S_x(f)$ [$m^2/Hz$] is the vibration spectrum of the tuning mechanism. In figure 4.22 the difference in tuning coefficient between measurements **b** ($\approx 3.5 \cdot 10^6$ Hz/mm) and **c** ($\approx 4 \cdot 10^7$ Hz/mm) is about 10. Thus there is a factor of approximately 20 dB difference in vibrational phase noise measured between **b** and **c**. Measurement **b** is limited by the mixer noise floor given by measurement **a**. From (4.5) the low frequency vibration level corresponding to -90 dBc/Hz at 10 Hz (**c**) can be calculated to be less than $10^{-12}$ $m/\sqrt{Hz}$. This level increased to -80 dBc/Hz when the tuning mechanism was coupled to the plunger, and to -70 dBc/Hz when the dewar was taken off the springs.

To avoid the corruption of synthesizer phase noise in the measurement, the T-SLOSC must be designed so a high Q mode may be tuned to the same frequency as a high Q mode in the SLOSC (see figure 4.1b). Both the T-SLOSC and the SLOSC used $TE_{61\delta}$ operational modes, the T-SLOSC mode was at 9.7485 GHz and the SLOSC mode was at 9.7328 GHz, with loaded Q values of $10^8$ and $2.5 \cdot 10^8$ respectively. Unfortunately when the T-SLOSC was tuned to the SLOSC at 9.7328 GHz, the T-SLOSC exhibited a tuning coefficient of $4 \cdot 10^7$ Hz/mm, and the measurement was limited by vibrations. Figure 4.22 curve c shows a phase noise level of -105 dBc/Hz at 1 kHz. In order to reduce this level to less than -140 dBc/Hz the required tuning coefficient must be less than $10^6$ Hz/mm (4.5). Limiting the tuning coefficient to $10^6$ Hz/mm restricts the useful tuning range to less than 5 MHz. Therefore the cavity must be designed to have a frequency to within a few MHz above the fixed frequency SLOSC oscillator so the tuning required is minimal. Precision calculations of resonant frequencies were presented in Chapter 2, which shows that this redesign goal is achievable but is beyond the scope of this work.

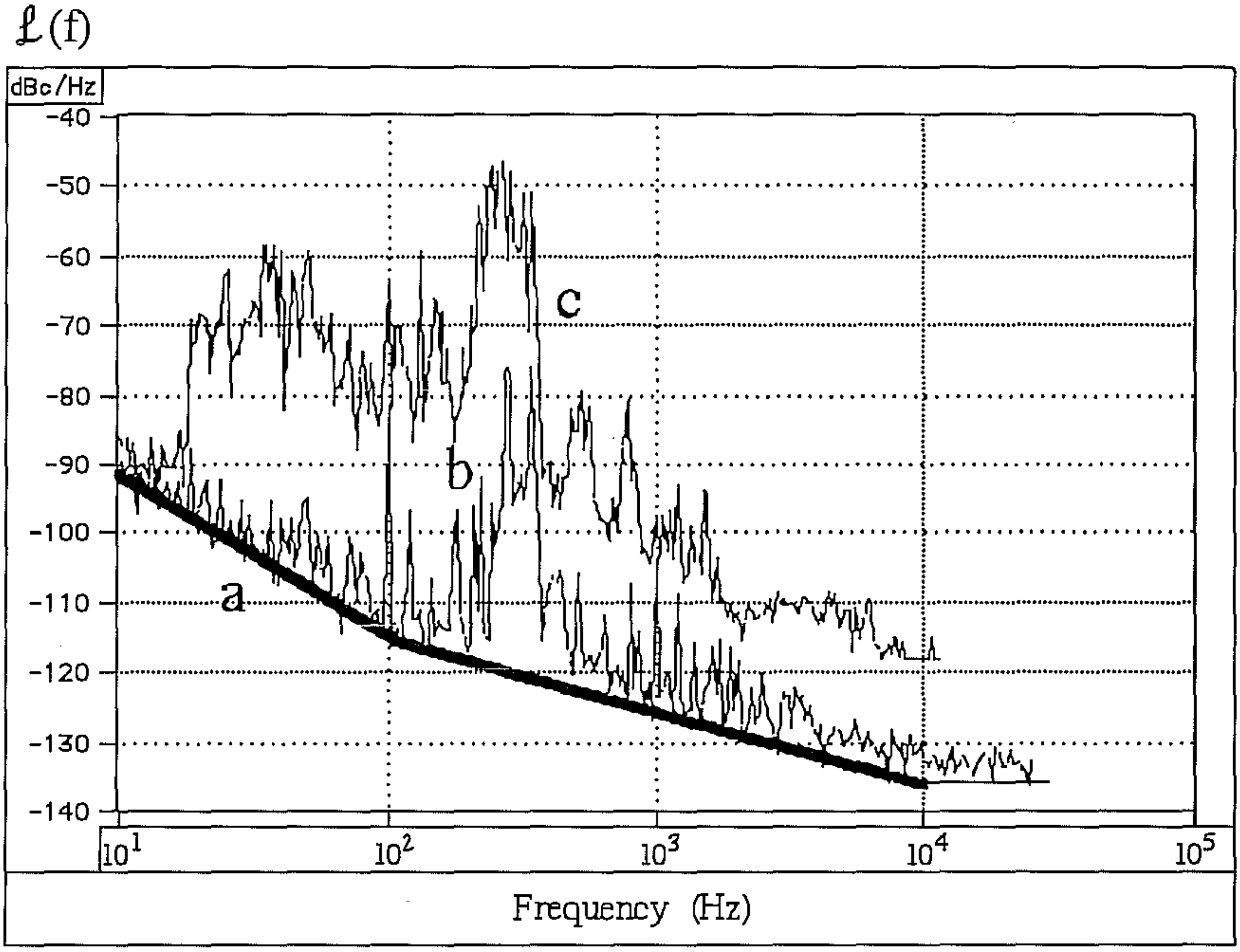

# Figure 4.22

T-SLOSC loop oscillator noise: **a**. Measurement noise floor: **b**. T-SLOSC oscillator SSB phase noise with a tuning coefficient of $3.5 \cdot 10^6$ Hz/mm: **c**. T-SLOSC oscillator SSB phase noise with a tuning coefficient of $4 \cdot 10^7$ Hz/mm

## 4.5 SLOSC OSCILLATOR PHASE NOISE

Because of the problems associated with the T-SLOSC, a phase noise measurement using the circuit in figure 4.1c was performed. Two quasi $TE_{61\delta}$ SLOSC oscillator modes at 9.7328 and 9.7338 GHz were heterodyned with a 1 MHz signal from a HP 8662A synthesiser. Mitec X-band room temperature amplifiers of the AMF series were used with the cryogenic phase shifter operated at room temperature. Only -10 dBm could be obtained from the filtered port, so the signal was taken from the unfiltered port, where up to 17 dBm was obtained. The noise from the active components can be calculated as $-102.5/f^{1.25}$ dBc/Hz from the oscillator characteristics shown in figure 4.23. This is very close to the measured room temperature phase noise of these amplifiers (Mann, 1992). Results are shown in figure 4.23, along with theoretical calculations of the phase noise from the filtered and unfiltered ports. The theoretical curves agree well with the data below 500 Hz, but above this frequency the noise is dominated by the noise from the 8662A synthesiser. At 1 kHz offset -140 dBc/Hz from the unfiltered port would be obtained if the synthesiser noise of -137.5 dBc did not interfere. These results imply that the phase noise from the filtered port was -175 dBc/Hz at 1 kHz offset.

## 4.6 OSCILLATOR COMPARISONS AND CONCLUSIONS

Past oscillator and amplifier measurements have determined the flicker noise component for similar active systems at liquid helium temperatures to be -87/f dBc/Hz (Giles et al. 1989), -90/f (dBc/Hz) (Lusher and Hardy 1989) and -91/f (dBc/Hz) (Tobar and Blair, May 1991). For the cryogenic T-SLOSC oscillator the limiting electronic component was the cryogenic AMF amplifier, which was more than 15 dB better at 1 Hz, and rolled off at a greater slope as; $-108.5/f^{1.25}$ (dBc/Hz). For the SLOSC oscillator with room temperature amplifiers the limiting components were the room temperature AMF amplifiers which had a combined active component noise of $-102.5/f^{1.25}$ dBc/Hz 6 dB worse than the cryogenic amplifier. Thus an oscillator configured with the cryogenic amplifier would be expected to operate with 6 dB improvement.

Figure 4.24 reveals that at present the best X-band phase noise performance for carrier offsets greater than 10 Hz with output power greater than a milliwatt is given by the cryogenic sapphire oscillators at the University of Western Australia. At the Jet Propulsion Laboratories a phase noise of $-80/f^3$ dBc/Hz was measured below 1 Hz, in a superconducting cavity maser (Dick and Wang, 1991). However their system has an output power of only a nanowatt and was incapable of measuring any noise beyond a few hertz. In the future using the circuit illustrated in figure 4.1b, we expect to measure

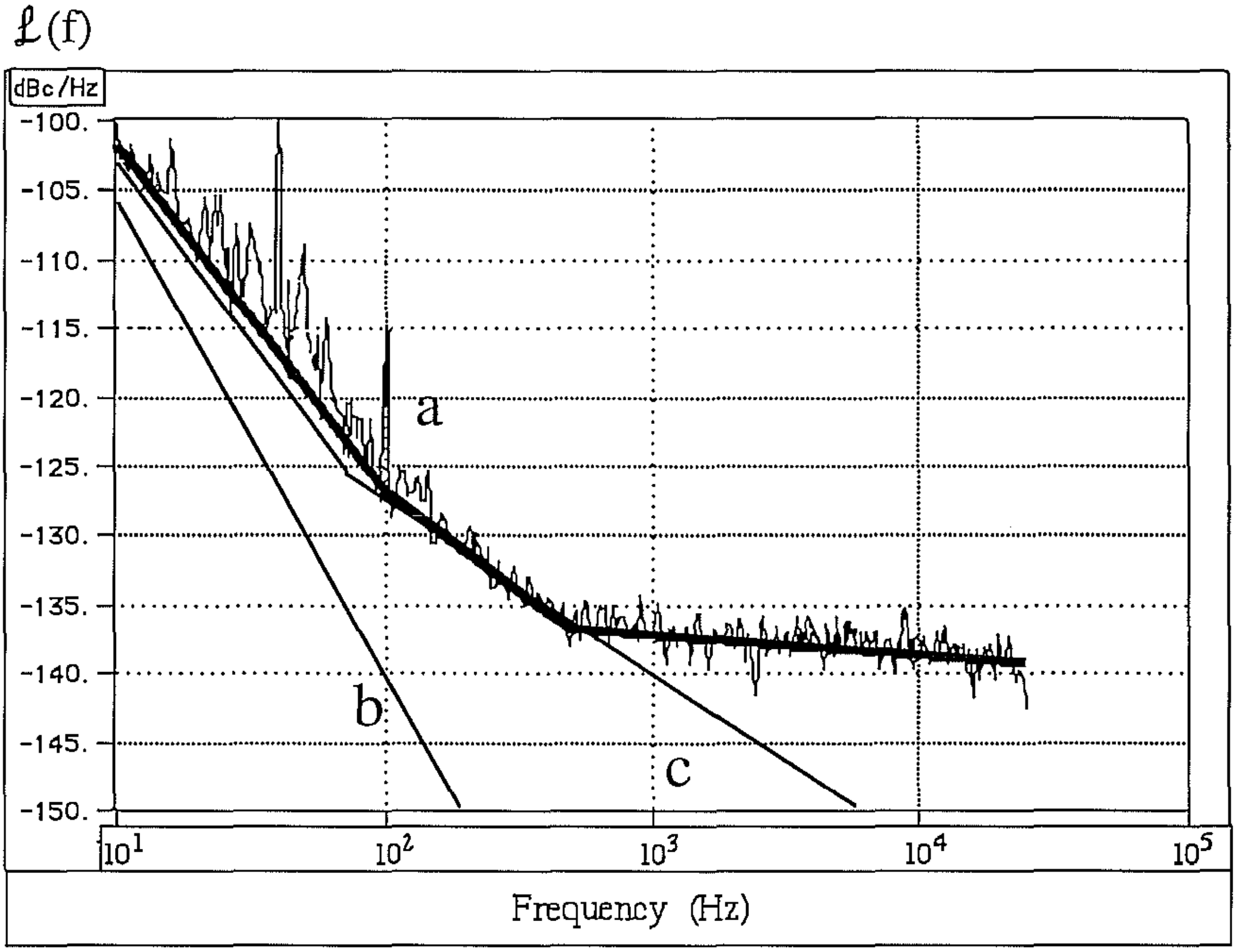

# Figure 4.23

**a**. SSB oscillator phase noise from the non-filtered port, measured with a high gain phase locked loop above 100 Hz and a low gain phase locked loop below 100 Hz. **b**. Theoretical calculation of the phase noise taken from the filtered port. **c**. Theoretical calculation for the phase noise taken from the unfiltered port, it follows closely to the measured value (bold line) below 500 Hz, above 500 Hz the noise characteristic of the 8662A synthesizer is measured

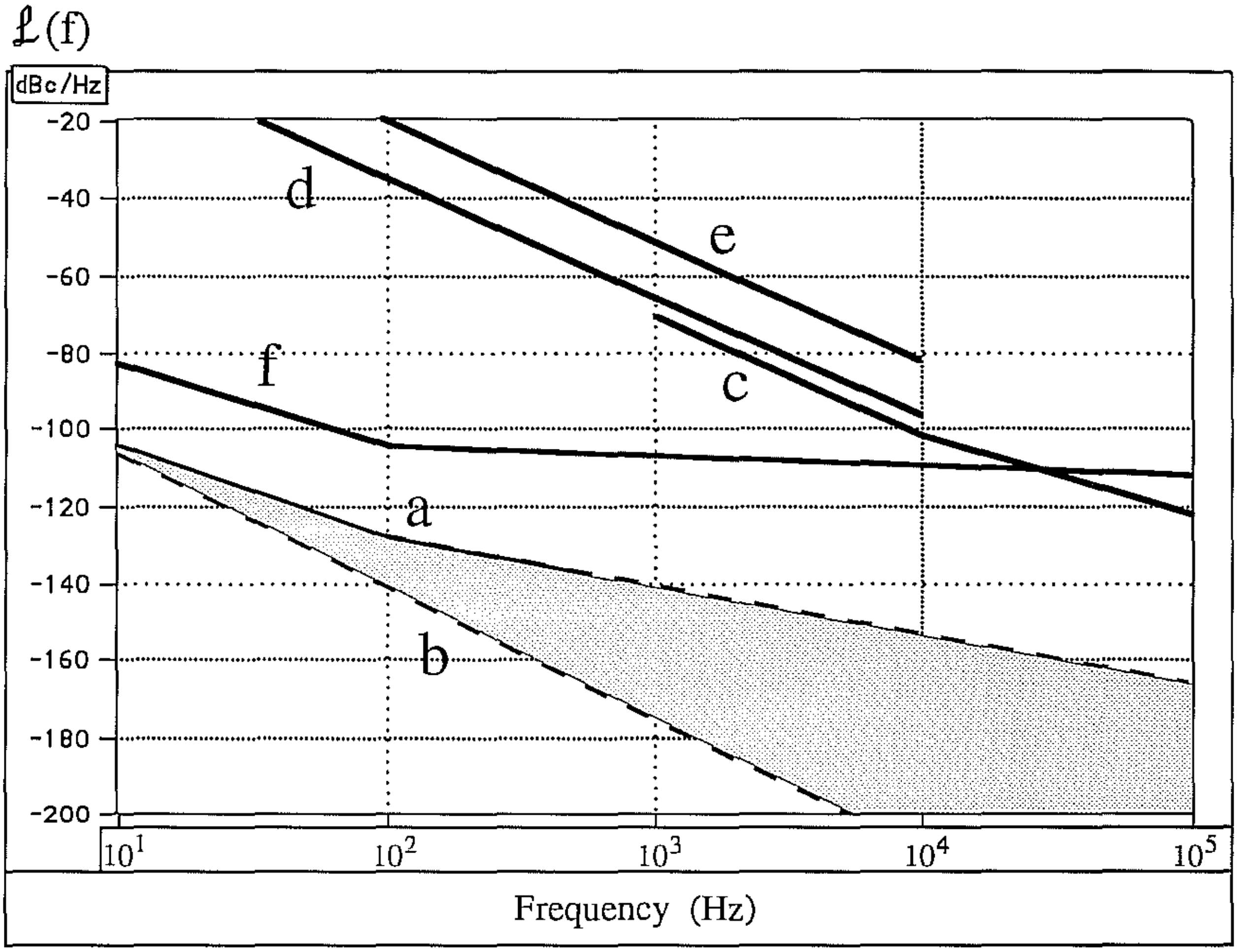

# Figure 4.24

SLOSC oscillator SSB phase noise: **a**. The bold line is the recently measured phase noise from the unfiltered port at liquid helium temperatures. Above 500 Hz the measurement was limited by the noise floor; the dashed continuation shows the calculated spectrum that would be observed if the noise floor was reduced: **b**. Predicted liquid heluim temperature phase noise of the signal taken from the filtered port: **c**. High Temperature Superconductor (HTS) oscillator phase noise (Khanna and Schmidt, 1991): **d**. Typical free running X-band DRO: **e**. Typical X-band GUNN oscillator: **f**. 5 MHz crystal oscillator performance at X-band.

phase noise in the shaded region of figure 4.24, depending on the residual phase noise of the SLOSC and T-SLOSC filters, as well as the off resonance feed through.

In conclusion we have successfully determined the optimum operating bias conditions to reduce flicker phase noise for the GaAs components in an all cryogenic SLOSC loop oscillator. A very low phase noise at the unfiltered port of -140 dBc/Hz at 1 kHz for a SLOSC oscillator with 17 dBm output has been determined. The results are consistent with the filtered port phase noise performance of -175 dBc/Hz but to directly measure this would require the use of a starved local oscillator mixer with a very low noise floor.

# CHAPTER 5

# NOISE PERFORMANCE OF THE RE-ENTRANT CAVITY PARAMETRIC TRANSDUCER

## 5.1 INTRODUCTION

Previous chapters have focused on the development of ultra-low noise sapphire resonators and oscillators, which potentially have many applications. At the University of Western Australia we require an ultra-low noise pump oscillator to operate in a low noise parametric transducer system, attached to the UWA gravitational wave detector. This and the remaining chapters deal with its inclusion into the parametric transducer system.

The preliminary system tested at the University of Western Australia (Veitch, 1986) is illustrated in figure 5.1. It consisted of a 1.5 tonne Nb bar which was supported near its centre by a low loss high isolation mechanical filter (Veitch, Jan 1991). A Nb bending flap was monitored by a microwave X-band re-entrant cavity parametric transducer (Mann, 1982; Linthorne and Blair, 1992), which is also bonded to the antenna. The pump power for the transducer was provided by an ultra stable low noise tunable oscillator, based on a Sapphire Loaded Superconducting Cavity (SLOSC).

This chapter focuses on some experimental noise results obtained when using the T-SLOSC oscillator as the pump oscillator in the UWA gravity wave detector system. Results reveal that the series noise level was limited by inadequate vibration isolation to the gravitational wave detector. This problem was identified in 1990 and now the dewar and vibration isolation system has been redesigned. An upper experimental limit for the displacement sensitivity per square root hertz has been determined, and is compared with the expected results given that certain improvements in the vibration isolation and parametric transducer are made.

## 5.2 EXPERIMENTAL RESULTS

The TSLOSC cavity configured as shown in figure 4.2a was set up as a loop oscillator and operated as the pump source for the re-entrant cavity parametric transducer. For better versatility in tuning range and resolution, the T-SLOSC loop oscillator was mixed with a HP-8662A, 0-1.2 GHz synthesiser as shown in figure 5.1. The vibrational modes presented in figure 4.3 to 4.6 were present in the phase and amplitude noise spectrum of the pump oscillator. However these modes could not be seen at the output of the gravity wave detector even when the T-SLOSC dewar was excited with a large impulse, showing that the vibrational series noise was well above the pump oscillator noise.

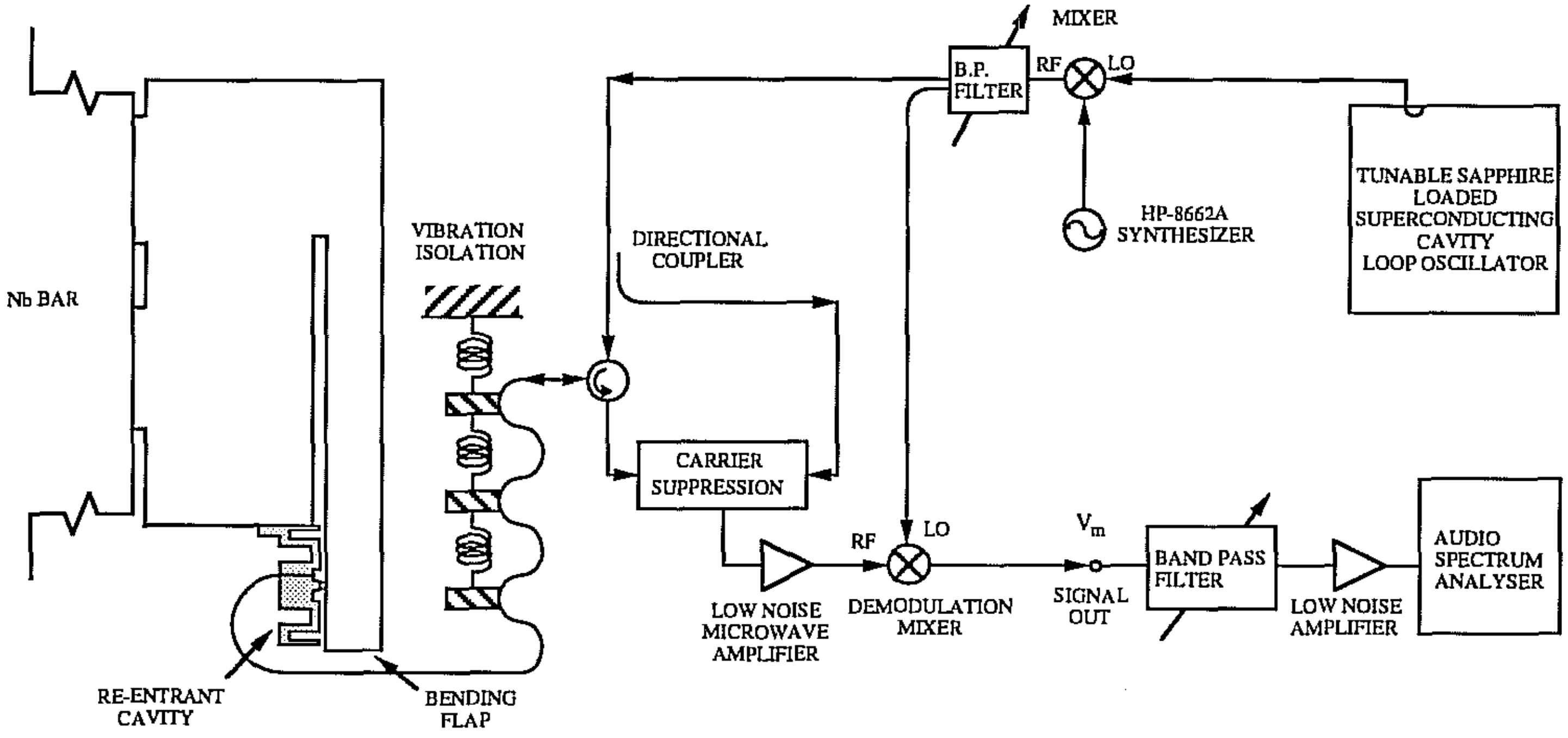

Figure 5.1

The UWA gravitational radiation antenna with a parametric re-entrant cavity transducer and contacting coaxial cable coupling. To analyse the series noise an off resonance frequency band is selected by a variable band pass filter at the output of the demodulation mixer, amplified by a low noise amplifier and then read by an audio spectrum analyser.

The series noise level was measured by using a variable band pass filter and a low noise amplifier at the output of the demodulation mixer, as shown in figure 5.1. The filter was needed to attenuate any unwanted peaks outside the frequency band of interest so the spectrum analyser would not overload when trying to measure the off resonance noise floor. The series voltage noise when referred to the mixer is plotted for various incident powers in figure 5.2. This voltage was seen to vary by over 10 dB depending on the level of excitation in the laboratory. No clear dependence on input power was determined in the presence of this changing level of vibration.

The smallest series noise measured at the mixer demodulation output at 700 Hz was $V_m = 2.6\cdot10^{-7}$ V/$\sqrt{\text{Hz}}$. The demodulation mixer sensitivity was $k_m \sim 10$ V/$\sqrt{\text{W}}$, the microwave amplifier gain was $G_{amp} = 30$ dB, the losses from the transducer to mixer RF port was L = 13 dB. From these values the effective noise power at the parametric transducer can be calculated to be;

$$N = \frac{V_m^2\,L}{k_m^2\,G_{amp}} = 1.9\cdot10^{-17}\ \text{W/Hz} \qquad (5.1).$$

The series noise component referred to the cavity displacement is given by;

$$S_{X2} = \sqrt{N}/S_{PX2} \quad \text{m}/\sqrt{\text{Hz}} \qquad (5.2),$$

where $S_{PX2}$ ($\sqrt{\text{W}}$/m), is the total forward transductance scattering parameter of the re-entrant cavity transducer, and is equal to the (modulated signal power)$^{1/2}$/(secondary mass displacement) and was $2\cdot10^7$ $\sqrt{\text{W}}$/m when this measurement was taken. The form of the forward transductance is given in appendix B section B.2.1. This noise component can thus be calculated to be approximately $2\cdot10^{-16}$ m/$\sqrt{\text{Hz}}$.

This level of series vibrational noise can also be calculated from calibrating the antenna (Veitch, 1991), where;

$$S_{X2} = \frac{dx}{dV}V_m = \frac{\Delta x}{2\,\Delta V}V_m \quad \text{m}/\sqrt{\text{Hz}} \qquad (5.3).$$

Here $\Delta x \sim 10^{-10}$, and is the position bandwidth of the re-entrant cavity, while $\Delta V = 50$ mV, and is the voltage bandwidth of the re-entrant cavity (Linthorne 1991). From this method $S_{X2}$ is calculated to be $2.6\cdot10^{-16}$ m/$\sqrt{\text{Hz}}$, which is in good agreement with (5.2).

## 5.3 DESIGN IMPROVEMENTS

The new design contains two new features to limit the excess vibrations exciting the antenna. Firstly, a new dewar system has been constructed (Blair et al, 1992) which eliminates the need for thermal grounds were thought to be a major source of excess

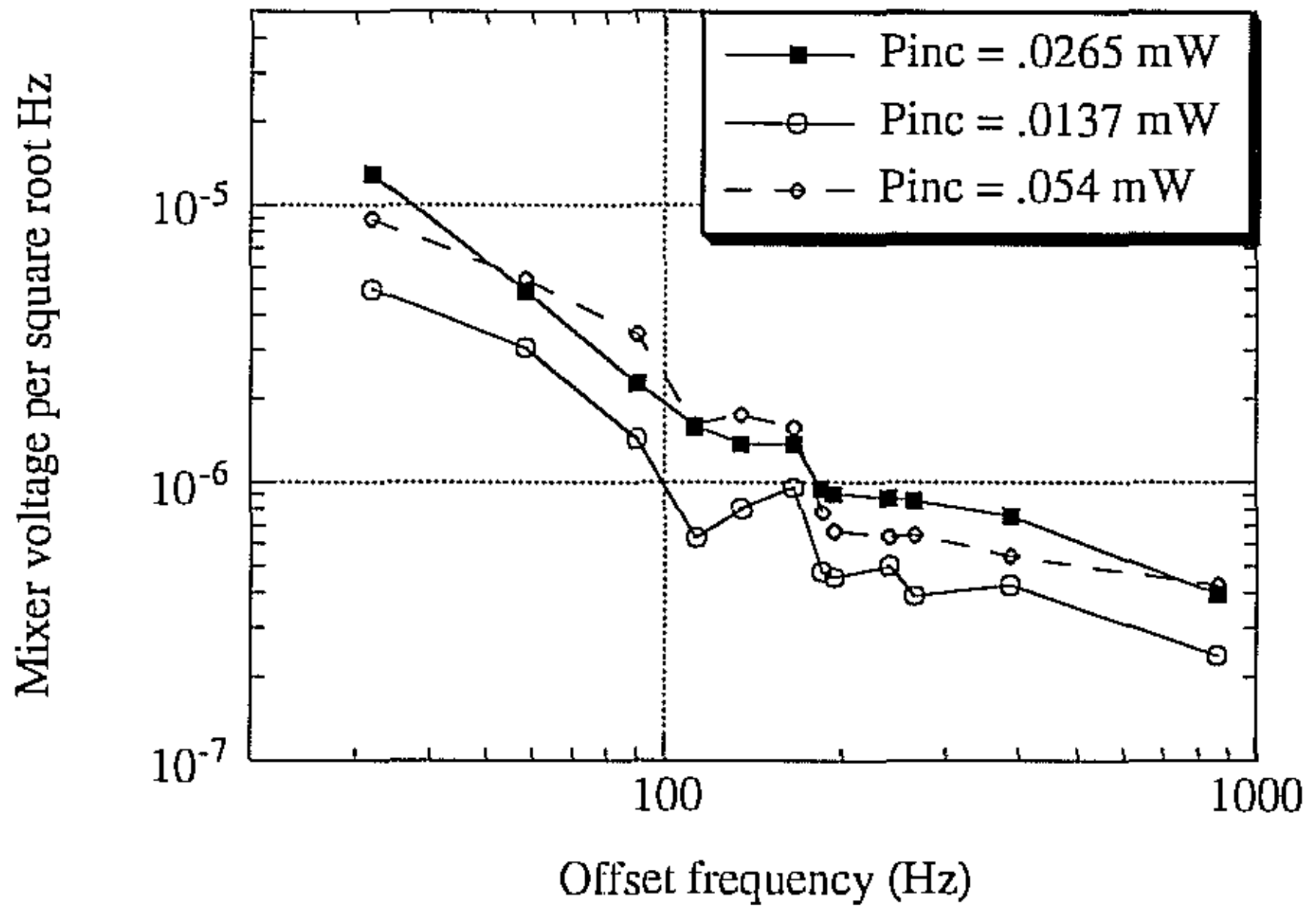

# Figure 5.2

Series noise measured in volts per square root Hz versus the demodulated offset frequency at the mixer output, for a varying re-entrant cavity incident power (Pinc). The normal modes of the resonant antenna occur at around 700 Hz.

noise. Secondly, the parametric transducer has been modified to include a non-contacting microwave coupling (Ivanov et al, 1993).

### 5.3.1. Dewar Design

The original cryostat arrangement has been described in detail by (Veitch, 1987). A schematic is shown in figure 5.3. The experimental can was supported by its pumping tube from an eight layer vibration isolation stack (alternating layers of rubber and lead discs) situated in the central tower. The dewar had a single thermal vacuum. The thermal conduction path between the 4 K shield, the liquid helium tanks and the experimental can was provided by 250 000 loosely bundled 0.08 mm diameter copper wires inside the thermal vacuum. Similar sheaves diverted heat conducted down the suspension tube into the 77 K, 30 K and 4 K thermal shields. This design allowed an economical cryostat to be constructed without a major internal vacuum vessel, and gave good cryogenic performance.

Thermal grounds were essential to this design to maintain the experimental chamber in thermal equilibrium with the small diameter helium tanks. Unfortunately, the mechanical isolation expected of the vibration isolation stack was partially short circuited by the presence of these thermal grounds. The isolation of the vibration isolation stack at 700 Hz has been calculated to be about 140 dB (Blair et al, 1991). Three stages of further isolation installed beneath the catherine wheel suspension for the niobium bar were insufficient to isolate against the level of excitation through the thermal grounds.

The new antenna system is shown in figure 5.4. Complete removal of the 4K shield has allowed the installation of a much larger experimental chamber within the existing dewar. This enables the isolation of noise sources ( such as boiling cryogens and structural vibrations) within the dewar from the antenna suspension. The antenna is now cooled by low pressure ($10^{-4}$ Torr) helium exchange gas in the experimental chamber, rather than via the sheaves of fine copper wires. The liquid helium is stored in the eccentric annulus formed by a double walled chamber of external diameter 970 mm, internal diameter 870 mm and length 3m.

The niobium antenna and its suspension have no mechanical contact with the cryostat as the entire assembly is supported from the isolation stack by a single Ti-6Al-4V alloy rod. The rod hangs in the centre of the 100 mm diameter suspension tube which supports the experimental chamber from the vibration isolation chamber. These two volumes form a single vacuum space separate from that of the main dewar vacuum. The experimental

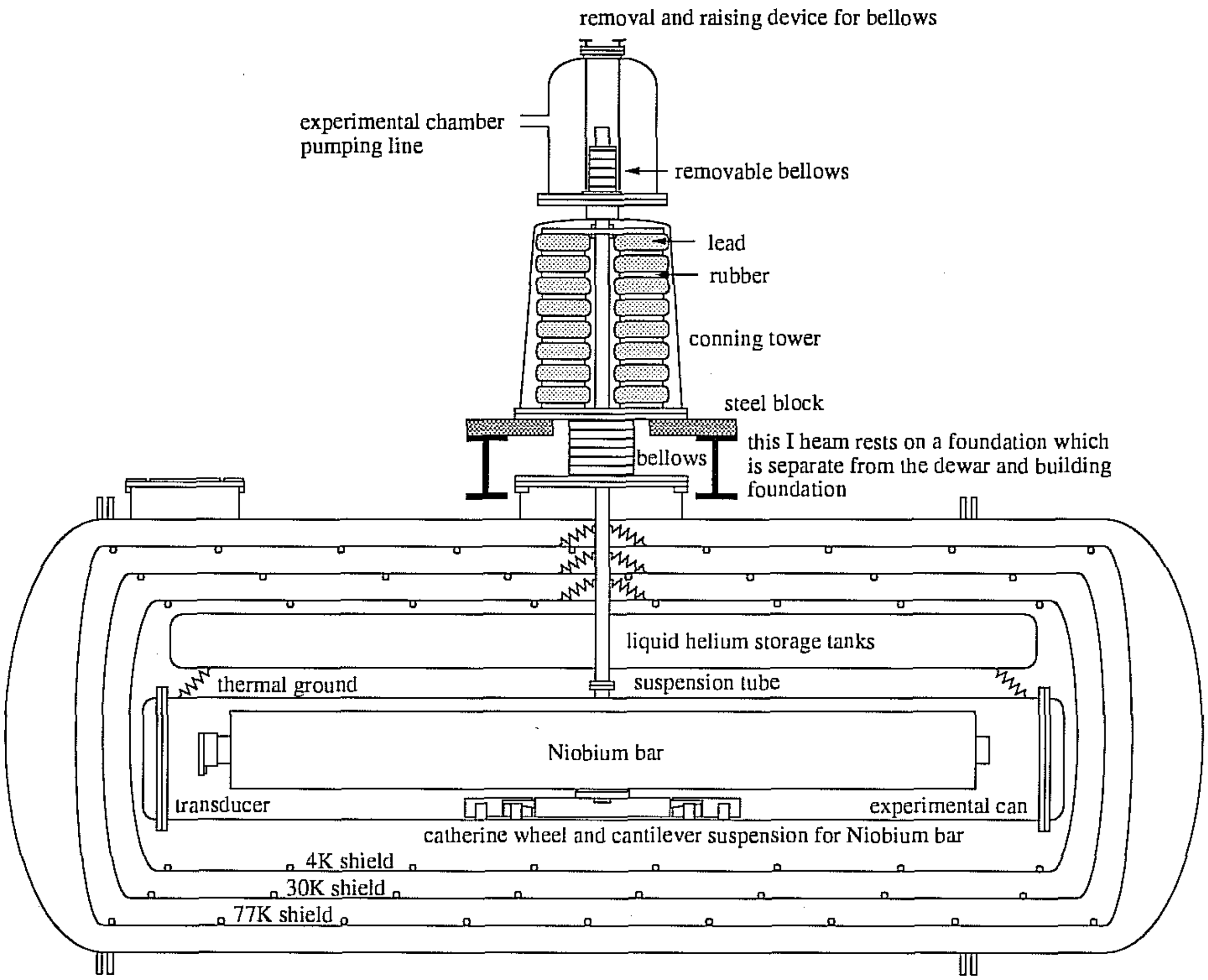

Figure 5.3

Schematic of the original cryostat arangement

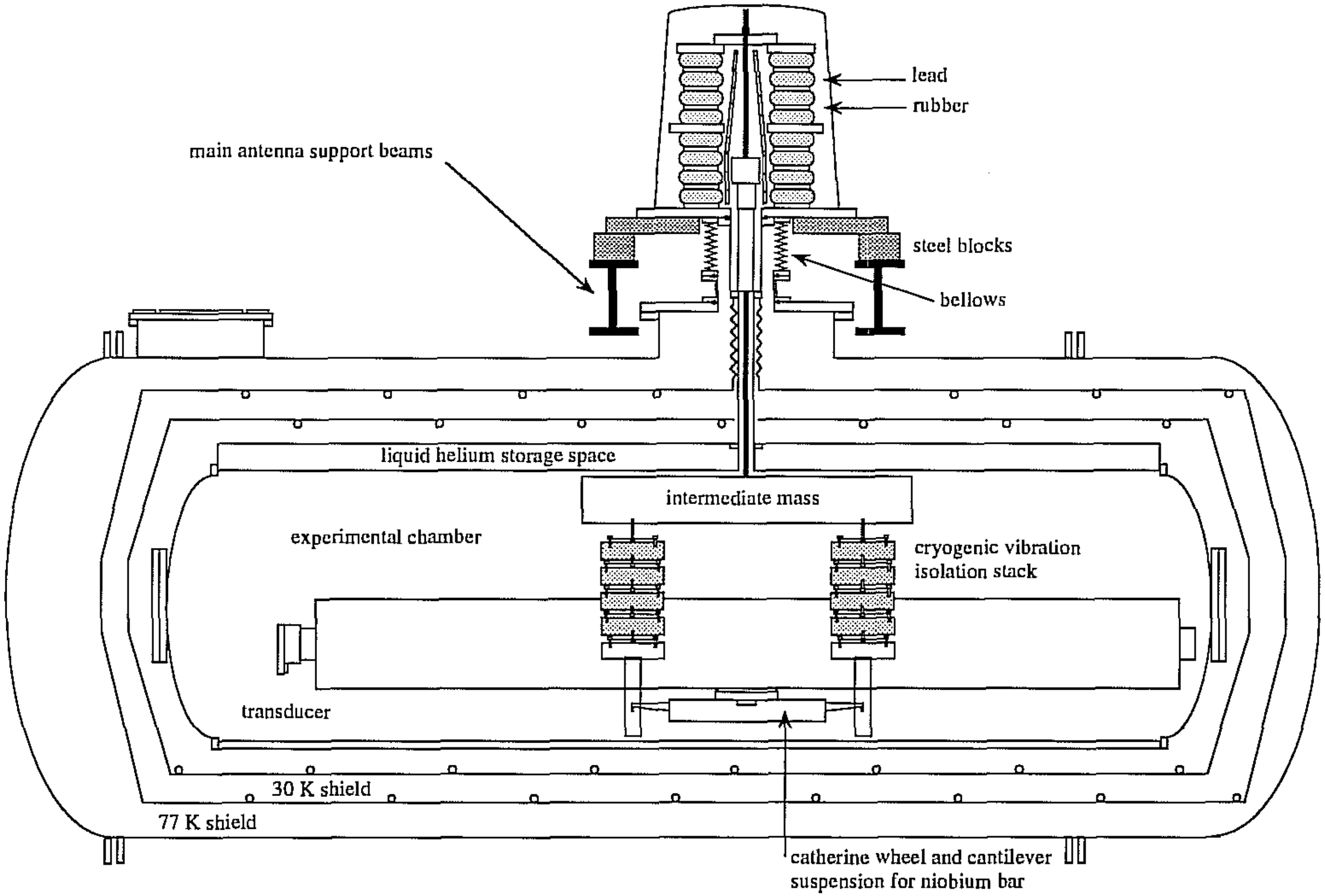

# Figure 5.4

Schematic of the new cryostat with side view of antenna.

chamber contains helium exchange gas to cool the niobium bar while the main dewar vacuum provides thermal isolation for the experiment.

### 5.3.2 Non-contacting transducer coupling

Direct connection of the external microwave pump oscillator to the transducer via a coaxial cable can degrade the antenna's acoustic Q and create an additional pathway for environmental seismic noise to enter the antenna. A cable connection must be long so it can pass through a mechanical vibration isolation stage. Also it must have a small diameter to reduce elastic stiffness. Such a cable has significant electrical losses even at 4.2 K, and attenuates the modulated gravity wave signal by about 6 dB before amplification, increasing the effective amplifier noise power when referred to the re-entrant cavity transducer.

To improve the noise performance of the antenna transducer system electromagnetically coupled microstrip antennas (transceivers) have been devised to realise a non-contacting microwave coupling to the re-entrant cavity transducer (Ivanov et al, 1993). The transceiver attached to the surface of the Nb bar must have minimal size and mass, have no mechanical modes near the bar resonant frequency (~710 Hz) and be attached in a manner that the intrinsically high acoustic Q of the antenna is not degraded. For this reason microwave horns (Blair et al, 1992) have been abandoned in favour of microstrip antennas. A rectangular patch antenna was chosen due to simplicity in design fabrication and adjustment. Details on design principle and how they operate are given in (Ivanov et al, 1993).

## 5.4 TRANSDUCER SENSITIVITY

The calculation of the gravitational wave detector sensitivity requires a detailed analysis of all narrow band and wide band noise sources along with the scattering parameters of the transducer system. This full analysis is presented in chapter 7. In this section the performance of the transducer only is considered. The transducer potentially supplies a narrow band noise component due to the a.m. noise of the pump oscillator causing back action, as well as additive electronic series noise. In this section only the improvement in the transducer wide band series noise due to the addition of the non-contacting microwave coupling, and a cryogenic amplifier are considered.

In the past the series noise component has been dominated by external vibrations. The new system should eliminate this noise source leaving electronic noise as the dominant series noise term in the system. A critical electronic noise source for a parametric transducer is the microwave phase noise, this has been alleviated in our system by the development of the ultra-low noise Sapphire Loaded Superconducting Cavity (SLOSC) oscillator. In the absence of this noise the dominant wide band component competing with the gravity wave signal is the noise power introduced by the first amplifier after reflection of the pump oscillator off the parametric transducer.

The effective energy per quadrature of noise introduced by the amplifier when referred to the re-entrant cavity transducer can be approximated by;

$$N \approx L_1 L_2 k_B T_{eff}/2 \quad \text{W/Hz} \qquad (5.4).$$

Assuming the noise energy is distributed equally to each quadrature (amplitude and phase). Here $k_B$ is Boltzman's constant and $T_{eff}$ is the effective noise temperature of the amplifier. As shown in figure 5.5, $L_1$ is the signal power loss from the transducer to the circulator and $L_2$ is the signal power loss from the circulator to the amplifier. Hence $L_1$, $L_2$ and $T_{eff}$ are the important parameters that must be minimised to decrease the wide band series noise term.

Previously the UWA antenna has been configured with a room temperature amplifier after the cryogenic carrier suppression system. This means that long lossy coaxial cables must transport the signal out of the cryogenic dewar before amplification. Incorporating a cryogenic amplifier reduces both $T_{eff}$ and $L_2$. A suitable amplifier that operates at X-band has been successfully tested and operated at a temperature of 8 K (see chapter 4). Carrier suppression ensures that the amplifier operates in the low input signal regime, where flicker noise (caused by non-linear up-conversion) is does not occur. In this regime the amplifier should operate at a noise temperature close to its physical temperature, but this is yet to be confirmed for this particular amplifier.

In the past direct coupling to the transducer was achieved with a long piece of coaxial cable between the circulator and re-entrant cavity transducer. This cable was about 1 meter long and 0.5 mm in diameter. It passed from the carrier suppression electronics through a mechanical low pass vibration isolation filter, and then was coupled directly to the re-entrant cavity transducer. Replacing this cable with the low loss microwave strip line antennas reduces $L_1$ by about 4 dB. Table 5.1 summarises the improvements in effective noise power per Hz, when introducing the cryogenic amplifier and microwave strip line coupling. It is evident that the greatest improvement is due to the addition of the cryogenic amplifier which reduces the effective noise power by 21 dB.

The displacement sensitivity is calculated by (5.2). If environmental vibrations were eliminated then the previous system would have been limited by electronic noise. The equivalent displacement sensitivity of the electronic noise is calculated to be to $S_{X2} = 5\cdot10^{-18}$ m/$\sqrt{\text{Hz}}$ which is 32 dB smaller than the measured value of $2\cdot10^{-16}$ m/$\sqrt{\text{Hz}}$

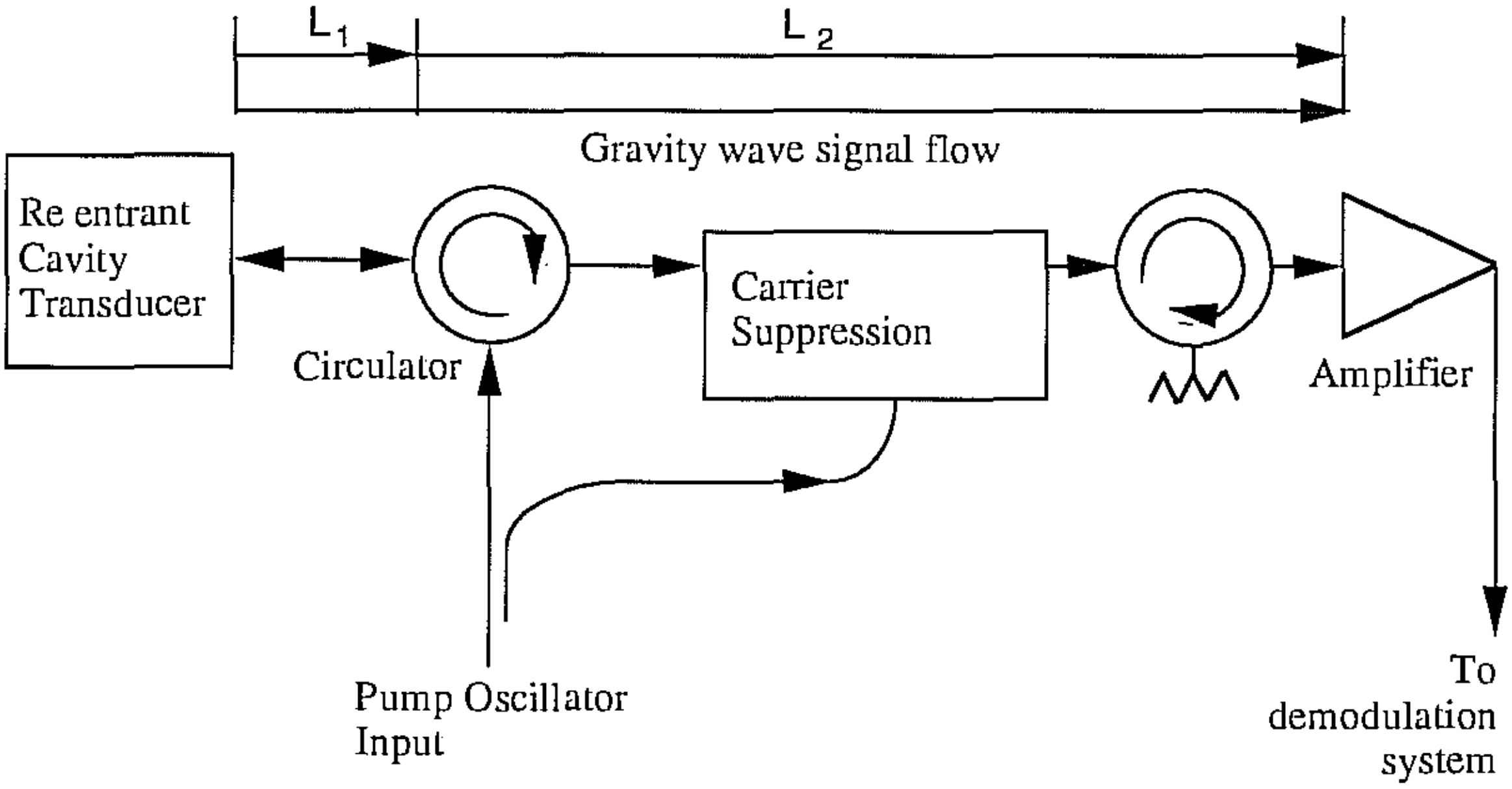

Figure 5.5

The losses as seen by a gravity wave signal are broken into two components; $L_1$ is the signal power loss from the transducer to the circulator; $L_2$ is the signal power loss from the circulator to the amplifier. $L_1$ is reduced by replacing a long piece of coaxial cable with the microstrip coupling. $L_2$ is reduced by including a cryogenic amplifier, eliminating signal loss before amplification via the long lossy cable that leads from the cryogenic dewar to the room temperature data acquisition.

| Configuration | $L_1$ (dB) | $L_2$ (dB) | $T_{eff}$ (K) | $L_1L_2T_{eff}$ (K) | Improvement (dB) |
|---|---|---|---|---|---|
| #1 | 6 | 7 | 300 | 6000 | 0 |
| #2 | 6 | 2 | 8 | 50.5 | 21 |
| #3 | 2 | 2 | 8 | 20 | 4 |

Table 5.1

Shows the change in the losses ( $L_1$ and $L_2$ ), the amplifier noise temperature ( $T_{eff}$ ), the effective amplifier noise temperature referred to the transducer output ( $L_1L_2T_{eff}$ ), and the overall improvement in noise power when compared to the previous antenna configuration (#1); after adding the cryogenic amplifier (#2); and then the microstrip coupling (#3).

presented in section 5.2. The new vibration isolation system should thus allow a 32 dB improvement of the series noise component. Further more the addition of the cryogenic amplifier should create a 21 dB improvement in effective noise, and the addition of the strip line coupling should create a 4 dB improvement in effective noise. These expected improvements should reduce $S_{X2}$ to $3 \cdot 10^{-19}$ m/$\sqrt{Hz}$ for the new system, corresponding to an overall reduction of 57 dB.

This chapter has looked at the re-entrant cavity transducer series noise sensitivity. The limiting effect in the past has been due to seismic vibrations. To calculate the gravity wave detector sensitivity the transfer function of a two-mode resonant bar coupled to a parametric transducer must be known, along with the effect of all series and narrow band noise sources. In chapter 6 the transfer function is determined, and in chapter 7 all series and narrow band noise sources are considered in conjunction with the transfer function derived, to determine the potential detector sensitivity.

# CHAPTER 6

# A GRAVITATIONAL RESONANT BAR ANTENNA COMBINED WITH A RESONANT PARAMETRIC TRANSDUCER

## 6.1 INTRODUCTION

Parametric transducers for resonant bar detectors are under development by the gravitational radiation research groups at Moscow (Braginsky et al, 1981), Louisiana (Oelfke and Hamilton, 1983; De Aguiar, 1990), Tokyo (Tsubono et al, 1986), Rome ( Bordoni et al, 1986), Western Australia (Veitch et al, 1987), and Rochester (Bocko et al, 1989; Fisher et al, 1991). A parametric transducer is a resonant circuit tuned to the same frequency as an external pump oscillator. The resonant bar attached to the transducer will cause a modulation of the transducer's resonant frequency, which will inturn modulate the incident pump signal. The modulated pump signal is then amplified by a low noise amplifier at the carrier frequency, and demodulated using part of the original pump signal as the reference. A schematic of the new parametric transducer system at the University of Western Australia is shown in figure 6.1.

In this chapter a general representation of the transfer function of a two-mode antenna interacting with a resonant parametric transducer capacitively is presented. The model was verified by applying it to the gravitational wave detector at the University of Western Australia. Assuming linear transfer function theory, a two-mode oscillator is of fourth order, and can be represented by the first order matrix equation given by (6.1).

$$\dot{\underline{x}} = \mathbf{A}\,\underline{x} + \mathbf{B}\,\underline{u} \tag{6.1}$$

Here **A** is a 4 by 4 matrix, $\underline{x}$ the state vector, $\underline{u}$ the input vector and **B** the matrix which relates the inputs to the states. Taking the Laplace transformation of (6.1) gives;

$$\underline{X}(s) = \mathbf{G}(s)\,\underline{U}(s) + \mathbf{H}(s)\,\underline{x}(t{=}0) \tag{6.2}$$

where $\mathbf{G}(s) = (s\mathbf{I}-\mathbf{A})^{-1}\,\mathbf{B}$ and $\mathbf{H}(s) = (s\mathbf{I}-\mathbf{A})^{-1}$ .

Here **G(s)** is the transfer function of the state vector per input force, and **H(s)** is the transfer function of the stste vector per initial condition. The addition of the parametric transducer modifies the matrix **A** and hence the transfer function, changing the normal mode Q values and resonant frequencies of the resonant bar antenna. This chapter explains these effects enabling a general form of the transfer function to be calculated. This allows optimisations based on this model to be made for the UWA system. In the

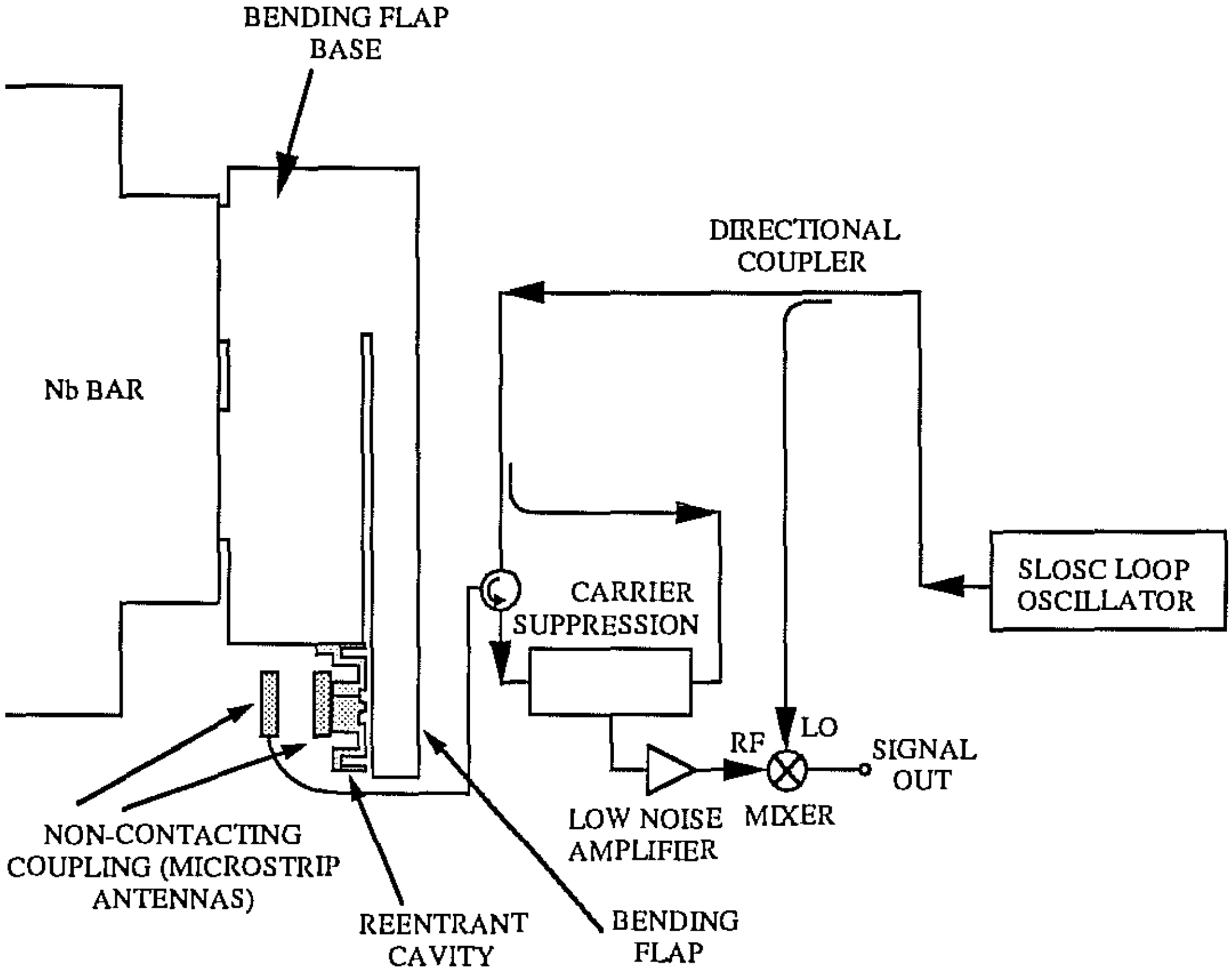

# Figure 6.1

Schematic diagram of the second generation 1.5 tonne niobium bar antenna, with a bonded bending flap and re-entrant cavity transducer.

next chapter a noise simulation program is presented based on this transfer function and is used to estimate the sensitivity of the detector at UWA.

By expanding the analysis of the UWA system to a three-mode model to include the bending flap base (shown in figure 6.1), the effect of the glue joint bonding the flap base to the bar was determined. This model enabled a better fit of experimental data, compared to the two-mode model.

## 6.2 TWO-MODE ANTENNA MODELS

A resonant bar antenna with a resonant secondary mass, may be modelled as a two-mode coupled harmonic oscillator. To solve this model, Paik (Paik, 1974) defined a detuning parameter based on the arithmetic mean of the uncoupled frequencies. This approach was used by (Veitch, 1991) and (Linthorne et al, Jan 1990) when considering the interaction of a re-entrant cavity parametric transducer with a two-mode resonant bar antenna. The equations given by Veitch are complicated first order approximations. To determine the detuning parameter and secondary mass, a cumbersome iterative technique was performed. By defining a detuning parameter based on the geometric mean of the uncoupled mode frequencies (Tobar et al, 1991), an analytical solution for the detuning parameter and secondary effective mass is obtained in this section. Taking the analysis further, a simple formula is derived which enables straight forward interpretation of the electro-mechanical couplings. Incorporating the interaction as a function of pump oscillator offset frequency, the transfer function at general offset frequencies is determined, correcting previous theory for a parametric transducer (Blair, 1980; Linthorne, 1991). This is verified experimentally.

### 6.2.1 Two-mode antenna model with no transducer influence

Resonant bar antennas usually have a smaller resonant mass attached to the bar. The secondary mass acts as a mechanical transformer, amplifying the bar's vibrations and thereby improving the impedance match of the transducer to the bar. The combined system forms a two-mode coupled harmonic oscillator as shown in figure 6.2.

The model has the following parameters (subscript 1 and 2 refer to the bar and secondary mass respectively); $m_{1,2}$ is the effective mass, $k_{1,2}$ is the spring constant, $h_{1,2}$ is the dissipation constant, $\omega_{1,2} = (k_{1,2}/m_{1,2})^{1/2}$ is the resonant frequency, $\tau_{1,2} = 2\ m_{1,2}/h_{1,2}$ is the amplitude decay time constant, and $Q_{1,2} = \tau_{1,2}\ \omega_{1,2}/2$ is the acoustic Q.

Initially the influence of the transducer is ignored. Letting $x_3 = \dot{x}_1$ and $x_4 = \dot{x}_2$ the equation of motion of the coupled oscillator can be written in the form given by (6.1) as;

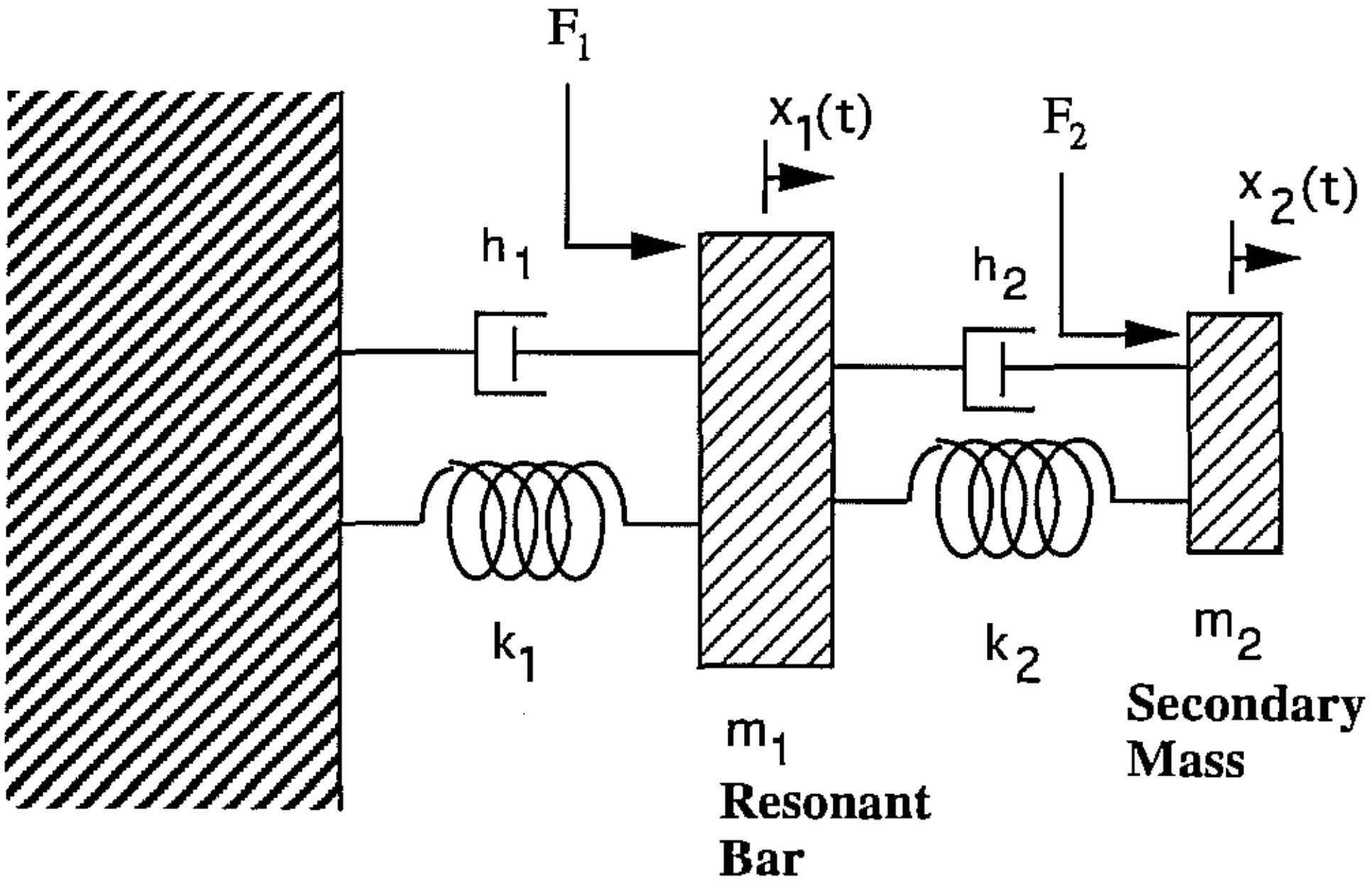

# Figure 6.2

Mechanical model of a two-mode resonant bar antenna

$$
\begin{bmatrix} \dot{x}_1 \\ \dot{x}_2 \\ \dot{x}_3 \\ \dot{x}_4 \end{bmatrix} =
\begin{bmatrix} 0 & 0 & 1 & 0 \\ 0 & 0 & 0 & 1 \\ -\omega_1^2-\frac{m_2}{m_1}\omega_2^2 & \frac{m_2}{m_1}\omega_2^2 & -\frac{2}{\tau_1}-\frac{2\,m_2}{\tau_2\,m_1} & \frac{2\,m_2}{\tau_2\,m_1} \\ \omega_2^2 & -\omega_2^2 & \frac{2}{\tau_2} & -\frac{2}{\tau_2} \end{bmatrix}
\begin{bmatrix} x_1 \\ x_2 \\ x_3 \\ x_4 \end{bmatrix}
+ \begin{bmatrix} 0 & 0 \\ 0 & 0 \\ 1/m_1 & 0 \\ 0 & 1/m_2 \end{bmatrix}
\begin{bmatrix} F_1 \\ F_2 \end{bmatrix} \qquad (6.3),
$$

where $F_1$ and $F_2$ are the input forces on mass $m_1$ and $m_2$ respectively. The characteristic equation of this matrix equation is;

$$
s^4+s^3\left(\frac{2}{\tau_1}+\frac{2}{\tau_2}\left(1+\frac{m_2}{m_1}\right)\right)+s^2\left(\omega_1^2+\omega_2^2\left(1+\frac{m_2}{m_1}\right)+\frac{4}{\tau_1\tau_2}\right)+s\left(\frac{2\omega_1^2}{\tau_2}+\frac{2\omega_2^2}{\tau_1}\right)+\omega_1^2\omega_2^2=0 \qquad (6.4).
$$

The characteristic equation for two independent second order systems with resonant frequencies $\omega_+$ and $\omega_-$, and amplitude decay time constants $\tau_+$ and $\tau_-$, is given by;

$$
s^4 + s^3\left(\frac{2}{\tau_+}+\frac{2}{\tau_-}\right)+s^2\left(\omega_+^2+\omega_-^2+\frac{4}{\tau_+\tau_-}\right)+s\left(\frac{2\omega_+^2}{\tau_-}+\frac{2\omega_-^2}{\tau_+}\right)+\omega_+^2\,\omega_-^2 = 0 \qquad (6.5).
$$

Here subscript + and - refer to the upper and lower tuned normal modes respectively. Equating coefficients of (6.4) and (6.5) gives the following relations between the normal mode and uncoupled mode parameters;

$$
\omega_+^2\,\omega_-^2 = \omega_1^2\,\omega_2^2 = \omega_0^4 \qquad (6.6),
$$

$$
\omega_+^2+\omega_-^2 = \omega_1^2+\omega_2^2\left(1+\frac{m_2}{m_1}\right)+\omega_0^2\left(\frac{1}{Q_1\,Q_2}-\frac{1}{Q_+\,Q_-}\right) \qquad (6.7),
$$

$$
\frac{2\omega_+^2}{\tau_-}+\frac{2\omega_-^2}{\tau_+} = \frac{2\omega_1^2}{\tau_2}+\frac{2\omega_2^2}{\tau_1} \qquad (6.8),
$$

$$
\frac{2}{\tau_+}+\frac{2}{\tau_-} = \frac{2}{\tau_1}+\frac{2}{\tau_2}\left(1+\frac{m_2}{m_1}\right) \qquad (6.9).
$$

If $Q_1$ and $Q_2 >> 1$, (6.6) and (6.7) are the product and sum respectively, of the solutions of the following polynomial in $\omega^2$.

$$
\omega^4-\omega^2\left(\omega_1^2+\omega_2^2\left(1+\frac{m_2}{m_1}\right)\right)+\omega_1^2\,\omega_2^2 = 0 \qquad (6.10).
$$

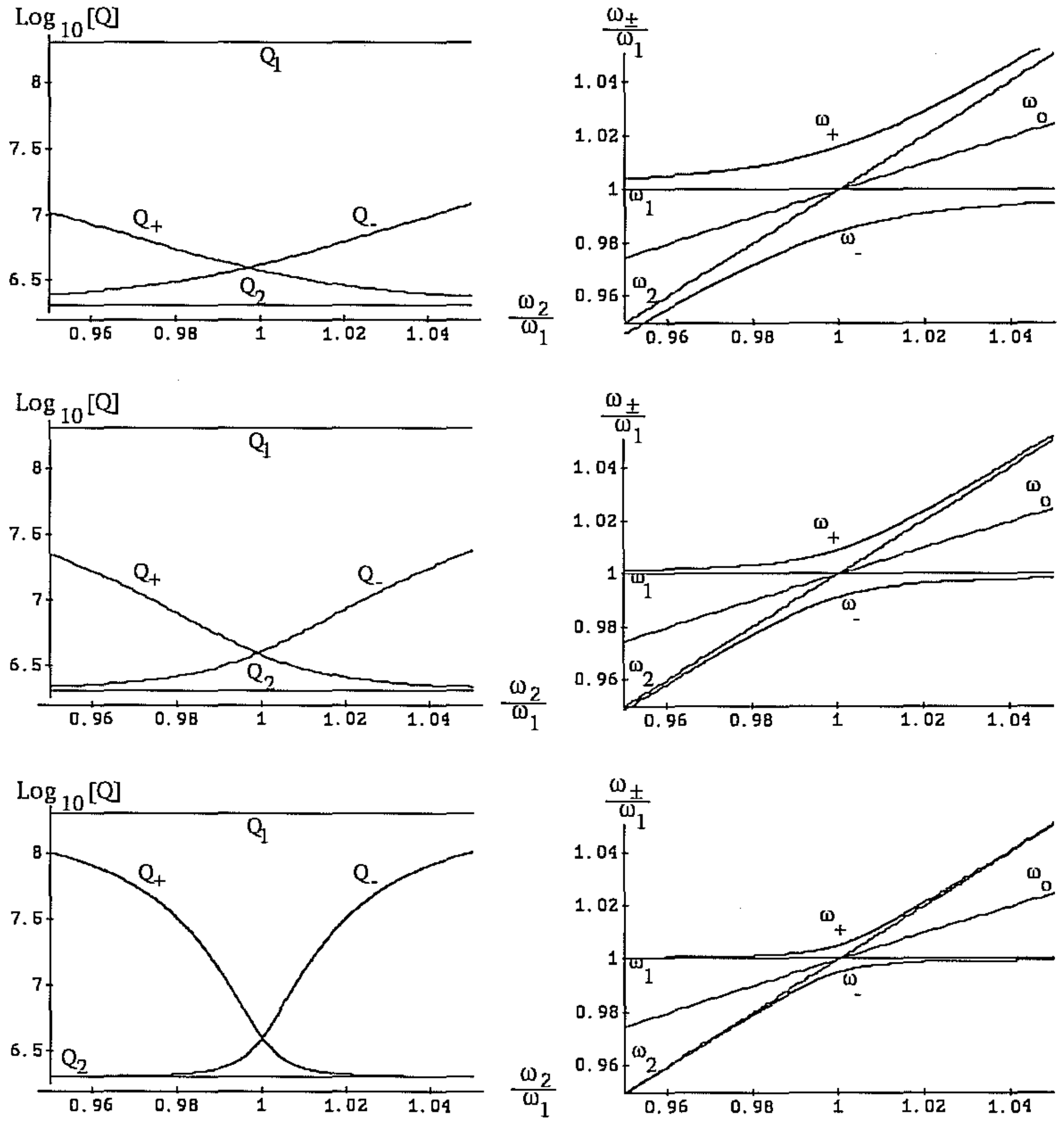

# Figure 6.3

Variation of normal mode Q values and resonant frequencies, versus secondary mass resonant frequency, with $m_2/m_1 = 10^{-3}$, $3 \cdot 10^{-3}$, $10^{-4}$ respectively.

When investigating the effect of tuning on a two-mode oscillator, Paik (Paik, 1974) defined a detuning parameter, $\delta'$, such that $\omega_{1,2} = \omega_0' (1 \pm \delta'/2)$, where $\omega_0'$ is the arithmetic mean of $\omega_1$ and $\omega_2$. Since the geometric mean is constant (6.6), a simplified solution in terms of the geometric mean is expected. Hence a new detuning parameter $\delta$ was defined (Tobar et al, 1991) such that;

$$\omega_{1,2} = \omega_0 (1+\delta/2)^{\pm 1} \qquad (6.11).$$

From (6.10) and (6.11) it can be shown that;

$$\omega_\pm^2 = \omega_0^2 \left( z \pm \sqrt{z^2-1} \right) \qquad (6.12).$$

where;

$$z = \frac{\omega_1^2 + \omega_2^2\left(1+\frac{m_2}{m_1}\right)}{2\,\omega_1\omega_2} = \frac{(1+\delta/2)^2}{2} + \frac{1+\frac{m_2}{m_1}}{2\,(1+\delta/2)^2} \qquad (6.13).$$

The frequency difference between the two normal modes is determined by both $m_2/m_1$ and $\delta$. The $m_2/m_1$ term determines the difference due to the inertial coupling between the bar and the secondary mass, while $\delta$ determines the component due to detuning. These effects are shown in figure 6.3 for selected $m_2/m_1$ ratios. Combining (6.8) and (6.9) we can solve for the uncoupled amplitude decay time constants, in terms of normal mode parameters;

$$\frac{1}{\tau_1} = \frac{\frac{1}{\tau_+\omega_+}\left(\omega_+(1+\delta/2)^2 - \omega_-\left(1+\frac{m_2}{m_1}\right)\right) + \frac{1}{\tau_-\omega_-}\left(\omega_-(1+\delta/2)^2 - \omega_+\left(1+\frac{m_2}{m_1}\right)\right)}{(1+\delta/2)^2 - (1+\delta/2)^{-2}\left(1+\frac{m_2}{m_1}\right)} \qquad (6.14),$$

$$\frac{1}{\tau_2} = \frac{\frac{1}{\tau_+\omega_+}\left(\omega_- - \omega_+(1+\delta/2)^{-2}\right) + \frac{1}{\tau_-\omega_-}\left(\omega_+ - \omega_-(1+\delta/2)^{-2}\right)}{(1+\delta/2)^2 - (1+\delta/2)^{-2}\left(1+\frac{m_2}{m_1}\right)} \qquad (6.15).$$

Solving (6.14) and (6.15) simultaneously we can determine the normal mode parameters. Figure 6.3 shows the variation of the normal mode acoustic Q values as a function of secondary mass resonant frequency for selected $m_2/m_1$ ratios, with $Q_1 = 2\cdot 10^8$ and $Q_2 = 2\cdot 10^6$.

### 6.2.2. Two-mode antenna model with a parametric transducer at small pump oscillator offsets

#### 6.2.2.1 Theory

A parametric transducer will change the resonant frequency of a mechanical oscillator. In most two-mode antenna, the parametric transducer will mostly act to change the resonant frequency of the second oscillator, since it has a much smaller effective mass and spring constant. For small perturbations in the normal mode frequencies, we assume that;

$$\omega_{\pm L} - \omega_{\pm} = \frac{d\,\omega_{\pm}}{d\,\omega_2}\left(\omega_{2L} - \omega_2\right)\Big]_{\omega_1 = \text{const}} \qquad (6.16),$$

where $\omega_{\pm L}$ are the loaded normal mode resonant frequencies, and $\omega_{2L}$ is the loaded secondary mass resonant frequency. For pump oscillator frequency offsets less than the transducer bandwidth, the loaded secondary mass resonant frequency is given by (Veitch, 1991);

$$\omega_{2L} \approx \omega_2\left(1 + \beta_0\, Q_e \frac{d\Omega}{\Omega_0}\right) \qquad (6.17),$$

where $\beta_0$ is the centre of resonance electromechanical coupling of the transducer to the secondary mass when $\delta\Omega=0$, $Q_e$ is the transducer loaded electrical Q, $\Omega_0$ is the transducer resonant frequency, and $\delta\Omega$ is the frequency offset of the pump from the transducer resonance. Combining (6.16) and (6.17), we can show that for small offsets from resonance;

$$K_{\pm} = \frac{d\omega_{\pm}}{d(\delta\Omega/\Omega_0)} \approx \frac{\omega_{\pm L} - \omega_{\pm}}{\delta\Omega/\Omega_0} = \left(Q_e\, \beta_0\, \omega_2\right)\frac{d\,\omega_{\pm}}{d\,\omega_2}\Big]_{\omega_1 = \text{const}} \qquad (6.18).$$

By implicitly differentiating (6.12) w.r.t. $\omega_2$, and substituting into (6.18), we can express the detuning parameter as;

$$(1 + \delta/2)^2 = \frac{\omega_+}{\omega_-}\left(\frac{1}{1+\beta_r}\right) + \frac{\omega_-}{\omega_+}\left(\frac{\beta_r}{1+\beta_r}\right); \quad \text{where} \quad \beta_r = \frac{K_+\,\omega_-}{K_-\,\omega_+} \qquad (6.19).$$

Combining (6.12) and (6.19) the mass ratio can be expressed as;

$$\frac{m_2}{m_1} = \beta_r\left(\frac{\omega_+^{\,2} - \omega_-^{\,2}}{(1+\beta_r)\,\omega_+\,\omega_-}\right)^2 \qquad (6.20).$$

Substituting (6.19) into $d\omega_{\pm}/d\omega_2$, and this term into (6.18), the total electromechanical coupling can be shown to be;

$$\beta_0 = \frac{K_+}{Q_e\,\omega_+} + \frac{K_-}{Q_e\,\omega_-} = \beta_{0+} + \beta_{0-} \qquad (6.21),$$

where $\beta_{0\pm}$ are the centre of resonance electromechanical couplings to the $\omega_{\pm}$ modes. Thus $\beta_r=\beta_{0+}/\beta_{0-}$, is the ratio of the electromechanical couplings. From a measurement of $K_+$, $K_-$, $Q_e$, $\omega_+$ and $\omega_-$, we can calculate $\delta$, $m_2/m_1$, $\beta_{0+}$ and $\beta_{0-}$. This then enables us to determine the uncoupled mechanical mode frequencies and Q factors.

If the tuning of the two masses are exact, ie $\delta = 0$, it can be shown from (6.19) that;

$$\beta_{0+} = \frac{\omega_+}{\omega_-} \beta_{0-} \Big]_{\delta=0} \qquad (6.22).$$

Here the frequency split will be solely due to the mass ratio of the resonant bar and secondary mass. The positively split mode has a slightly higher coupling, which is associated with a slightly lower Q value (figure 6.3). Hence the + mode will have slightly more of the secondary mass eigenmode associated with it, and the - mode will have slightly more of the resonant bar eigenmode associated with it. If the resonant bar and secondary mass are detuned, the coupled modes will start to become more identifiable with either the resonant bar or secondary mass. The mode associated with the secondary mass will increase its coupling. This is highlighted in figures 6.5 and 6.6, where the secondary mass flapish mode (+ mode) has a larger slope than the barish mode (-mode), these slopes are directly proportional to the components of coupling as given by (6.21). In the extreme case, when the resonant bar and secondary mass are totally detuned, the electro-mechanical coupling due to the bar mode will approach zero, and the secondary mass mode approach $\beta_0$ (assuming (6.16) is true).

#### 6.2.2.2 Experimental Verification

In 1987 and 1989 the secondary oscillator (bending flap #1) was tuned above the bar frequency. The base of this bending flap was glued to the resonant bar to create a coupled mode oscillator. In 1991 a new optimised secondary oscillator was built (bending flap #2) (Linthorne, 1991), and was tuned below the bar frequency. Table 6.1 and 6.2 compare these three experimental runs of the UWA system with the two-mode model. Table 6.1 summarises the important measured parameters, details on how these measurements are achieved are presented in (Linthorne, 1991). Table 6.2 summarises some calculated parameters based on the two-mode model and measurements presented in table 6.1.

From table 6.2 it can be seen that the calculated bar frequencies from the two-mode model are slightly less than the expected value. Also only the 1991 experiment with the redesigned flap gave a meaningful result with regards to the Q value of the resonant bar. However the rest of the model gave good results, determining the bending flap effective mass to be about .53 kg and .45 kg for bending flap #1 and #2 respectively, very close to estimated values (Linthorne, 1991). According to this model bending flap #1 decreased its resonant frequency by 5 Hz when it was rebonded between 1987 and 1989. This was due to the different boundary conditions provided by the glue joint in 1987 and 1989. The resonant frequency of bending flap #1 was highly dependent on this bond. In section 6.4 it is shown that the bending flap acoustic Q is also highly dependent on this bond when a three-mode

| measured parameter | 1987 | 1989 | 1991 |
|---|---|---|---|
| $\omega_+/2\pi$ (Hz) | 725.62 | 722.014 | 711.334 |
| $\omega_-/2\pi$ (Hz) | 703.15 | 702.000 | 688.859 |
| $Q_+$ ($10^6$) | 2.5±.1 | .83±.03 | 16.0±.5 |
| $Q_-$ ($10^6$) | 9.0±.3 | 11.3±.05 | 3.7±.1 |
| $\Omega_0/2\pi$ (GHz) | 10.432736 | 10.3741909 | 9.773200 |
| $Q_1$ | ---------------------------- | $(2.3 \pm .3)\ 10^8$ | ---------------------------- |
| $\omega_1/2\pi$ (Hz) | ---------------------------- | 709.7 | ---------------------------- |
| $m_1$ (kg) | ---------------------------- | 755 | ---------------------------- |

# Table 6.1

Measured parameters for the resonant bar-bending flap gravitational wave antenna with re-entrant cavity transducer at UWA. Measurements for three seperate experiments in 1987, 1989 and 1991 are presented.

| calculated parameter | 1987 | 1989 | 1991 |
|---|---|---|---|
| $\omega_1/2\pi$ (Hz) | 708.4 | 708.6 | 707.5 |
| $\omega_2/2\pi$ (Hz) | 720.3 | 715.3 | 692.6 |
| $Q_1$ ($10^8$) | - | - | 2.0±.1 |
| $Q_2$ ($10^6$) | 1.85±.05 | 0.45±.02 | 3.1±.1 |
| $m_2$ (kg) | 0.53 | 0.53 | 0.45 |
| $\beta_r$ | 3.37 | 2.07 | 0.213 |

# Table 6.2

Calculated parameters for the resonant bar-bending flap gravitational wave antenna with re-entrant cavity transducer at UWA. Calculations for three seperate experiments in 1987, 1989 and 1991 are presented.

model is considered. Bending flap #2 was designed to reduce this interaction, section 6.4 reveals that this interaction was reduced, but still could do with further reduction.

Bending flap #2 is still not an optimum design. For a well tuned resonant bar and bending flap, $\beta_r$ should be approximately one. This is essential to maximise the useful electro-mechanical coupling. In section 6.3 it is shown how bending flap #2 can be optimised to achieve this.

### 6.2.3. Two-mode antenna model with a parametric transducer at general pump oscillator offsets

#### 6.2.3.1 Theory

The effect of a parametric transducer on a resonant bar antenna was first analysed by (Giffard and Paik, 1977). Some properties of their impedance matrix description, have been analysed by (Blair, 1980), (Veitch, 1986) and then (Linthorne, 1991). However the impedance matrix description assumes that the pump oscillator input voltage at the transducer is constant as a function of offset frequency (Giffard and Paik, 1977). For an upconverting parametric transducer this is not true, it is the power incident on the cavity that is constant, resulting in errors in (Blair, 1980) and (Linthorne, 1991). For this reason a scattering matrix description is more appropriate than a impedance matrix description.

The scattering matrix introduced in this section is incorporated with the input impedance of the resonant bar to the parametric transducer. The input impedance varies as a function of pump offset frequency, and acts to change the resonant frequencies and acoustic Q values of the resonant bar system. In this section these effects are theoretically analysed and compared with experiment. In appendix B all components of the scattering matrix are derived, and in chapter 7 they are incorporated into a noise analysis program.

A schematic of a resonant bar interacting with a parametric transducer is shown in figure 6.4, the matrix description accompanying this schematic is given by (6.23). Here **a** is the incident signal ($\sqrt{\text{watt}}$), **b** is the reflected signal ($\sqrt{\text{watt}}$) off the transducer, **u** is the input velocity to the transducer and **f** is the force on the transducer. Rather than analysing the incident and reflected signal as upper and lower side bands (Giffard and Paik 1977), the signal is analysed in terms of the observable a.m and p.m quadrature components, which gives a more meaningful description when analysing the sensitivity of the system. For each normal mode sensed by the transducer, the following matrix is defined; (subscript + and - refer to the positively and negatively tuned normal modes respectively)

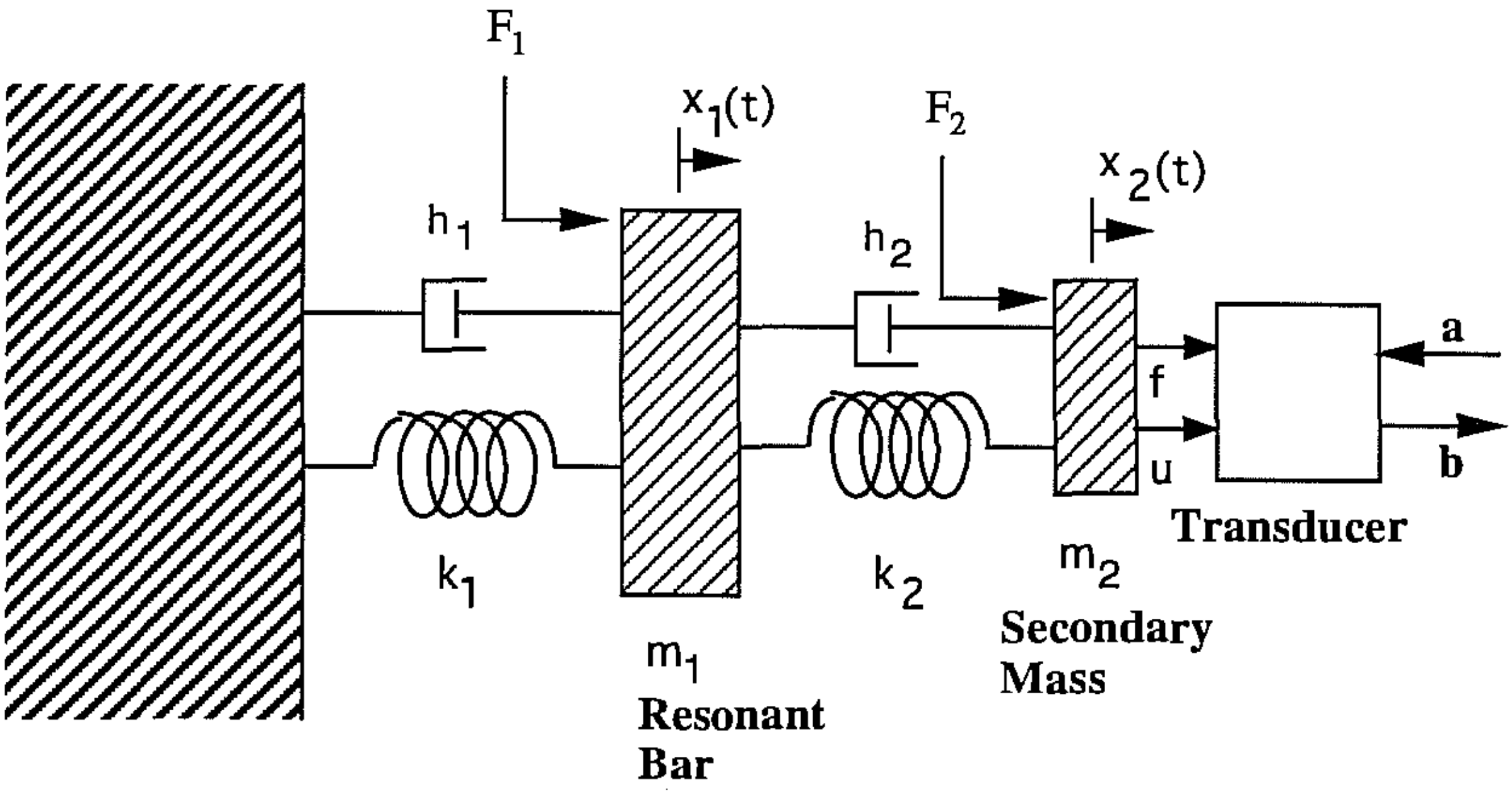

# Figure 6.4

Model of a resonant bar interacting with a parametric transducer

$$\begin{bmatrix} b_{pm}(\omega_\pm) \\ f(\omega_\pm) \\ b_{am}(\omega_\pm) \end{bmatrix} = \begin{bmatrix} S_{pp\pm} & S_{pu\pm} & S_{pa\pm} \\ S_{fp\pm} & Z_{in\pm} & S_{fa\pm} \\ S_{ap\pm} & S_{au\pm} & S_{aa\pm} \end{bmatrix} \begin{bmatrix} a_{pm}(\omega_\pm) \\ u(\omega_\pm) \\ a_{am}(\omega_\pm) \end{bmatrix} \qquad (6.23).$$

The transducer scattering components for each normal mode are defined as; $S_{pa\pm}$, ratio of reflected p.m converted from incident a.m when **u**=0 and $\mathbf{a}_{pm}$=0; $S_{ap\pm}$, ratio of reflected a.m converted from incident p.m when **u**=0 and $\mathbf{a}_{am}$=0; $S_{pp\pm}$, ratio of reflected p.m to incident p.m when **u**=0 and $\mathbf{a}_{am}$=0; $S_{aa\pm}$, ratio of reflected a.m to incident a.m when $\mathbf{a}_{pm}$=0 and **u**=0; $S_{pu\pm}$, p.m transductance scattering parameter; $S_{au\pm}$, a.m transductance scattering parameter; $S_{fp\pm}$, p.m reverse transductance scattering parameter; $S_{fa\pm}$, a.m reverse transductance scattering parameter; $Z_{in\pm}$, the mechanical input impedance. This method naturally includes a.m to p.m and p.m to a.m conversions.

Defining $Z_m$ as the mechanical input impedance of the transducer in units (seconds)$^{-1}$, where $Z_m = Z_{m+}+Z_{m-}$, $Z_{m\pm} = Z_{in\pm}/m_2$, the equation of motion given by (6.3) is modified to;

$$\begin{bmatrix} \dot{x}_1 \\ \dot{x}_2 \\ \dot{x}_3 \\ \dot{x}_4 \end{bmatrix} = \begin{bmatrix} 0 & 0 & 1 & 0 \\ 0 & 0 & 0 & 1 \\ -\omega_1{}^2-\frac{m_2}{m_1}\omega_2{}^2 & \frac{m_2}{m_1}\omega_2{}^2 & -\frac{2}{\tau_1}-\frac{2\,m_2}{\tau_2\,m_1} & \frac{2\,m_2}{\tau_2\,m_1} \\ \omega_2{}^2 & -\omega_2{}^2 & \frac{2}{\tau_2} & -\frac{2}{\tau_2}+Z_m \end{bmatrix} \begin{bmatrix} x_1 \\ x_2 \\ x_3 \\ x_4 \end{bmatrix} + \begin{bmatrix} 0 & 0 \\ 0 & 0 \\ 1/m_1 & 0 \\ 0 & 1/m_2 \end{bmatrix} \begin{bmatrix} F_1 \\ F_2 \end{bmatrix} \qquad (6.24).$$

Appendix B shows the input impedance to be of the form;

$$\mathrm{Re}[Z_{in\pm}] = \frac{\beta_{0\pm}\,(\omega_2^2/\omega_\pm)(1+4Q_e^2\omega_\pm^2/\Omega_0^2)}{2\left(1+4Q_e^2\,\delta^2\right)}\left(\frac{1}{(2Q_e\,\delta_{U\pm})^2+1} - \frac{1}{(2Q_e\,\delta_{L\pm})^2+1}\right) \quad \text{(B25a)},$$

$$\mathrm{Im}[Z_{m\pm}] = \frac{-\beta_{0\pm}\,(\omega_2^2/\omega_\pm)(1+4Q_e^2\omega_\pm^2/\Omega_0^2)}{2\left(1+4Q_e^2\,\delta^2\right)}\left(\frac{2Q_e\,\delta_{U\pm}}{(2Q_e\,\delta_{U\pm})^2+1} + \frac{2Q_e\,\delta_{L\pm}}{(2Q_e\,\delta_{L\pm})^2+1}\right) \quad \text{(B25b)},$$

where;
$$\delta_{U\pm} = \frac{\delta\Omega+\omega_\pm}{\Omega_0}, \quad \delta_{L\pm} = \frac{\delta\Omega-\omega_\pm}{\Omega_0}, \quad \delta = \frac{\delta\Omega}{\Omega_0}.$$

The input impedance has an extra filtering term outside the brackets compared to previous derivations (Blair, 1980) (Linthorne, 1991), this is due to the voltage across the cavity not being a constant as the pump oscillator is tuned across the transducer resonance. The input impedance is substituted into (6.24), and the normal mode resonant frequencies are calculated from its characteristic equation. For small electro-mechanical couplings the coupled mode behaviour can be ignored and the normal modes can be treated independently, for this case the loaded normal mode frequencies and Q values can be expresses as (Veitch, 1991);

$$\omega_{\pm L} \approx \omega_{\pm}\left(1 - \frac{\mathrm{Im}[Z_{m\pm}]}{\omega_{\pm}}\right)^{1/2} \qquad (6.26a)$$

$$Q_{\pm L}^{-1} \approx Q_{\pm}^{-1} + \mathrm{Re}[Z_{m\pm}]/\omega_{\pm} \qquad (6.26b)$$

By differentiating (6.26a&b) and assuming the normal mode modulation frequencies are small compared with the halfbandwidth ($\Delta\Omega_{hbw}$) of the transducer, the turning points for the normal mode frequency and Q characteristics are respectively calculated to occur at;

$$\delta\Omega_{ftp} = \pm\Delta\Omega_{hbw}/\sqrt{3} \qquad (6.27a),$$

$$\delta\Omega_{Qtp} = \pm\Delta\Omega_{hbw}/\sqrt{5} \qquad (6.27b).$$

This result is general for any system comprising of a resonant bar with a parametric transducer. This result also explains the discrepancy between experiment and theory presented in (Linthorne, 1991), where the frequency turning points were calculated to occur at $\delta\Omega_{ftp} = \pm\Delta\Omega_{hbw}$.

#### 6.2.3.2 Experimental Verification

This section compares the theory discussed above with the experimental results taken in 1991 involving the UWA resonant bar antenna. Comparisons of experiment with theory (6.26a&b) are presented in figure 6.5 and 6.6 for small electro-mechanical couplings. To increase the coupling the incident power was increased. Figure 6.7 shows the theoretical effects when couplings of greater than $10^{-2}$ are achieved. In this regime it is evident that (6.26a&b) are insufficient to describe the behaviour. The behaviour was analysed by plotting the roots of the characteristic equation (6.24), coupled mode behaviour is observed as the pump frequency is varied. Changing the pump frequency causes a change in secondary mass flap frequency which has the notable affect of tuning or detuning the flap from the resonant bar at high couplings. This effect begins to become evident around couplings of $10^{-2}$, and is extremely evident at couplings greater than $10^{-1}$. Figure 6.7 shows how normal mode identities change as the pump frequency tunes the bending flap frequency across the resonant bar frequency.

Originally it was planned to operate the transducer with the pump on the centre of resonance, but this has proved to be impossible due to low frequency drifts of the cavity

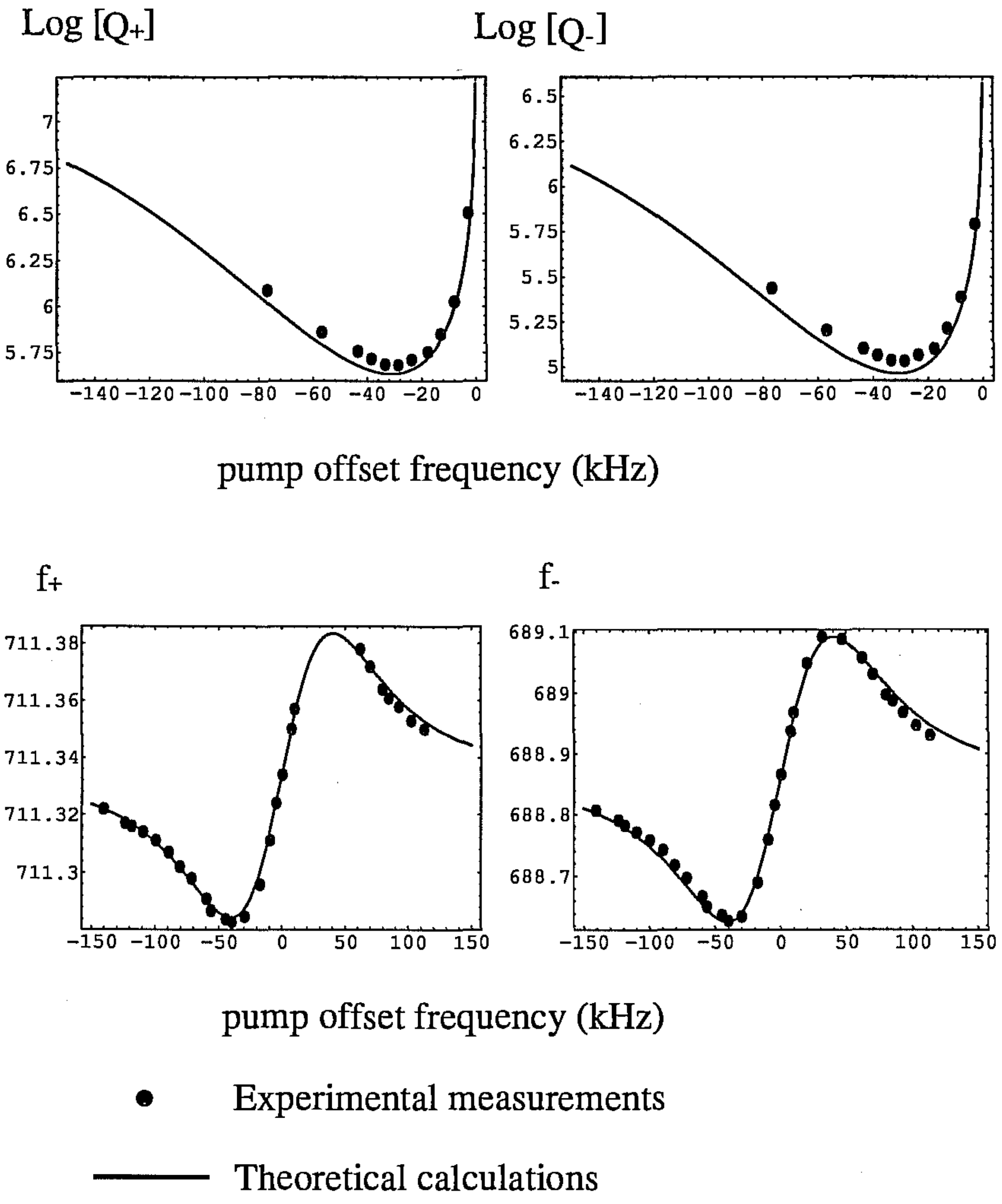

# Figure 6.5

Parametric transducer effects on antenna frequency and Q values as a function of pump offset frequency, with $P_{inc} = 30\ \mu w$ and $\beta_o = 2.5\ 10^{-3}$

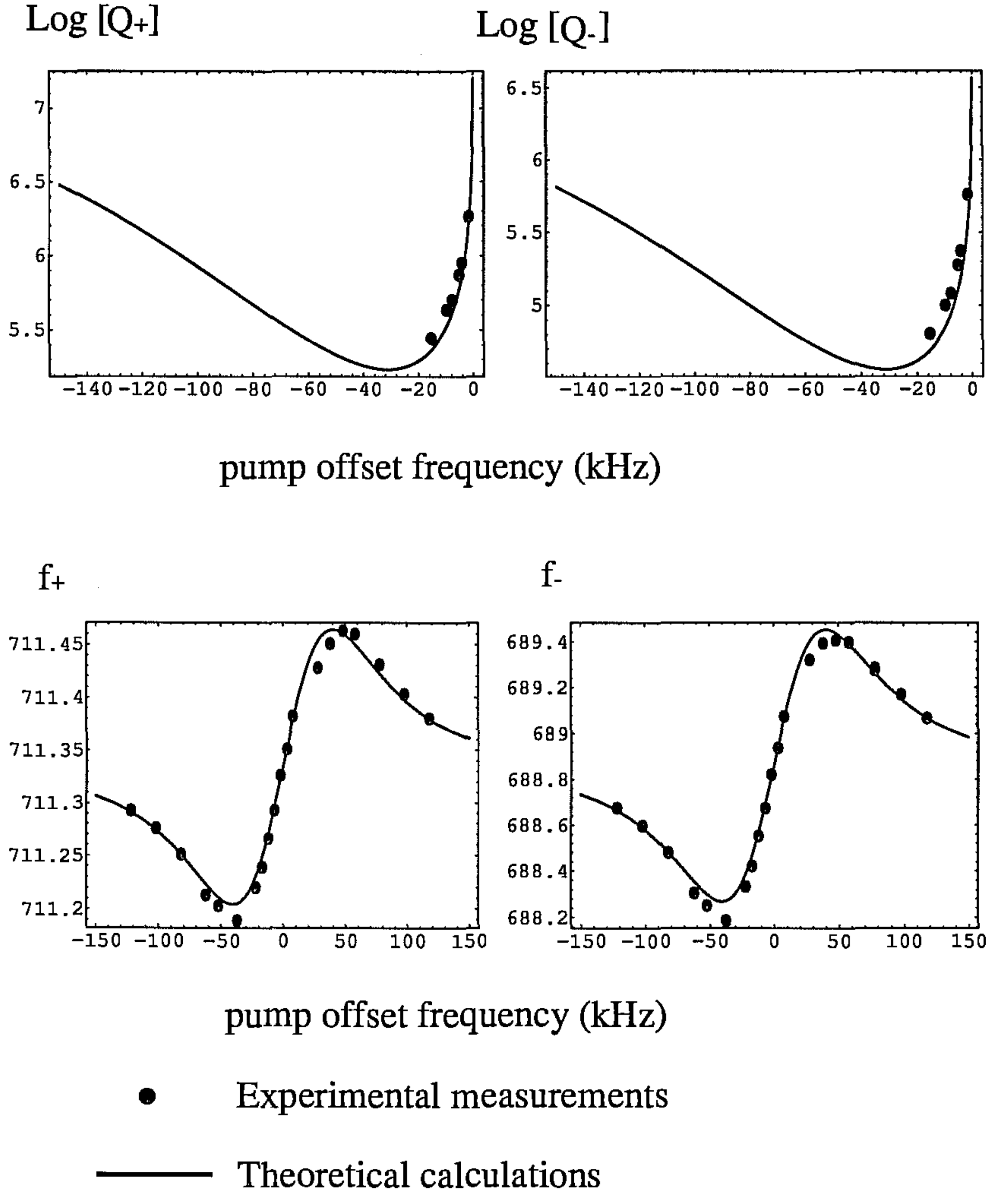

# Figure 6.6

Parametric transducer effects on antenna frequency and Q values as a function of pump offset frequency, with $P_{inc} = 77\ \mu w$ and $\beta_b = 6.4\ 10^{-3}$

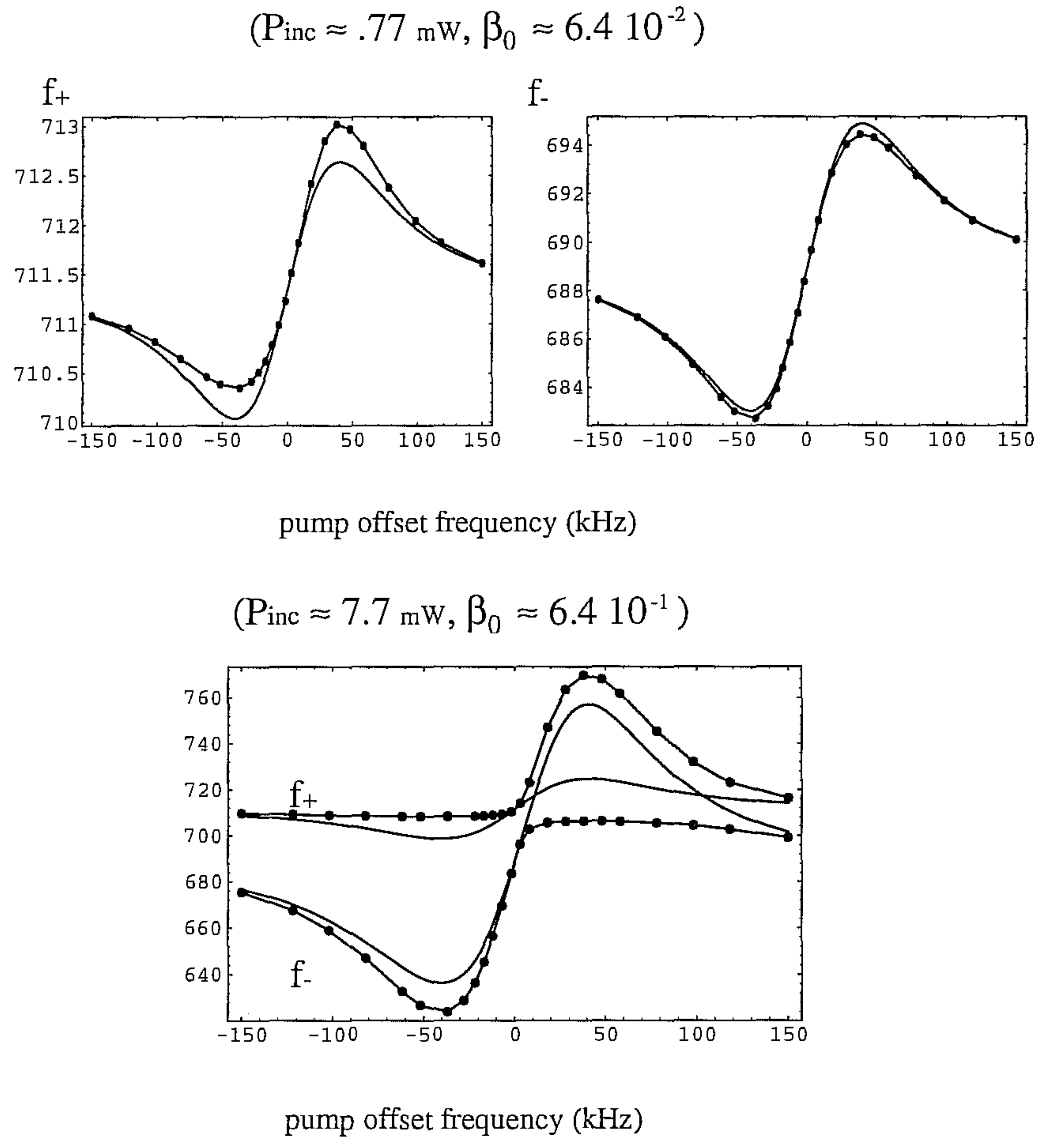

—●— Locus of characteristic equation (6.24)

—— Theoretical calculations from equations (6.26a&b)

# Figure 6.7

Parametric transducer effects on antenna frequency for increased electro-mechanical couplings.

and pump oscillator. This caused the pump to occasionally drift above resonance which had the effect of parametrically exciting the resonant bar. The transducer must operate in the cold damped regime below the centre of resonance (see figures 6.5 and 6.6). Q values are only shown below resonance as above resonance parametric excitations cause the Q values to be infinite. Also, increasing the coupling causes the slope near the centre of resonance to increase (6.21). This is a source of possible upconversions since a low frequency modulation will cause a change in antenna Q and frequency. It is therefore desirable to operate on the lower turning points so the coupling may be increased without increasing to first order the amount of upconversions.

Independent measurements of the Q of the re-entrant cavity reveal that the halfbandwith is 69.8 kHz. This means that the frequency turning point occurs at ±40.3 kHz pump frequency offset and the Q turning point occurs at ±31.2 kHz pump frequency offset, this is verified in figures 6.5 and 6.6. The turning points of the Q and frequency characteristics do not occur at the same point. However at the lower frequency turning point the Q characteristic is still quite flat. Also modulations of the frequency will perturb the antenna phase while modulations of the Q will perturb the amplitude, since a phase demodulation scheme is in operation, operation on the frequency turning point is more desirable.

#### 6.2.3.3 Electro-Mechanical Coupling

If the pump oscillator is not on the centre of resonance the electro-mechanical coupling will be reduced compared to the centre of resonance coupling. The coupling must be calculated as a function of pump offset frequency. Veitch (Veitch, 1991) uses a definition which depends on the reverse and forward transductances of the transducer. This definition is not appropriate when determining the signal to noise ratio of the antenna system. The alternative definition that only depends on the forward transductance is used here (Paik, 1974);

$$\beta_{\pm} = \frac{\text{signal energy in side bands due to the} \pm \text{resonant mode}}{\text{signal energy in antenna}} \qquad (6.28).$$

In terms of the transducer electrical cavity parameters the total electro-mechanical coupling as derived in appendix B can be shown to be;

$$\beta = \frac{2\beta_e P_{inc}}{(1+\beta_e)m_2\omega_2^2\Omega_0}\left(\frac{2Q_e df}{f_o\ dx}\right)^2\left(\frac{1}{1+4Q_e^2\delta^2}\right)\left(\frac{1}{1+4Q_e^2\delta_U^2} + \frac{1}{1+4Q_e^2\delta_L^2}\right) \qquad (6.29)$$

where; $\beta_e$ is the electrical cavity coupling, $P_{inc}$ is the incident pump power on the cavity, df/dx is the transducer rate of resonant frequency change with respect to displacement,

($P_{inc} \approx 77\ \mu W$, $\beta_0 \approx 6.4\ 10^{-3}$)

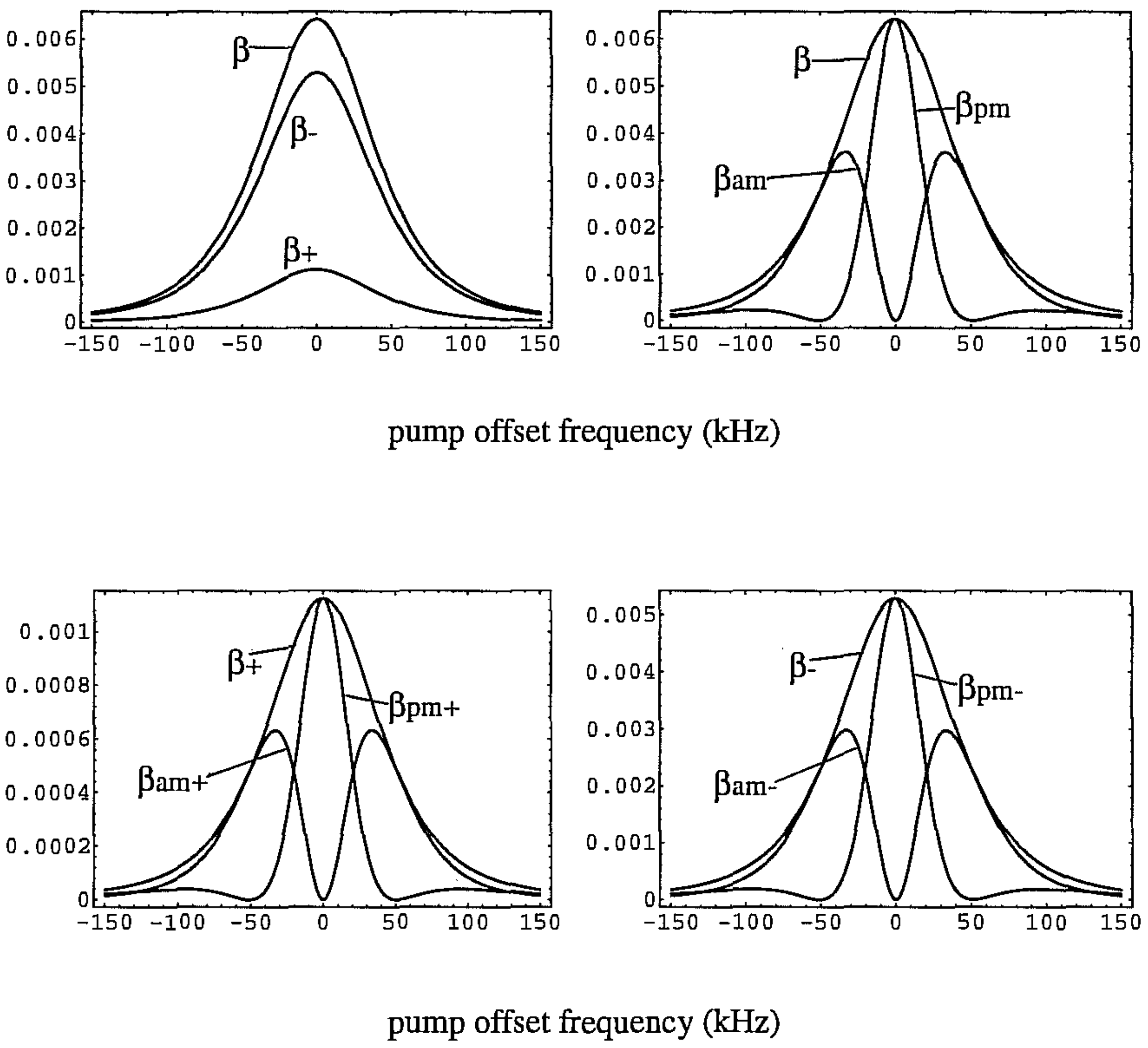

$$\beta = \beta_+ + \beta_- = \beta_{pm} + \beta_{am}$$

(on resonance $\beta = \beta_0$)

# Figure 6.8

Electro-mechanical coupling components as a function of pump offset frequency when the electrical coupling to the transducer is $\beta_e = 0.3$. The significant component of electro-mechanical coupling for signal to noise calculations is from the barish + mode. The transducer is operated with the pump offset on the lower frequency turning point, here the useful coupling is; $\beta_+ = 6.33\ 10^{-4}$ $f_0=\Omega_0/2\pi$, $\delta_U = (\delta\Omega+\omega_m)/\Omega_0$ and $\delta_L = (\delta\Omega-\omega_m)/\Omega_0$, where $\omega_m$ is the resonant bar modulation frequency. From (6.29) the center of resonance coupling can be shown to be;

$$\beta_0 = \frac{4\beta_e P_{inc}}{(1+\beta_e)m_2\omega_2^2\Omega_0}\left(\frac{2Q_{fac}}{f_o}\frac{df}{dx}\right)^2 \qquad (6.30),$$

where

$$Q_{fac}^2 = \left(\frac{Q_e^2}{1+4Q_e^2\omega_m^2/\Omega_0^2}\right) \qquad (6.31).$$

From (6.30) it is evident that the electro-mechanical coupling is proportional to the incident power. In the last experimental run it has been estimated that $\beta_0 \sim 0.012\ P_{inc}$, implying from (6.30) that $\beta_e \sim 1$. Actually $\beta_e$ is measured to be 0.3 to within 50% accuracy, the discrepancy is probably due to the inaccuracy in the estimations of transducer df/dx and incident power, $P_{inc}$.

Figure 6.8 illustrates the components of the electro-mechanical coupling as a function of pump offset frequency for the data presented in figure 6.6. The total electro-mechanical coupling is broken into four components. First it is broken down to components associated with either the + or - normal modes by relating $\beta_0$ to $\beta_{0\pm}$ via (6.21) and substituting $\omega_m$ for either $\omega_\pm$. Then by defining electro-mechanical coupling components that are proportional to the foward transductance scattering parameters $S_{pu\pm}$ and $S_{au\pm}$ respectively, the coupling components may be broken down further into p.m and a.m quadrature components. The four components are calculated in appendix B to be;

$$\beta_{pm\pm} = \beta_{0\pm}F_{pm\pm} \quad \text{and} \quad \beta_{am\pm} = \beta_{0\pm}F_{am\pm} \qquad (6.32a\&b),$$

where $F_{pm\pm}$ and $F_{am\pm}$ are defined in appendix B.

Even though the total coupling at the center of resonance is $\beta_0 = 6.4\ 10^{-3}$, the useful coupling component with regards to gravity wave signal to noise ratio is mainly determined by the bar-ish + mode component. At the turning point given by (6.27a), this component is only $\beta_+=6.33\ 10^{-4}$. To increase this component the resonant bar and bending flap frequencies must be tuned more closely, the detuning is calculated to be 14.9 Hz for the 1991 experimental run (see table 6.2).

To obtain better signal to noise performance the coupling needs to be increased. At a constant incident power and secondary oscillator detuning the coupling may be increased by increasing $Q_{fac}^2$ or $(df/dx)^2$. A sapphire transducer has been proposed (Blair and Hong, 1991) which will increase $Q_{fac}^2$ at the expense of $(df/dx)^2$. Table 6.3 compares these values between the current re-entrant cavity transducer and proposed sapphire

| parameter | re-entrant cavity | sapphire transducer |
|---|---|---|
| $\frac{df}{dx}$ (Hz/m) | $3 \cdot 10^{14}$ | $10^{12}$ |
| $Q_e$ | $10^5$ | $10^8$ |
| $Q_{fac}$ | $10^5$ | $7 \cdot 10^6$ |

# Table 6.3

Re-entrant cavity and sapphire transducer parameters.

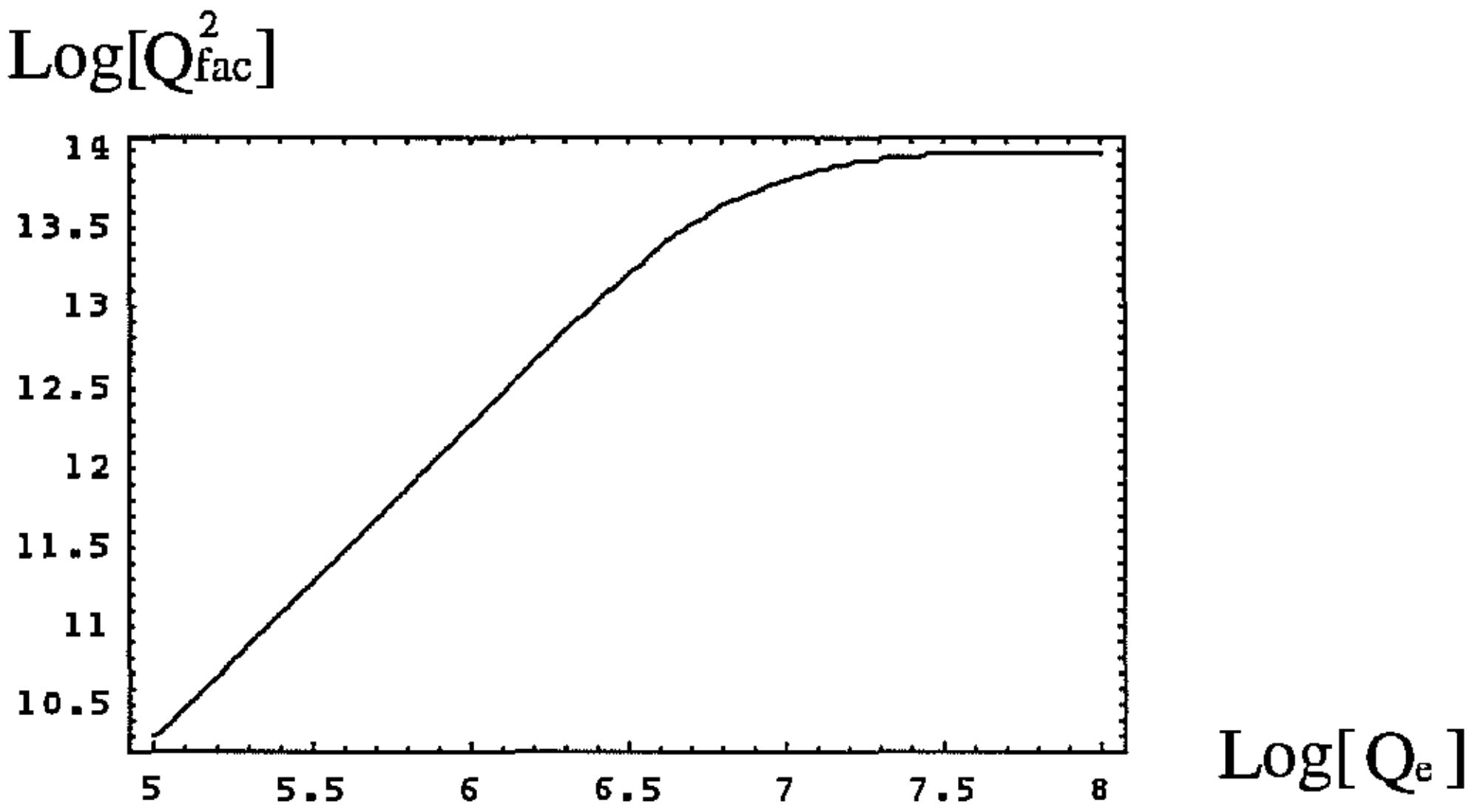

# Figure 6.9

$Q_{fac}$ is equal to $Q_e$ if the half bandwidth is greater than the modulation frequency ($\omega_m$), and equal to $\Omega_o/(2\omega_m)$ if the half bandwidth is less than the modulation frequency. The modulation frequency of the resonant bar at UWA is about 710 Hz, while the pump frequency ($\Omega_o$) is about 9.7 GHz, this causes the break point in the above graph to occur at a Q value of about $10^7$.

transducer. These values imply that for a constant input power the ratio of the center of resonance electro-mechanical coupling is $\beta_{0re\text{-}ent}/\beta_{0spph} \approx 18$. To achieve a comparable coupling to the re-entrant cavity the sapphire transducer must be able to take 18 times the incident power. However the increased power necessary to maintain the same electro-mechanical coupling will set a stricter limit on the phase and amplitude noise of the pump oscillator.

Figure 6.9 shows the Log[$Q_{fac}^2$] versus Log[$Q_e$] characteristic. This reveals that the coupling is proportional to $Q_e^2$ if the half bandwith of the transducer is greater than the resonant bar modulation frequency (ie. $Q_{fac} \approx Q_e$). As the half bandwidth decreases (Q increases) below the modulation frequency, $Q_{fac} \rightarrow \Omega_0/(2\omega_m)$, and the centre of resonance coupling, $\beta_0$, becomes independent of transducer Q. However increasing $Q_e$ still affects the coupling as a function of pump offset frequency, due to the filtering nature of the resonant cavity transducer. For example, the turning points of the normal mode frequency and Q characteristics will occur at a smaller pump offset frequency as the Q increases, because the turning points depend on the bandwith of the transducer. For a higher Q cavity the pump oscillator will thus have a narrower bandwidth to operate within, setting a stricter limit on the pump oscillator frequency stability.

The model presented in this section enables predictions on how a sapphire transducer will interact parametrically with a resonant bar antenna. For details of its proposed construction see (Blair and Hong, 1991). Figure 6.10 illustrates the expected parametric effects for this type of transducer, when configured to achieve the same centre of resonance coupling as the re-entrant cavity data presented in figure 6.6. Comparing these results, the sapphire transducer causes a greater cold damping effect and there are four turning points due to the modulation frequency being large compared to the half bandwidth of the resonantor. It is also evident from the coupling curves in figure 6.10 that coupling is possible outside the transducer bandwidth at a pump oscillator offset equal to the resonant bar modulation frequency.

## 6.3 DESIGN OPTIMISATION WITH RESPECT TO THE TWO-MODE MODEL

To optimise the band width and signal to noise ratio the secondary oscillator must be tuned to the resonant bar. Figure 6.7 highlights the fact that the pump oscillator can be used to tune the bending flap. The larger the incident pump power and hence electro-mechanical coupling, the greater the frequency tuning of the bending flap. When operating on the lower frequency turning point, the bending flap will have its resonant frequency detuned down in frequency, with the magnitude of detuning depending on the electro-mechanical coupling. It is evident from table 6.2 that bending flap #2 is 14.9 Hz below the bar frequency already. It was decided to tune the flap frequency just above the bar resonance, so at a unique coupling when operating on the lower frequency turning point, the flap and bar will be perfectly tuned.

Assuming that the pump oscillator only perturbs the bending flap frequency and the flap modulation frequency is much smaller than the cavity bandwidth, the loaded bending flap frequency can be approximated by the following formula;

$$f_{2L} \approx f_2 \left(1 + \frac{2\beta_0 Q_e \delta}{1+4Q_e^2\,\delta^2}\right)^{1/2} \qquad (6.33).$$

Thus the loaded bending flap frequency when the pump frequency is tuned to the lower frequency turning point, is given by;

$$f_{ftp} \approx f_2 \left(1 - \frac{3\sqrt{3}\beta_0}{16}\right)^{1/2} \qquad (6.34)$$

When the antenna is operating with the pump oscillator on the lower frequency turning point, for optimum operation the bar and flap should be tuned with $f_1 = f_{ftp}$. This means that the flap must be tuned above the bar frequency, with (6.34) defining a unique electro-mechanical coupling needed to achieve this optimum condition. From (6.34) and defining the frequency detuning as $\Delta f = f_2 - f_1$, the centre of resonance electro-mechanical coupling needed to achieve $f_1 = f_{ftp}$ is given by;

$$\beta_0 = \frac{16(2f_1+\Delta f)}{3\sqrt{3}(f_1+\Delta f)^2}\Delta f \qquad (6.35)$$

The useful coupling is however reduced at this lower turning point compared to the centre of resonance coupling. From (6.32) it can be shown that $\beta_{ftp} = 9/16\ \beta_0$ which can be broken into a.m. and p.m. components. It is possible to utilise the total coupling by demodulating the pump signal with the right phase as described in appendix B. Figure 6.11 shows the value of the electro-mechanical coupling needed to achieve $f_1 = f_{ftp}$, as a function of frequency detuning ($\Delta f$).

To tune the flap resonant frequency from 14.9 Hz below the resonant bar ($\Delta f$=-14.9) to above the resonant bar, the length of the bending flap must be reduced (Linthorne, 1991). The frequency is inversely proportional to the length squared. To increase the frequency, mass is etched off the end of the bending flap. The bending flap mass is proportional to the length of the bending flap, therefore the frequency of the bending flap

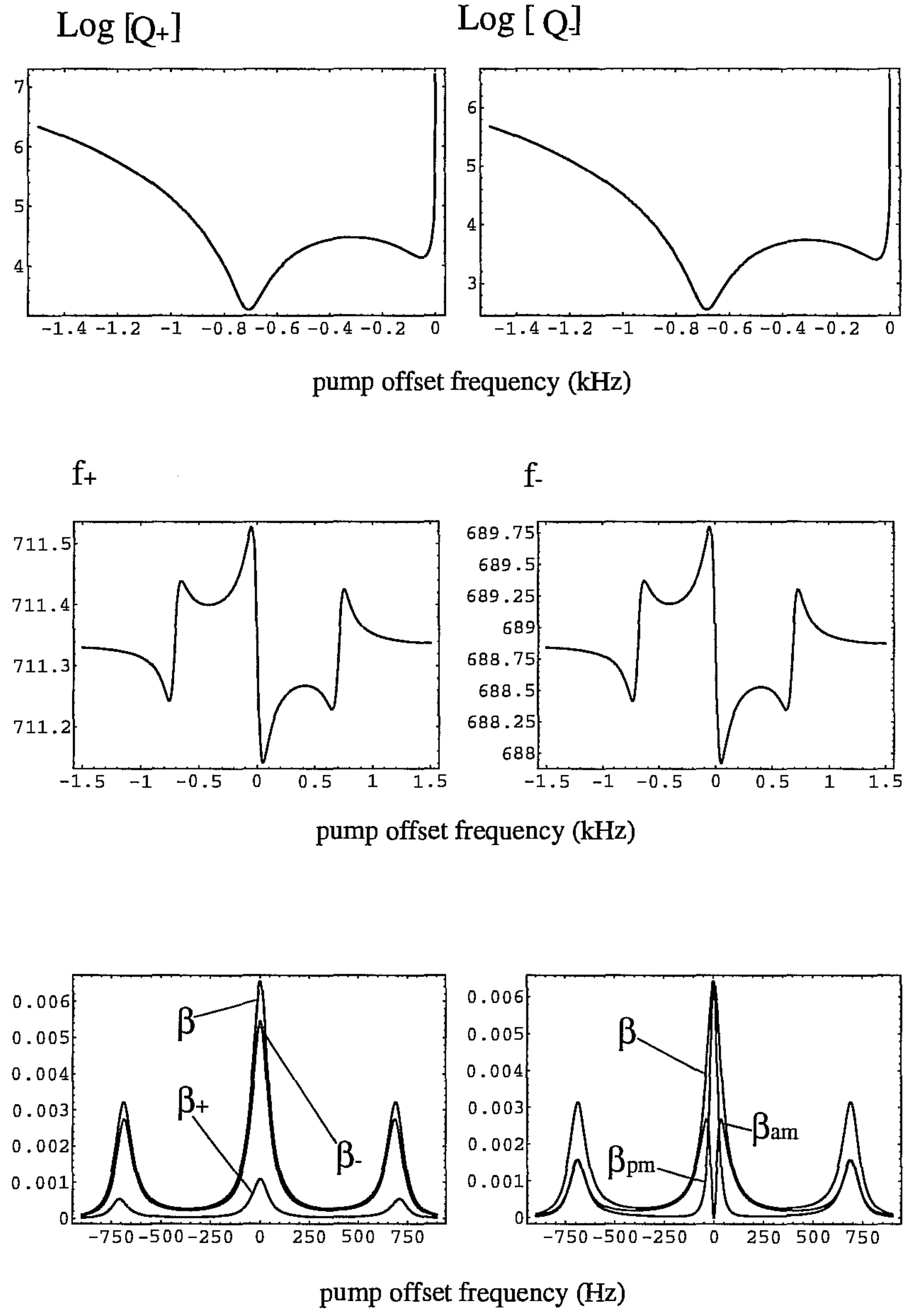

# Figure 6.10

Predicted parametric effects for a sapphire transducer attached to a resonant bar gravitational wave antenna, showing antenna frequency, Q values and coupling as a function of pump offset frequency, with $P_{inc} = 1.4$ mw, $\beta_o = 6.4\ 10^{-3}$ and $\beta_e = 0.3$.

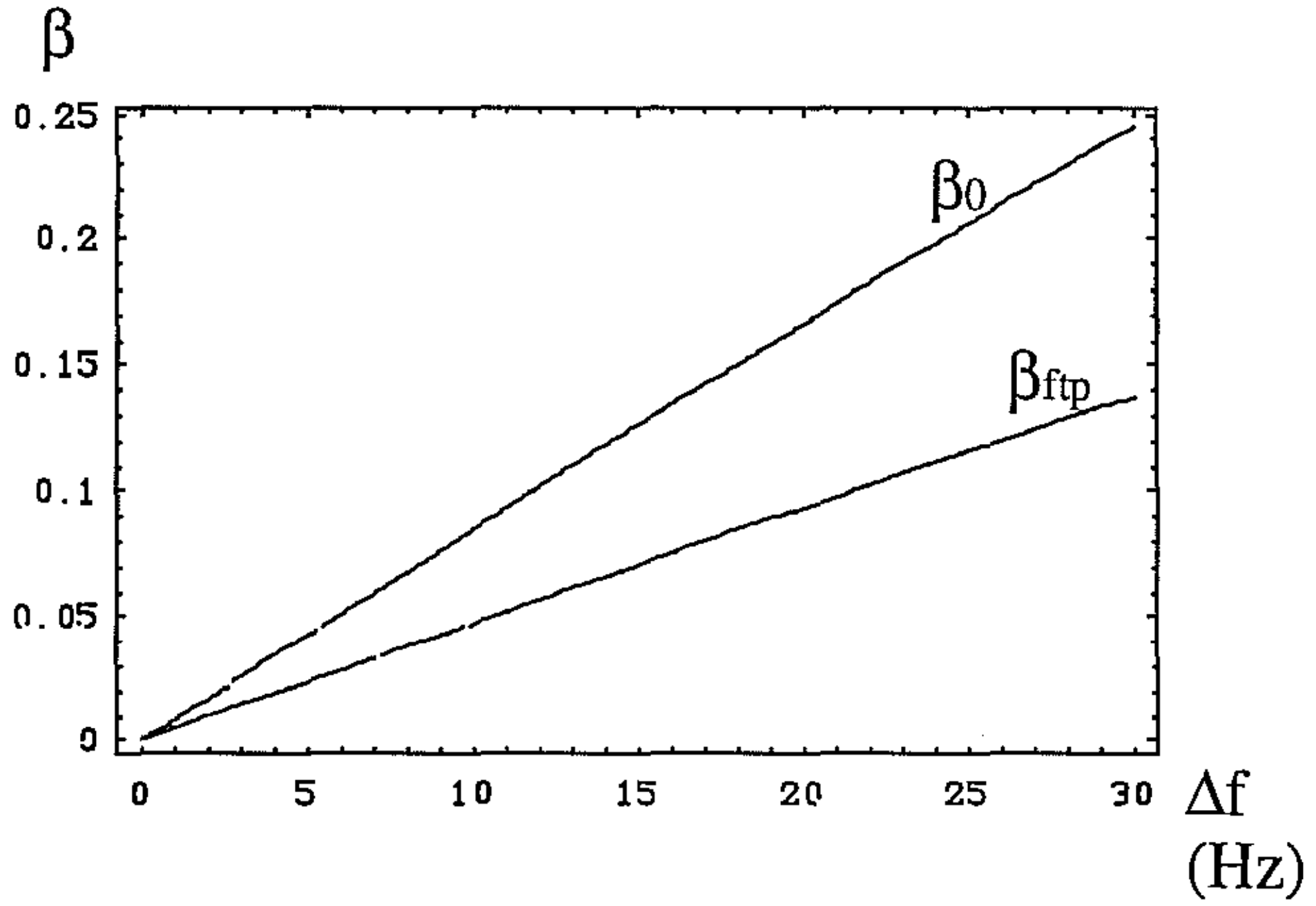

# Figure 6.11

The electro-mechanical coupling (β) required to tune bending flap #2 to the bar when operating on the lower turning point of the normal mode versus pump offset frequency characteristic. This is plotted as a function of bending flap detuning with respect to the resonant bar (Δf).

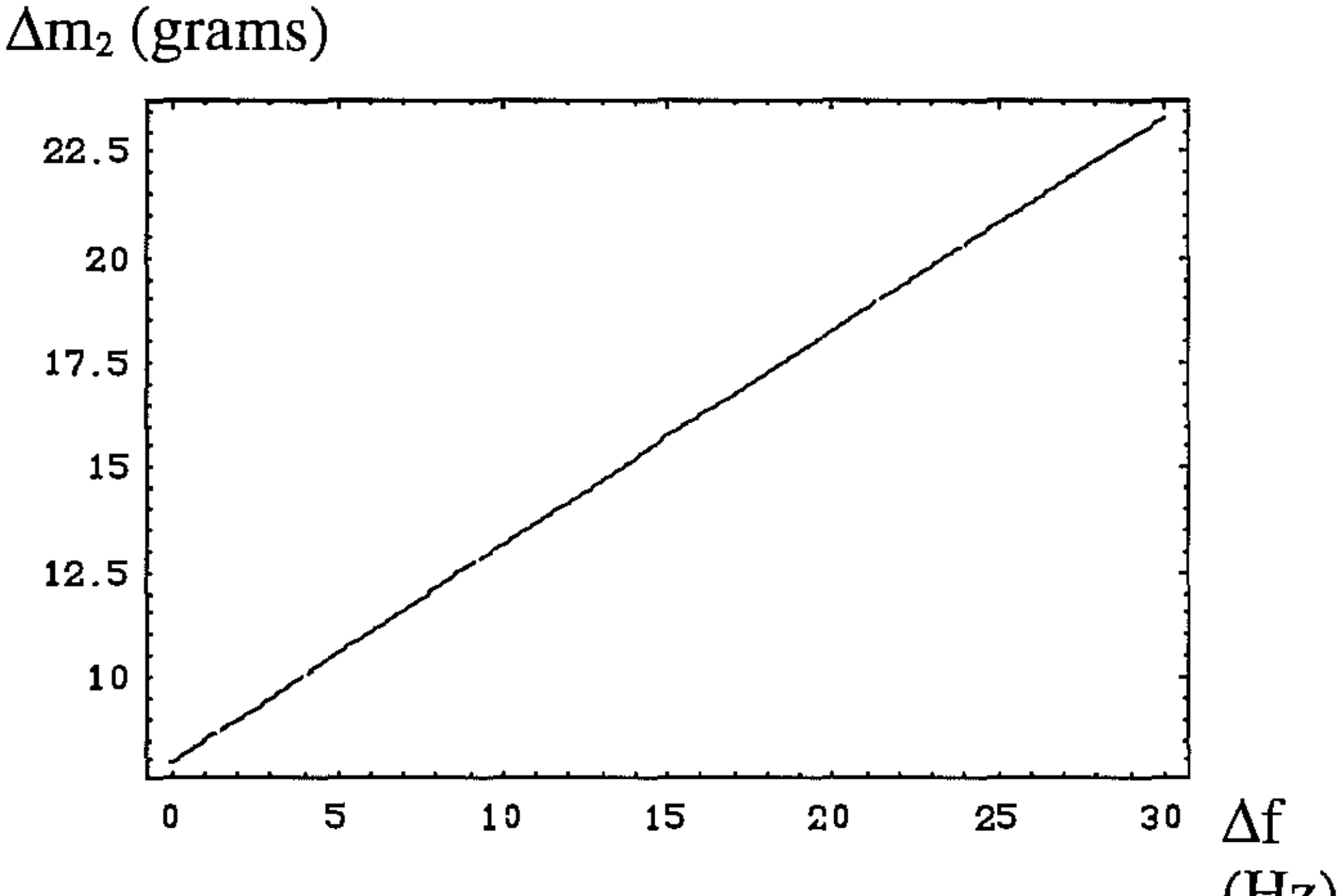

# Figure 6.12

Bending flap #2 mass reduction required ($\Delta m_2$) to obtain a specific detuning, with respect to the resonant bar antenna (Δf).

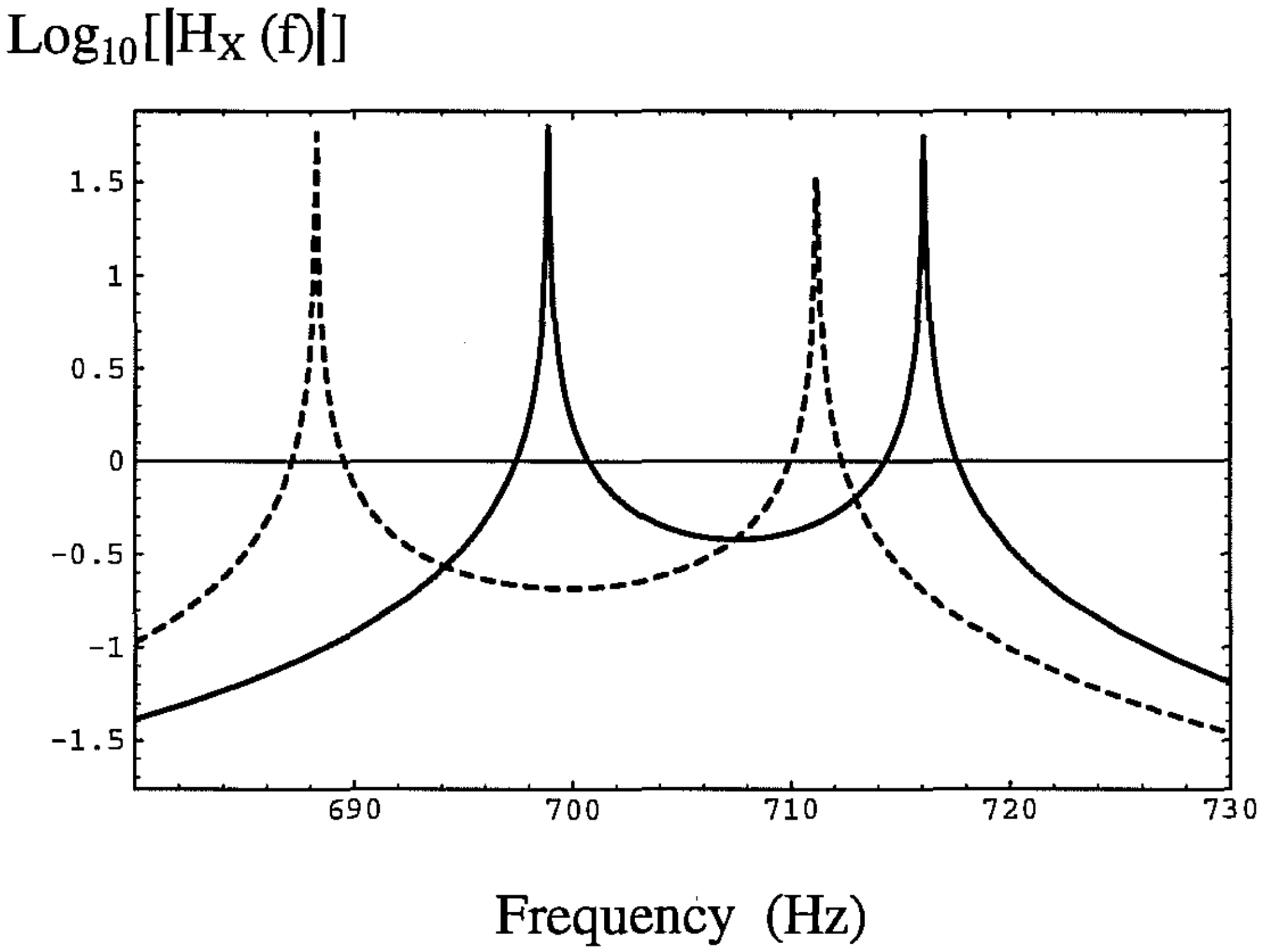

# Figure 6.13

Displacement transfer function of bending flap response per initial impulse displacement at the resonant bar. Bold line represents the optimised transfer function, while the dashed line represents the previous unoptimised transfer function.

will be inversely proportional to the mass of the bending flap squared. Empirically it has been determined that the mass of the bending flap is equal to 1.7 times the effective mass (Linthorne, 1991). From this information it is possible to calculate the amount of mass needed to be etched away from the end of the bending flap as a function of Δf;

$$\Delta m_{bf} = 1.7 \times m_2^* \left( 1 - \sqrt{\frac{f_2^*}{f_1 + \Delta f}} \right) \quad (6.36).$$

Here $m_2^*$ =.45, $f_2^*$ = 692.6 and $f_1$ = 707.5 are the values given in table 6.2 when Δf= -14.9. Figure 6.12 shows (6.36) plotted as a function of Δf. The flap will need approximately 10 gm etched of the length to tuned it to 711 Hz, 3.5 Hz above the bar frequency. With this modification the tuning requirement from 6.35 should enable a coupling of $\beta_0$=0.03, $\beta_{ftp}$ = 0.017 and a flap effective mass of $m_2$ = 0.444.

The optimised transfer function $H_X(f)$ for an initial impulse displacement at the resonant bar, $x_1(t=0)$, to displacement frequency response at the bending flap, $x_2(f)$, is shown in figure 6.13, and is compared to the transfer function before optimisation. Note that this transfer function is maximised in the optimised configuration, with the peak displacement amplification occuring at the resonant peaks.

## 6.4 THREE MODE ANTENNA MODELS

The U.W.A antenna and transducer system is more accurately modelled as the three mode oscillator shown in Figure 6.14 (Tobar et al, 1991). The model includes the base of the bending flap ($m_b$), as a low Q oscillator between the flap and the bar. The spring constant, $k_b$, and damping constant, $h_b$, are governed by the properties of the glue which bonds the base of the flap to the bar. The glue joint is treated as a spring between the bar and the base of the bending flap, with the resonance frequency given by;

$$\omega_b = \sqrt{\frac{E\,A}{m_b\,t}} \quad (6.37),$$

where A and t are the area and thickness of the glue joint, and E the Young's modulus of the epoxy glue. For 24 hour Araldite E ≈ $8 \cdot 10^9$ N/m$^2$ at 5K (Touloukian and Ho,1970). In 1987 parameters for the glue joint were; A ≈ 80 cm$^2$, t ≈ 135 μm and $m_b$ ≈ 3 kg, giving rise to $\omega_b/2\pi$ ≈ 60 kHz. In 1991 parameters for the glue joint were; A ≈ 10 cm$^2$, t ≈ 130 μm and $m_b$ ≈ 5 kg, giving rise to $\omega_b/2\pi$ ≈ 20 kHz. The Q factor ($Q_b$) of the glue joint is the last parameter to be estimated. This value will depend on many factors including the quantity of glue in the joint. Bending flap #2 has been designed with less

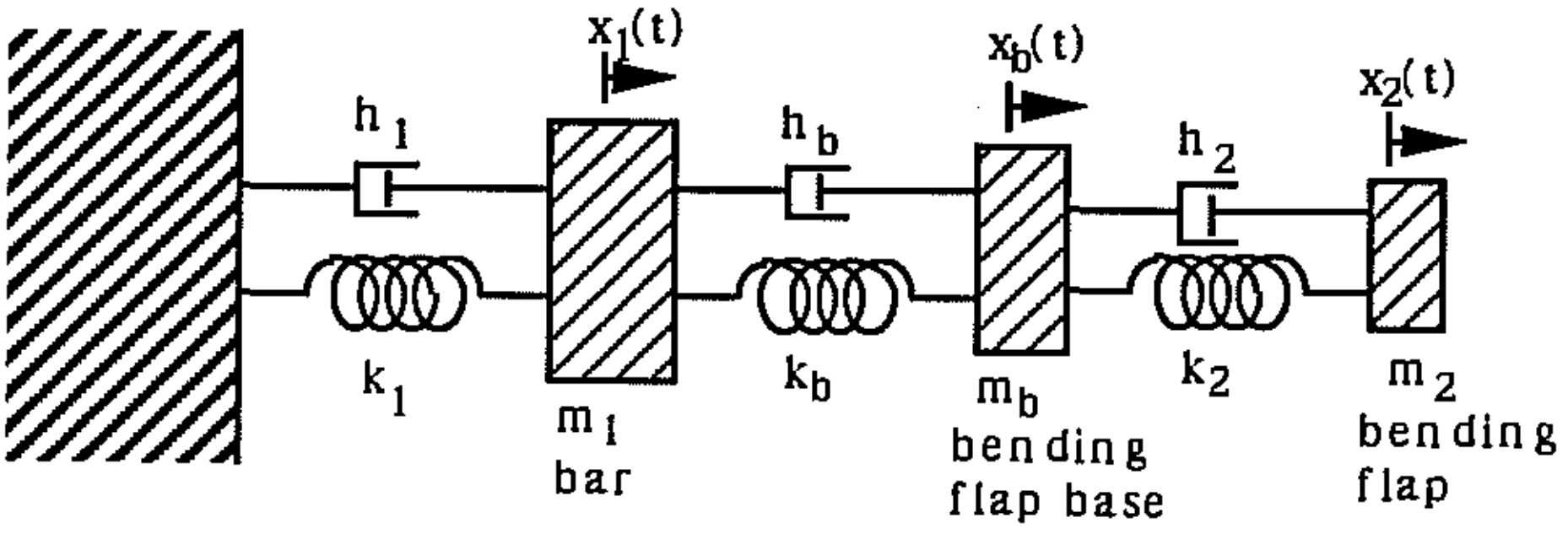

# Figure 6.14

Mechanical model of a three mode resonant bar antenna contact area so the glue joint will contribute less dissipation. Also the flap is further from the vacinity of the glue joint to further reduce interactions.

Considering the third longitudinal mode of the resonant bar and the bending flap base interacting as a two-mode coupled oscillator, the value of $Q_b$ can be estimated. It is assumed that the Q of this third mode is of the order $10^8$, with a measured frequency of 2103.8 Hz (Veitch, 1987). The normal mode values of the third mode with the bending flap attached have been measured to be f = 2098.9 (Hz) and $Q = 5.5\cdot10^7$ in 1987, and f = 2095.2 (Hz) and $Q = 3\cdot10^7$ in 1991. Assuming that the discrepancies between these values and resonant bar values are due to the resonant bar coupling to the bending flap base as a two-mode oscillator, $Q_b$ may be calculated. This model predicts that the frequency of the third mode of the resonant bar should be 2103 Hz and 2102.5 Hz in 1987 and 1991 respectively, which is close to the measured value of 2103.8 Hz. Assuming that the Q is independent of frequency, the values of $Q_b$ are predicted to be 70 and 800 during 1987 and 1991 respectively. Thus the new bending flap has an improved value of $Q_b$.

Equating poles as in the two-mode model, and assuming $Q_1, Q_2 >> 1$ and $Q_b \geq 1$, we obtain a cubic equation for the resonant frequencies, and a matrix equation for the amplitude decay time constants;

$$\omega^6 - \left(\omega_1{}^2 + \omega_2{}^2\left(1+\frac{m_2}{m_b}\right) + \omega_b{}^2\left(1+\frac{m_b}{m_1}\right)\right)\omega^4 +$$

$$\left(\omega_1{}^2\omega_2{}^2\left(1+\frac{m_2}{m_b}\right) + \omega_1{}^2\omega_b{}^2 + \omega_2{}^2\omega_b{}^2\left(1+\frac{m_2+m_b}{m_1}\right)\right)\omega^2 - \omega_1{}^2\omega_2{}^2\omega_b{}^2 = 0 \quad (6.38),$$

$$\begin{bmatrix} 1 & 1+\dfrac{m_2}{m_b} & 1+\dfrac{m_b}{m_1} \\ \omega_2{}^2\left(1+\dfrac{m_2}{m_b}\right)+\omega_b{}^2 & \omega_1{}^2\left(1+\dfrac{m_2}{m_b}\right)+\omega_b{}^2\left(1+\dfrac{m_2+m_b}{m_1}\right) & \omega_1{}^2+\omega_2{}^2\left(1+\dfrac{m_2+m_b}{m_1}\right) \\ \omega_2{}^2\,\omega_b{}^2 & \omega_1{}^2\,\omega_b{}^2 & \omega_1{}^2\,\omega_2{}^2 \end{bmatrix} \begin{bmatrix} \dfrac{2}{\tau_1} \\ \dfrac{2}{\tau_2} \\ \dfrac{2}{\tau_b} \end{bmatrix}$$

$$= \begin{bmatrix} 1 & 1 & 1 \\ \omega_-{}^2+\omega_{bn}{}^2 & \omega_+{}^2+\omega_{bn}{}^2 & \omega_+{}^2+\omega_-{}^2 \\ \omega_-{}^2\omega_{bn}{}^2 & \omega_+{}^2\omega_{bn}{}^2 & \omega_+{}^2\omega_-{}^2 \end{bmatrix} \begin{bmatrix} \dfrac{2}{\tau_+} \\ \dfrac{2}{\tau_-} \\ \dfrac{2}{\tau_{bn}} \end{bmatrix} \quad (6.39),$$

| measured parameter | 1987 | 1989 | 1991 |
|---|---|---|---|
| $\omega_1/2\pi$ (Hz) | 709.5 | 709.8 | 709.8 |
| $\omega_2/2\pi$ (Hz) | 721.0 | 715.5 | 692.6 |
| $Q_1$ $(10^8)$ | >0.54 | - | >1.7 |
| $Q_2$ $(10^6)$ | 2.0±.1 | - | 3.1 |

# Table 6.4

Three-mode model; calculated parameters for the resonant bar-bending flap gravitational wave antenna with re-entrant cavity transducer at UWA. Calculations for three seperate experiments in 1987, 1989 and 1991 are presented.

where $\omega_{bn}$ and $\tau_{bn}$ are the resonant frequency and amplitude decay time constant of the low Q oscillator normal mode. These equations are solved using Mathematica (Wolfram, 1991).

For three different experiments in 1987, 1989 and 1991 table 6.4 summarises calculations from the three mode model above. The three mode model gives better agreement with experimental data than does the two-mode model. The resonant frequency of the bar obtained with the three mode model is in excellent agreement with the measured value of 709.7 Hz (Veitch 1982). The calculated resonant frequency of the bending flaps are similar to those obtained with the two-mode model. However the calculated bending flap Q values are lower than expected; $Q_2 = 2{\cdot}10^6$ for bending flap #1 in 1987 and $Q_2 = 3.1{\cdot}10^6$ for bending flap #2 in 1991. In 1989 the model was not successful in predicting the Q values, the measurement was done in a high vibration environment which could have effected the accuracy of Q measurements. The bending flaps were expected to have a Q of greater than $10^7$, with bending flap #2 having a measured lower limit of $5{\cdot}10^6$ (Linthorne 1991).

Even though $Q_b$ has been improved in the new flap design, calculations show that it has virtually no effect on degrading the normal mode resonant Q values. The calculated Q values of the bending flaps are insensitive to changes in $Q_1$, $Q_b$, $\omega_b$ and $m_b$, indicating that the low Q oscillator comprising the bending flap base and the glue joint does not significantly degrade the normal mode Q values. This is emphasised by a order of magnitude increase in $Q_b$ not significantly increasing the bending flap Q value between designs. The observed degradation arises from a dissipative coupling to the flap from the glue joint, which is probably due to residual bending stresses at the glue joint. The modest increase in bending flap Q between designs is probably due to the flap being further removed from the vicinity of the glue joint. The Q should be further improved by making the bending flap base larger to separate the glue joint and the flap even further. This is the same principal that the Rochester group have used to eliminate bonding effects of their transducer system (Fisher et al 1991).

# CHAPTER 7

# DETECTION SENSITIVITY ANALYSIS OF A TWO-MODE RESONANT BAR ANTENNA WITH A PARAMETRIC TRANSDUCER

## 7.1 INTRODUCTION

This chapter describes sensitivity simulations of a two-mode resonant bar gravitational wave detector with a parametric transducer. Simulations are achieved using the commercial algebraic manipulation program "Mathematica". This program supersedes a Fortran program written for a single mode antenna at UWA (Veitch, 1986). An optimal filter approach is adopted here, similar to the methods used at Stanford University (Michelson and Taber, 1981), Louisiana State University (Xu et al, 1989) and Maryland University (Richard, 1986). Previous analysis conducted at UWA used an approximation to the optimal filter by considering a sampling time approach (Veitch, 1986; Linthorne, 1991).

In this chapter we present the theory of optimal filters applied to resonant-mass gravitational wave detectors. The theory is expanded on by defining the noise sources with respect to the detector in terms of strain sensitivity per root Hertz. Simple formulae are derived that calculate the optimal detectable strain sensitivity and effective bandwidth in terms of the spectral strain sensitivity. The model is then applied to the gravitational wave detector at the University of Western Australia. Previously the sensitivity of the detector was limited by environmental vibrations. Assuming a vibration free environment major noise sources are estimated and the potential sensitivity of the detector determined.

## 7.2 OPTIMAL FILTER APPROACH

A standard result in signal detection theory (Wainstein and Zubakov, 1962) states that the signal to noise ratio is optimised by a filter which has a transfer function proportional to the complex conjugate of the signal Fourier transform divided by the total noise spectral density. Expressing all noise components and the signal as a displacement at the transducer, the signal to noise ratio when utilising an optimal filter is given by;

$$\mathrm{SNR} = \frac{1}{2\pi}\int_{-\infty}^{\infty}\frac{|G_{21}(j\omega)|^2|F_1(\omega)|^2}{S_{x2}(\omega)}\,d\omega = 4\int_{0}^{\infty}\frac{|G_{21}(j2\pi f)|^2|F_1(f)|^2}{S_{x2}^{+}(f)}\,df \qquad (7.1).$$

Here $G_{21}(j\omega)$ is the Fourier transform of the impulse response (or transfer function) of displacement sensed by the transducer per force input at the resonant bar, $F_1(\omega)$ is the Fourier transform of the force input, and $S_{x2}(\omega)$ is the double sided spectral density of all noise components referred to a displacement at the transducer. The single sided spectral density is related to the double sided spectral density by, $S^+_{x2}(\omega) = 2S_{x2}(\omega)$ (Reif, 1965).

### 7.2.1 Gravity Wave Signal

Resonant bar detectors will only be sensitive to the most violent gravitational events occurring in space, like supernova explosions. Gravity waves from these events will be short bursts, of duration much less than the ring down time of the antenna. Typically the burst duration is expected to be about 1 ms, corresponding to a broad bandwidth of about 1 kHz. It is therefore valid to assume that the signal force is an impulse given by $F_1(t)=F_g\delta(t)$ [N], which Fourier transforms to $F_1(f)=F_g$ [N/Hz].

To calculate the signal force density of the gravitational wave ($F_g$) on the resonant bar in terms of the impulse energy and resonant bar effective mass ($m_1$) the mean energy for a harmonic oscillator may be considered [13];

$$<E> = \frac{<p^2>}{2m_1} + \frac{k_1<x_1^2>}{2} \quad (7.2).$$

Here the first term is the mean kinetic energy and the second term is the mean potential energy. From the equipartition theorem both these terms are equivalent and equal to kT/2 or $(m_1\omega_1^2<x_1>^2)/2$, where k is Boltzman's constant and T is the surrounding temperature that is in equilibrium with the harmonic oscillator. Hence it follows that;

$$<p> = \sqrt{<E>m_1} = \sqrt{kTm_1} = m_1\omega_1<x_1> \quad (7.3).$$

The peak momentum of a mechanical oscillator is related to the signal force density by;

$$p_{peak} = \sqrt{2}<p> = \int_{-\infty}^{\infty} F_g\,\delta(t)\,dt = F_g \quad (7.4).$$

By equating the impulse energy (E) of a gravitational wave to <E>, from (7.3) and (7.4) the signal force density of a gravitational wave on the resonant bar is given by;

$$F_g = \sqrt{2\,E\,m_1} \quad (7.5)$$

Thus substituting (7.5) in (7.1) the signal to noise ratio can be written as;

$$SNR = \frac{E\,m_1}{\pi}\int_{-\infty}^{\infty}\frac{|G_{21}(j\omega)|^2}{S_{x2}(\omega)}\,d\omega = 8\,E\,m_1\int_{0}^{\infty}\frac{|G_{21}(j2\pi f)|^2}{S^+_{x2}(f)}\,df \quad (7.6),$$

as calculated previously (Michelson and Taber, 1981).

### 7.2.2 Transfer Function

Assuming the harmonic oscillator model from chapter 6, the equation of motion of a resonant bar detector with force inputs at the resonant bar and secondary mass can be written as a matrix equation of the form (Tobar and Blair, 1993);

$$\dot{\underline{x}}(t) = A\underline{x}(t) + \mathbf{B}\underline{f}(t) \qquad (7.7),$$

where A is a 4×4 matrix and B is the matrix that relates force inputs, $\underline{f}(t)$, to states, $\underline{x}(t)$. The state vector contains primary and secondary oscillator displacements ($x_1$ and $x_2$) and velocities ($x_3=\dot{x}_1$ and $x_4=\dot{x}_2$). Defining the 4×4 matrix in Laplace space $\mathbf{H}(s) = (s\mathbf{I}-A)^{-1}$, the Laplace transformation of the equation of motion (7.7) can be written as;

$$\underline{X}(s) = \mathbf{H}(s)\,\mathbf{B}\,\underline{F}(s) + \mathbf{H}(s)\,\underline{x}(t{=}0) \qquad (7.8)$$

The first term in (7.8) is the response due to the force inputs and the second term is the response due to the initial state of the antenna. Substituting s=jω the zero state transfer function of the displacement sensed by the transducer per gravitational impulse force at the resonant bar is given by;

$$G_{21}(j\omega) = \frac{X_2(j\omega)}{F_1} = \frac{H_{23}(j\omega)}{m_1} \qquad (7.9),$$

where $H_{23}(j\omega)$ is a matrix element of $\mathbf{H}(j\omega)$. The zero input transfer function of displacement density at the transducer per initial impulse displacement of the bar is given by;

$$H_x(j\omega) = \frac{X_2(j\omega)}{x_1(t{=}0)} = H_{21}(j\omega) \approx j\omega H_{23}(j\omega) \qquad (7.10),$$

where $H_x(j\omega)$ has the dimension of 1/Hz. From (7.9) and (7.10) we can show;

$$|G_{21}(j\omega)| \approx \frac{|H_x(j\omega)|}{\omega\, m_1} \qquad (7.11).$$

For the UWA detector $H_x(j\omega)$ has been calculated from the equation of motion (Tobar and Blair, 1993; also see chapter 6), and is used in the forth coming noise simulations presented in this paper. This approach can be easily expanded to a multi-mode antenna by expanding the matrix equation (7.7) to a higher order matrix depending on the number of modes.

### 7.2.3 Effective Noise Temperature

In a noisy environment the magnitude of the signal density must be compared with the magnitude of the noise density as a function of frequency. Thus we define;

$$S^{+}_{ns}(f) = \frac{\sqrt{S^{+}_{x2}(f)}}{|H_x(j\omega)|} \qquad (7.12),$$

which has the dimension $[m\sqrt{Hz}]$, and is a spectral form of the displacement noise referred to a displacement at the resonant bar. The effective noise temperature ($T_n$) of the detector is calculated by calculating the effective temperature of the impulse signal required to give a signal to noise ratio of one. Thus by substituting SNR=1 and <E> = $kT_n$ in (7.6), the effective noise temperature can be shown to be;

$$T_n = \frac{\pi}{km_1}\left(\int_{-\infty}^{\infty}\frac{|G_{21}(j\omega)|^2}{S_{x2}(\omega)}\,d\omega\right)^{-1} = \frac{m_1\,\pi^2\,f_1^2}{2\,k}\left(\int_{0}^{\infty}\left(\frac{f_1}{f}\right)^2\frac{1}{S_{ns}^{+}(f)^2}\,df\right)^{-1} \qquad (7.13)$$

as calculated previously (Michelson and Taber, 1981).

### 7.2.4 Signal Strain Density and Spectral Strain Sensitivity

Detailed analysis of resonant bar gravitational wave detectors have been presented previously (Astone et al, 1991; Pallottino and Pizzella, 1991). In this section we expand on the analysis by defining the spectral strain sensitivity which is independent of the incident signal.

It has been shown previously (Pizzella, 1975) that the rms displacement at the end of a resonant bar is related to the signal strain density $\{H_s(f_1)$ [strain/Hz]$\}$ by;

$$H_s(f_1) = \frac{\sqrt{2}\,\pi <x_1>}{4\,f_1\,L} \qquad (7.14)$$

where L is the length of the resonant bar. The rms displacement of an oscillator of effective mass $m_1$ and in equilibrium at temperature $T_n$ is independent of Q and given by;

$$<x_1> = \sqrt{\frac{kT_n}{m_1\omega_1^2}} \qquad (7.15)$$

Thus combining (7.14) and (7.15) we obtain the same formula as derived in (Astone et al, 1991);

$$H_s(f_1) = \frac{\sqrt{2}\,\pi}{4\,f_1\,L}\sqrt{\frac{kT_n}{m_1\omega_1^2}} \qquad (7.16),$$

which relates the detectible signal strain density to the effective noise temperature of the detector. The signal strain density is the natural quantity measured by the detector, to determine the detected strain of the gravitational wave the signal bandwidth must be known ($\Delta f_s$), and is given by;

$$h = H_s(f_1)\,\Delta f_s \qquad (7.17)$$

If the signal bandwidth is greater than the detector bandwidth then $\Delta f_s$ will not be known and can only be estimated, this will always be true for a gravitational burst. Most gravitational wave groups assume a burst duration of 1 ms equivalent to $\Delta f_s$ = 1 kHz.

The resonant bar detector at the University of Western Australia has a large cylindrical niobium bar as the fundamental resonant-mass, with an effective mass of 755 kg (half the bar mass), length of 2.75 m and a fundamental resonant frequency of 710 Hz. Substituting these values in (7.17) gives $H_s(f_1) = 5.45 \cdot 10^{-22}\sqrt{T_n(mK)}$ which is equivalent to $h_{1ms} = 5.45 \cdot 10^{-19}\sqrt{T_n(mK)}$.

For a broad band free-mass detector (interferometer) with a constant spectral density $\{S_x^+\}$ of noise, the optimum SNR is given by;

$$SNR = \frac{4}{S_x^+}\int_0^\infty X(f)^2 \, df \qquad (7.18).$$

Here X(f) is the Fourier transform of the gravitational wave signal in [m/Hz] between the free test masses and $S_x^+$ is the single sided spectral displacement density of noise in [m²/Hz]. Substituting SNR=1 into (7.18) we can write;

$$h^+ = \frac{\sqrt{S_x^+}}{L} = \frac{2}{L}\left(\int_0^\infty X(f)^2 \, df\right)^{1/2} \qquad (7.19)$$

Assuming that the signal density is constant and equal to $X_0$ [m/Hz] over the bandwidth $\Delta f_s$, then the integrated displacement is given by $x = X_0 \Delta f_s$. Substituting the assumed signal density into (7.19) we can show that the optimal detectable strain sensitivity is related to the spectral strain sensitivity of an interferometer by;

$$h = X_0 \Delta f_s / L = \sqrt{\Delta f_s}\, h^+/2 \qquad (7.20)$$

For a resonant bar the optimal detectable signal density can be calculated by substituting (7.13) into (7.16), rearranging we can show that;

$$\frac{1}{H_s(f_1)^2} = 4\int_0^\infty \frac{1}{h^+(f)^2} df \qquad (7.21),$$

where

$$h^+(f) = \frac{\pi\, S_{ns}^+(f)}{4\, f\, L} \qquad (7.22),$$

assuming $f_1$ is the frequency of maximum signal to noise density. Note that $h^+(f)$ is related to $S_{ns}^+(f)$ in exactly the same way as $H_s(f_1)$ is related to the peak value of $x_1$ in (7.14). Here $h^+(f)$ is the spectral strain sensitivity [strain/√Hz] of the resonant bar detector akin to the spectral strain sensitivity of an interferometer detector. This can be proved by calculating the optimal detectable strain sensitivity of a resonant bar antenna when the signal bandwidth is less than the detector bandwidth ($\Delta f_d$), from (7.21) it can be shown that;

$$h = H_s(f_1)\, \Delta f_s = \sqrt{\Delta f_s}\, h^+(f_1)/2 \qquad (7.23),$$

which is exactly the same as expected (7.20). If the detector bandwidth is less than the signal bandwidth then (7.21) can be shown to give;

$$h = H_s(f_1)\,\Delta f_s = \frac{\Delta f_s\, h^+(f_1)}{2\sqrt{\Delta f_d}} \qquad (7.24),$$

Equation (7.23) and (7.24) give a means of converting from spectral strain sensitivity to strain sensitivity.

For a narrow band detector (including resonant bars) it is more meaningful to quote the spectral strain sensitivity $\{h^+(f_1)\}$ and bandwidth $\{\Delta f_d\}$, rather than the strain sensitivity $\{h\}$ when comparing detectors, because the comparison is independent of any incoming signal. This is exactly what is done when comparing narrow and broad band configurations of an interferometer detector (Vinet et al, 1988).

The total strain sensitivity per $\sqrt{Hz}$ due to the sum of all noise components can be calculated by;

$$h^+(f) = \sqrt{\sum_i h_i^+(f)^2} \qquad (7.25).$$

This is a convenient way to plot noise components as they can be directly compared to spectral plots for laser interferometer gravitational wave detectors.

### 7.2.5 Effective Bandwidth

To calculate the effective detector bandwidth the signal to noise ratio may be written as;

$$SNR = \int_0^\infty Snr(f)\, df \approx Snr(f_0)\, \Delta f_{eff} \qquad (7.26),$$

where Snr(f) is the signal to noise ratio density, with the maximum value occurring at $f_0$ denoted by $Snr(f_0)$. Thus the effective bandwidth is given by;

$$\Delta f_{eff} = \frac{SNR}{Snr(f_0)} = h^+(f_0)^2 \int_0^\infty \frac{1}{h^+(f)^2}\, df = \frac{h^+(f_0)^2}{4\, H_s(f_0)^2} \qquad (7.27).$$

From any signal to noise ratio or spectral strain density plot for any gravitational wave detector, (7.27) determines an easy method for calculating the effective bandwidth.

## 7.3 NOISE COMPONENTS

This section analyses all the significant noise components that can occur in a two-mode resonant bar antenna with a parametric transducer. In the next section a detailed analysis of the optimisation processes with numerical calculations of expected improvements are presented.

### 7.3.1 Acoustic Thermal Noise

The acoustic losses in a two mode resonant bar antenna cause a white spectral density of force to excite the resonant system according to Nyquist's theorem (Reif, 1965). The single sided spectral density of force in (Newtons)$^2$ per Hz, exciting a mechanical oscillator in equilibrium with the surroundings of temperature T, is given by;

$$Sf_n^+ = 4\,k\,T\,h_n = 4\,k\,T\,\omega_n\,m_n/Q_n \qquad (7.28).$$

Here n = 1 refers to the resonant bar and n = 2 refers to the secondary mass in the two-mode mechanical oscillator system.

The UWA system consists of a 1.5 tonne niobium resonant bar as the primary oscillator and a niobium bending flap with 0.45 kg effective mass as the secondary oscillator. In the past the secondary mass Q value has been limited to $3\cdot10^6$ by dissapative effects due to the glue joint between the oscillators and gas damping of the secondary mass. Assuming that these limitations can be overcome a bending flap and resonant bar Q value of around $5\cdot10^7$ and $2\cdot10^8$ should be achieved respectively.

From (7.28) it is clear that the level of noise is proportional to temperature and inversely proportional to acoustic Q value. This is the reason that high acoustic Q materials cooled to low temperatures are in use in the majority of resonant bar detectors around the world. It is planned to operate the antenna at UWA in the cold damped regime below resonance, in this regime both the Q value and the effective temperature are reduced, however nothing happens to the ratio of T/Q so the level of (7.28) is unchanged (Tsubono, 1991). However the transducer does supply a smaller resistive component of damping that contributes mostly to the acoustic loss of the secondary mass (Veitch, 1991). The equivalent Q of this component is given by;

$$\frac{1}{Q_r} = \frac{\beta\,\omega_2}{\Omega_0} \qquad (7.29),$$

where $\beta$ is the total electro-mechanical coupling given by (B18) in appendix B.

For the UWA operating conditions given in section 6.3, $\beta$ = 0.017 and hence $Q_r$ = $8\cdot10^8$. This equation can be used to calculate the maximum coupling before the resistive damping starts to affect the thermal performance. This will occur when $Q_r \sim 5\cdot10^7$, which limits the electro-mechanical coupling to $\beta \sim 0.25$.

The displacement at the transducer due to the spectral density of force given by (7.28) will be filtered by the transfer function of the system. Thus the single sided displacement density at the transducer will be;

$$S^+_{xn} = Sf^+_n \, |G_{2n}(j2\pi f)|^2 \qquad (7.30),$$

where $G_{2n}(j2\pi f)$ is the transfer function of force at oscillator n to displacement at the transducer. Substituting (7.30) into (7.22) the strain sensitivity per $\sqrt{Hz}$ is given by;

$$h^+_{th\,n}(f) = \frac{i\,\pi}{4\,L\,f}\frac{|G_{2n}(j2\pi f)|}{|H_x(j2\pi f)|}\sqrt{Sf^+_n} \qquad (7.31)$$

For the resonant bar when n=1, (7.31) reduces to;

$$h^+_{th\,1}(f) = \frac{\pi}{2\,L\,f}\sqrt{\frac{k\,T}{\omega_1\,Q_1\,m_1}} \qquad (7.32).$$

Figure 7.1 compares the thermal noise components at 5 K, in terms of displacement sensitivity (7.30) at the transducer, when the acoustic Q value of the bending flap was limited to $3{\cdot}10^6$ and then assumed to be $5{\cdot}10^7$. These components are narrow band with respect to the broad band series electronic noise when referred to the transducer. If the resonant bar thermal noise was the only noise component the signal to noise density (SNR/Hz) is independent of frequency, which highlights the potential of using a resonant bar detector as a broad band detector (Michelson and Taber, 1984). However the addition of the secondary mass (bending flap) limits the broad band nature of the thermal noise. Figure 7.2 and 7.3 illustrate that the peak value of the SNR/Hz due to the thermal noise is limited by the resonant bar, while the bandwidth is dependent on the Q of the secondary mass. When the Q value of the secondary mass is $3{\cdot}10^6$ the spectral strain sensitivity is $3.6{\cdot}10^{22}$ with a detector bandwidth of 3.3 Hz (figure 7.2). If the bending flap Q increases to $5{\cdot}10^7$ the spectral strain sensitivity remains constant as it is limited by the thermal noise in the bar, however the detector bandwidth will increase to 12.1 Hz.

Thus the secondary mass provides amplitude amplification of the resonant bar reducing any broad band additive noise when referred to the secondary mass at the expense of degrading the narrow band noise by limiting the detector bandwidth.

### 7.3.2 Electronic Noise

Figure 7.4 shows a schematic of the parametric transducer system at UWA. It consists of an ultra-low noise pump oscillator incident on a re-entrant cavity transducer. The

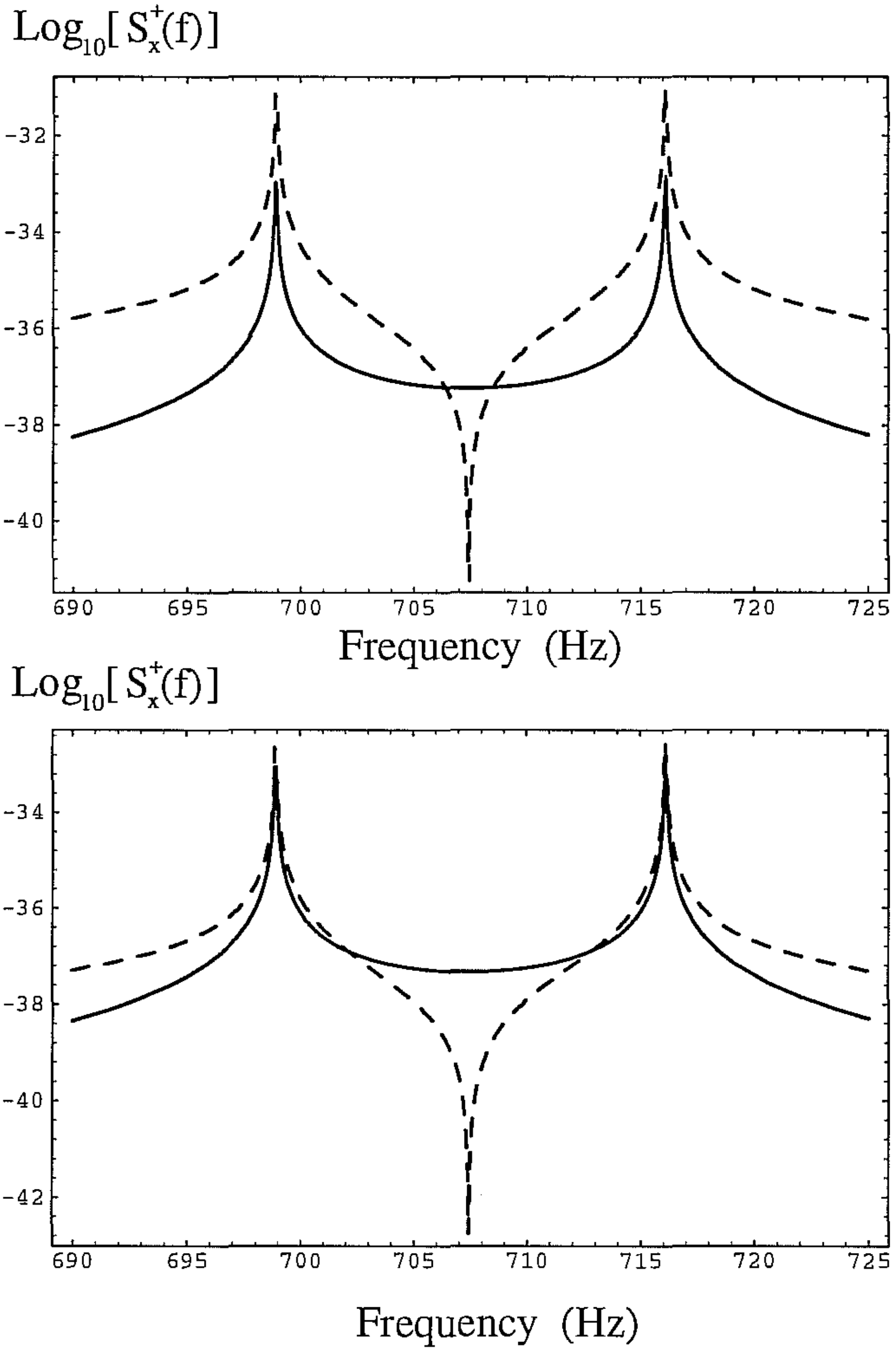

# Figure 7.1

Themal noise components referred as a spectral displacement density in square meters per hertz at the transducer, the dashed line is the dissipative thermal noise due to the secondary mass and the bold line is the dissipative thermal noise due to the resonant bar (Q=2 $10^8$): Above, when the secondary mass Q value was limited to 3 $10^6$ due to a dissipative effect: Below, assuming a bending flap Q value of 5 $10^7$.

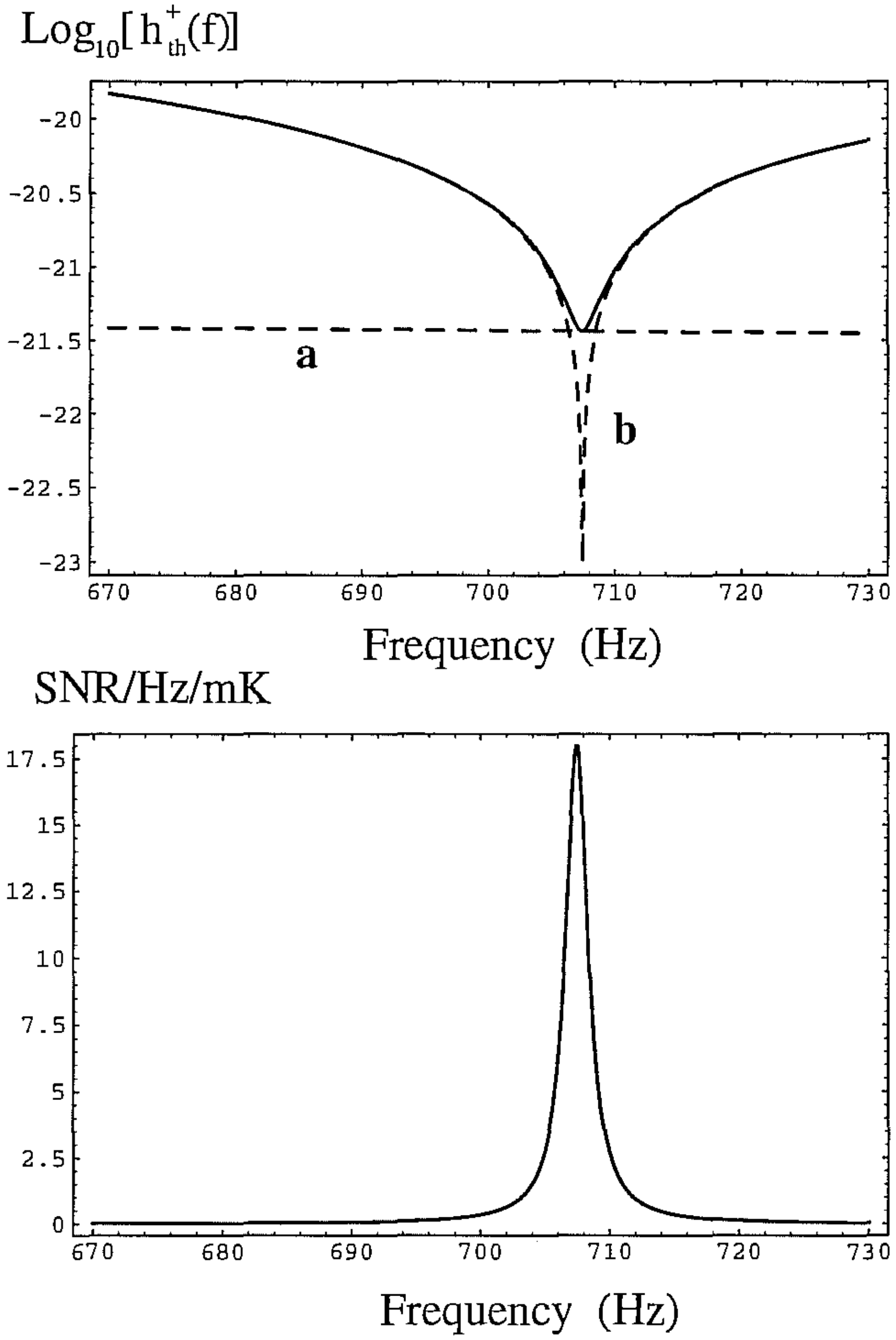

# Figure 7.2

Thermal noise components when the bending flap at 5 K is limited to a Q value of 3 $10^6$ and the resonant bar Q value is 2 $10^8$ : Above, strain sensitivity per root Hertz. The dashed curves **a** and **b** are the components due to the resonant bar and the bending flap respectively, while the bold line is the total thermal component of strain sensitivity per root Hertz: Below, signal to noise ratio per Hertz per mK impulse, if the thermal noise was the only noise component.

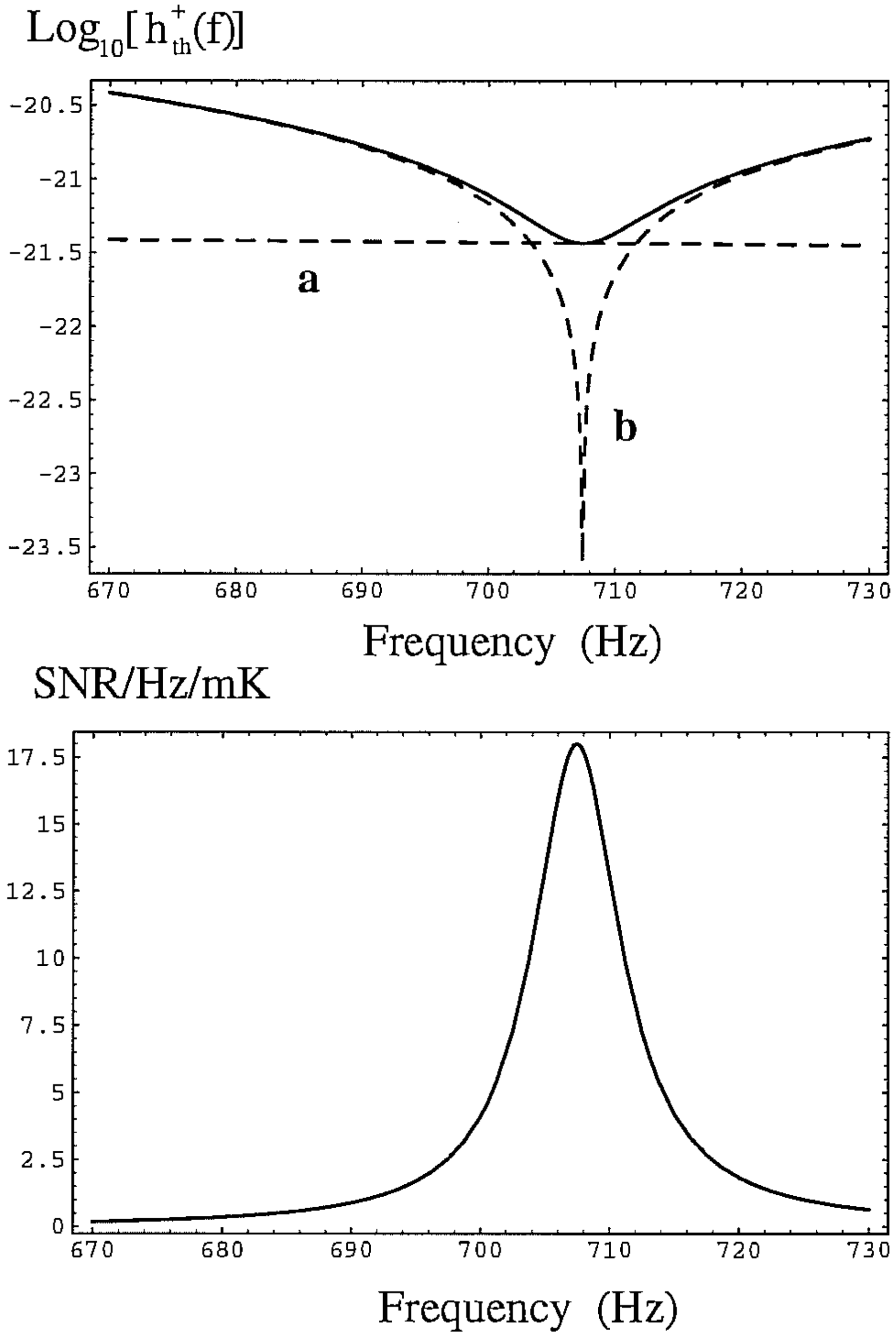

# Figure 7.3

Thermal noise components when the bending flap has a Q value of 5 $10^7$ and the resonant bar Q value is 2 $10^8$ at 5 K: Above, strain sensitivity per root Hertz. The dashed curves **a** and **b** are the components due to the resonant bar and bending flap respectively, while the bold line is the total thermal component of strain sensitivity per root Hertz: Below, signal to noise ratio per Hertz per mK impulse, if the thermal noise was the only noise component.

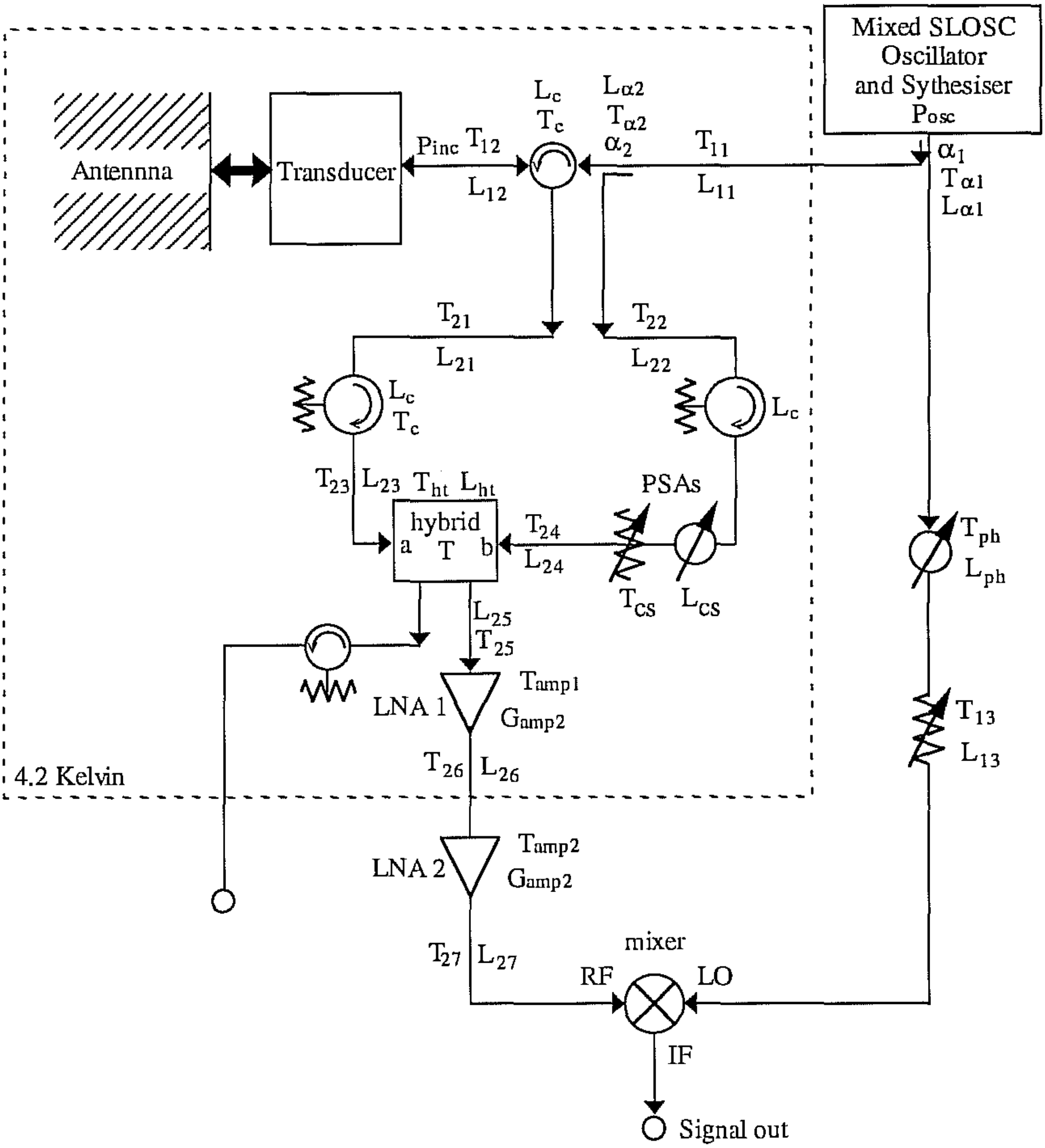

# Figure 7.4

The UWA signal demodulation circuit. Symbols representing the effective noise temperature (T), losses (L) and amplifier gain are shown along side each element. The symbols $\alpha_1$ and $\alpha_2$ represent the directional coupler power splitting ratio. $P_{inc}$ represents the power incident on the transducer, and $P_{osc}$ the power available from the oscillator. $G_{amp1}$ and $G_{amp2}$ are the gains of amplifiers 1 and 2 respectively.

reflected pump signal is modulated by the two-mode resonant bar. This signal is demodulated by first suppressing the carrier then amplifying the signal at 4.2 K before mixing down the signal to base band. In this section the form of the significant electronic noise sources supplied by the transducer are determined.

#### 7.3.2.1 Series Demodulation Noise

After the pump signal is reflected off the re-entrant cavity transducer the signal will be attenuated by losses in the non-contacting coupling, co-axial cable, circulator and hybrid T carrier suppression system, before the first Low Noise Amplifier (LNA1) is encountered in the demodulation system. These losses will increase the effective noise temperature of this amplifier as discussed in chapter 5, due to the attenuation of the signal before amplification. Also these devices will supply their own component of Nyquist noise due to the dissipative nature of the losses (Robins, 1984). The Nyquist noise of the components after LNA1 and the noise temperature of LNA2 are rendered less significant due to the amplification of the signal in LNA1. The carrier suppression is needed to decrease the input power into LNA1 and hence suppress any non-linear upconversions that may enhance the amplifier flicker noise spectral density. The second amplifier (LNA2) may be necessary to supply enough signal power into the RF port of the mixer in order to overcome the mixer flicker noise. However the input power level into LNA2 will be greater than LNA1, and may be large enough to create non-linear upconversions. It must be experimentally determined whether this level of flicker noise is less than the mixer noise level without LNA2. In the following simulations LNA2 is assumed to operate in the white noise regime like LNA1.

The carrier suppression involves a hybrid T to reduce the RF carrier level before amplification by adding some carrier signal from the incident power. This must be adjusted to have nearly the same amplitude and opposite phase, cancelling the carrier but leaving the reflected signal unaffected. It has been shown in (Veitch, 1991) that a badly designed carrier suppression system can cause the in phase component of the phase noise to decorrelate, and thus not cancel in the double balance mixer. However the suppression system must be designed well enough to ensure that the amplifier operates in the low input power regime, where flicker noise is eliminated. To achieve the desired flicker noise suppression the carrier suppression must be well designed a priori. Hence the effect of the carrier suppression decorrelating the phase noise will not be considered in this thesis as it should be negligible in a well designed system. The system at UWA has been shown to operate for long periods of time with better than 60 dB pump signal suppression (Ivanov, 1993), which is good enough to disregard amplifier flicker noise. The suppression system will however add an extra component of Nyquist noise due to the extra arm used to cancel the modulated carrier.

The Nyquist single sided noise that is generated in a passive device with loss L and temperature T is equal to kT(L-1) Watts/Hz when referred to the input of the passive device. Here we assume that a half of this energy is present in each quadrature component of amplitude and phase. Referring the noise to the transducer output, the Nyquist noise per quadrature (N) for the devices in the output line from the transducer to the mixer can be written as;

$$
\begin{aligned}
N_{12a} &= k\,T_{12}\,(L_{12}-1)/2 \\
N_{ca} &= k\,T_c\,L_{12}\,(L_c-1)/2 \\
N_{21} &= k\,T_{21}\,L_{12}\,L_c\,(L_{21}-1)/2 \\
N_{cb} &= k\,T_c\,L_{12}\,L_c\,L_{21}\,(L_c-1)/2 \\
N_{23} &= k\,T_{23}\,L_{12}\,L_c\,L_{21}\,L_c\,(L_{23}-1)/2 \\
N_{hta} &= k\,L_{23}\,L_{12}\,L_c\,L_{21}\,L_cT_{ht}\,(L_{ht}-1)/2 \\
N_{htb} &= k\,L_{23}\,L_{12}\,L_c\,L_{21}\,L_c/2\ \times \\
&\quad \left(T_{22}(1-1/L_{22})/(L_cL_{24}L_{cs})+T_c(1-1/L_c)/(L_{24}L_{cs})+T_{cs}(1-1/L_{cs})/(L_{24})+T_{24}(1-1/L_{24})\right) \\
N_{25} &= k\,T_{25}\,L_{ht}\,L_{23}\,L_{12}\,L_cL_{21}\,L_c\,(L_{25}-1)/2 \\
N_{ampl} &= k\,T_{ampl}\,L_{25}\,L_{ht}\,L_{23}\,L_{12}\,L_cL_{21}\,L_c/2 \\
N_{26} &= k\,T_{26}\,L_{25}\,L_{ht}\,L_{23}\,L_{12}\,L_cL_{21}\,L_c\,(L_{26}-1)/(2\,G_{ampl}) \\
N_{amp2} &= k\,T_{amp2}\,L_{26}\,L_{25}\,L_{ht}\,L_{23}\,L_{12}\,L_cL_{21}\,L_c/(2\,G_{ampl}) \\
N_{27} &= k\,T_{27}\,L_{26}\,L_{25}\,L_{ht}\,L_{23}\,L_{12}\,L_cL_{21}\,L_c\,(L_{27}-1)/(2\,G_{ampl}\,G_{amp2})
\end{aligned}
\qquad (7.33).
$$

Here the subscripts match the subscripts of the symbols representing the elements shown in figure 7.4. Note there are two terms for the Nyquist noise produced by the hybrid T, $N_{hta}$ is the Nyquist noise produced by the losses in the hybrid T and $N_{htb}$ is the Nyquist noise produced in the extra arm that cancels the carrier from the modulated signal.

The last component of noise in the demodulation chain is the series noise due to the double balanced mixer (DBM) (Veitch, 1991). The two major components of noise that are considered here are the mixer 1/f noise and local oscillator amplitude noise. A real DBM has an output voltage spectral density due to the local oscillator amplitude noise of;

$$
\begin{aligned}
S_{V\,LOAM} &= 400\,N_{LOAM}/I \quad \text{volts}^2/\text{Hz} \qquad (7.34), \\
N_{LOAM} &\approx S_{am}\,P_{osc}\,(1-1/\alpha_1)/(L_{ph}\,L_{13})
\end{aligned}
$$

where I is the amplitude isolation of the LO port which is typically about 36 dB for a low noise X-band mixer. Here it is assumed that the Nyquist noise produced in this arm is insignificant, which will always be true when the local oscillator power is greater than a milliwatt with an amplitude noise greater than -175 dBc/Hz. The output mixer flicker noise for a low noise DBM is of the order $S_{V\,1/f}(f) \approx 1.9 \cdot 10^{-14}/f$ volts$^2$/Hz.

In order to refer the mixer voltage noise to the transducer, the sensitivity of the mixer ($k_m$ volts/$\sqrt{W}$) from RF port to output must be determined. The sensitivity of the mixer is related to the single side band conversion loss ($L_{ssb}$) by;

$$k_m^2 = 380/L_{ssb} \tag{7.35}$$

and the single sided conversion loss is related to the local oscillator input power. For a low noise X-band Watkins Johnson DBM the single sided conversion loss is given by;

$$L_{ssb}(dB) = 6.78 - 0.525\, P_{LO}(dBm) + 0.0645\, P_{LO}^2(dBm) \tag{7.36}$$

This gives a minimum conversion loss of $L_{ssb}(dB) = 5.7$ dB when the local oscillator power is $P_{LO}(dBm) = 4.0$ dBm. Thus the maximum sensitivity of the mixer can be calculated to be $k_m = 10$ volts/$\sqrt{W}$. As the local oscillator power varies from the optimum, the mixer sensitivity will decrease and hence the mixer noise floor will increase when referred to the RF port of the mixer. The effective mixer noise power when referred to the output of the transducer can be written as;

$$N_m = S_v\, L_{27}\, L_{26}\, L_{25}\, L_{ht}\, L_{23}\, L_{12}\, L_c L_{21}\, L_c / (k_m^2\, G_{ampl}\, G_{amp2}) \tag{7.37}$$

Here $S_v = S_{V\,LOAM} + S_{V\,1/f}$, and is the total mixer output noise in volts squared per Hz.

To convert all these components of series noise discussed in this sub-section to a displacement noise at the transducer it follows that;

$$S_x = \sqrt{\sum_i N_i / S_{PX2}}\ \ m/\sqrt{Hz} \tag{7.38}$$

which is similar to equation (5.2) in chapter 5. Here $S_{PX2}$ is the forward transductance of the transducer in $\sqrt{W}$/m.

### 7.3.2.2 Incident Series Noise

The incident series noise is comprised of the oscillator amplitude and phase noise ($S_{am}(f)$ & $S_{pm}(f)$), and the Nyquist noise produced by the losses in the pump oscillator path that couples the energy from the oscillator to the transducer. In the demodulation system shown in figure 7.4 there will be noise cancellation in the mixer and carrier suppression circuit. Appendix C describes noise side band behaviour and cancellation in the mixer when reflecting of a resonant transducer cavity and (Veitch, 1991) describes the noise cancellation in the carrier suppression system. Both cancel correlated in phase components. Therefore the carrier suppression will not suppress any more oscillator noise than the mixer does. However it will suppress the Nyquist noise produced by the co-axial cable loss $L_{11}$ as it is not correlated at the mixer LO and RF ports, but is correlated at the hybrid T carrier suppression ports. The Nyquist noise per quadrature produced by the passive elements connecting the oscillator to the transducer, when referred to the input of the transducer are given by;

$$
\begin{aligned}
N_{\alpha 1} &= k\,T_{\alpha 1}\,(1-1/L_{\alpha 1})(1-1/\alpha_2)/(2\,L_{11}\,L_{12}\,L_c\,L_{\alpha 2}\,\alpha_1)\\
N_{11} &= k\,T_{11}\,(1-1/L_{11})(1-1/\alpha_2)/(2\,L_{12}\,L_c\,L_{\alpha 2})\\
N_{\alpha 2} &= k\,T_{\alpha 2}\,(1-1/L_{\alpha 2})(1-1/\alpha_2)/(2\,L_{12}\,L_c)\\
N_{cc} &= k\,T_c\,(1-1/L_c)(1-1/\alpha_2)/(2\,L_{12})\\
N_{12b} &= k\,T_{12}\,(1-1/L_{12})(1-1/\alpha_2)/2 \qquad (7.39).
\end{aligned}
$$

When the phase of the demodulation bridge circuit is set to maximise the mixer sensitivity (and hence electro-mechanical coupling) at a certain pump oscillator offset frequency, the phase noise power that is reflected off the transducer and is not cancelled in the mixer is given by (see appendix C equation (C8));

$$N_q = \left((S_{pm}\,P_{osc})(1-1/\alpha_2)/(L_{11}\,L_{12}\,L_c\,L_{\alpha 1}\,L_{\alpha 2}\,\alpha_1) + N_{\alpha 1}\right)\left(|S_{pp}|_q^2 + |S_{ap}|^2\right) \quad (7.40),$$

when referred to the output of the transducer. Here $|S_{pp}|_q^2$ is the fraction of quadrature phase noise reflected off a resonant cavity and $|S_{ap}|^2$ is the fraction of phase noise reflected as amplitude noise. Equation (7.40) calculates the component due to the oscillator phase noise and Nyquist noise in the first power splitter. The phase noise produced by the rest of the incident line that is not suppressed by the carrier suppression system, is given by;

$$N_{nyq} = (\varepsilon^2(N_{11} + N_{\alpha 2}) + N_{cc} + N_{12b})\left(|S_{pp}|^2 + |S_{ap}|^2\right) \quad (7.41),$$

when referred to the output of the transducer. Here $|S_{pp}|^2$ is the fraction of phase noise reflected off the transducer and $\varepsilon^2$ is the power carrier suppression factor (Veitch, 1991).

To calculate the effective net series noise when referred to a displacement at the transducer, these incident noise components given by (7.40) and (7.41) are summed up with the series noise components of the demodulation system derived in the last sub-section using (7.38).

### 7.3.2.3 Back Action Noise

The back action noise creates a force spectral density at the secondary mass due to the amplitude noise inside the resonant parametric transducer. This noise component has a similar nature to the thermal noise produced in the secondary mass, and is narrow band when referred to the displacement of the transducer. The back action force spectral density acting on the secondary mass caused by the incident noise is given by;

$$S^{+}_{f\,ba} = |S_{fp}|^2\, N_{pm} + |S_{fa}|^2\, N_{am} \qquad (7.42),$$

where $N_{pm}$ and $N_{am}$ are the double side band (or single sided) phase and amplitude noise side band power incident on the transducer. The back action scattering parameters $S_{fp}$ and $S_{fa}$ are described in appendix B. This noise source limits the available electro-mechanical coupling. The electro-mechanical coupling is increased by increasing the incident power, which will inturn increase the power in the noise side bands. The back action noise is proportional to both the electro-mechanical coupling and the power in the noise side bands, hence both act to cause an increase in the spectral density of force acting on the secondary mass.

The spectral density of displacement at the transducer due to the spectral density of force is;

$$S^{+}_{x\,ba} = S^{+}_{f\,ba}\, |G_{22}(j2\pi f)|^2 \qquad (7.43).$$

This component of noise is narrow band and similar to the secondary mass thermal noise given by (7.30) when n=2.

### 7.3.2.4 Correlated Back Action and Series Noise

There will be correlated noise components due to the reflected series noise and the back action noise. Defining the following scattering parameters referred to the displacement at the transducer; $S^{x}_{fa} = G_2(j2\pi f)S_{fa}$, $S^{x}_{fp} = G_2(j2\pi f)S_{fp}$, $S^{x}_{ap} = S_{ap}/S_{PX2}$ and $S^{x}_{pp} = S_{pp}/S_{PX2}$, the effective complex displacement when referred to the transducer is;

$$x = S^{x}_{fa}\, N_{am} + S^{x}_{fp}\, N_{pm} + S^{x}_{pa}\, N_{am} + S^{x}_{pp}\, N_{pm} + S^{x}_{aa}\, N_{am} + S^{x}_{ap}\, N_{pm} \qquad (7.44).$$

When operating the demodulation system to maximise the mixer sensitivity means that $S_{pa}+S_{aa}=0$ (see appendix C). Therefore (7.44) reduces to;

$$x = S^{x}_{fa}\, N_{am} + S^{x}_{fp}\, N_{pm} + (S^{x}_{pp}+S^{x}_{ap})\, N_{pm} \qquad (7.45)$$

The first two terms are back action terms and the last is the reflected series noise. To calculate the noise spectral density (7.45) must be multiplied by its complex conjugate. The cross terms between the back action and series noise will give the correlated components, while the rest of the terms will give the back action and series terms as discussed in the last two sub-sections. Hence the correlated noise can be written from (7.45) as;

$$S_{x\,cor} = \left(S_{fp}^{x\,*}(S_{pp}^{x}+S_{ap}^{x}) + (S_{pp}^{x}+S_{ap}^{x})^{*}S_{fp}^{x}\right) N_{pm} \qquad (7.46).$$

This component is only dependent on the phase noise and must be further broken down into a component that is correlated in the mixer and a component that is not. For the component that is correlated in the mixer only the quadrature component will be important (or Imaginary), thus the effective correlated noise component is given by;

$$S_{x\,cor\,q} = \left(S_{fp}^{x\,*}(S_{pp}^{x}+S_{ap}^{x})+(S_{pp}^{x}+S_{ap}^{x})^{*}S_{fp}^{x}\right) N_{pm\,unc} + \mathrm{Im}\left[\left(S_{fp}^{x\,*}(S_{pp}^{x}+S_{ap}^{x})+(S_{pp}^{x}+S_{ap}^{x})^{*}S_{fp}^{x}\right)\right] N_{pm\,cor} \qquad (7.47),$$

where $N_{pm} = N_{pm\,unc} + N_{pm\,cor}$, and $N_{pm\,unc}$ is the uncorrelated noise power per Hz at the RF port of the mixer referred to the output of the transducer, and $N_{pm\,cor}$ is the correlated noise power per Hz at the RF port of the mixer referred to the output of the transducer. It turns out that for the level of noise components in the following simulations, this component of noise is insignificant.

**7.3.2.5 Resonant Transducer Nyquist Noise**

There is kT/2 Watts/Hz of Nyquist noise produced per quadrature (amplitude and phase) in the transducer resonant circuit. At the output of the transducer the series noise will be filtered and is given by;

$$N_{sn} = |T(j\omega)|^{2}\, kT/2 \quad ; \quad T(j\omega) = \frac{2\sqrt{\beta_e}}{(1+\beta_e)(1+j2Q_e(\omega-\Omega_o)/\Omega_o)} \qquad (7.48),$$

where $T(j\omega)$ is the complex transmission coefficient of the resonant circuit.

The Nyquist noise that is not at the output of the transducer will remain in the cavity. The noise power in the amplitude quadrature will create back action noise and is given by;

$$N_{bn} = |\Gamma(j\omega)|^{2}\, kT/2 \quad ; \quad |\Gamma(j\omega)|^{2} = 1-|T(j\omega)|^{2} \qquad (7.49),$$

where $\Gamma(j\omega)$ is the complex reflection coefficient of the resonant circuit.

### 7.3.3 Seismic Modulation Noise

It was first proposed by Veitch to operate a parametric transducer with the pump oscillator frequency just below the resonant frequency of the transducer (Veitch, 1986). This was to avoid pump oscillator drift above resonance which causes parametric excitation of the resonant bar detector. As the pump oscillator frequency is varied below resonance the interaction between the transducer and the antenna causes the vibrational frequency and amplitude to change. This effect was shown in chapter 6 (see figures 6.5 - 6.7). Low frequency seismic noise modulates the resonant frequency of the cavity, thereby changing the pump frequency offset, and hence modulating the antenna frequency and mode temperature. Close to resonance this effect is maximised and is proportional to the electro-mechanical coupling.

Previous analysis shows that the seismic effect will limit the electro-mechanical coupling to $10^{-3}$ (Linthorne, 1991). It has therefore been proposed in this thesis that the transducer be operated with the pump oscillator incident on the lower turning point of the antenna frequency versus pump offset characteristic, calculated in chapter 6. All the following calculations of the UWA antenna sensitivity are calculated with the pump incident on the turning lower point. This is at the expense of limiting the electro-mechanical coupling to 9/16 times its value on resonance, for a given incident pump power. This is however a small price to pay as operating on the lower turning point will eliminate to first order the seismic upconversions, allowing electro-mechanical couplings larger than $10^{-3}$.

## 7.4 ANTENNA SENSITIVITY AND OPTIMISATION

In this section the sensitivity of the UWA detector in its previous configuration during 1991 is estimated using the optimal filter theory. Then by making incremental improvements in the detector a step by step optimisation process is undertaken to estimate the sensitivity for the next experimental operation of the UWA detector. Further simulations are presented to consider the possibility of approaching the thermal noise limits calculated in section 7.3.1.

Table 7.1a shows a summary of the major noise components for a variety of configurations of the UWA gravitational radiation detector. The tabulated noise components are broken into two categories: 1. $S_x$, Series noise components which are given as a constant spectral density in $m^2$/Hz at 710 Hz, when referred to displacement at the transducer: 2. $S_{F2}$, Narrow band noise components acting on the secondary mass which are given as a constant spectral density of force in $N^2$/Hz, when referred to the force driving the secondary oscillator. The last three columns estimate the spectral

| Simulation | $S_x$ Amp | $S_x$ HTb | $S_x$ Nyq | $S_x$ Osc' | $S_x$ Tot' | $S_{f2}$ Therm' | $S_{f2}$ B.A. | $S_{f2}$ Dis' | $S_{f2}$ Tot' | $h^+(f)$ | Bandwidth (Hz) | $h_{rms}$ pulse ~1ms |
|---|---|---|---|---|---|---|---|---|---|---|---|---|
| **1** | — | — | — | — | — | $1.8\times10^{-25}$ | — | — | **$1.8\times10^{-25}$** | $3.7\times10^{-22}$ | 3.3 | $7.1\times10^{-20}$ |
| **2** | — | — | — | — | — | $1.1\times10^{-26}$ | — | — | **$1.1\times10^{-26}$** | $3.7\times10^{-22}$ | 12.1 | $3.7\times10^{-20}$ |
| **3** | — | — | — | — | **$4.3\times10^{-32}$** | — | — | — | **$3.5\times10^{-23}$** | $1.8\times10^{-20}$ | 0.84 | $7.1\times10^{-18}$ |
| **4** | $9.5\times10^{-35}$ | $4.5\times10^{-37}$ | $3.1\times10^{-35}$ | $1.5\times10^{-37}$ | **$1.3\times10^{-34}$** | $1.8\times10^{-25}$ | $4.8\times10^{-26}$ | $1.8\times10^{-28}$ | **$2.3\times10^{-25}$** | $1.5\times10^{-21}$ | 1.06 | $5.0\times10^{-19}$ |
| **5** | $9.5\times10^{-35}$ | $4.5\times10^{-37}$ | $3.1\times10^{-35}$ | $1.5\times10^{-37}$ | **$1.3\times10^{-34}$** | $1.8\times10^{-25}$ | $8.3\times10^{-28}$ | $1.8\times10^{-28}$ | **$1.8\times10^{-25}$** | $1.3\times10^{-21}$ | 0.95 | $4.7\times10^{-19}$ |
| **6** | $3.3\times10^{-37}$ | $1.2\times10^{-37}$ | $1.1\times10^{-37}$ | $1.5\times10^{-37}$ | **$7.2\times10^{-37}$** | $1.8\times10^{-25}$ | $4.8\times10^{-26}$ | $1.8\times10^{-28}$ | **$2.3\times10^{-25}$** | $1.3\times10^{-21}$ | 10.7 | $1.4\times10^{-19}$ |
| **7** | $3.3\times10^{-37}$ | $1.2\times10^{-37}$ | $1.1\times10^{-37}$ | $1.5\times10^{-37}$ | **$7.2\times10^{-37}$** | $1.8\times10^{-25}$ | $8.3\times10^{-28}$ | $1.8\times10^{-28}$ | **$1.8\times10^{-25}$** | $1.3\times10^{-21}$ | 12.0 | $1.3\times10^{-19}$ |
| **8** | $3.3\times10^{-37}$ | $1.3\times10^{-37}$ | $1.1\times10^{-37}$ | $1.5\times10^{-37}$ | **$7.2\times10^{-37}$** | $1.1\times10^{-26}$ | $4.8\times10^{-26}$ | $1.9\times10^{-28}$ | **$5.9\times10^{-26}$** | $1.3\times10^{-21}$ | 20.2 | $1.1\times10^{-19}$ |
| **9** | $3.3\times10^{-37}$ | $1.3\times10^{-37}$ | $1.1\times10^{-37}$ | $1.5\times10^{-37}$ | **$7.2\times10^{-37}$** | $1.1\times10^{-26}$ | $8.3\times10^{-28}$ | $1.9\times10^{-28}$ | **$1.2\times10^{-26}$** | $7.9\times10^{-22}$ | 15.7 | $7.1\times10^{-20}$ |
| **10** | $7.0\times10^{-38}$ | $2.6\times10^{-38}$ | $2.3\times10^{-38}$ | $1.5\times10^{-37}$ | **$2.7\times10^{-37}$** | $1.8\times10^{-25}$ | $1.0\times10^{-24}$ | $8.9\times10^{-28}$ | **$1.2\times10^{-24}$** | $8.7\times10^{-22}$ | 3.0 | $1.8\times10^{-19}$ |
| **11** | $7.0\times10^{-38}$ | $2.6\times10^{-38}$ | $2.3\times10^{-38}$ | $1.5\times10^{-37}$ | **$2.7\times10^{-37}$** | $1.8\times10^{-25}$ | $1.8\times10^{-26}$ | $8.9\times10^{-28}$ | **$2.0\times10^{-25}$** | $8.7\times10^{-22}$ | 7.4 | $1.1\times10^{-19}$ |
| **12** | $7.0\times10^{-38}$ | $2.6\times10^{-38}$ | $2.3\times10^{-38}$ | $1.5\times10^{-37}$ | **$2.7\times10^{-37}$** | $1.1\times10^{-26}$ | $1.0\times10^{-24}$ | $8.9\times10^{-28}$ | **$1.1\times10^{-24}$** | $8.6\times10^{-22}$ | 3.2 | $1.7\times10^{-19}$ |
| **13** | $7.0\times10^{-38}$ | $2.6\times10^{-38}$ | $2.3\times10^{-38}$ | $1.5\times10^{-37}$ | **$2.7\times10^{-37}$** | $1.1\times10^{-26}$ | $1.8\times10^{-26}$ | $8.9\times10^{-28}$ | **$2.9\times10^{-26}$** | $8.6\times10^{-22}$ | 18.4 | $7.1\times10^{-20}$ |
| **14** | $7.0\times10^{-38}$ | $2.6\times10^{-38}$ | $2.3\times10^{-38}$ | $1.5\times10^{-37}$ | **$2.7\times10^{-37}$** | $1.1\times10^{-26}$ | $5.4\times10^{-27}$ | $8.0\times10^{-28}$ | **$1.6\times10^{-26}$** | $8.3\times10^{-22}$ | 21.6 | $6.3\times10^{-20}$ |
| **15** | $7.0\times10^{-38}$ | $2.6\times10^{-38}$ | $2.3\times10^{-38}$ | $1.5\times10^{-37}$ | **$2.7\times10^{-37}$** | $1.1\times10^{-26}$ | $1.4\times10^{-28}$ | $8.0\times10^{-28}$ | **$1.1\times10^{-26}$** | $7.5\times10^{-22}$ | 21.5 | $5.6\times10^{-20}$ |
| **16** | $7.0\times10^{-38}$ | $2.6\times10^{-38}$ | $2.3\times10^{-38}$ | — | **$1.2\times10^{-37}$** | $1.1\times10^{-26}$ | $1.4\times10^{-28}$ | $8.0\times10^{-28}$ | **$1.1\times10^{-26}$** | $6.3\times10^{-22}$ | 20.1 | $4.9\times10^{-20}$ |
| **17** | $2.0\times10^{-38}$ | $7.5\times10^{-39}$ | $6.7\times10^{-39}$ | $1.5\times10^{-37}$ | **$1.8\times10^{-37}$** | $1.1\times10^{-26}$ | $3.9\times10^{-27}$ | $2.7\times10^{-27}$ | **$1.8\times10^{-26}$** | $7.3\times10^{-22}$ | 20.5 | $5.7\times10^{-20}$ |
| **18** | $2.0\times10^{-38}$ | $7.5\times10^{-39}$ | $6.7\times10^{-39}$ | — | **$3.5\times10^{-38}$** | $1.1\times10^{-26}$ | $3.9\times10^{-27}$ | $2.7\times10^{-27}$ | **$1.8\times10^{-26}$** | $4.6\times10^{-22}$ | 13.5 | $4.5\times10^{-20}$ |

# Table 7.1a

Summary of the major noise components for simulation 1-18. The noise components are broken into two categories: $S_x[m^2/Hz]$, series noise components referred to a displacement at the transducer: $S_{F2}[m^2/Hz]$, narrow band noise components referred to a force driving the bending flap. Here (refer to figure 7.4): **Column Ampl** gives the noise due to LNA1: **Column HTb** gives the Nyquist noise added at port b in the hybrid T carrier suppression: **Column Nyq** gives the Nyquist noise added between the transducer and LNA1: **Column Osc'** gives the noise due to the pump oscillator phase noise: **Column Therm'** gives the noise due to acoustic losses in the bending flap: **Column B.A.** gives the back action noise due to the pump oscillator: **Column Dis'** gives the resistive damping component due to the pump. The total series and narrow band noise are given in the columns $S_x$ **Tot'** and $S_{F2}$ **Tot'** respectively. The final 3 columns estimate the spectral strain sensitivity, bandwidth and strain sensitivity for a 1 millisecond gravitational pulse.

| Simu-lation | $Q_2$ | $S_{am}$ dBc/Hz | $S_\phi$ dBc/Hz | $\beta_0$ | $Q_e$ | $T_{ampl}$ K | detuning Hz | SNR/ mK | $T_{eff}$ mK |
|---|---|---|---|---|---|---|---|---|---|
| **4** | $3\times10^6$ | -142 | -135 | .0064 | $7\times10^4$ | 300 | -14 | 1.2 | .85 |
| **5** | $3\times10^6$ | -160 | -135 | .0064 | $7\times10^4$ | 300 | -14 | 1.3 | .75 |
| **6** | $3\times10^6$ | -142 | -135 | .0064 | $7\times10^4$ | 8 | 0 | 14.8 | .065 |
| **7** | $3\times10^6$ | -160 | -135 | .0064 | $7\times10^4$ | 8 | 0 | 16.6 | .060 |
| **8** | $5\times10^7$ | -142 | -135 | .0064 | $7\times10^4$ | 8 | 0 | 28.2 | .036 |
| **9** | $5\times10^7$ | -160 | -135 | .0064 | $7\times10^4$ | 8 | 0 | 58.6 | .017 |
| **10** | $3\times10^6$ | -142 | -135 | .03 | $7\times10^4$ | 8 | 0 | 9.6 | .11 |
| **11** | $3\times10^6$ | -160 | -135 | .03 | $7\times10^4$ | 8 | 0 | 23.8 | .042 |
| **12** | $5\times10^7$ | -142 | -135 | .03 | $7\times10^4$ | 8 | 0 | 10.4 | .095 |
| **13** | $5\times10^7$ | -160 | -135 | .03 | $7\times10^4$ | 8 | 0 | 59.6 | .017 |
| **14** | $5\times10^7$ | -142 | -135 | .03 | $10^6$ | 8 | 0 | 76 | .013 |
| **15** | $5\times10^7$ | -160 | -135 | .03 | $10^6$ | 8 | 0 | 90 | .011 |
| **16** | $5\times10^7$ | -160 | —— | .03 | $10^6$ | 8 | 0 | 122 | .008 |
| **17** | $5\times10^7$ | -160 | -135 | .1 | $10^6$ | 8 | 0 | 92 | .011 |
| **18** | $5\times10^7$ | -160 | —— | .1 | $10^6$ | 8 | 0 | 154 | .007 |

# Table 7.1b

The major variables that can be optimised to maximise the detector sensitivity for simulations 4-18. Here: $Q_2$ is the bending flap acoustic Q: $S_{am}$ the pump oscillator amplitude noise at 710 Hz: $S_\phi$ the pump oscillator phase noise at 710 Hz: $\beta_0$ the electro-mechanical coupling when the oscillator is incident on the transducer centre of resonance: $Q_e$ the re-entrant cavity electrical Q: $T_{ampl}$ the effective noise temperature of LNA1: Detuning, the amount of frequency detuning between the bending flap and the resonant bar. The final two columns compare the sensitivity of each simulation in terms of integrated signal to noise ratio per mK impulse (SNR/mK) and effective noise temperature ($T_{eff}$).

strain sensitivity, bandwidth and strainsensitivity for a 1 millisecond pulse. The strain sensitivity/$\sqrt{Hz}$ and SNR/Hz per mK impulse are plotted for simulations 3 to 18 in figures 7.5 to 7.20.

The first two simulations in table 7.1a show the thermal noise limit at 5K as discussed in 7.3.1, when the bending flap has a Q value of $3 \cdot 10^6$ and $5 \cdot 10^7$ respectively. Simulation 3 estimates the UWA detector sensitivity in the previous experimental run. In this experiment both the broad band series noise and the narrow band noise were limited by vibrations. This experiment was operating with an electro-mechanical coupling of 0.0064, with the broad band vibrational noise determined in chapter 5 to be $4.3 \cdot 10^{-32}$ $m^2/Hz$. The mode temperature was measured to be in the order of $10^3$ K, which is equivalent to $S_{F2} = 3.5 \cdot 10^{-23}$ $N^2/Hz$, when the Q value of the bending flap was $3 \cdot 10^6$. Utilising the optimum filter theory it can be determined that this detector was operating with a bandwidth of 0.84 Hz with a 1 millisecond pulse strain sensitivity of approximately $7.1 \cdot 10^{-18}$, which is much greater than the sub $10^{-19}$ thermal limit. Chapter 5 discusses the reasons for the vibration corruption from the environment and how it has now been eliminated. Figure 7.5 shows the strain sensitivity/$\sqrt{Hz}$ and SNR/Hz per mK impulse for this simulation.

What if vibrations were not the limiting effect in the previous experiments? What would be the sensitivity that could be achieved and how could it be improved? In principle could the sensitivity limited by the thermal noise be approached? Simulations 4-18 deal with these questions by using reasonable estimates of the limiting components when predicting improvements. Table 7.1b summarise the major variables that influence the sensitivity and their values assumed in each simulation.

The phase noise of the pump oscillator that drives the parametric transducer has been measured and was shown in figure 4.23 to be $S_\phi$ = -135 dBc/Hz at 710 Hz. A lower phase noise could be achieved if the synthesiser was eliminated and the signal taken from the filtered port of the SLOSC resonator. However this is not practical as a tunable oscillator is needed with the ability for frequency and phase control (Ivanov, 1993). The amplitude noise of the oscillator is assumed to be limited by a series low noise amplifier as measured in figure 4.18, which is $S_{am}$ = -142 dBc/Hz at 710 Hz. The amplitude noise of the oscillator driving this measurement was not seen above the amplifier noise. If enough power can be extracted from the pump oscillator without a series amplifier, the feed back amplifier in the loop oscillator will determine the amplitude noise. This amplifier is always operating in a saturated state and hence the oscillator amplitude noise would be expected to be below the small signal amplitude noise of the series amplifier. It is assumed for the following simulations that the oscillator amplitude noise is in the order of -160 dBc/Hz in this state.

Simulations 4-18 investigate what happens when the secondary mass Q value is increased from $3 \cdot 10^6$ to $5 \cdot 10^7$; when the oscillator amplitude noise is improved from -142 to -160 dBc/Hz; when the oscillator phase noise is -135 dBc/Hz and then cancelled using a secondary cavity in an interferometric configuration; when the electro-mechanical coupling is increased from .0064 to .03 to .1; when the re-entrant cavity Q value is increased from $7 \cdot 10^4$ to $10^6$; and a cryogenic amplifier for LNA1 is included instead of a room temperature amplifier as discussed in chapter 5. Table 7.1b shows assumed changes of key parameters for given optimisations, while table 7.1a shows the calculated noise contributions that arise from these changes.

For each simulation the dominant narrow and braod band noise components as a function of frequency are plotted (figures 7.6 to 7.20) in terms of spectral strain sensitivity. Also, the total signal to noise density per mK impulse is plotted. If the series noise is dominant over the spectrum the SNR/Hz versus frequency characteristic will have two humps, distinguishing the two normal modes. In this case the narrow band noise will only determine the magnitude of the SNR/Hz at the resonant frequencies (see figures 7.5,7.6,7.7 and 7.11). Otherwise if the narrow band secondary mass noise is dominant the SNR/Hz characteristic will have one hump centred between the acoustic normal modes. In this case the series noise will only determine the maximum magnitude of the SNR/Hz (see figure 7.12 and 7.14). Figures 7.16, 7.18 and 7.19 give good examples of what the SNR/Hz is like when both dominate.

To minimise the effect of a constant level of broad band series noise the electro-mechanical coupling (or signal sideband power) should be increased. To increase the electro-mechanical coupling either the incident power or transducer Q must be increased. As long as the frequency of signal modulation remains within the transducer bandwidth (see chapter 6) the effect of the series noise added by the phase noise of the pump oscillator remains unchanged. This is because the amount of quadrature phase noise increases by the same ratio with power and Q as does the electro-mechanical coupling. However the relative level of Nyquist and amplifier noise will be reduced with respect to the enhanced signal. Other ways to reduce the series noise is by making direct improvements to the device noise, ie. by the addition of a cryogenic amplifier or by cancelling the oscillator phase noise using a dummy cavity combined with the transducer in an interferometer system.

To decrease the narrow band back action noise the electro-mechanical coupling and incident power must be reduced. Thus a high electrical Q transducer is necessary so less power is needed to obtain the desired electro-mechanical coupling. The reduced power means that back action effects will be reduced. To decrease the narrow band thermal noise the acoustic Q of the bending flap must be maximised.

The UWA gravitational detector will now be tuned to operate with an electro-mechanical coupling of 0.03 (see chapter 6), with a cryogenic amplifier in the demodulation chain. If the pump oscillator noise, bending flap Q and transducer Q remained the same as in the previous experiment, the increased coupling would actually make the detector less sensitive ( compare simulation 6 to simulation 10). Even by improving the bending flap acoustic Q only a very minimal effect on the sensitivity is obtained (compare simulation 10 to simulation 12). This is because the limiting noise component is the narrow band back action noise. To decrease the effect of the back action noise the amplitude noise power in the oscillator side bands incident on the cavity must be reduced. This can be done by either reducing the amplitude noise of the oscillator or increasing the Q of the transducer. The re-entrant cavity transducer has been shown to achieve an electrical Q value of $6 \cdot 10^5$ (Linthorne and Blair, 1991). By assuming optimistic values for the amplitude noise and Q value of the re-entrant cavity, simulation 13 to 15 show that it may be possible to improve the strain sensitivity to less than $10^{-19}$ for a 1 millisecond pulse.

It has been reported that it is possible to operate this type of transducer with an electro-mechanical coupling of about 0.1 (Linthorne and Blair, 1992). Simulation 17 shows that the series noise would be reduced slightly at the expense of a slightly increased narrow band noise level. However when comparing this with simulation 15, there has been no significant increase in integrated sensitivity. The main component of series noise for this case is due to the phase noise of the pump oscillator. Thus at this high level of coupling an interferometer system using the transducer with a matched dummy cavity can be used to cancel phase noise. In this case the detector sensitivity would be significantly increased. Simulation 18 predicts a 1 millisecond pulse strain sensitivity of $4.5 \cdot 10^{-20}$, which is approaching the thermal limit of $3.7 \cdot 10^{-20}$.

## 7.5 CONCLUSION

This chapter has reviewed the theory of optimal filters as applied to resonant bar gravitational wave detectors. Strain sensitivity/$\sqrt{\text{Hz}}$ was defined and related to the strain sensitivity. This theory was applied to the detector at UWA. It has been determined that it should operate with a 1 millisecond pulse strain sensitivity of less than $10^{-19}$, with a bandwidth of about 18 Hz. With further improvement it seems possible that the detector could operate with a 1 millisecond pulse strain sensitivity of around $6 \cdot 10^{-20}$.

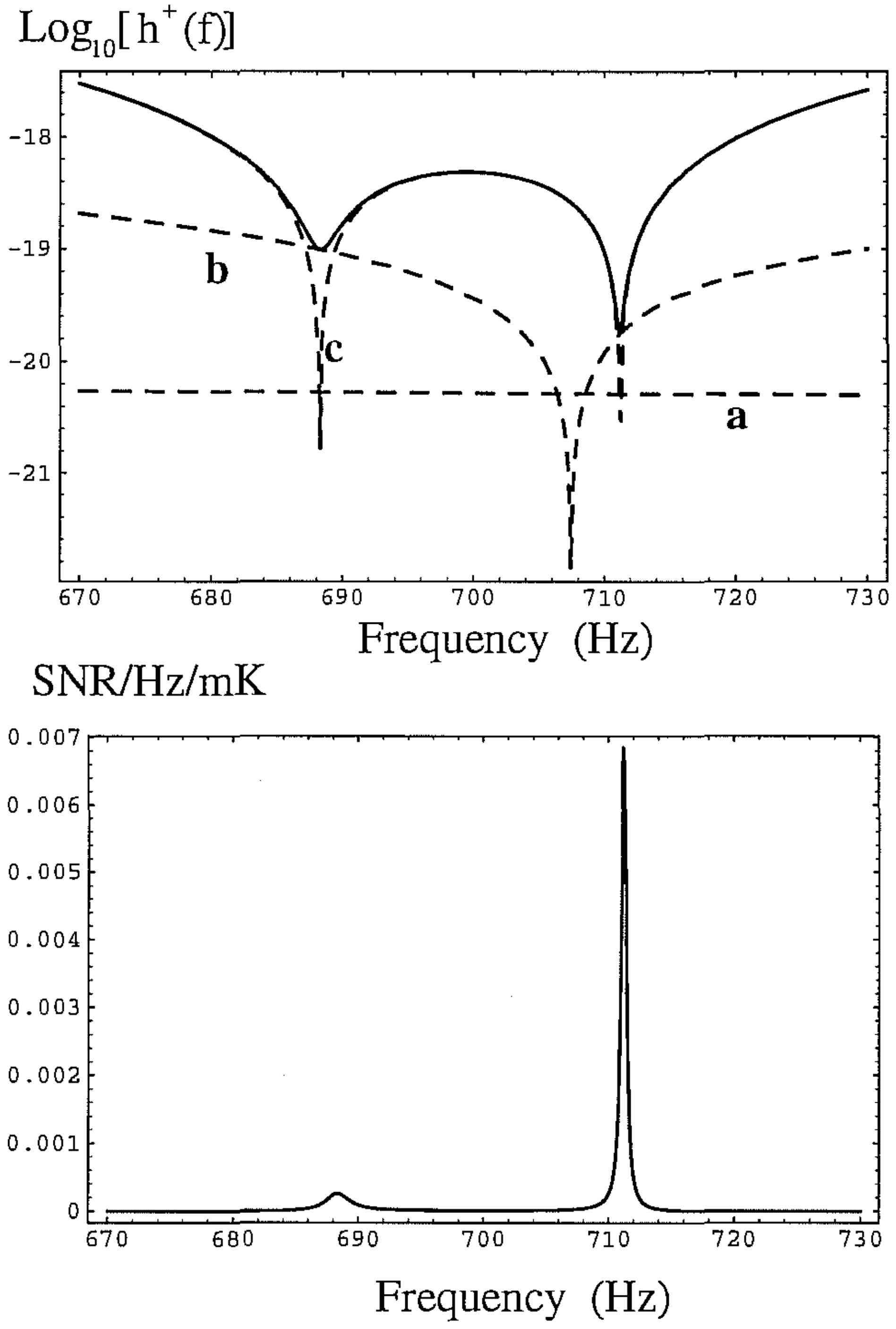

# Figure 7.5

<u>Simulation 3</u>, sensitivity of the UWA detector as a function of frequency in the a previous experimental run (see table 7.1a&b): Above, strain sensitivity per root Hertz. The dashed curves **a** and **b** are the narrow band components due to the resonant bar and secondary mass respectively while **c** is the broad band series noise. The bold line is the total strain sensitivity per root Hertz: Below, signal to noise ratio per Hertz per mK impulse.

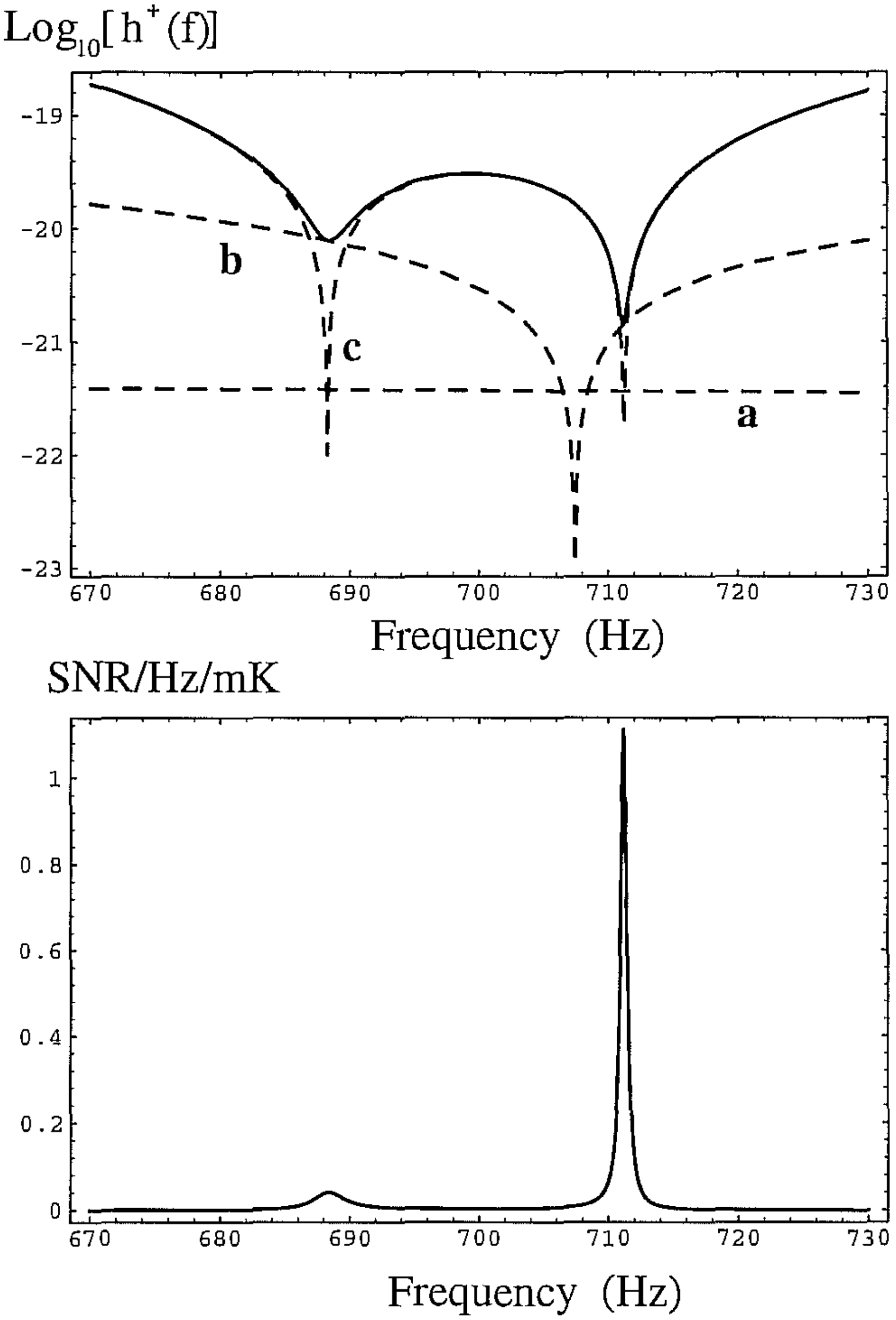

# Figure 7.6

Simulation 4, predicted sensitivity of the UWA detector as a function of frequency if excitations by the surrounding environment were eliminated in the last experimental configuration (see table 7.1a&b): Above, strain sensitivity per root Hertz. The dashed curves **a** and **b** are the narrow band components due to the resonant bar and secondary mass respectively while **c** is the broad band series noise. The bold line is the total strain sensitivity per root Hertz: Below, signal to noise ratio per Hertz per mK impulse.

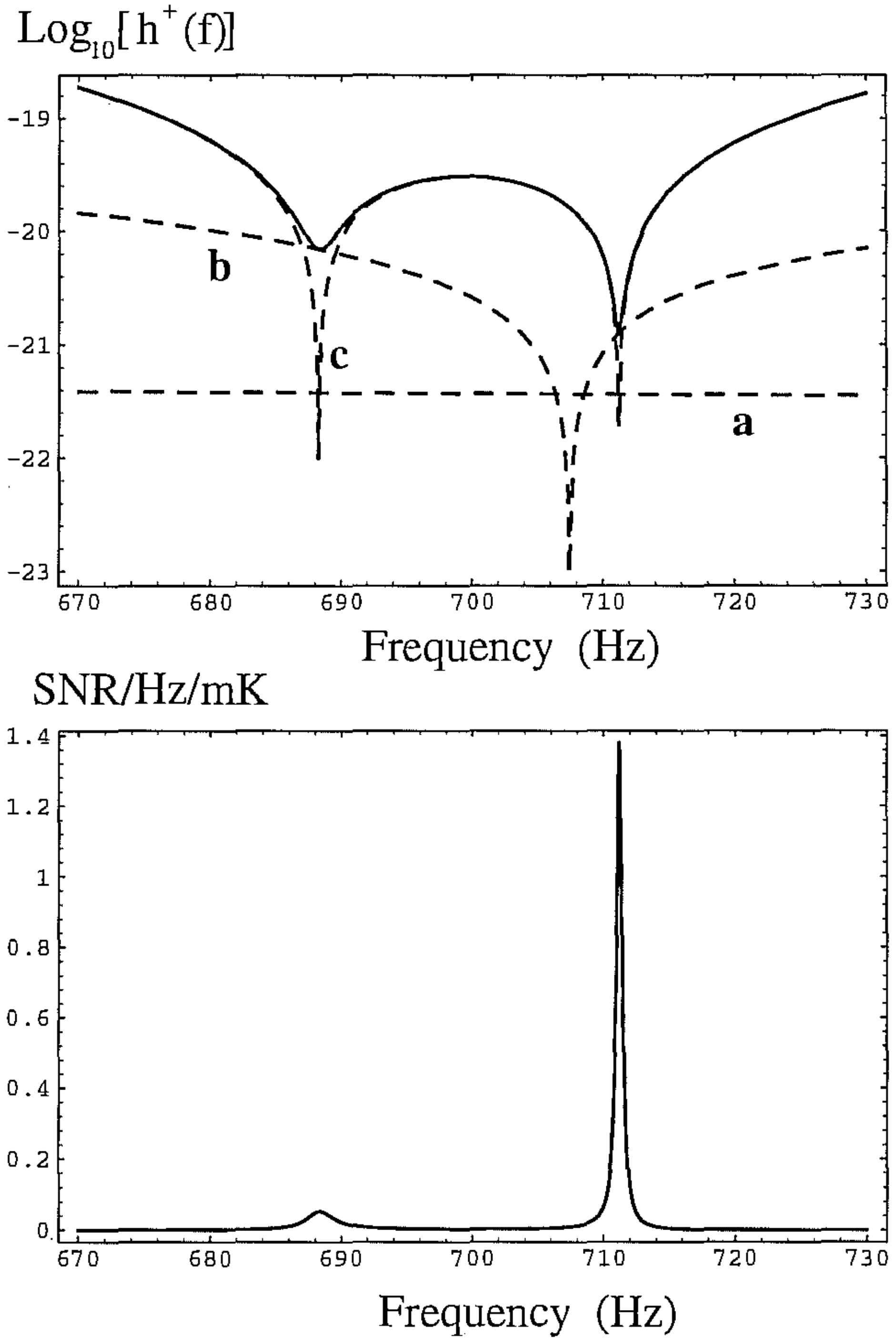

# Figure 7.7

<u>Simulation 5</u>, predicted sensitivity of the UWA detector as a function of frequency if excitations by the surrounding environment were eliminated in the last experimental configuration and the oscillator amplitude noise was reduced (see table 7.1a&b): Above, strain sensitivity per root Hertz. The dashed curves **a** and **b** are the narrow band components due to the resonant bar and secondary mass respectively while **c** is the broad band series noise. The bold line is the total strain sensitivity per root Hertz: Below, signal to noise ratio per Hertz per mK impulse.

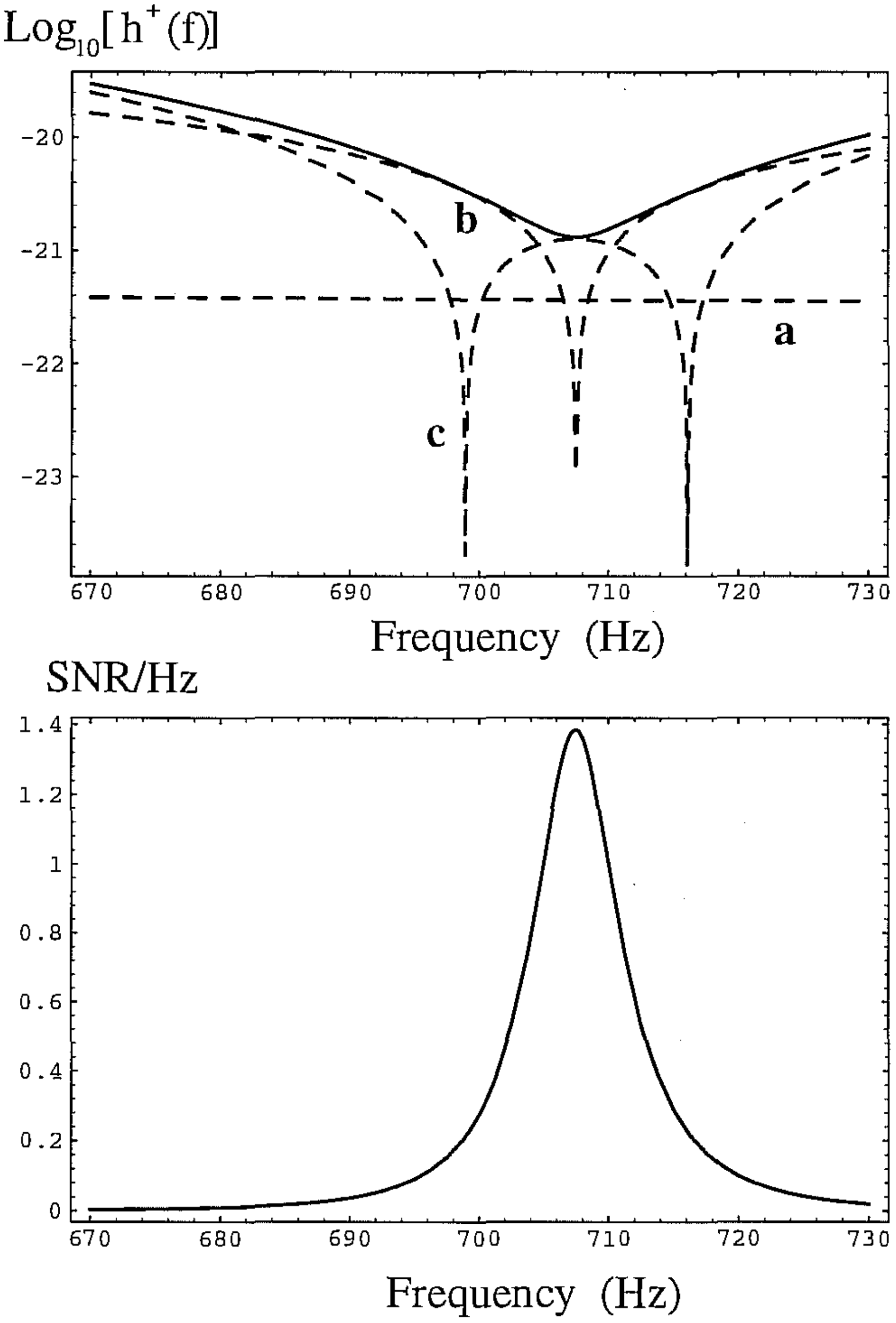

# Figure 7.8

**Simulation 6**, predicted sensitivity of the UWA detector as a function of frequency if excitations by the surrounding environment were eliminated with the bending flap tuned to the resonant bar and a cryogenic amplifier in the demodulation chain (see table 7.1a&b): Above, strain sensitivity per root Hertz. The dashed curves **a** and **b** are the narrow band components due to the resonant bar and secondary mass respectively while **c** is the broad band series noise. The bold line is the total strain sensitivity per root Hertz: Below, signal to noise ratio per Hertz per mK impulse.

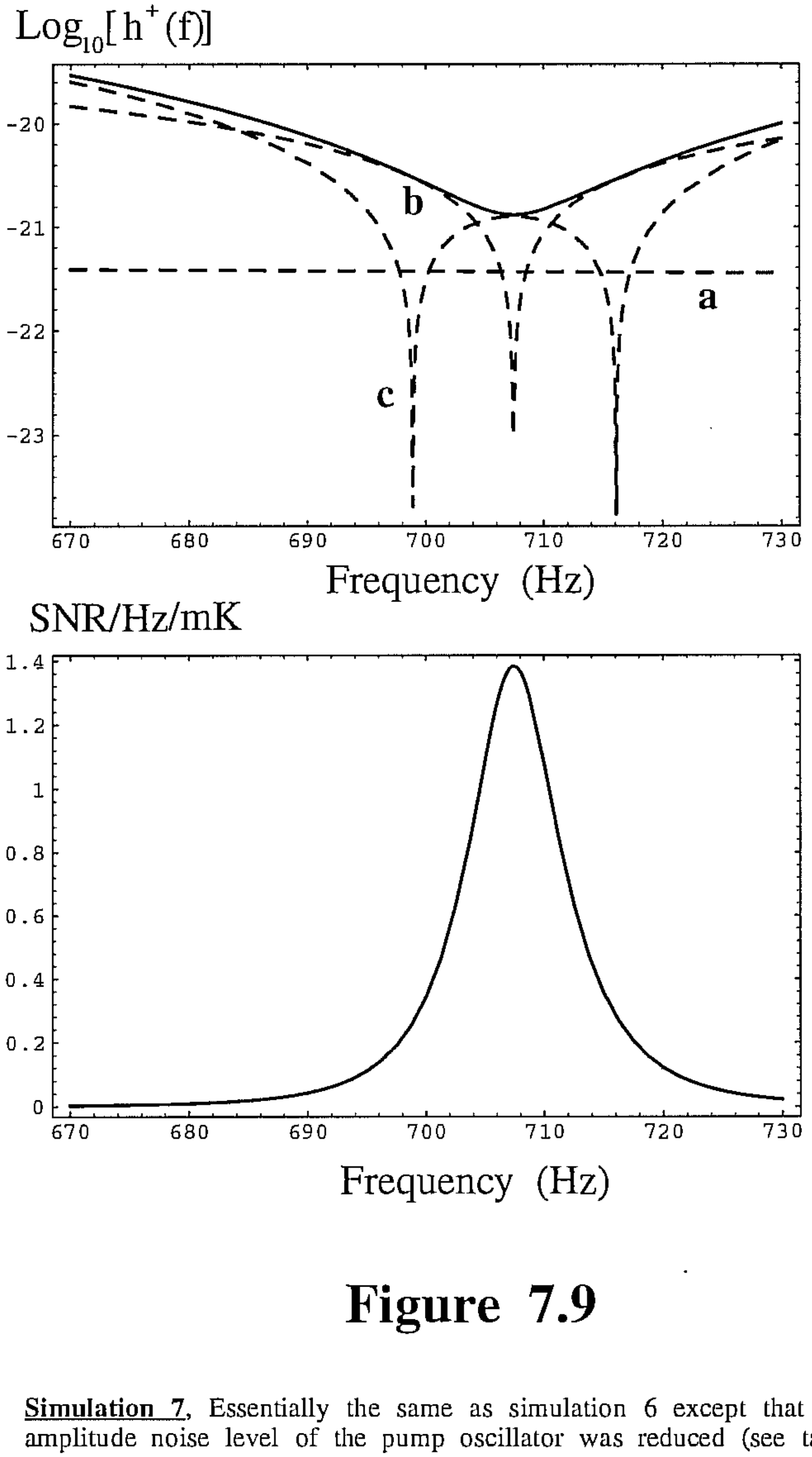

# Figure 7.9

__Simulation 7__, Essentially the same as simulation 6 except that the amplitude noise level of the pump oscillator was reduced (see table 7.1a&b): Above, strain sensitivity per root Hertz. The dashed curves **a** and **b** are the narrow band components due to the resonant bar and secondary mass respectively while **c** is the broad band series noise. The bold line is the total strain sensitivity per root Hertz: Below, signal to noise ratio per Hertz per mK impulse.

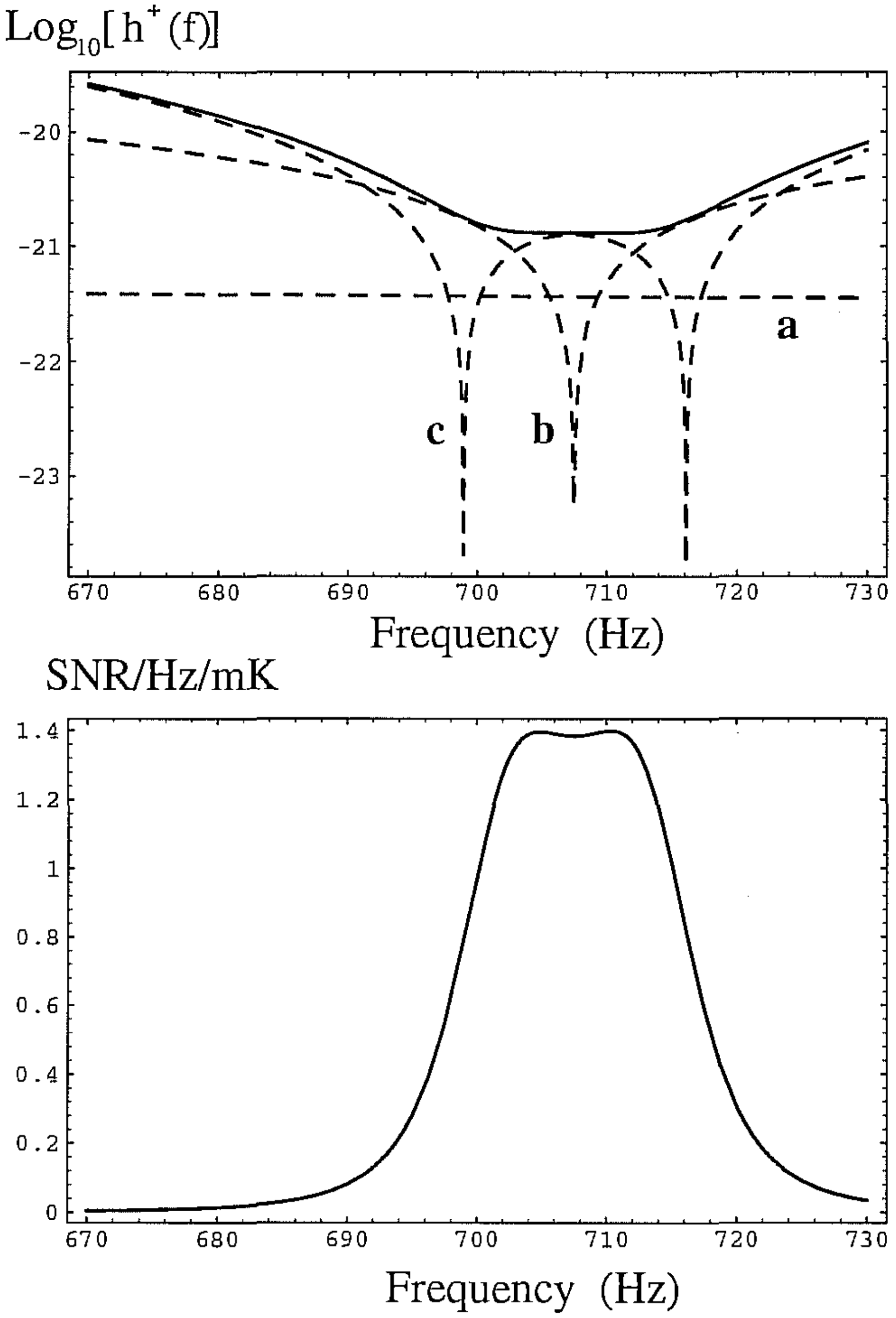

# Figure 7.10

**Simulation 8**, Essentially the same as simulation 6 except that the bending flap Q value was increased (see table 7.1a&b): Above, strain sensitivity per root Hertz. The dashed curves **a** and **b** are the narrow band components due to the resonant bar and secondary mass respectively while **c** is the broad band series noise. The bold line is the total strain sensitivity per root Hertz: Below, signal to noise ratio per Hertz per mK impulse.

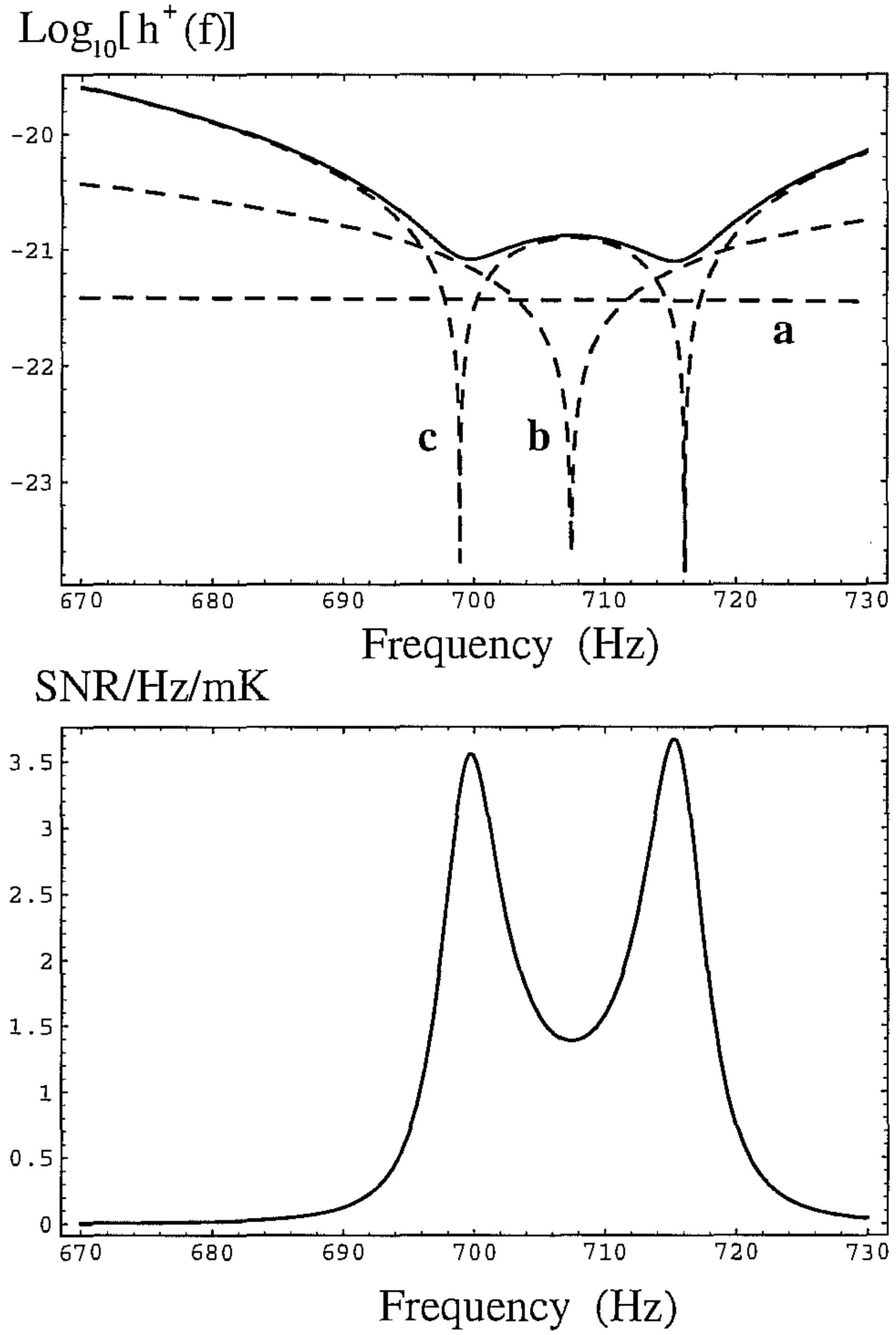

# Figure 7.11

__Simulation 9__, Essentially the same as simulation 7 except that the bending flap Q value was increased (see table 7.1a&b): Above, strain sensitivity per root Hertz. The dashed curves **a** and **b** are the narrow band components due to the resonant bar and secondary mass respectively while **c** is the broad band series noise. The bold line is the total strain sensitivity per root Hertz: Below, signal to noise ratio per Hertz per mK impulse.

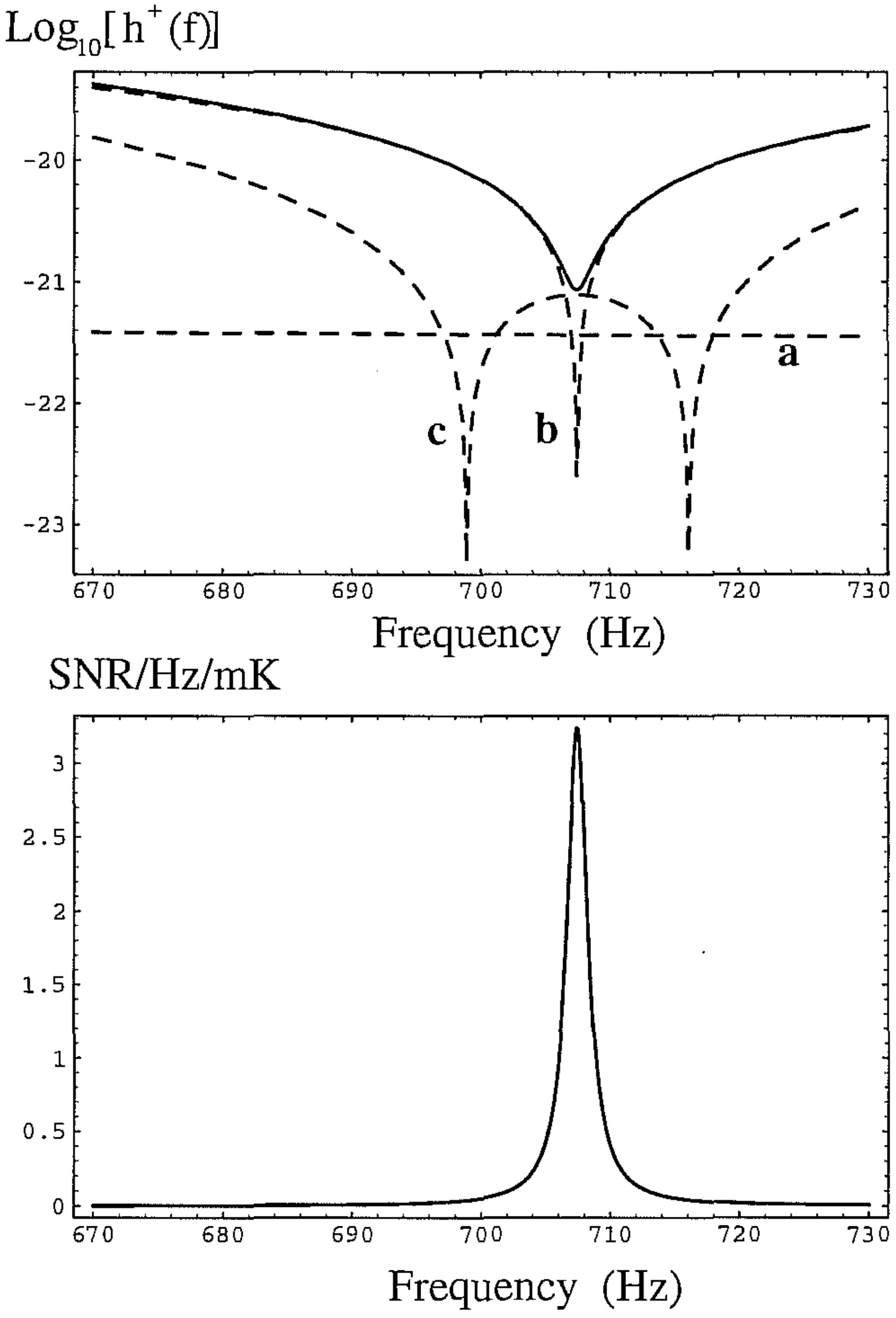

# Figure 7.12

**<u>Simulation 10</u>**, Essentially the same as simulation 6 except that the electro-mechanical coupling was increased by a factor of 4.7 (see table 7.1a&b): Above, strain sensitivity per root Hertz. The dashed curves **a** and **b** are the narrow band components due to the resonant bar and secondary mass respectively while **c** is the broad band series noise. The bold line is the total strain sensitivity per root Hertz: Below, signal to noise ratio per Hertz per mK impulse.

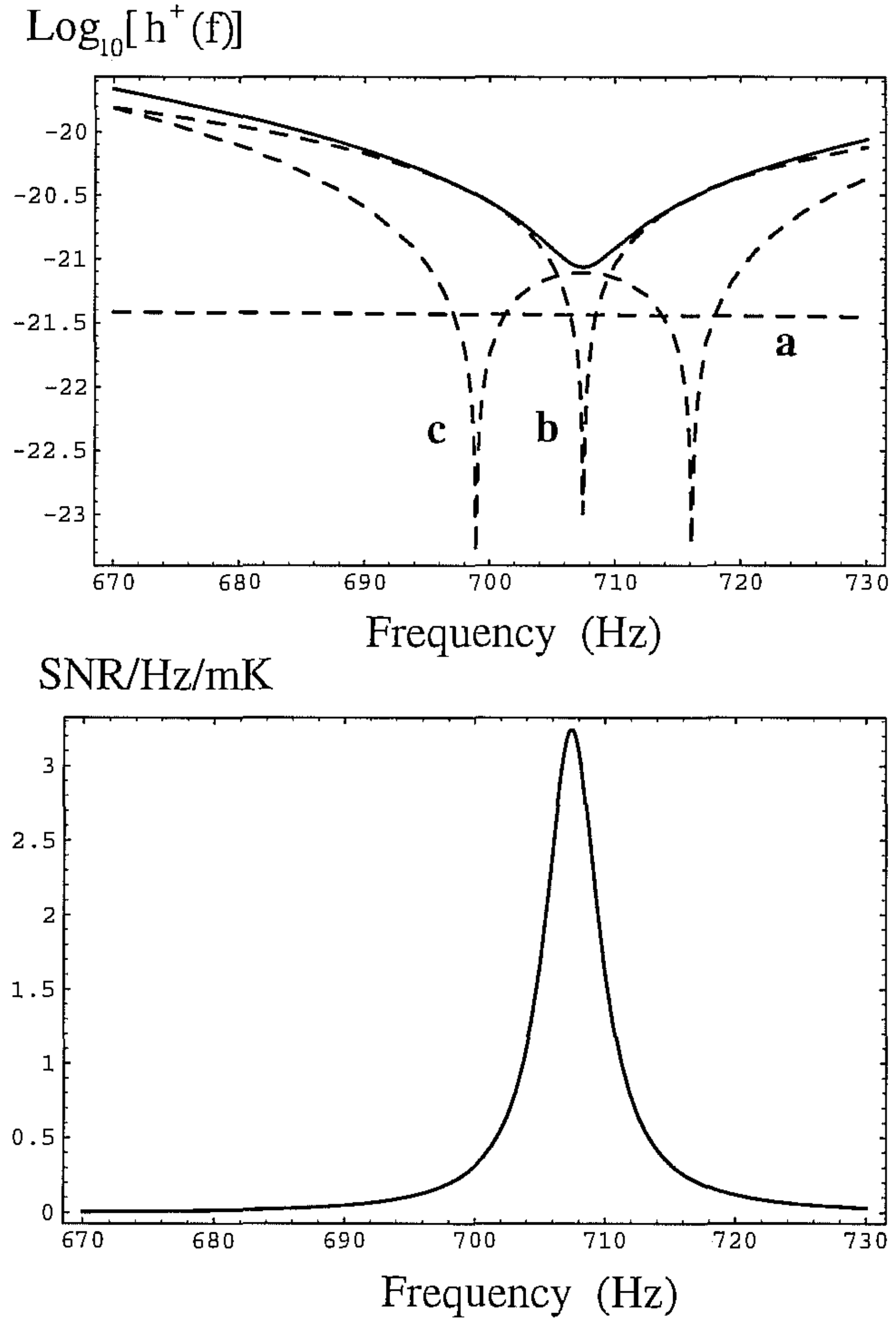

# Figure 7.13

<u>Simulation 11</u>, Essentially the same as simulation 7 except that the electro-mechanical coupling was increased by a factor of 4.7 (see table 7.1a&b): Above, strain sensitivity per root Hertz. The dashed curves **a** and **b** are the narrow band components due to the resonant bar and secondary mass respectively while **c** is the broad band series noise. The bold line is the total strain sensitivity per root Hertz: Below, signal to noise ratio per Hertz per mK impulse.

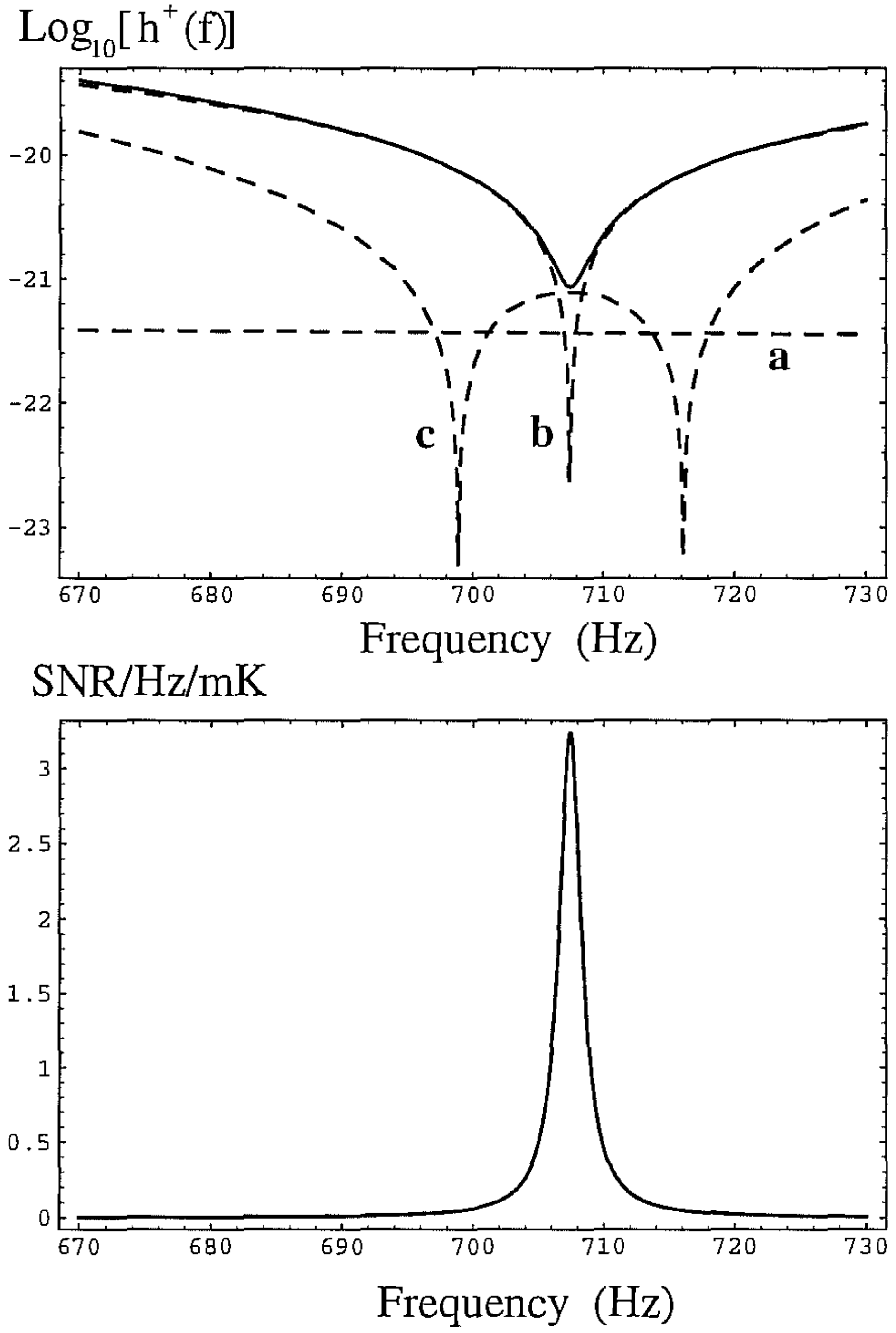

# Figure 7.14

<u>Simulation 12</u>, Essentially the same as simulation 8 except that the electro-mechanical coupling was increased by a factor of 4.7 (see table 7.1a&b): Above, strain sensitivity per root Hertz. The dashed curves **a** and **b** are the narrow band components due to the resonant bar and secondary mass respectively while **c** is the broad band series noise. The bold line is the total strain sensitivity per root Hertz: Below, signal to noise ratio per Hertz per mK impulse.

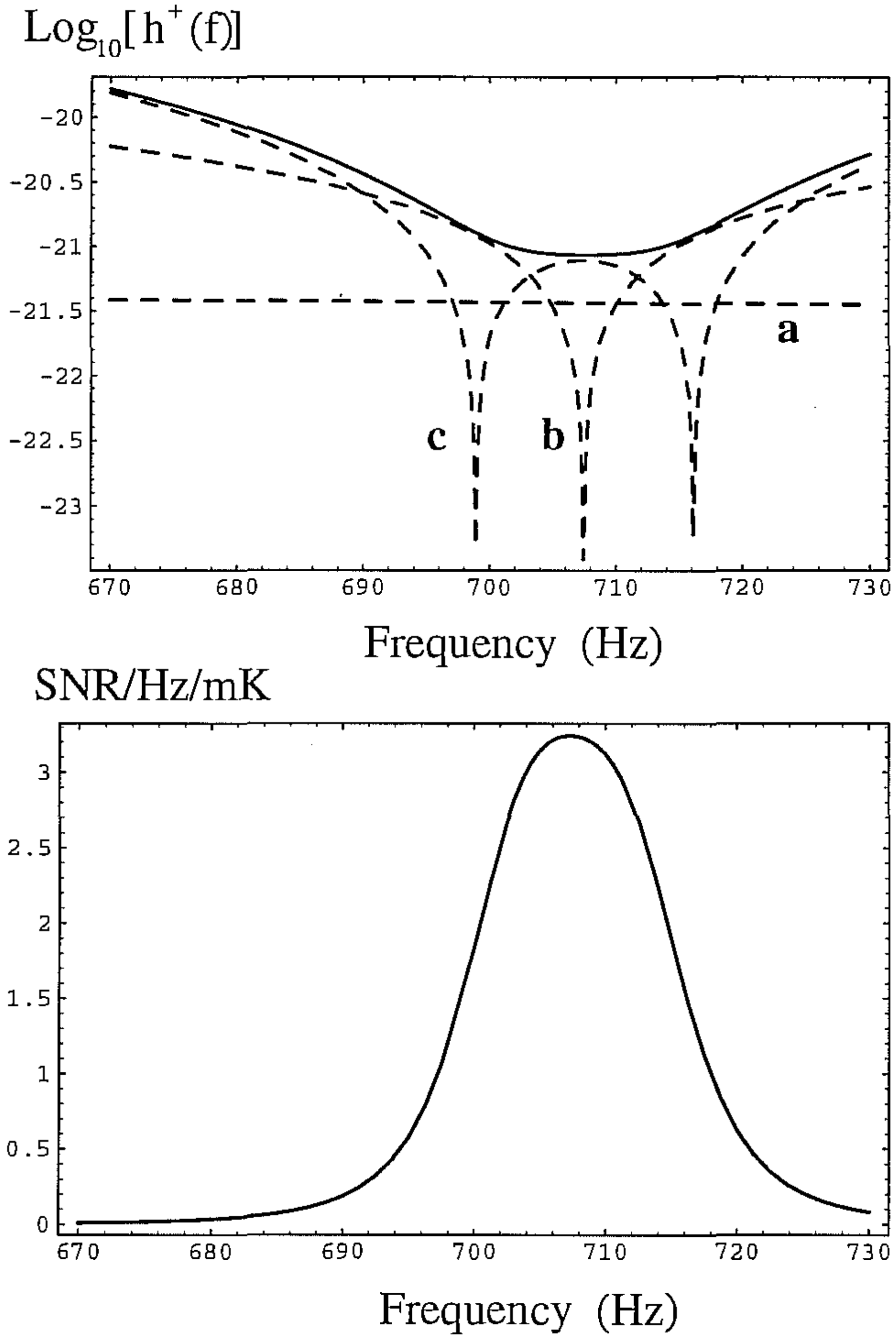

# Figure 7.15

<u>**Simulation 13**</u>, Essentially the same as simulation 9 except that the electro-mechanical coupling was increased by a factor of 4.7 (see table 7.1a&b): Above, strain sensitivity per root Hertz. The dashed curves **a** and **b** are the narrow band components due to the resonant bar and secondary mass respectively while **c** is the broad band series noise. The bold line is the total strain sensitivity per root Hertz: Below, signal to noise ratio per Hertz per mK impulse.

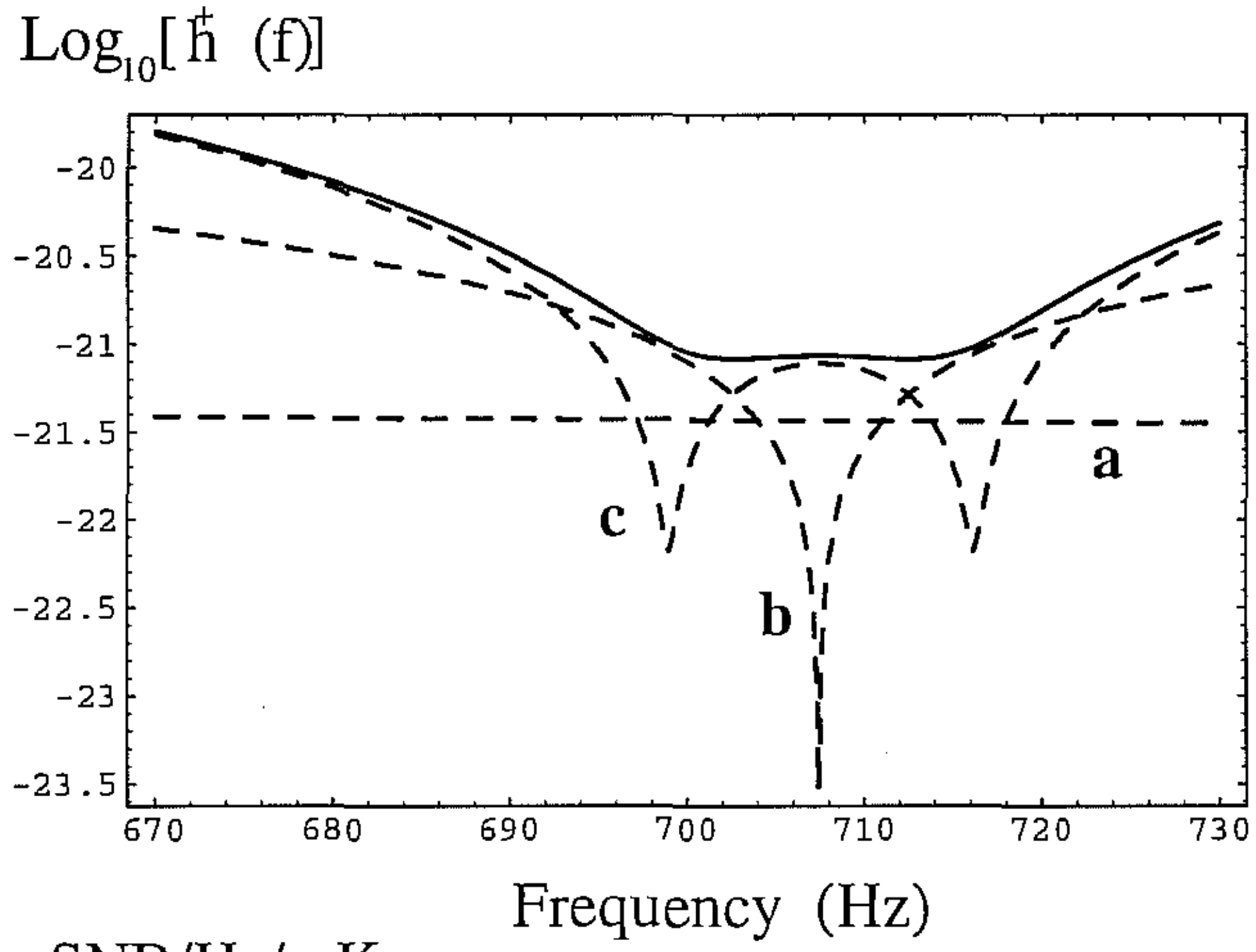

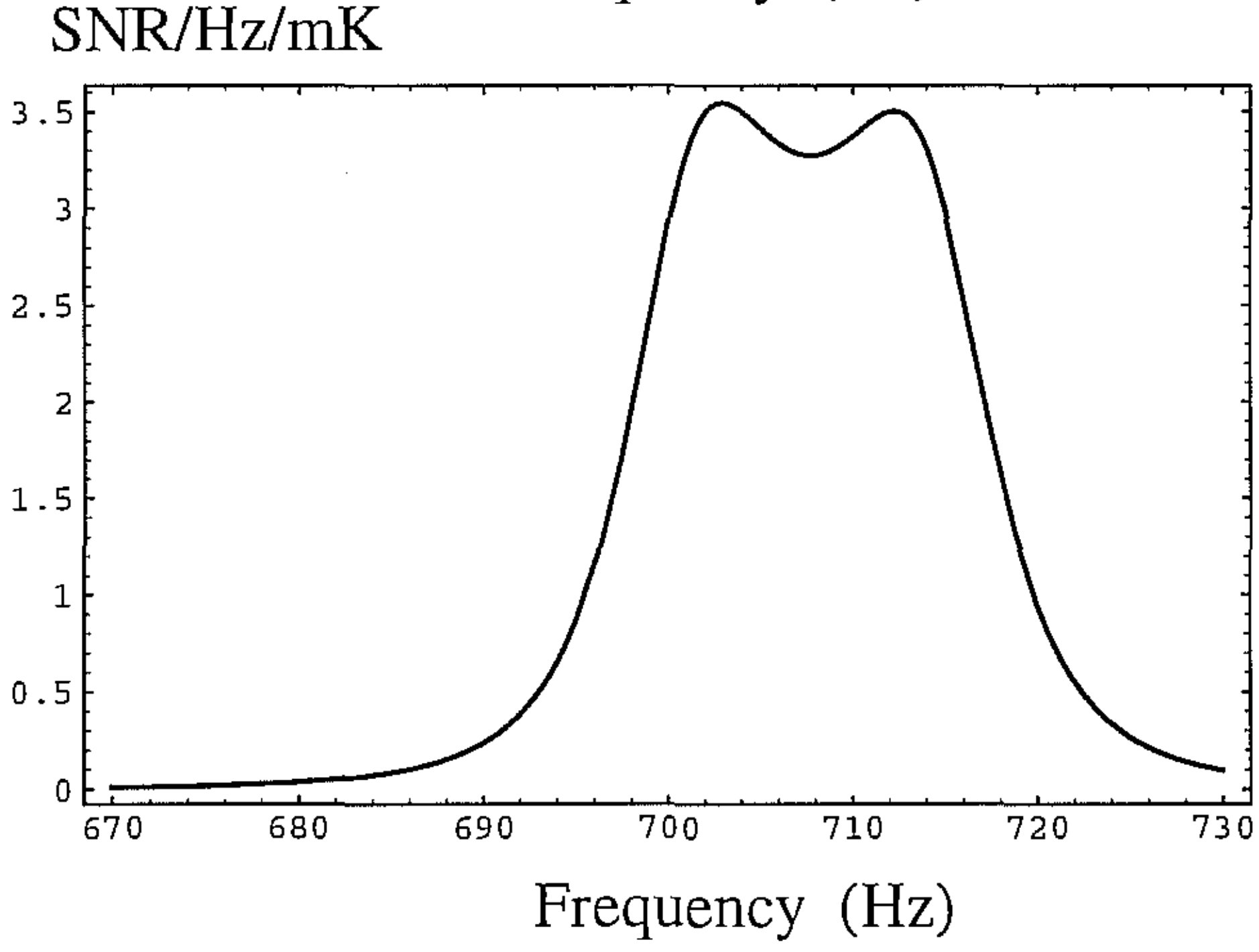

# Figure 7.16

<u>**Simulation 14**</u>, Essentially the same as simulation 12 except that the transducer Q was increased (see table 7.1a&b): Above, strain sensitivity per root Hertz. The dashed curves **a** and **b** are the narrow band components due to the resonant bar and secondary mass respectively while **c** is the broad band series noise. The bold line is the total strain sensitivity per root Hertz: Below, signal to noise ratio per Hertz per mK impulse.

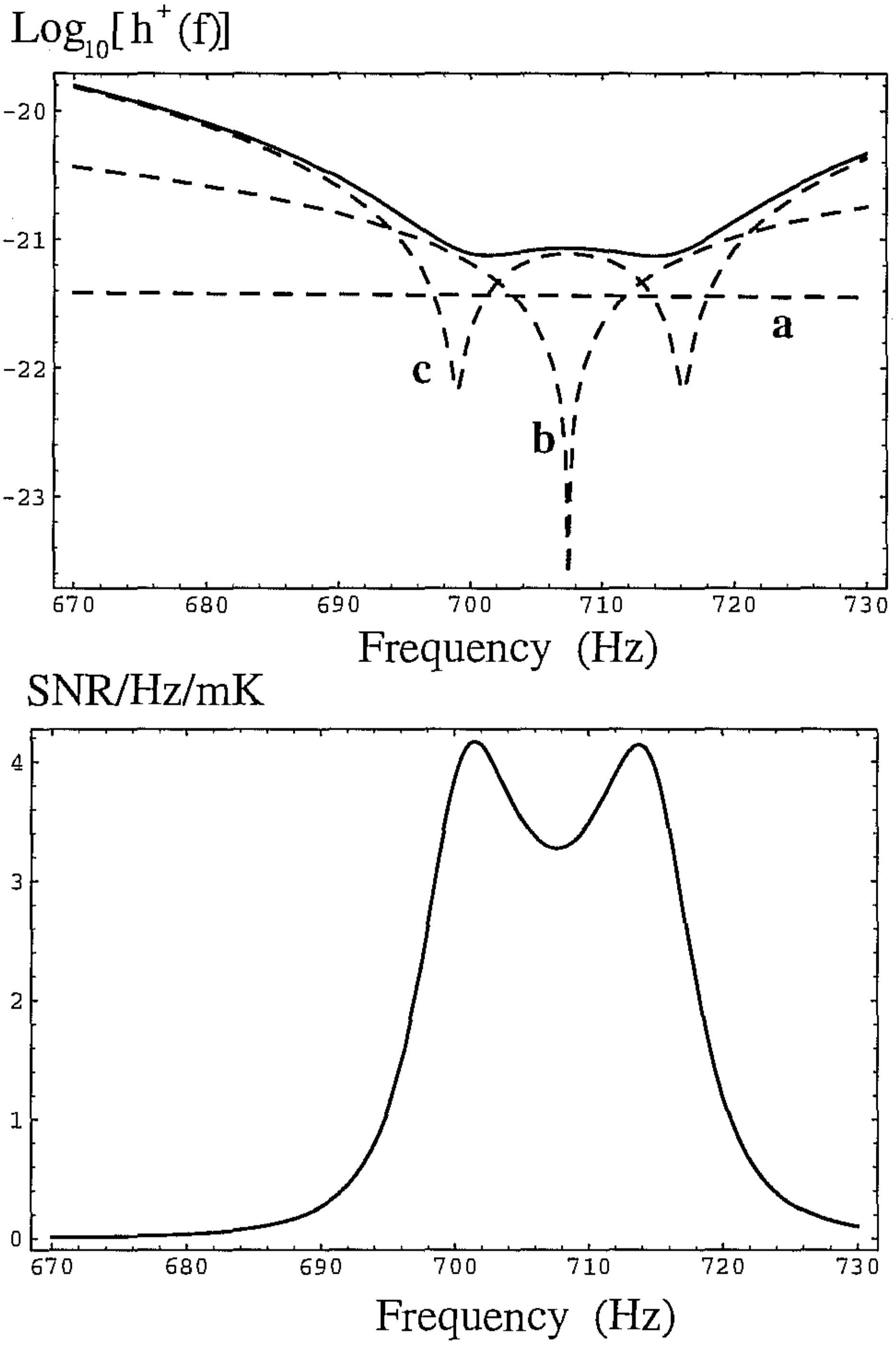

# Figure 7.17

<u>**Simulation 15**</u>, Essentially the same as simulation 13 except that the transducer Q was increased (see table 7.1a&b): Above, strain sensitivity per root Hertz. The dashed curves **a** and **b** are the narrow band components due to the resonant bar and secondary mass respectively while **c** is the broad band series noise. The bold line is the total strain sensitivity per root Hertz: Below, signal to noise ratio per Hertz per mK impulse.

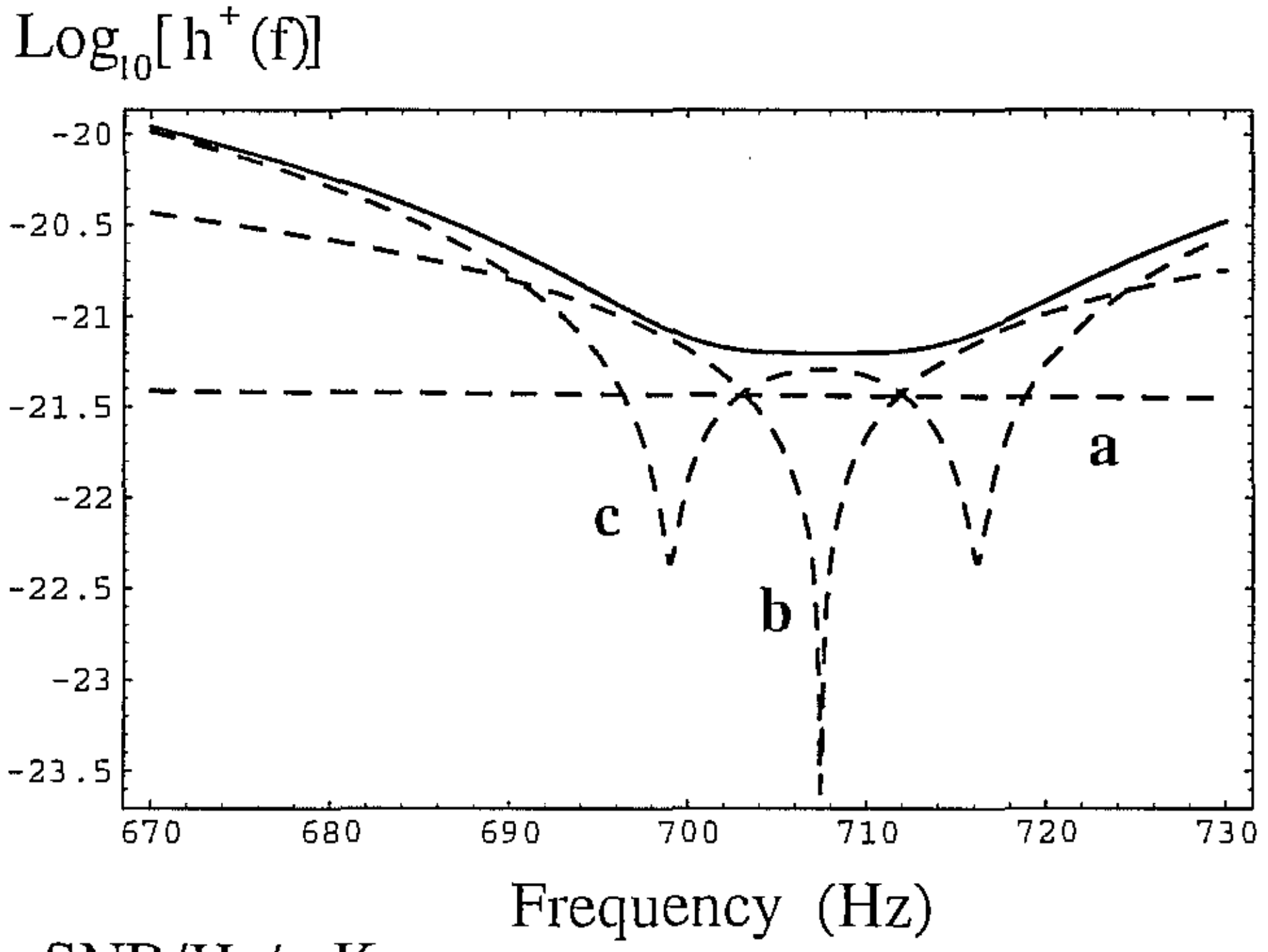

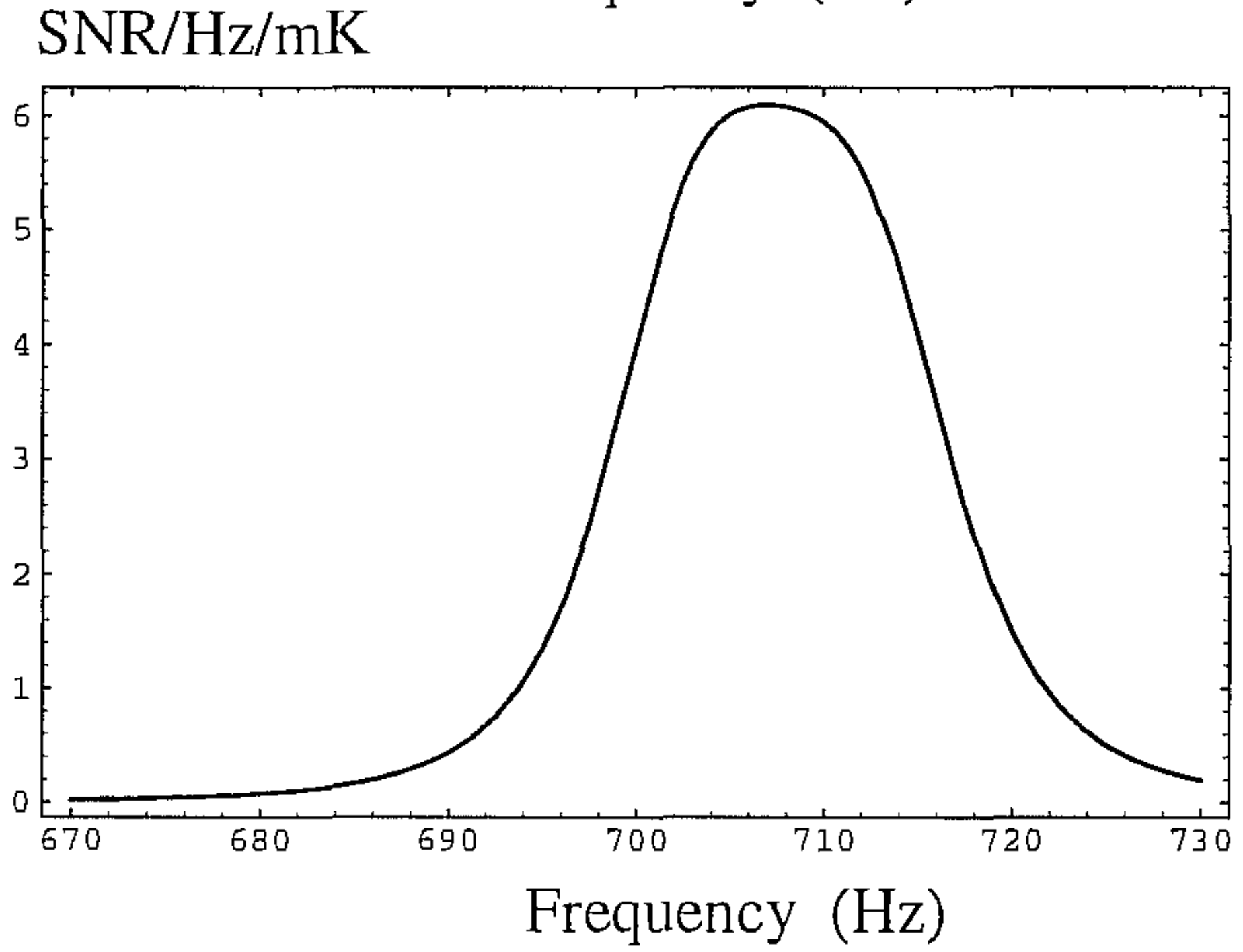

# Figure 7.18

**<u>Simulation 16</u>**, Essentially the same as simulation 15 except all the oscillator phase noise was assumed to be cancelled using a matched dummy cavity and the transducer in an interferometer system (see table 7.1a&b): Above, strain sensitivity per root Hertz. The dashed curves **a** and **b** are the narrow band components due to the resonant bar and secondary mass respectively while **c** is the broad band series noise. The bold line is the total strain sensitivity per root Hertz: Below, signal to noise ratio per Hertz per mK impulse.

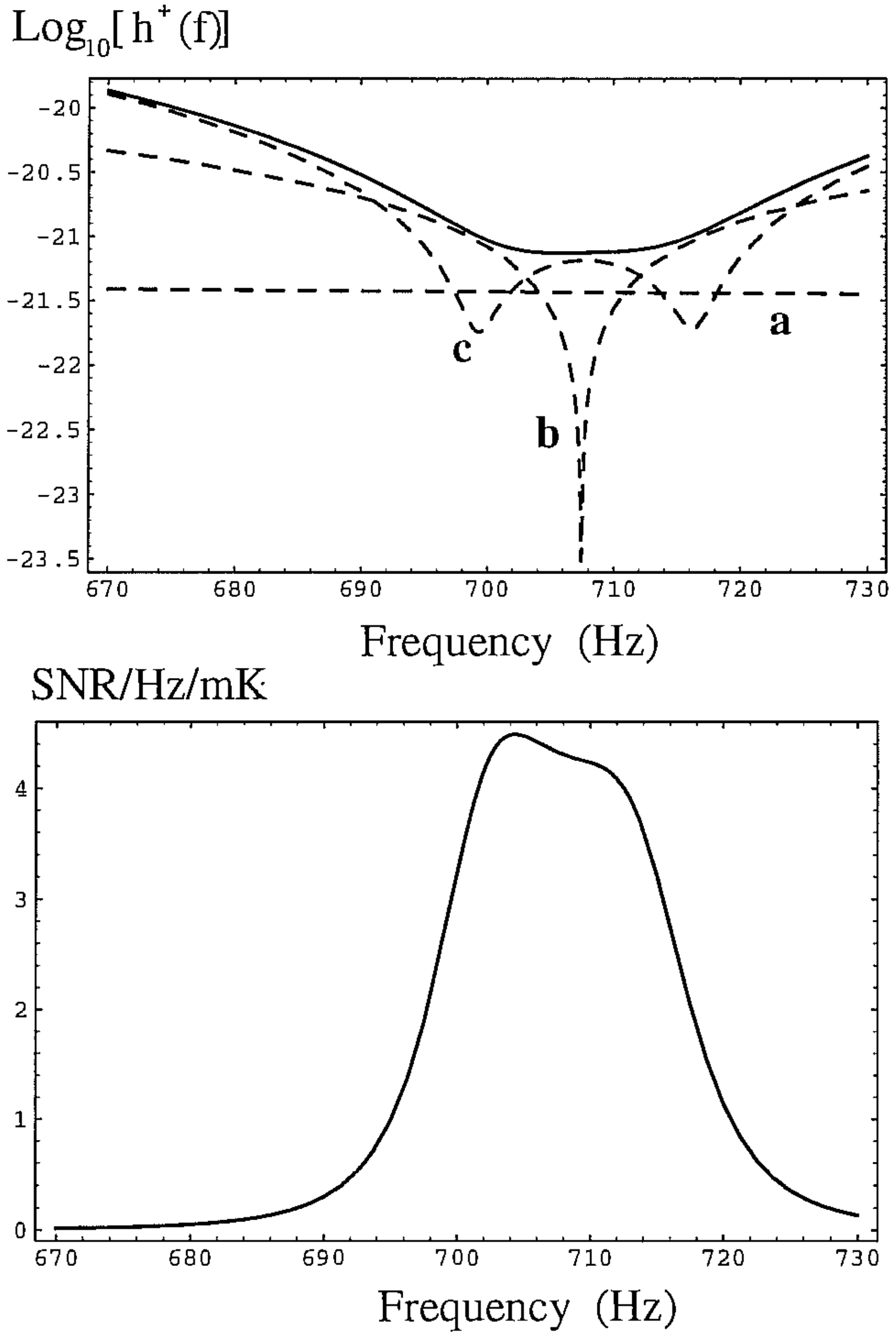

# Figure 7.19

<u>**Simulation 17**</u>, Essentially the same as simulation 15 except that the electro-mechanical coupling was further increased by a factor of 3.3 (see table 7.1a&b): Above, strain sensitivity per root Hertz. The dashed curves **a** and **b** are the narrow band components due to the resonant bar and secondary mass respectively while **c** is the broad band series noise. The bold line is the total strain sensitivity per root Hertz: Below, signal to noise ratio per Hertz per mK impulse.

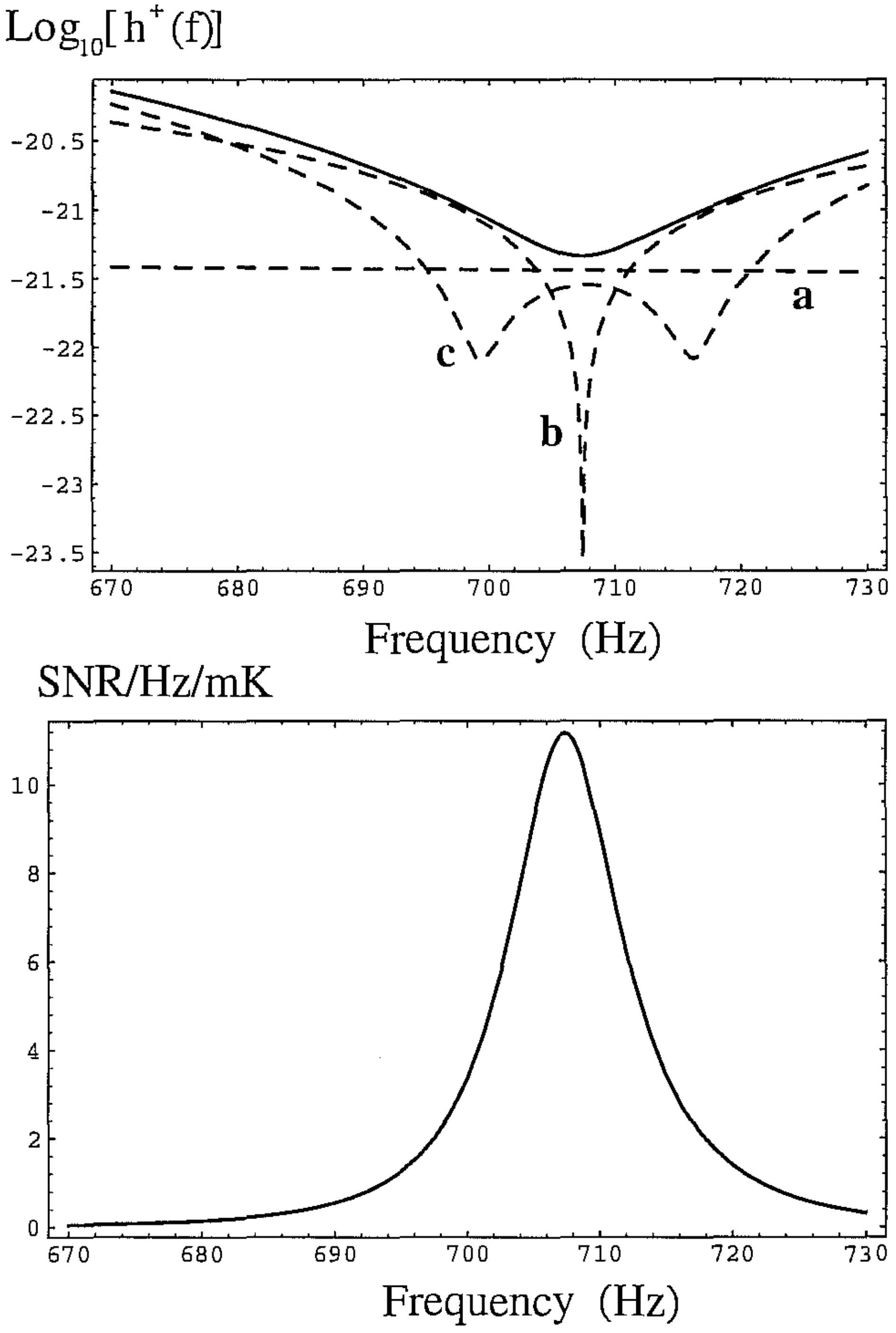

# Figure 7.20

**Simulation 18**, Essentially the same as simulation 17 except all the oscillator phase noise was assumed to be cancelled using a matched dummy cavity and the transducer in an interferometer system (see table 7.1a&b): Above, strain sensitivity per root Hertz. The dashed curves **a** and **b** are the narrow band components due to the resonant bar and secondary mass respectively while **c** is the broad band series noise. The bold line is the total strain sensitivity per root Hertz: Below, signal to noise ratio per Hertz per mK impulse.

# CONCLUSION

This thesis has described the development and multi-mode characterisation of a high Q Tunable Sapphire Loaded Superconducting Cavity (T-SLOSC) resonator over X-band, and its inclusion as the resonator in a tunable resonator oscillator. This type of resonant cavity is very sensitive to vibrations, the greater the tuning range the more sensitive it becomes. The phase noise of the T-SLOSC oscillator was dominated by vibrations. Vibrational effects were reduced to below the measurement noise floor with the aide of three stages of vibration isolation and by restricting the tuning range to less than 5 MHz. For this type of cavity to be useful with a larger tuning range, a more elaborate vibration isolation system would be needed.

An easier way to create a tunable low noise oscillator is by mixing a fixed frequency SLOSC oscillator with a tunable low noise oscillator (a 0-1.2 GHz HP-8662A synthesiser was used). The draw back in this method was that the phase noise of the synthesiser dominated at offset frequencies above 500 Hz. However the SLOSC oscillator phase noise between 10 to 500 Hz is the best ever directly measured at X-band. At 1 kHz these results imply that the single sideband phase noise from the non-filtered port and filtered port of the oscillator was -140 and -175 dBc/Hz respectively.

The SLOSC-synthesiser oscillator has been included into the parametric transducer system of the UWA resonant bar gravitational wave antenna. It has been determined that the current transducer is adequate to enable the detector to operate with a strain sensitivity of less than $10^{-19}$. This is comparable to the Louisiana and Rome detectors which are now in operation, and is sufficient to observe predicted gravitational fluxes from stellar collapses to neutron stars and black holes (supernova) at a distance of up to 10 kiloparsecs.

# APPENDIX A

# CALCULATION OF FIELD COMPONENTS IN AN ANISOTROPIC CYLINDRICAL DIELECTRIC RESONATOR

## A.1 BASIC EQUATIONS

This appendix deals with the basic time-harmonic electromagnetic field equations in a linear anisotropic medium. Assuming the dielectric medium is anisotropic with no off diagonal terms in the permittivity tensor, lossless and in a source free region, the electric and magnetic field satisfy Maxwell's two curl equations;

$$\nabla \times \mathbf{H} = j\omega\mathbf{D} \qquad (A.1),$$

$$\nabla \times \mathbf{E} = j\omega\mu\mathbf{H} \qquad (A.2),$$

with

$$\mathbf{D} = \begin{bmatrix} \varepsilon_r & 0 & 0 \\ 0 & \varepsilon_\phi & 0 \\ 0 & 0 & \varepsilon_z \end{bmatrix} \varepsilon_0 \mathbf{E} \qquad (A.3),$$

together with Gauss's law in a charge free region;

$$\nabla.\mathbf{D} = 0 \qquad (A.4).$$

The phasors **E**, **H** and **D** are the electric field, magnetic field and electric flux density vectors respectively. The constant scalars $\varepsilon_r$, $\varepsilon_\phi$ and $\varepsilon_z$ are the radial, azimuthal and axial relative permittivities respectively, and the constant scalars $\varepsilon_0$ and $\mu$ are the permittivity of free space and the isotropic permeability of the medium respectively.

When solving these equations with respect to an anisotropic medium it is assumed that the cylindrical crystal is orientated with its c-axis parallel to the z-axis of the cylinder. Hence the relative permittivity parallel to the c-axis is given by $\varepsilon_z$ and perpendicular to the c-axis is given by $\varepsilon_r$ where $\varepsilon_\phi = \varepsilon_r$ as well.

The amount of difficulty can be greatly reduced by realising that only two field components directed along a constant direction are needed to derive all other components (Auda and Kajfez, 1986). In cylindrical co-ordinates the easiest direction to solve for is in the z direction.

Taking the curl of (A.1) it becomes;

$$\nabla(\nabla.\mathbf{H}) - \nabla^2\mathbf{H} = j\omega\, \nabla \times \mathbf{D} \qquad (A.5)$$

Utilising the fact that $\varepsilon_\phi = \varepsilon_r$ the z component of $\nabla \times \mathbf{D}$ is equal to the z component of $\varepsilon_r\varepsilon_0\nabla \times \mathbf{E}$ which from (A2) equals $j\omega\varepsilon_z\varepsilon_0\mu H_z$. Also the magnetic field is solenoidal, hence $\nabla.\mathbf{H} = 0$. Substituting this into (A.5), the z component of (A.5) can be written as;

$$\nabla^2 H_z + k_r^2 H_z = 0 \qquad \text{(A.6)},$$

where $k_r = \omega(\varepsilon_r\varepsilon_0\mu)^{1/2}$.

Taking the curl of (A.2) it becomes;

$$\nabla(\nabla.\mathbf{E}) - \nabla^2\mathbf{E} = -j\omega\,\nabla \times \mathbf{H} \qquad \text{(A.7)}$$

To calculate the z component of (A.7) is more complicated because the divergence of the electric field is not constant for an anisotropic dielectric. In cylindrical co-ordinates the z component of $\nabla(\nabla.\mathbf{E})$ can be written as;

$$\nabla(\nabla.\mathbf{E})\big|_z = \left(\frac{1}{r}\frac{\partial E_r}{\partial z} + \frac{\partial^2 E_r}{\partial z\partial r} + \frac{1}{r}\frac{\partial^2 E_\phi}{\partial z\partial\phi}\right) + \frac{\partial^2 E_z}{\partial z^2} \qquad \text{(A.8)}$$

However the divergence of the electric flux density is zero, hence utilising (A.3) the z component of $\nabla(\nabla.\mathbf{D})$ is equal to zero and can be written as;

$$\nabla(\nabla.\mathbf{D})\big|_z = \varepsilon_r\left(\frac{1}{r}\frac{\partial E_r}{\partial z} + \frac{\partial^2 E_r}{\partial z\partial r} + \frac{1}{r}\frac{\partial^2 E_\phi}{\partial z\partial\phi}\right) + \varepsilon_z\frac{\partial^2 E_z}{\partial z^2} = 0 \qquad \text{(A.9)}$$

Eliminating the term in brackets from (A.8) and (A.9) the z component of $\nabla(\nabla.\mathbf{E})$ becomes;

$$\nabla(\nabla.\mathbf{E})\big|_z = \frac{\partial^2 E_z}{\partial z^2}\left(1 - \frac{\varepsilon_z}{\varepsilon_r}\right) \qquad \text{(A.10)}$$

Hence by substituting (A.1) and (A.10) into (A.7) the z component of (A.7) can be written as;

$$\nabla^2 E_z + \frac{\partial^2 E_z}{\partial z^2}\left(\frac{\varepsilon_z}{\varepsilon_r} - 1\right) + k_z^2 E_z = 0 \qquad \text{(A.11)},$$

where $k_z = \omega(\varepsilon_z\varepsilon_0\mu)^{1/2}$.

The axial field components in the resonator, $E_z$ and $H_z$, can thus be obtained by solving (A.6) and (A.11). Once these components have been solved, the following relations derived from Maxwell's equations are used to calculate the rest (Auda and Kajfez, 1986);

$$\left(\frac{\partial^2}{\partial z^2} + k_r^2\right)H_r = \frac{j\omega\varepsilon_r}{r}\frac{\partial E_z}{\partial \phi} + \frac{\partial^2 H_z}{\partial z \partial r} \quad \text{(A.12)}$$

$$\left(\frac{\partial^2}{\partial z^2} + k_r^2\right)E_\phi = j\omega\mu\frac{\partial H_z}{\partial r} + \frac{1}{r}\frac{\partial^2 E_z}{\partial z \partial \phi} \quad \text{(A.13)}$$

$$\left(\frac{\partial^2}{\partial z^2} + k_r^2\right)E_r = \frac{-j\omega\mu}{r}\frac{\partial H_z}{\partial \phi} + \frac{\partial^2 E_z}{\partial z \partial r} \quad \text{(A.14)}$$

$$\left(\frac{\partial^2}{\partial z^2} + k_r^2\right)H_\phi = -j\omega\varepsilon_r\frac{\partial E_z}{\partial r} + \frac{1}{r}\frac{\partial^2 H_z}{\partial z \partial \phi} \quad \text{(A.15)}$$

The solution of $E_z$ and $H_z$ is readily accomplished using the separation of variables method.

## A.2 SOLUTION OF THE ANISOTROPIC WAVE EQUATION IN CYLINDRICAL CO-ORDINATES

In this section, the method of separation of variables is used to obtain solutions for the anisotropic wave equations given by (A.6) and (A.11).

### A.2.1 Solution of $H_z$

Equation (A.6) is of the same form for the isotropic case and is given by;

$$\frac{1}{r}\frac{\partial}{\partial r}\left(r\frac{\partial H_z}{\partial r}\right) + \frac{1}{r^2}\frac{\partial^2 H_z}{\partial \phi^2} + \frac{\partial^2 H_z}{\partial z^2} + k_r^2 H_z = 0 \quad \text{(A.16).}$$

The z component of the magnetic field can not see the permittivity parallel to the c-axis, and hence (A.16) is independent of $\varepsilon_z$, with the effective dielectric constant equal to $\varepsilon_r$.

The method of separation of variables seeks a solution to (A.16) of the form,

$$H_z = R_H(r)\,F_H(\phi)\,Z_H(z) \quad \text{(A.17)}$$

Dividing (A.16) by $H_z$ then substituting in (A.17), it becomes;

$$\frac{1}{rR_H}\frac{d}{dr}\left(r\frac{dR_H}{dr}\right) + \frac{1}{r^2F_H}\frac{d^2F_H}{d\phi^2} + \frac{1}{Z_H}\frac{d^2Z_H}{dz^2} + k_r^2 = 0 \quad \text{(A.18)}$$

The third term is explicitly independent of r and $\phi$. It is also necessarily independent of z, if (A.18) is to sum to zero for all $(r,\phi,z)$. Thus;

$$\frac{1}{Z_H}\frac{d^2Z_H}{dz^2} = -\beta_H^2 \qquad \text{(A.19)},$$

where $\beta_H$ is a constant. Substituting (A.19) into (A.18), then multiplying through by $r^2$, it becomes;

$$\frac{r}{R_H}\frac{d}{dr}\left(r\frac{dR_H}{dr}\right)+\frac{1}{F_H}\frac{d^2F_H}{d\phi^2}+\left(k_r^2 - \beta_H^2\right) r^2 = 0 \qquad \text{(A.20)}$$

The second term is a function of $\phi$ only, and by the same argument;

$$\frac{1}{F_H}\frac{d^2F_H}{d\phi^2} = -m^2 \qquad \text{(A.21)},$$

where m is constant. Substituting (A.21) into (A.20) and multiplying throughout by $R_H$, (A.20) becomes;

$$r\frac{d}{dr}\left(r\frac{dR_H}{dr}\right)+\left((k_H\, r)^2 - m^2\right) R_H = 0 \qquad \text{(A.22)},$$

where,

$$k_H^2 = k_r^2 - \beta_H^2 \qquad \text{(A.23)}$$

The wave equation is thus separated into three equations, each of which determines one of the functions $R_H$,$F_H$ or $Z_H$. The first two, (A.19) and (A.21), are harmonic equations, whose solutions are harmonic functions. The last equation (A.22) is a Bessel equation of the mth order, whose solutions are Bessel functions.

### A.2.2 Solution of $E_z$

Equation (A.11) is given by;

$$\frac{1}{r}\frac{\partial}{\partial r}\left(r\frac{\partial E_z}{\partial r}\right)+\frac{1}{r^2}\frac{\partial^2 E_z}{\partial \phi^2}+\frac{\varepsilon_z}{\varepsilon_r}\frac{\partial^2 E_z}{\partial z^2}+k_z^2\, E_z = 0 \qquad \text{(A.24)}.$$

The method of separation of variables seeks a solution to (A.24) of the form,

$$E_z = R_E(r)\, F_E(\phi)\, Z_E(z) \qquad \text{(A.25)}$$

Dividing (A.24) by $E_z$ then substituting in (A.25), it becomes;

$$\frac{1}{rR_E}\frac{d}{dr}\left(r\frac{dR_E}{dr}\right)+\frac{1}{r^2F_E}\frac{d^2F_E}{d\phi^2}+\frac{\varepsilon_z}{\varepsilon_r}\frac{1}{Z_E}\frac{d^2Z_E}{dz^2}+k_z^2 = 0 \qquad \text{(A.26)}$$

The third term is explicitly independent of r and $\phi$. It is also necessarily independent of z, if (A.26) is to sum to zero for all (r,$\phi$,z). Thus;

$$\frac{\varepsilon_z}{\varepsilon_r}\frac{1}{Z_E}\frac{d^2Z_E}{dz^2} = -\beta_E^2 \qquad \text{(A.27)},$$

where $\beta_E$ is a constant. Substituting (A.27) into (A.26), then multiplying through by $r^2$, it becomes;

$$\frac{r}{R_E}\frac{d}{dr}\left(r\frac{dR_E}{dr}\right)+\frac{1}{F_E}\frac{d^2F_E}{d\phi^2}+\left(k_z^2 - \beta_E^2\right) r^2 = 0 \qquad \text{(A.28)}$$

The second term is a function of $\phi$ only, and by the same argument;

$$\frac{1}{F_E}\frac{d^2F_E}{d\phi^2} = -m^2 \qquad \text{(A.29)},$$

where m is constant. Substituting (A.29) into (A.28) and multiplying throughout by $R_E$, (A.28) becomes;

$$r\frac{d}{dr}\left(r\frac{dR_E}{dr}\right)+\left((k_E\, r)^2 - m^2\right) R_E = 0 \qquad \text{(A.30)},$$

where,

$$k_E^2 = k_z^2 - \beta_E^2 \qquad \text{(A.31)}$$

The wave equation is thus separated into three equations, each of which determines one of the functions $R_E$,$F_E$ or $Z_E$. The first two, (A.27) and (A.29), are harmonic equations, whose solutions are harmonic functions. The last equation (A.30) is a Bessel equation of the mth order, whose solutions are Bessel functions.

## A.3 FIELD COMPONENTS

Consider the cylindrical dielectric resonator shown in figure A.1. The field components $E_z$ and $H_z$ are periodic inside the dielectric and evanescent outside, also they must satisfy the separation of variables solutions derived in the previous section. These requirements are met by choosing $E_z$ and $H_z$ in the form;

$$E_{z1} = A\, J_m(k_E r) \begin{matrix}\text{Cos}(m\phi)\\ \text{Sin}(m\phi)\end{matrix} \begin{matrix}\text{Cos}\left((\varepsilon_r/\varepsilon_z)^{1/2}\beta_E\, z\right)\\ \text{Sin}\left((\varepsilon_r/\varepsilon_z)^{1/2}\beta_E\, z\right)\end{matrix} \qquad \text{(A.32a)}$$

$$E_{z2} = C\, K_m(k_{out} r) \begin{matrix}\text{Cos}(m\phi)\\ \text{Sin}(m\phi)\end{matrix} \begin{matrix}\text{Cos}\left((\varepsilon_r/\varepsilon_z)^{1/2}\beta_E\, z\right)\\ \text{Sin}\left((\varepsilon_r/\varepsilon_z)^{1/2}\beta_E\, z\right)\end{matrix} \qquad \text{(A.32b)}$$

$$E_{z3} = E\, J_m(k_E r) \begin{matrix}\text{Cos}(m\phi)\\ \text{Sin}(m\phi)\end{matrix} \exp(-\alpha z) \qquad \text{(A.32c)}$$

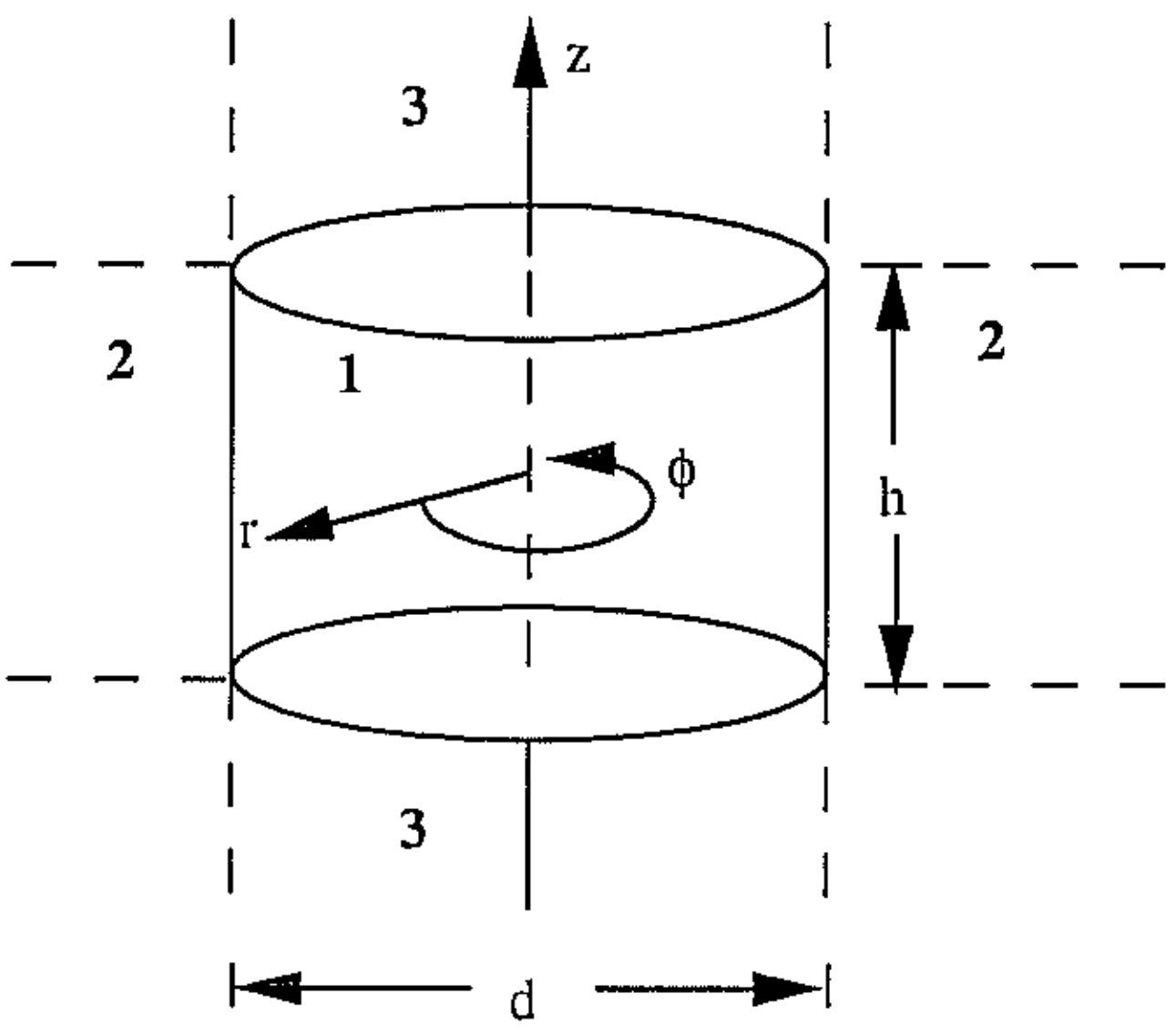

# Figure A.1

The open dielectric crystal is analyzed in cylindrical coordinates $\{r,\phi,z\}$. Resonant frequencies are solved by matching tangential fields between regions 1 and 2, and regions 1 and 3.

$$H_{z1} = B\, J_m(k_H r) \begin{matrix} \mathrm{Sin}(m\phi) \\ \mathrm{Cos}(m\phi) \end{matrix} \begin{matrix} \mathrm{Sin}(\beta_H\, z) \\ \mathrm{Cos}(\beta_H\, z) \end{matrix} \tag{A.32d}$$

$$H_{z2} = D\, K_m(k_{out} r) \begin{matrix} \mathrm{Sin}(m\phi) \\ \mathrm{Cos}(m\phi) \end{matrix} \begin{matrix} \mathrm{Sin}(\beta_H\, z) \\ \mathrm{Cos}(\beta_H\, z) \end{matrix} \tag{A.32e}$$

$$H_{z3} = F\, J_m(k_H r) \begin{matrix} \mathrm{Sin}(m\phi) \\ \mathrm{Cos}(m\phi) \end{matrix} \exp(-\alpha z) \tag{A.32f}$$

For the infinite cylindrical rod problem, region 3 does not exist and the z dependence of the magnetic and electric fields are the same, essentially reducing the problem to a two dimensional problem. For this case the arguments for the z dependence between the **E** and **H** field must be equal, and therefore $\beta_H = (\varepsilon_r/\varepsilon_z)^{1/2}\beta_E$. However the finite extent of the radial dimension causes the anisotropy to affect the radial wave numbers $k_E$ and $k_H$ differently, for the $E_z$ and $H_z$ field respectively. Likewise when the z dimension is finite it is expected that the periodic arguments for the z dependence are not necessarily equal due to the different permittivities seen by the $E_z$ and $H_z$ field components.

To simplify the solutions of the resonant frequencies presented in chapter 2 of this thesis, it has been assumed that $\beta_H = \beta_E = \beta$, with the $E_z$ and $H_z$ components assumed to have the same dependence in the z direction given by;

$$E_{z1} = A\, J_m(k_E r) \begin{matrix} \mathrm{Cos}(m\phi) \\ \mathrm{Sin}(m\phi) \end{matrix} \left(P_1\exp(-j\beta z) + P_2\exp(+j\beta z)\right) \tag{A.33a}$$

$$E_{z2} = C\, K_m(k_{out} r) \begin{matrix} \mathrm{Cos}(m\phi) \\ \mathrm{Sin}(m\phi) \end{matrix} \left(P_1\exp(-j\beta z) + P_2\exp(+j\beta z)\right) \tag{A.33b}$$

$$E_{z3} = E\, J_m(k_E r) \begin{matrix} \mathrm{Cos}(m\phi) \\ \mathrm{Sin}(m\phi) \end{matrix} \exp(-\alpha z) \tag{A.33c}$$

$$H_{z1} = B\, J_m(k_H r) \begin{matrix} \mathrm{Sin}(m\phi) \\ \mathrm{Cos}(m\phi) \end{matrix} \left(P_1\exp(-j\beta z) + P_2\exp(+j\beta z)\right) \tag{A.33d}$$

$$H_{z2} = D\, K_m(k_{out} r) \begin{matrix} \mathrm{Sin}(m\phi) \\ \mathrm{Cos}(m\phi) \end{matrix} \left(P_1\exp(-j\beta z) + P_2\exp(+j\beta z)\right) \tag{A.33e}$$

$$H_{z3} = F\, J_m(k_H r) \begin{matrix} \mathrm{Sin}(m\phi) \\ \mathrm{Cos}(m\phi) \end{matrix} \exp(-\alpha z) \tag{A.33f}$$

These approximations are the same as assuming the divergence of **E** is not dependent on z, so the second term in (A.11) is equal to zero and can be ignored. Here m is the azimuthal mode number, $\beta$ the axial propagation constant inside the dielectric, $k_0$ the free space wave number, $k_{out}$ the radial propagation constant outside the dielectric, and $k_E$ and $k_H$ the radial propagation constants inside the dielectric for TM and TE modes respectively. Here $k_E^2 = \varepsilon_z k_0^2 - \beta^2$, $k_H^2 = \varepsilon_r k_0^2 - \beta^2$ and $k_{out}^2 = \beta^2 - k_0^2$.

To calculate the other field components equations (A.12) to (A.15) are used. To calculate the resonant frequencies one follows the methods used in (Kajfez and Guillon,1986) and (Garault and Guillon, 1976). Tangential field components are matched at the boundary of regions 1 and 2, and regions 1 and 3, to obtain two coupled transcendental eigen-value equations, to solve for the resonant frequencies. Chapter 2 shows the comparisons between experiment and theory in anisotropic cylindrical sapphire resonators.

# APPENDIX B

# DERIVATIONS OF ELECTRICAL AND MECHANICAL TRANSDUCER PARAMETERS FOR A MECHANICALLY MODULATED RESONANT PARAMETRIC TRANSDUCER

## B.1 TRANSDUCER MODEL

The operation of a parametric transducer has been considered from various points of view (Braginsky, 1970; Giffard and Paik, 1977; Blair, 1980; Darling et al, 1982; Veitch 1991). The approach presented here differs from previous work, in that it considers the transducer in scattering parameter form, with the modulation considered in terms of a.m. and p.m. quadrature components, rather than an impedance formulation that analyses upper and lower side bands. Thus an augmented scattering matrix is presented which includes all a.m. and p.m. components, which simplifies calculations of a.m. to p.m. and p.m. to a.m. conversions.

Figure B.1 shows an equivalent circuit for a capacitance modulated transducer. The transducer is modelled as a parallel LCR circuit coupled with a transmission line of characteristic impedance $Z_o$, and pumped at a frequency $\Omega_p$. The relative motion of the resonant bar antenna is assumed to modulate the capacitance. For a harmonic time varying capacitor plate displacement of, $x(t) = x_o\,(1 + \alpha\,\mathrm{Sin}(\omega_m t))$, and hence the time harmonic velocity of, $u(t) = u_p \mathrm{Sin}(\omega_m t)$ ($u_p = x_o \omega_m \alpha$), the capacitance is of the form $c(t) = C_o\,(1 - \alpha \mathrm{Sin}(\omega_m t))$, assuming $\alpha \ll 1$ which is true for a gravitational wave detector.

The matrix description accompanying figure B.1 is given by (B1). Here **a** is the incident signal ($\sqrt{\mathrm{watt}}$), **b** is the reflected signal ($\sqrt{\mathrm{watt}}$) off the transducer, **u** is the input velocity to the transducer and **f** is the force on the transducer. For each normal mode sensed by the transducer, the following matrix is defined; (subscript + and - refer to the positively and negatively tuned normal modes respectively)

$$\begin{bmatrix} \mathbf{b}_{pm}(\omega_\pm) \\ \mathbf{f}(\omega_\pm) \\ \mathbf{b}_{am}(\omega_\pm) \end{bmatrix} = \begin{bmatrix} S_{pp\pm} & S_{pu\pm} & S_{pa\pm} \\ S_{fp\pm} & Z_{in\pm} & S_{fa\pm} \\ S_{ap\pm} & S_{au\pm} & S_{aa\pm} \end{bmatrix} \begin{bmatrix} \mathbf{a}_{pm}(\omega_\pm) \\ \mathbf{u}(\omega_\pm) \\ \mathbf{a}_{am}(\omega_\pm) \end{bmatrix} \qquad \text{(B1).}$$

The transducer scattering components for each normal mode are defined as; $S_{pa\pm}$, ratio of reflected p.m converted from incident a.m when **u**=0 and $\mathbf{a}_{pm}=0$; $S_{ap\pm}$, ratio of reflected a.m converted from incident p.m when **u**=0 and $\mathbf{a}_{am}=0$; $S_{pp\pm}$, ratio of reflected p.m to incident p.m when **u**=0 and $\mathbf{a}_{am}=0$; $S_{aa\pm}$, ratio of reflected a.m to incident a.m when

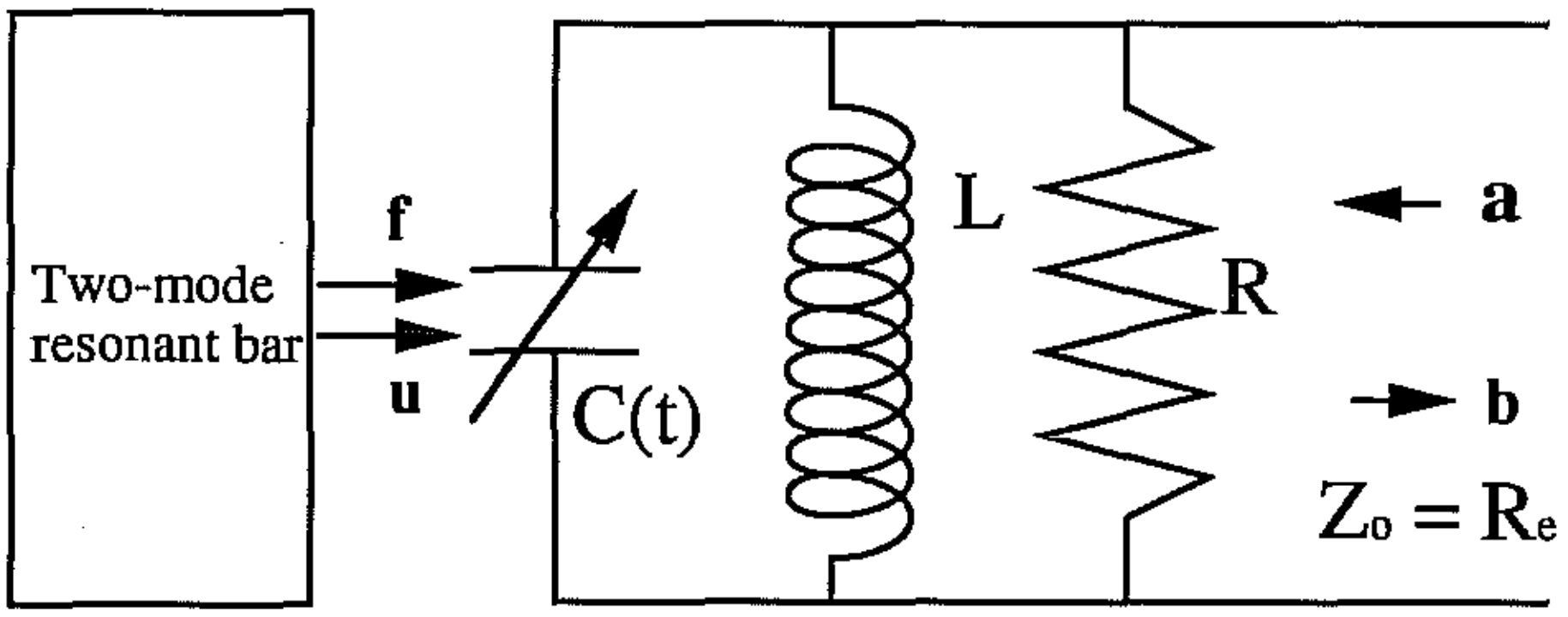

# Figure B.1

Parametric upconverting transducer, modelled as a capacitance modulated LCR circuit. here **u** and **f** are the input velocity and force applied by the resonant bar to the transducer, and **a** and **b** are the electrical incident and reflected signal in $(\text{watt})^{1/2}$ $a_{pm}$=0 and u=0; $S_{pu\pm}$, p.m transductance scattering parameter; $S_{au\pm}$, a.m transductance scattering parameter; $S_{fp\pm}$, p.m reverse transductance scattering parameter; $S_{fa\pm}$, a.m reverse transductance scattering parameter; $Z_{in\pm}$, the mechanical input impedance. This method naturally includes a.m to p.m and p.m to a.m conversions.

## B.2 CALCULATION OF SCATTERING MATRIX COMPONENTS

### B.2.1 Forward transductance

To calculate the forward transductance scattering parameters $S_{pu}$ and $S_{au}$, the amplitude and phase sidebands are calculated in the presence of an ideal pump source with no noise side bands incident on the cavity (ie $a_{am}(\omega_m) = a_{pm}(\omega_m) = 0$) at the modulation frequency $\omega_m$. Hence the pump signal can be represented by the following phasor;

$$A(t) = \sqrt{P_{inc}}\,\mathrm{Exp}(j\Omega_p t) \qquad \text{(B2)},$$

where $P_{inc}$ and $\Omega_p$ are the incident pump power and frequency respectively. The signal that enters the transducer cavity will be modulated, since $|\alpha| << 1$ most of the signal will be in the first order sidebands. Thus the modulated signal will be of the form;

$$\sqrt{P_c(t)} = \sqrt{P_o}\left[\mathrm{Cos}(\Omega_p t) + g_U \mathrm{Sin}(\Omega_U t) + h_U \mathrm{Cos}(\Omega_U t) + g_L \mathrm{Sin}(\Omega_L t) + h_L \mathrm{Cos}(\Omega_L t)\right] \qquad \text{(B3)},$$

where $\Omega_U = \Omega_p + \omega_m$ and $\Omega_L = \Omega_p - \omega_m$. Equation (B.3) can be written in terms of amplitude and phase modulation indices;

$$\sqrt{P_c(t)} = \sqrt{P_o}\left[\mathrm{Cos}(\Omega_p t)\left(1 + m_{am}\mathrm{Cos}(\omega_m t + \theta_{am})\right) + \mathrm{Sin}(\Omega_p t)\left(m_{pm}\mathrm{Cos}(\omega_m t + \theta_{pm})\right)\right] \text{(B4)}$$

where;

$$m_{am}^2 = (g_L - g_U)^2 + (h_L + h_U)^2 \qquad \text{(B5a)}$$
$$\mathrm{Tan}(\theta_{am}) = (g_L - g_U)/(h_L + h_U)$$

$$m_{pm}^2 = (h_U - h_L)^2 + (g_U + g_L)^2 \qquad \text{(B5b)}$$
$$\mathrm{Tan}(\theta_{pm}) = (h_U - h_L)/(g_U + g_L)$$

The values of $g_{L,U}$ and $h_{L,U}$ have been calculated previously (Giffard and Paik, 1977; Veitch, 1986) to be;

$$g_{L,U} = \pm\frac{\alpha\,\Omega_{L\,U}}{\Omega_o}\,\frac{Q_e^2\delta_{L\,U}}{1+4Q_e^2\delta_{L\,U}^2} \qquad \text{(B6a)}$$

$$h_{L,U} = \pm\frac{\alpha\,\Omega_{L\,U}}{2\Omega_o}\,\frac{Q_e}{1+4Q_e^2\delta_{L\,U}^2} \qquad \text{(B6b)},$$

where $\Omega_o = (LC_o)^{-1/2}$, $Q_e = R/((1+\beta_e)(\Omega_o L))$, $\beta_e = R/R_e$ and $\delta_{U,L} = (\Omega_{U,L}-\Omega_o)/\Omega_o$.

The modulation coefficients given by (B5) are with respect to the carrier signal in the cavity. The observed modulation coefficients $m^{\#}_{am}{}^2$ and $m^{\#}_{pm}{}^2$ must be calculated with respect to the reflected carrier signal. Figure B.2 shows the phasor diagram of (B3) when the carrier is phase shifted. From this diagram the rotated values ($g^{\#}_{L,U}$ and $h^{\#}_{L,U}$) may be written in terms of the old by (note phasor g lags h);

$$g^{\#}_{L,U} = g_{L,U}\mathrm{Cos}[\gamma] + h_{L,U}\mathrm{Sin}[\gamma] \qquad \text{(B7a)}$$

$$h^{\#}_{L,U} = h_{L,U}\mathrm{Cos}[\gamma] - g_{L,U}\mathrm{Sin}[\gamma] \qquad \text{(B7b)}$$

Substituting the rotated values from (B7) into (B5) the amplitude and phase components with respect to the rotated carrier may be written as;

$$m^{\#}_{am}{}^2 = ((g_L-g_U)\mathrm{Cos}[\gamma] + (h_L-h_U)\mathrm{Sin}[\gamma])^2 + ((h_L+h_U)\mathrm{Cos}[\gamma]-(g_L+g_U)\mathrm{Sin}[\gamma])^2 \qquad \text{(B8a)}$$

$$m^{\#}_{pm}{}^2 = ((h_U-h_L)\mathrm{Cos}[\gamma] + (g_L-g_U)\mathrm{Sin}[\gamma])^2 + ((g_L+g_U)\mathrm{Cos}[\gamma]+(h_L+h_U)\mathrm{Sin}[\gamma])^2 \qquad \text{(B8b)}$$

The angle ($\gamma$) between the carrier inside and reflected from the cavity is given by;

$$\gamma = \mathrm{Arg}[\Gamma(j\delta)/T(j\delta)] \; ; \; \Gamma(j\delta) = \frac{(\beta_e-1)/(\beta_e+1)-j2Q_e\delta}{1+j2Q_e\delta} \; ; \; T(j\delta) = \frac{2\sqrt{\beta_e}/(\beta_e+1)}{1+j2Q_e\delta} \qquad \text{(B9)}$$

Here $\Gamma(j\delta)$ and $T(j\delta)$ are the reflection and transmission coefficient of the carrier respectively and $\delta = (\Omega_p-\Omega_o)/\Omega_o$. Thus the sine and cosine of the angle $\gamma$ are given by;

$$\mathrm{Cos}[\gamma] = \mathrm{Abs}[T(j\delta)/\Gamma(j\delta)]\,\mathrm{Re}[\Gamma(j\delta)/T(j\delta)] = \frac{(\beta_e-1)/(\beta_e+1)}{\sqrt{(\beta_e-1)/(\beta_e+1)^2+4Q_e^2\delta^2}} \qquad \text{(B10a)}$$

$$\mathrm{Sin}[\gamma] = \mathrm{Abs}[T(j\delta)/\Gamma(j\delta)]\,\mathrm{Im}[\Gamma(j\delta)/T(j\delta)] = \frac{-2Q_e\delta}{\sqrt{(\beta_e-1)/(\beta_e+1)^2+4Q_e^2\delta^2}} \qquad \text{(B10b)}$$

By substituting (B6) and (B10) into (B8) $m^{\#}_{am}{}^2$ and $m^{\#}_{pm}{}^2$ may be calculated with respect to cavity parameters.

The carrier power is determined by calculating the total power dissipated in the equivalent circuit illustrated in figure B.1. The carrier power may be split into two components;

$$P_d = V_p^2/(2R_e) + V_p^2/(2R) = P_{de} + P_{dcav} \qquad \text{(B11)}$$

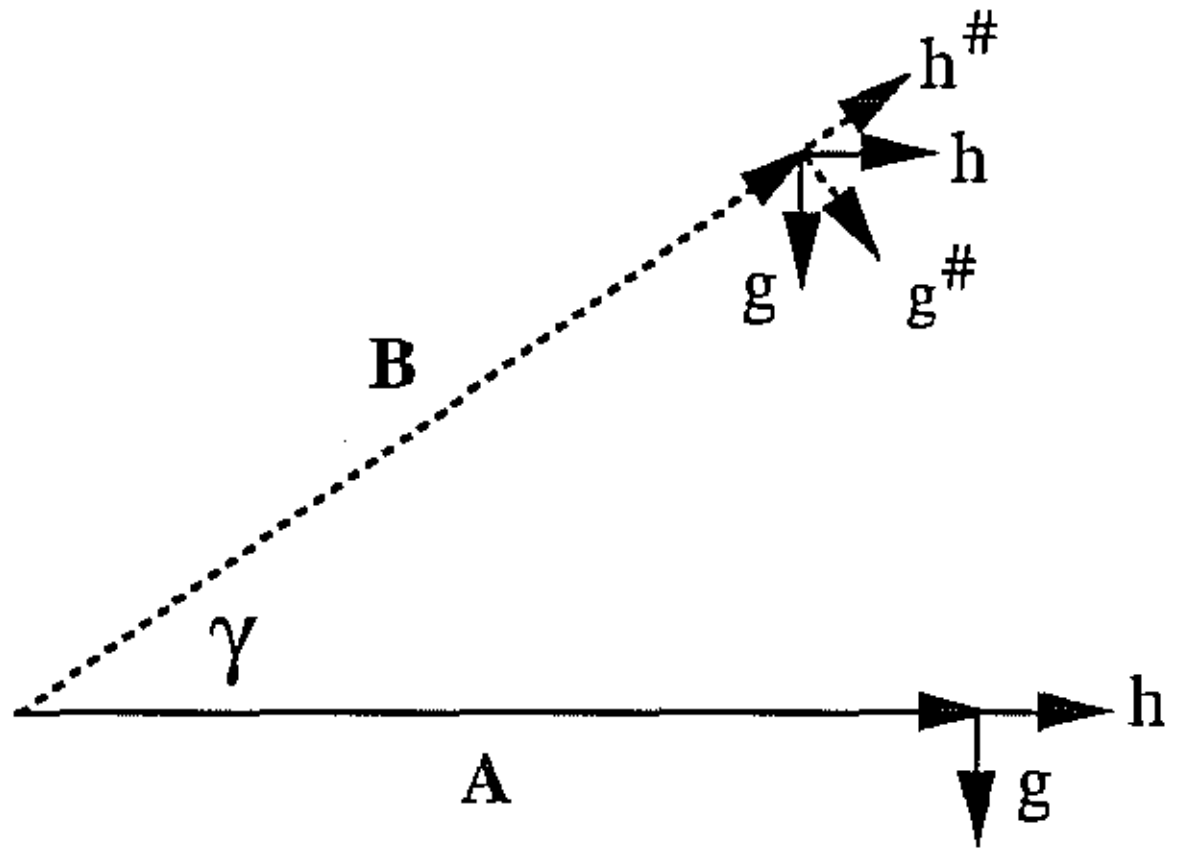

# Figure B.2

Phasor A is a representation of equation (B3). Phasor B shows what happens to the modulation components with respect to the carrier when the carrier is phase shifted. Eequations (B7a&b) were deduced from this diagram.

where $V_p$ is the peak voltage across the parallel circuit. The first term in (B11) represents the carrier power in the coupling circuit ($P_{de}$), and the second term represents the power dissipated inside the cavity $P_{dcav}$. From (B11) it can be shown that;

$$P_{de} = \frac{\beta_e P_d}{\beta_e+1} \text{ and } P_{dcav} = \frac{P_d}{\beta_e+1} \qquad \text{(B12a\&b)}$$

Also the carrier power inside the cavity can be related to the incident pump power by $P_{dcav} = |T(j\delta)|^2 P_{inc}$, equating this to (B12b) the total carrier power can be calculated to be;

$$P_d = \frac{4\beta_e P_{inc}}{(\beta_e+1)(1+4Q_e^2\delta^2)} \qquad \text{(B13)}$$

The fraction of power that is coupled to is given by the carrier in the coupling circuit, substituting (B13) into (B12a) obtains;

$$P_{de} = \frac{4\beta_e^2 P_{inc}}{(\beta_e+1)^2(1+4Q_e^2\delta^2)} \qquad \text{(B14)}$$

The peak power in the sidebands transmitted to the coupling circuit is given by; $|\mathbf{b}_{am}|^2 = P_{de}\, m_{am}^{\#\,2}$ and $|\mathbf{b}_{pm}|^2 = P_{de}\, m_{pm}^{\#\,2}$ ; Thus the forward transductance may be calculated from the ratio of the sideband peak power to $u_p^2$. Therefore by multiplying (B8) by $P_{de}/u_p^2$ ( with $x_o = f_o/(2\ df/dx)$ ) and assuming that $|\Omega_p - \Omega_o| << \Omega_o$, the forward transductance is calculated to be;

$$|S_{au}|^2 = |S_{fwd}|^2 F_{am} \quad \text{and} \quad |S_{pu}|^2 = |S_{fwd}|^2 F_{pm} \qquad \text{(B15a\&b)},$$

where;

$$|S_{fwd}|^2 = \frac{4P_{inc}\beta_e^2}{\omega_m^2(\beta_e+1)^2(1+4Q_e^2\omega_m^2/\Omega_o^2)}\left(\frac{2Q_e df}{f_o\ dx}\right)^2$$

$$F_{am} = \frac{1+4Q_e^2\omega_m^2/\Omega_o^2}{4(1+4Q_e^2\delta^2)((\beta_e-1)^2/(\beta_e+1)^2+4Q_e^2\delta^2)} \times$$

$$\left(\left(\frac{1}{1+4Q_e^2\delta_L^2} - \frac{1}{1+4Q_e^2\delta_U^2}\right)\left(\frac{\beta_e-1}{\beta_e+1}\right) + \left(\frac{2Q_e\delta_U}{1+4Q_e^2\delta_U^2} - \frac{2Q_e\delta_L}{1+4Q_e^2\delta_L^2}\right)2Q_e\delta\right)^2 +$$

$$\left(-\left(\frac{1}{1+4Q_e^2\delta_L^2} + \frac{1}{1+4Q_e^2\delta_U^2}\right)2Q_e\delta + \left(\frac{2Q_e\delta_U}{1+4Q_e^2\delta_U^2} + \frac{2Q_e\delta_L}{1+4Q_e^2\delta_L^2}\right)\left(\frac{\beta_e-1}{\beta_e+1}\right)\right)^2$$

$$F_{pm} = \frac{1+4Q_e^2\omega_m^2/\Omega_o^2}{4(1+4Q_e^2\delta^2)((\beta_e-1)^2/(\beta_e+1)^2+4Q_e^2\delta^2)} \times$$

$$\left(\left(\frac{1}{1+4Q_e^2\delta_L^2} + \frac{1}{1+4Q_e^2\delta_U^2}\right)\left(\frac{\beta_e-1}{\beta_e+1}\right) + \left(\frac{2Q_e\delta_U}{1+4Q_e^2\delta_U^2} + \frac{2Q_e\delta_L}{1+4Q_e^2\delta_L^2}\right)2Q_e\delta\right)^2 +$$

$$\left(\left(\frac{1}{1+4Q_e^2\delta_L^2} - \frac{1}{1+4Q_e^2\delta_U^2}\right)2Q_e\delta + \left(\frac{2Q_e\delta_L}{1+4Q_e^2\delta_L^2} - \frac{2Q_e\delta_U}{1+4Q_e^2\delta_U^2}\right)\left(\frac{\beta_e-1}{\beta_e+1}\right)\right)^2$$

Here $S_{fwd}$ is the forward transductance when the pump oscillator is incident on the centre of resonance. Thus for the first time non-quasi static formulae for the modulation indices of a parametric transducer operating in reflection have been calculated. If $\omega_m$ -> 0 these formulae reduce to the quasi static formulae as derived in (Veitch 1991). The total forward transductance may be calculated from the sum of (B15a) and (B15b) as;

$$|S_{tu}|^2 = |S_{au}|^2 + |S_{pu}|^2 = \frac{|S_{fwd}|^2(1+4Q_e^2\omega_m^2/\Omega_o^2)}{2(1+4Q_e^2\delta^2)}\left(\frac{1}{1+4Q_e^2\delta_U^2} + \frac{1}{1+4Q_e^2\delta_L^2}\right) \quad \text{(B16)}$$

To extend (B15a&b) and (B16) to the two-mode system $\omega_\pm$ is substituted for $\omega_m$ to calculate $|S_{au\pm}|^2$ and $|S_{pu\pm}|^2$. The forward transductance in terms of the square root phase modulation power per displacement as presented in section 5.2, may be calculated by $S_{PX2} = \omega_m|S_{tu}|$.

Transducers are usually compared by the electro-mechanical coupling, which can be considered as a normalised forward transductance. Hence from (B15a&b) the electro-mechanical coupling is calculated in the following sub-section.

### B.2.2 Electro-mechanical Coupling

If the pump oscillator is not on the centre of resonance the electro-mechanical coupling will be reduced compared to the centre of resonance coupling. The coupling must be calculated as a function of pump offset frequency. Veitch (Veitch, 1991) uses a definition which depends on the reverse and forward transductances of the transducer. This definition is not appropriate when determining the signal to noise ratio of the antenna system (Paik, 1974; Chapter 3), as the strength of the signal only depends on the forward transductance. Hence it is appropriate to define the electro-mechanical coupling as;

$$\beta = P_s/P_a \quad \text{(B17),}$$

where $P_s$ is the total electrical side band signal power in the cavity and coupling circuit due to the mechanical resonant bar modes, and $P_a$ is the upconverted antenna signal power = $\Omega_p \times E_a$, where $E_a$ is the mechanical antenna energy. A resonant bar antenna with a secondary mass attached to one end will only couple to half the antenna energy, therefore substituting in $E_a = 2 \times$ secondary mass energy = $m_2\omega_2^2\alpha^2x_o^2$ and $P_s = \omega_2^2\alpha^2x_o^2P_dm_\phi^2 = P_{inc}\omega_2^2\alpha^2x_o^2\left(|S_{au}|^2+|S_{pu}|^2\right)(1+\beta_e)/\beta_e$ in (B15) one obtains;

$$\beta = \frac{2\beta_eP_{inc}}{(1+\beta_e)m_2\omega_2^2\Omega_0}\left(\frac{2Q_e}{f_o}\frac{df}{dx}\right)^2\left(\frac{1}{1+4Q_e^2\delta^2}\right)\left(\frac{1}{1+4Q_e^2\delta_U^2}+\frac{1}{1+4Q_e^2\delta_L^2}\right) \quad \text{(B18)},$$

where the rate of resonant frequency change with respect to displacement for a capacitance modulated transducer is given by $df/dx = f_o/(2x_o)$. The centre of resonance coupling, when $\Omega_p = \Omega_o$ can thus be shown to be;

$$\beta_0 = \frac{4\beta_eP_{inc}}{(1+\beta_e)m_2\omega_2^2\Omega_0}\left(\frac{2Q_{fac}}{f_o}\frac{df}{dx}\right)^2 \quad \text{(B19)},$$

where

$$Q_{fac}^2 = \left(\frac{Q_e^2}{1+4Q_e^2\omega_m^2/\Omega_0^2}\right) \quad \text{(B20)}.$$

The centre of resonance electro-mechanical coupling $\beta_0$, has been previously defined as $(C_oV_p^2Q_e)/(2m_2\omega_2^2x_o^2)$ (Veitch, 1991; Linthorne 1991), and is consistent with the above definition. This value may be split into two components, $\beta_0 = \beta_{0e}+\beta_{0cav}$, where $\beta_{0e} = \beta_e\beta_0/(1+\beta_e)$ is the component of electro-mechanical coupling in the coupling circuit, and $\beta_{0cav} = \beta_0/(1+\beta_e)$ is the component of electro-mechanical coupling inside the cavity transducer (not coupled to). In this thesis I refer to the total electro-mechanical coupling $\beta_0$ to be consistent with others.

The total electro-mechanical coupling can be broken into four components when considering a two-mode resonant bar. Firstly two components are associated with either the + or - normal modes by relating $\beta_0$ to $\beta_{0\pm}$ via (6.21) and substituting $\omega_m$ for either $\omega_\pm$. Then by defining electro-mechanical coupling components that are proportional to the forward transductance scattering parameters $|S_{pu\pm}|^2$ and $|S_{au\pm}|^2$ respectively, the coupling components may be broken down further into p.m and a.m quadrature components. Thus the four components are given by;

$$\beta_{pm\pm} = \beta_{0\pm}F_{pm\pm} \text{ and } \beta_{am\pm} = \beta_{0\pm}F_{am\pm} \quad \text{(B21a\&b)}.$$

If the pump oscillator driving the transducer is incident on resonance (ie $\delta$=0) all the coupling energy will be in the p.m. side bands. Figure B.3 shows the basic demodulation scheme which is phase sensitive when the phase shifter supplies a $\pi/2$ phase shift at the LO port with respect to RF port. If the pump oscillator is offset from resonance some of the coupling energy will manifest in the a.m. side bands. It is therefore clear the electro-mechanical coupling can be maximised by offsetting the phase by $\theta_{off}$, such that;

$$\beta_{pm} = \beta\, \mathrm{Cos}^2[\theta_{off}] \;\; ; \;\; \beta_{am} = \beta\, \mathrm{Sin}^2[\theta_{off}] \;\; ; \;\; \mathrm{Tan}^2[\theta_{off}] = \beta_{am}/\beta_{pm} \qquad \text{(B22)}$$

This is equivalent to adjusting the phase to maximise the mixer sensitivity at the required pump oscillator offset.

When the output of the mixer is set to be phase sensitive on resonance, the cavity frequency response will be proportional to the imaginary part of the reflection coefficient, as given by (B23) when $\theta_{max} = 0$;

$$V_{mix} = \mathrm{Abs}[\,\Gamma(j\delta)]\mathrm{Sin}[\mathrm{Arg}[\,\Gamma(j\delta)]+\theta_{max}] \qquad \text{(B23)},$$

Here $\theta_{max}$ is the phase that has to be added to the imaginary response to obtain maximum electro-mechanical coupling at the required pump oscillator offset $\delta_{off}$. Thus $\theta_{max}$ must supply the following phase shift;

$$\theta_{max} = \theta_{off} - \mathrm{Arg}[\,\Gamma(j\delta_{off})] \qquad \text{(B24)}$$

It has been considered in chapter 6 that it may be necessary to operate at a pump offset of $1/\sqrt{3}$ of the lower half bandwidth (6.26a) to limit seismic noise. At this offset $\theta_{max} = \pi/3$ independent of cavity parameters. Figure B.4 shows the normalised mixer voltage when $\beta_e = .3$, with $\theta_{max} = 0$ and $\pi/3$, which corresponds to a maximum sensitivity on resonance and at $1/\sqrt{3}$ of the lower half bandwidth respectively. The ratio of the maximum slope squared is equal to 9/16 the same as the ratio of $\beta$ on resonance to $\beta$ at $1/\sqrt{3}$ of the lower half bandwidth as expected. Thus it is possible to couple to the total electro-mechanical coupling given by (B18) by setting the phase of the demodulation system to maximise the voltage sensitivity at the required offset.

The electro-mechanical coupling components have been analysed with respect to the gravity wave detector at UWA in chapter 6.

### B.2.3 Input impedance

The input impedance is the proportionality constant relating the force on the antenna at $\omega_m$ to the input velocity in the absence of input pump power side bands (ie $a_{am}(\omega_m)$ =

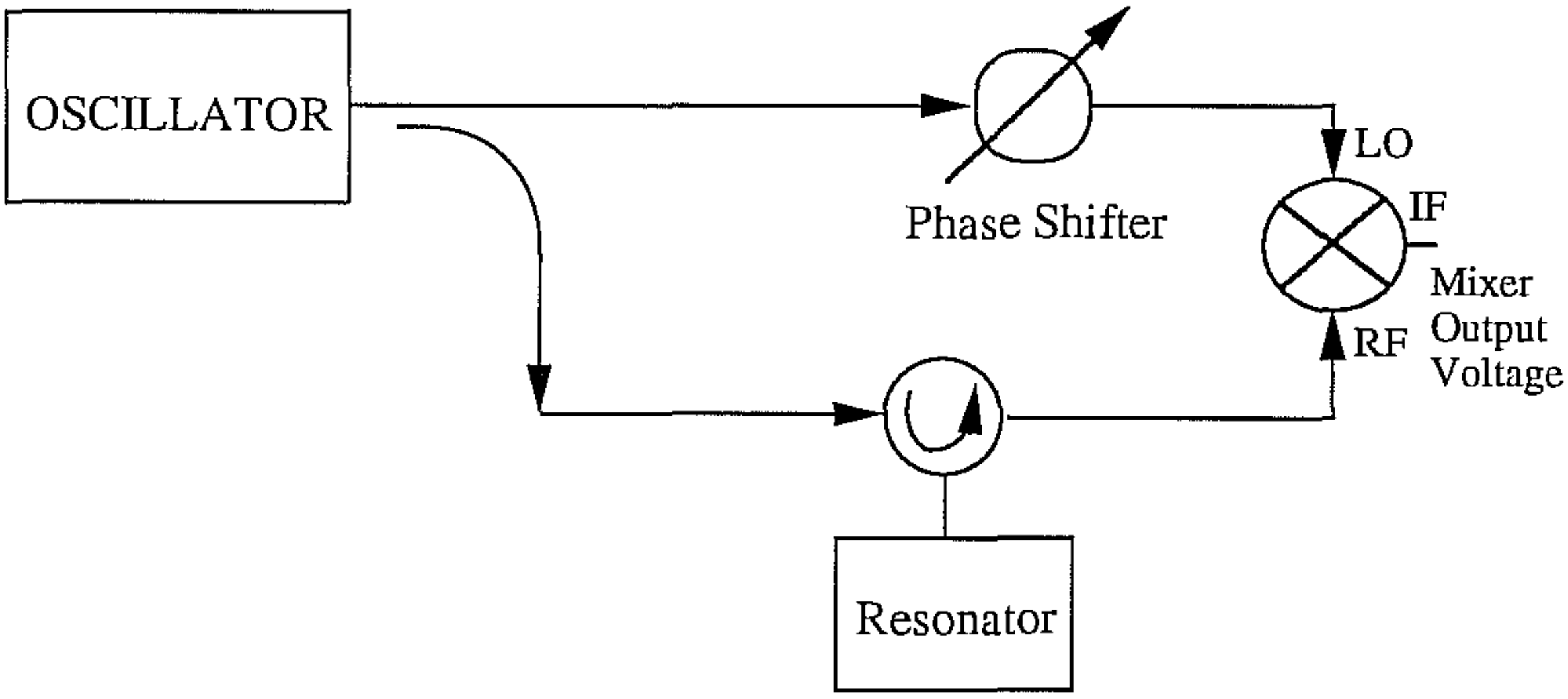

# Figure B.3

Standard bridge circuit to detect a cavity resonator/transducer. When the phase shifter is set so the LO port is in quadrature to the RF port on resonance, then the imaginary part of the reflection coefficient will be traced out as the oscillator is swept across resonant cavity. If the phase is adjusted then a combination of the real and imaginary parts will be observed as the oscillator is swept across resonance.

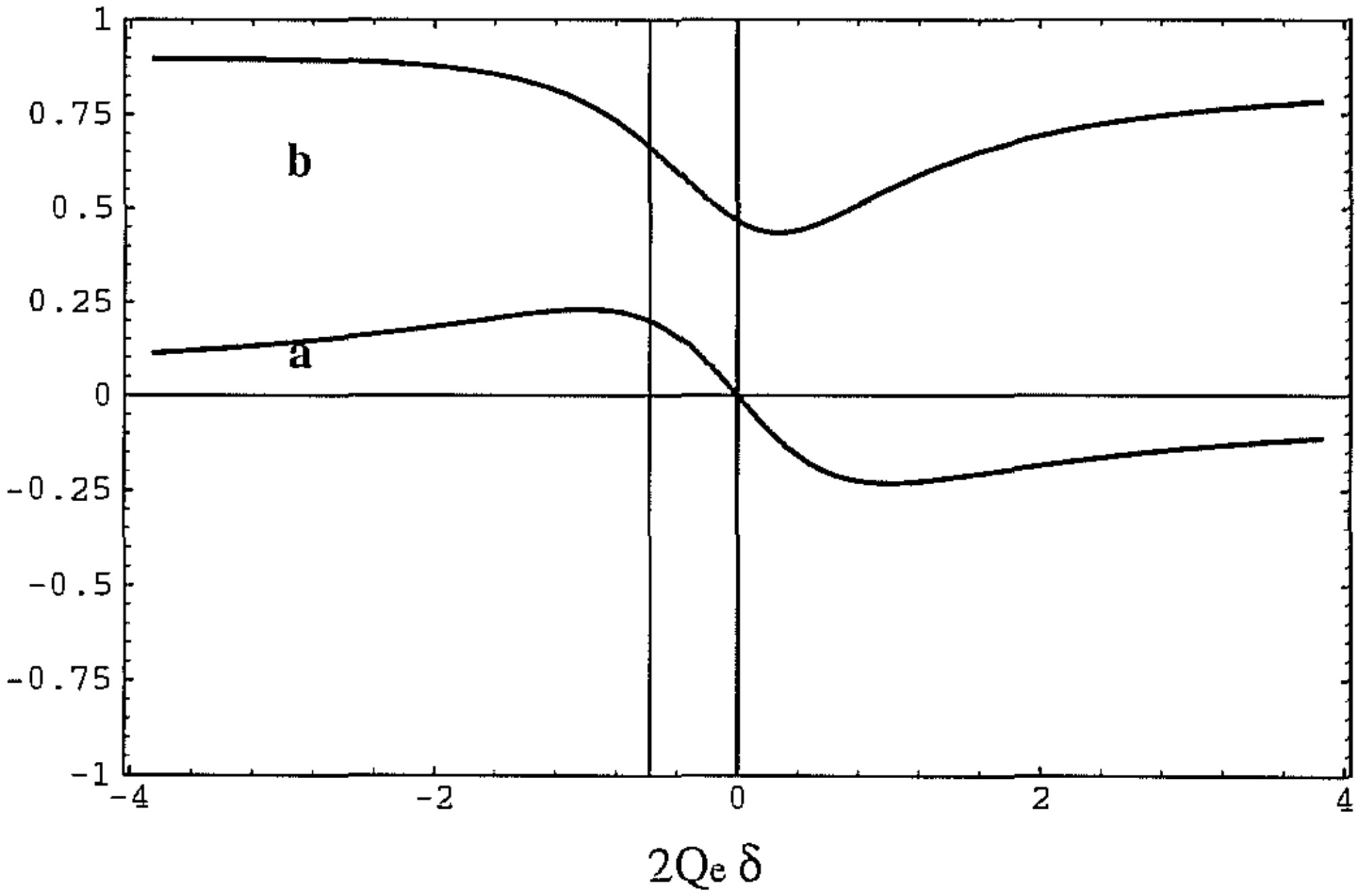

# Figure B.4

Normalised output voltage as a function of pump oscillator frequency;
**a**. When the LO and RF port are in quadrature on resonance, here the maximum mixer sensitivity occurs at resonance.
**b**. When an extra phase shift of $\pi/3$ is added to maximise the mixer sensitivity when the pump is offset to $(1/3)^{1/2}$ of the lower half bandwidth

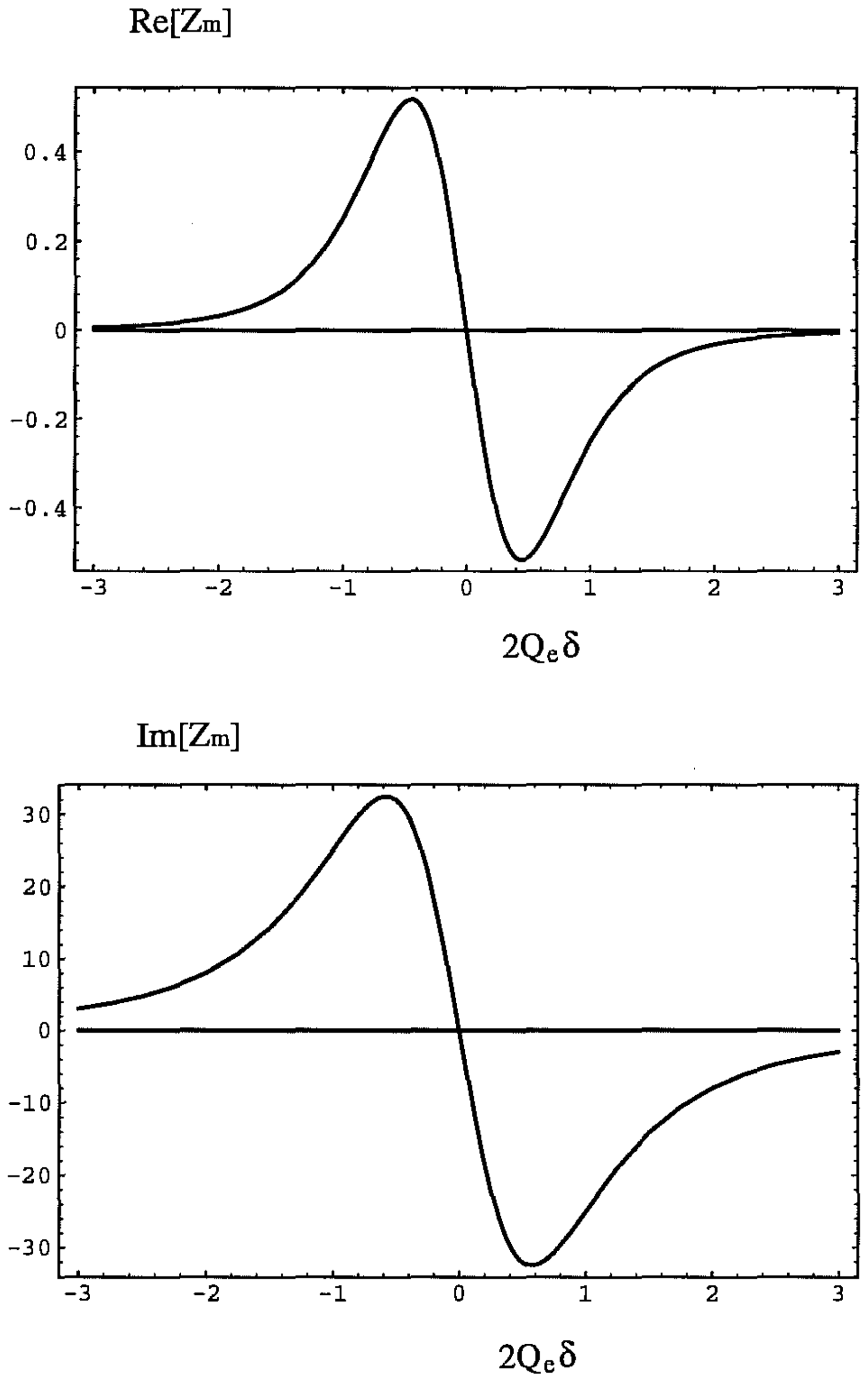

# Figure B.5

Real and imaginary components of the input impedance per secondary oscillator mass, ($Z_m = Z_{in}/m2$), as a function of pump oscillator frequency ($2Q_e\,\delta$). Here $\beta_o(\omega_2^2/\omega_m) = 100$ and $\omega_m/\Delta\omega = .005$, where $\Delta\omega = \Omega_o/Q_e$ is the bandwidth of the transducer.

$a_{pm}(\omega_m) = 0$). Assuming that $|\Omega_p - \Omega_o| << \Omega_o$, the form of the input impedance has been shown to be (Giffard and Paik, 1977; Veitch, 1982);

$$\mathrm{Re}[Z_{in}] = \frac{C_o V_p^2 Q_e}{4x_o^2 \omega_m}\left(\frac{1}{1+4Q_e^2\,\delta_U^2} - \frac{1}{1+4Q_e^2\,\delta_L^2}\right) \quad \text{(B25a)},$$

$$\mathrm{Im}[Z_{in}] = \frac{C_o V_p^2 Q_e}{4x_o^2 \omega_m}\left(\frac{2Q_e\,\delta_U}{1+4Q_e^2\,\delta_U^2} + \frac{2Q_e\,\delta_L}{1+4Q_e^2\,\delta_L^2}\right) \quad \text{(B25b)},$$

where $V_p$ is the peak pump oscillator voltage across the capacitor. The pump power inside the resonator is $P_{dcav} = V_p^2/(2\ R)$, where $R = Q_e/(\Omega_o C_o)$ with $P_{inc}$ related to the pump power inside the cavity by $P_{dcav} = |T(j\delta)|^2\ P_{inc}$. Combining these equations to eliminate $C_o V_p^2$ in (B25) it can easily be shown that;

$$\mathrm{Re}\big[Z_m\big] = \frac{\beta_0\,(\omega_2^2/\omega_m)(1+4Q_e^2\omega_m^2/\Omega_0^2)}{2\left(1+4Q_e^2\,\delta^2\right)}\left(\frac{1}{(2Q_e\,\delta_U)^2+1} - \frac{1}{(2Q_e\,\delta_L)^2+1}\right) \quad \text{(B26a)},$$

$$\mathrm{Im}\big[Z_m\big] = \frac{-\beta_0\,(\omega_2^2/\omega_m)(1+4Q_e^2\omega_m^2/\Omega_0^2)}{2\left(1+4Q_e^2\,\delta^2\right)}\left(\frac{2Q_e\,\delta_U}{(2Q_e\,\delta_U)^2+1} + \frac{2Q_e\,\delta_L}{(2Q_e\,\delta_L)^2+1}\right) \quad \text{(B26b)},$$

where $Z_m = Z_{in}/m_2$. Figure B.5 shows how the real and imaginary parts of the input impedance given by (B26) behave as a function of pump offset, $2Q_e\,\delta$, when $\beta_0\,(\omega_2^2/\omega_m) = 100$, and $\omega_m/\Delta\omega = .005$, where $\Delta\omega = \Omega_o/Q_e$ is the half bandwidth.

Because the input impedance is proportional to the centre of resonance electro-mechanical coupling, it may be easily separated into the following components associated with the two normal modes of a two-mode resonant bar system;

$$\mathrm{Re}\big[Z_{m\pm}\big] = \frac{\beta_{0\pm}\,(\omega_2^2/\omega_\pm)(1+4Q_e^2\omega_\pm^2/\Omega_0^2)}{2\left(1+4Q_e^2\,\delta^2\right)}\left(\frac{1}{(2Q_e\,\delta_{U\pm})^2+1} - \frac{1}{(2Q_e\,\delta_{L\pm})^2+1}\right) \quad \text{(B27a)},$$

$$\mathrm{Im}\big[Z_{m\pm}\big] = \frac{-\beta_{0\pm}\,(\omega_2^2/\omega_\pm)(1+4Q_e^2\omega_\pm^2/\Omega_0^2)}{2\left(1+4Q_e^2\,\delta^2\right)}\left(\frac{2Q_e\,\delta_{U\pm}}{(2Q_e\,\delta_{U\pm})^2+1} + \frac{2Q_e\,\delta_{L\pm}}{(2Q_e\,\delta_{L\pm})^2+1}\right) \quad \text{(B27b)},$$

How these values of the input impedance effect the normal mode properties of a two-mode resonant bar are discussed in chapter 6.

### B.2.4 Reflection scattering parameters

A resonant parametric transducer will affect the pump oscillator noise side bands when reflecting off the transducer. In the following analysis the incident signal is considered first with a phase modulation to calculate $S_{pp}$ and $S_{ap}$, and then with an amplitude modulation to calculate $S_{aa}$ and $S_{pa}$. The phasors representing the incident signal for the two respective cases, with a modulation at ω rads/sec, can be written mathematically as;

$$\sqrt{P_{inc}}\,Exp\left(j\Omega_p t\right)\left[1 - \frac{m_{pm}}{2}Exp\left(-j(\omega t+\theta_{pm})\right) + \frac{m_{pm}}{2}Exp\left(j(\omega t+\theta_{pm})\right)\right] \quad (B28a)$$
$$= \sqrt{P_{inc}}\,Exp\left(j\Omega_p t\right)\left[1 + jm_{pm}Sin\left(\omega t+\theta_{pm}\right)\right]$$

$$\sqrt{P_{inc}}\,Exp\left(j\Omega_p t\right)\left[1 + \frac{m_{am}}{2}Exp\left(-j(\omega t+\theta_{am})\right) + \frac{m_{am}}{2}Exp\left(j(\omega t+\theta_{am})\right)\right] \quad (B28b)$$
$$= \sqrt{P_{inc}}\,Exp\left(j\Omega_p t\right)\left[1 + m_{am}Cos\left(\omega t+\theta_{am}\right)\right]$$

Figure B.6 shows a diagram of these phasors when $\theta_{pm}$ and $\theta_{am}$ are zero. The following analysis examine what happens to the pump oscillator noise side bands with respect to the pump signal, when reflected off a resonant cavity. Similar analysis was presented by (Veitch, 1986) but for only the specific case when $\beta_e < 1$ and $\Omega_p \approx \Omega_o$. The transducer does not necessarily operate in these regimes and here a generalised theory is presented.

### B.2.4.1 Incident signal with phase modulation

When the incident pump signal contains phase modulation sidebands as described in (B28a), the reflected signal can be defined in terms of the reflection coefficient as;

$$\sqrt{P_{inc}}\,Exp\left(j\Omega_p t\right)\Gamma(j\delta)\left[1+Exp(jy)\left(\frac{m_{pm}\Gamma_U}{2}Exp(jx) - \frac{m_{pm}\Gamma_L}{2}Exp(-jx)\right)\right] \quad (B29),$$

where $y = (\theta_U+\theta_L)/2$, $x = \omega t+\theta_{pm}+(\theta_U-\theta_L)/2$, $\Gamma_U = |\Gamma(j\delta_U)/\Gamma(j\delta)|$, $\Gamma_L = |\Gamma(j\delta_L)/\Gamma(j\delta)|$, $\theta_U = Arg[\Gamma(j\delta_U)/\Gamma(j\delta)]$, $\theta_L = Arg[\Gamma(j\delta_L)/\Gamma(j\delta)]$. Expanding (B21) one obtains;

$$\sqrt{P_{inc}}\,Exp\left(j\Omega_p t\right)\Gamma(j\delta) \quad \times \quad (B30).$$
$$[\,1 + \frac{m_{pm}}{2}Cos(y)(\Gamma_U-\Gamma_L)Cos(x) - \frac{m_{pm}}{2}Sin(y)(\Gamma_U+\Gamma_L)Sin(x) +$$
$$j\frac{m_{pm}}{2}Sin(y)(\Gamma_U-\Gamma_L)Cos(x) + j\frac{m_{pm}}{2}Cos(y)(\Gamma_U+\Gamma_L)Sin(x)\,]$$

Using standard trigonometry (B30) can be rearranged as;

$$\sqrt{P_{inc}}\,Exp\left(j\Omega_p t\right)\Gamma(j\delta)\left[1+m^{\dagger}_{am}Cos\left(\omega t+\theta_{pm}+\theta^{\dagger}_{am}\right)+jm^{\dagger}_{pm}Sin\left(\omega t+\theta_{pm}+\theta^{\dagger}_{pm}\right)\right] \quad (B31),$$

where

$$m^{\dagger}_{pm} = \frac{m_{pm}}{2}\left[\Gamma_U^2+\Gamma_L^2+2\Gamma_U\Gamma_L Cos(2y)\right]^{1/2}$$
$$m^{\dagger}_{am} = \frac{m_{pm}}{2}\left[\Gamma_U^2+\Gamma_L^2-2\Gamma_U\Gamma_L Cos(2y)\right]^{1/2}$$

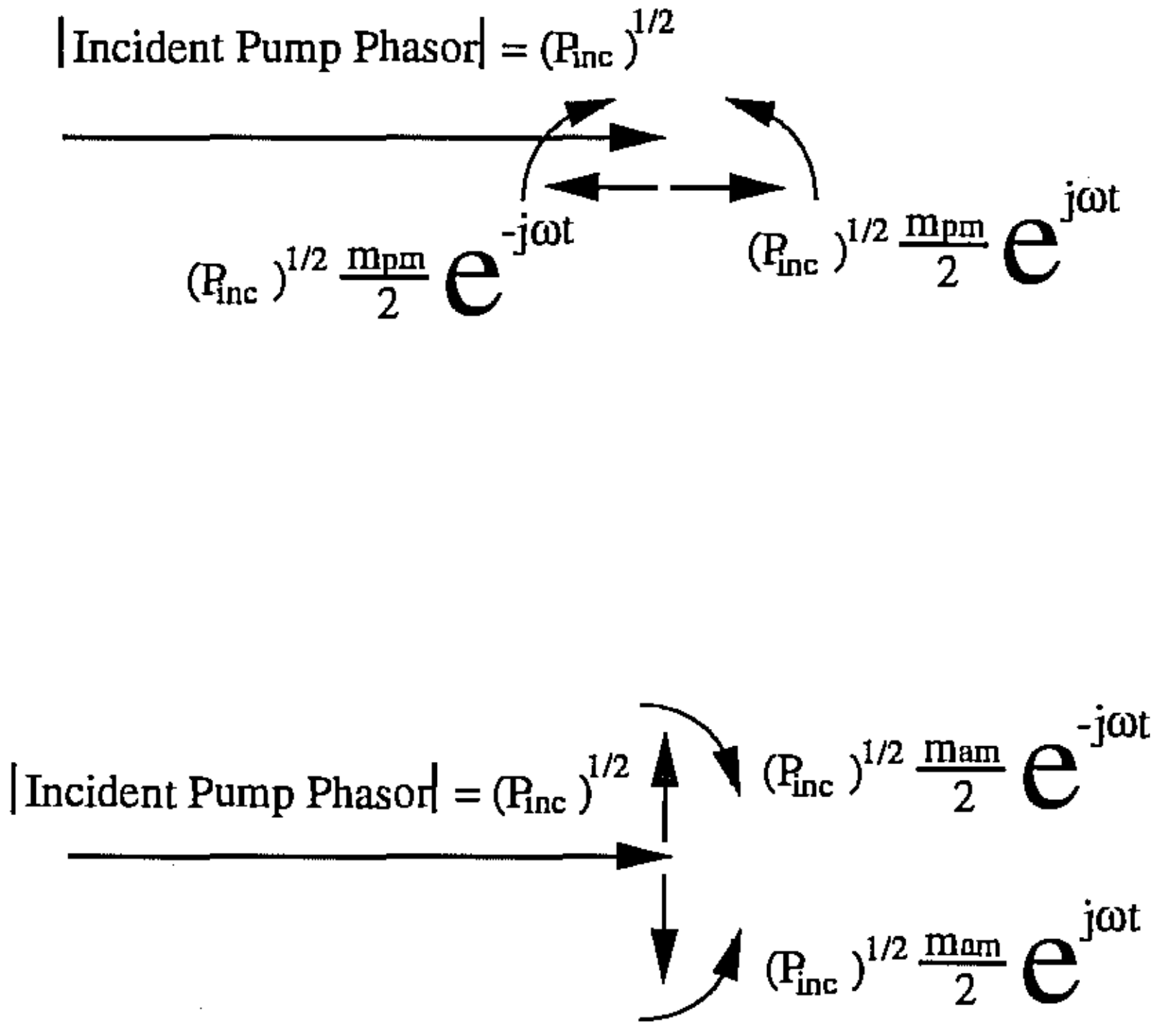

# Figure B.6

Top phasor diagram illustrates equation (B.20a), showing the phasor of the incident signal with rotating phase modulation phasors at t=0 with $\theta_{pm}=0$. Bottom phasor diagram illustrates equation (B.20b), showing the phasor of the incident signal with rotating amplitude modulation phasors at t=0 with $\theta_{am}=0$.

$$\theta^{\dagger}_{pm} = \theta_{\Gamma 1} = \frac{\theta_U-\theta_L}{2} + \mathrm{ArcTan}\left[\frac{(\Gamma_U-\Gamma_L)\mathrm{Sin}(y)}{(\Gamma_U+\Gamma_L)\mathrm{Cos}(y)}\right]$$

$$\theta^{\dagger}_{am} = \theta_{\Gamma 2} = \frac{\theta_U-\theta_L}{2} + \mathrm{ArcTan}\left[\frac{(\Gamma_U+\Gamma_L)\mathrm{Sin}(y)}{(\Gamma_U-\Gamma_L)\mathrm{Cos}(y)}\right]$$

The above modulation indices are with respect to the reflected pump signal $\sqrt{P_{inc}}\mathrm{Exp}(j\Omega_p t)\Gamma(j\delta)$. From (B31) the following scattering parameters with respect to the pump signal for the transducer are calculated to be;

$$S_{pp} = \frac{|\Gamma(j\delta)|}{2}\left[\Gamma_U^2+\Gamma_L^2+2\Gamma_U\Gamma_L\mathrm{Cos}(2y)\right]^{1/2}\mathrm{Exp}(j\theta_{\Gamma 1}) \qquad \text{(B32)}$$

$$S_{ap} = \frac{|\Gamma(j\delta)|}{2}\left[\Gamma_U^2+\Gamma_L^2-2\Gamma_U\Gamma_L\mathrm{Cos}(2y)\right]^{1/2}\mathrm{Exp}(j\theta_{\Gamma 2}) \qquad \text{(B33)}$$

These scattering parameters are used in chapter 7 to calculate the series noise components due to the incident pump oscillator's amplitude and phase noise.

### B.2.4.2 Incident signal with amplitude modulation

When the incident pump signal contains amplitude modulation sidebands as described in (B28b), the reflected signal can be defined in terms of the reflection coefficient as;

$$\sqrt{P_{inc}}\,\mathrm{Exp}(j\Omega_p t)\Gamma(j\delta)\left[1+\mathrm{Exp}(jy)\left(\frac{m_{am}\Gamma_U}{2}\mathrm{Exp}(jx) + \frac{m_{am}\Gamma_L}{2}\mathrm{Exp}(-jx)\right)\right] \qquad \text{(B34)},$$

where in this case $x = \omega t+\theta_{am}+(\theta_U-\theta_L)/2$. Expanding (B34) one obtains;

$$\sqrt{P_{inc}}\,\mathrm{Exp}(j\Omega_p t)\Gamma(j\delta) \quad \times \qquad \text{(B35)}.$$
$$[\,1 + \frac{m_{am}}{2}\mathrm{Cos}(y)(\Gamma_U+\Gamma_L)\mathrm{Cos}(x) - \frac{m_{am}}{2}\mathrm{Sin}(y)(\Gamma_U-\Gamma_L)\mathrm{Sin}(x) +$$
$$j\frac{m_{am}}{2}\mathrm{Cos}(y)(\Gamma_U-\Gamma_L)\mathrm{Sin}(x) + j\frac{m_{am}}{2}\mathrm{Sin}(y)(\Gamma_U+\Gamma_L)\mathrm{Cos}(x)\,]$$

Using standard trigonometry (B30) can be rearranged as;

$$\sqrt{P_{inc}}\,\mathrm{Exp}(j\Omega_p t)\Gamma(j\delta)\left[1+m^{\dagger}_{am}\mathrm{Cos}\left(\omega t+\theta_{am}+\theta^{\dagger}_{am}\right)+jm^{\dagger}_{pm}\mathrm{Sin}\left(\omega t+\theta_{am}+\theta^{\dagger}_{pm}\right)\right] \qquad \text{(B36)},$$

where

$$m^{\dagger}_{am} = \frac{m_{am}}{2}\left[\Gamma_U^2+\Gamma_L^2+2\Gamma_U\Gamma_L\mathrm{Cos}(2y)\right]^{1/2}$$

$$m^{\dagger}_{pm} = \frac{m_{am}}{2}\left[\Gamma_U^2+\Gamma_L^2-2\Gamma_U\Gamma_L\mathrm{Cos}(2y)\right]^{1/2}$$

$$\theta^{\dagger}_{am} = \theta_{\Gamma 1} = (\theta_U-\theta_L)/2 + \mathrm{ArcTan}\left[\frac{(\Gamma_U-\Gamma_L)\mathrm{Sin}(y)}{(\Gamma_U+\Gamma_L)\mathrm{Cos}(y)}\right]$$

$$\theta^{\dagger}_{pm} = \theta_{\Gamma 2} = (\theta_U-\theta_L)/2 + \mathrm{ArcTan}\left[\frac{(\Gamma_U+\Gamma_L)\mathrm{Sin}(y)}{(\Gamma_U-\Gamma_L)\mathrm{Cos}(y)}\right]$$

The above modulation indices are with respect to the reflected pump signal $\sqrt{P_{inc}}\mathrm{Exp}(j\Omega_p t)\Gamma(j\delta)$. From (B36) the following scattering parameters for the transducer are calculated to be;

$$S_{aa} = \frac{|\Gamma(j\delta)|}{2}\left[\Gamma_U^2+\Gamma_L^2+2\Gamma_U\Gamma_L\mathrm{Cos}(2y)\right]^{1/2}\mathrm{Exp}(j\theta_{\Gamma 1}) \qquad (B37)$$

$$S_{pa} = \frac{|\Gamma(j\delta)|}{2}\left[\Gamma_U^2+\Gamma_L^2-2\Gamma_U\Gamma_L\mathrm{Cos}(2y)\right]^{1/2}\mathrm{Exp}(j\theta_{\Gamma 2}) \qquad (B38)$$

These scattering parameters are also used in chapter 7 to calculate the series noise components due to the incident pump oscillator's amplitude and phase noise.

### B.2.5 Reverse transductance

The power that is not reflected off the transducer enters the resonant cavity and consists of the transmitted pump signal and amplitude and phase side bands that are either dissipated or stored inside the resonant cavity. The amplitude noise inside the resonant cavity will be dissipated and react back on the transducer, creating a time varying force on the capacitor plates, which will inturn excite the two-mode resonant bar at the secondary mass. Since the resonant bar is a high Q mechanical filter this noise source is thus inherently narrow band akin to the thermal noise. The phase noise inside the cavity will create no back action because it is the reactive power component, and causes no real power fluctuations.

The reverse transductance scattering parameters that determine the strength of the back action are $S_{fp}$ and $S_{fa}$. Previously (Veitch, 1986) did not consider the possibility of phase to amplitude and amplitude to phase conversions. This must occur to the transmitted signal in the same way as the reflected signal analysed in B.2.3. Thus $S_{fp}$ will depend on the p.m. to a.m. conversion factor $T_{ap}$, and $S_{fa}$ will depend on the transmitted a.m. factor $T_{aa}$. Assuming the same incident signal as described in (B28), the transmitted signal can be defined in terms of the transmission coefficient (B9), and by defining $y = (\theta_U+\theta_L)/2$, $x = \omega t+\theta_{pm}+(\theta_U-\theta_L)/2$, $T_U = |T(j\delta_U)/T(j\delta)|$, $T_L = |T(j\delta_L)/T(j\delta)|$, $\theta_U = \mathrm{Arg}[T(j\delta_U)/T(j\delta)]$, $\theta_L = \mathrm{Arg}[T(j\delta_L)/T(j\delta)]$.

The transmitted signal due to incident p.m. side bands is given by;

$$\sqrt{P_{inc}}\mathrm{Exp}(j\Omega_p t)T(j\delta)\left[1+m^{\dagger}_{am}\mathrm{Cos}\left(\omega t+\theta_{pm}+\theta^{\dagger}_{am}\right)+jm^{\dagger}_{pm}\mathrm{Sin}\left(\omega t+\theta_{pm}+\theta^{\dagger}_{pm}\right)\right] \qquad (B39),$$

where

$$m^{\dagger}_{pm} = \frac{m_{pm}}{2}\left[T_U^2+T_L^2+2T_UT_L\mathrm{Cos}(2y)\right]^{1/2}$$

$$m^{\dagger}_{am} = \frac{m_{pm}}{2}\left[T_U^2+T_L^2-2T_UT_L\mathrm{Cos}(2y)\right]^{1/2}$$

$$\theta^{\dagger}_{pm} = \theta_{T1} = (\theta_U-\theta_L)/2 + \mathrm{ArcTan}\left[\frac{(T_U-\Gamma_L)\mathrm{Sin}(y)}{(T_U+T_L)\mathrm{Cos}(y)}\right]$$

$$\theta^{\dagger}_{am} = \theta_{T2} = (\theta_U-\theta_L)/2 + \mathrm{ArcTan}\left[\frac{(T_U+T_L)\mathrm{Sin}(y)}{(T_U-T_L)\mathrm{Cos}(y)}\right]$$

The above modulation indices are with respect to the stored pump signal $\sqrt{P_{inc}}\mathrm{Exp}(j\Omega_p t)T(j\delta)$. From (B39) the following transmission parameters for the transducer are calculated to be;

$$T_{pp} = \frac{|T(j\delta)|}{2}\left[T_U^2+T_L^2+2T_UT_L\mathrm{Cos}(2y)\right]^{1/2}\mathrm{Exp}(j\theta_{T1}) \qquad \text{(B40)}$$

$$T_{ap} = \frac{|T(j\delta)|}{2}\left[T_U^2+T_L^2-2T_UT_L\mathrm{Cos}(2y)\right]^{1/2}\mathrm{Exp}(j\theta_{T2}) \qquad \text{(B41)}$$

Defining $x = \omega t+\theta_{am}+(\theta_U-\theta_L)/2$, the transmitted signal due to incident a.m. side bands is given by;

$$\sqrt{P_{inc}}\,\mathrm{Exp}(j\Omega_p t)T(j\delta)\left[1+m^{\dagger}_{am}\mathrm{Cos}\left(\omega t+\theta_{am}+\theta^{\dagger}_{am}\right)+jm^{\dagger}_{pm}\mathrm{Sin}\left(\omega t+\theta_{am}+\theta^{\dagger}_{pm}\right)\right] \qquad \text{(B42)},$$

where

$$m^{\dagger}_{am} = \frac{m_{am}}{2}\left[T_U^2+T_L^2+2T_UT_L\mathrm{Cos}(2y)\right]^{1/2}$$

$$m^{\dagger}_{pm} = \frac{m_{am}}{2}\left[T_U^2+T_L^2-2T_UT_L\mathrm{Cos}(2y)\right]^{1/2}$$

$$\theta^{\dagger}_{am} = \theta_{T1} = (\theta_U-\theta_L)/2 + \mathrm{ArcTan}\left[\frac{(T_U-T_L)\mathrm{Sin}(y)}{(T_U+T_L)\mathrm{Cos}(y)}\right]$$

$$\theta^{\dagger}_{pm} = \theta_{T2} = (\theta_U-\theta_L)/2 + \mathrm{ArcTan}\left[\frac{(T_U+T_L)\mathrm{Sin}(y)}{(T_U-T_L)\mathrm{Cos}(y)}\right]$$

The above modulation indices are with respect to the transmitted pump signal $\sqrt{P_{inc}}\mathrm{Exp}(j\Omega_p t)T(j\delta)$. From (B42) the following transmission parameters for the transducer are calculated to be;

$$T_{aa} = \frac{|T(j\delta)|}{2}\left[T_U^2+T_L^2+2T_UT_L\mathrm{Cos}(2y)\right]^{1/2}\mathrm{Exp}(j\theta_{T1}) \qquad \text{(B43)}$$

$$T_{pa} = \frac{|T(j\delta)|}{2}\left[T_U^2+T_L^2-2T_UT_L\mathrm{Cos}(2y)\right]^{1/2}\mathrm{Exp}(j\theta_{T2}) \qquad \text{(B44)}$$

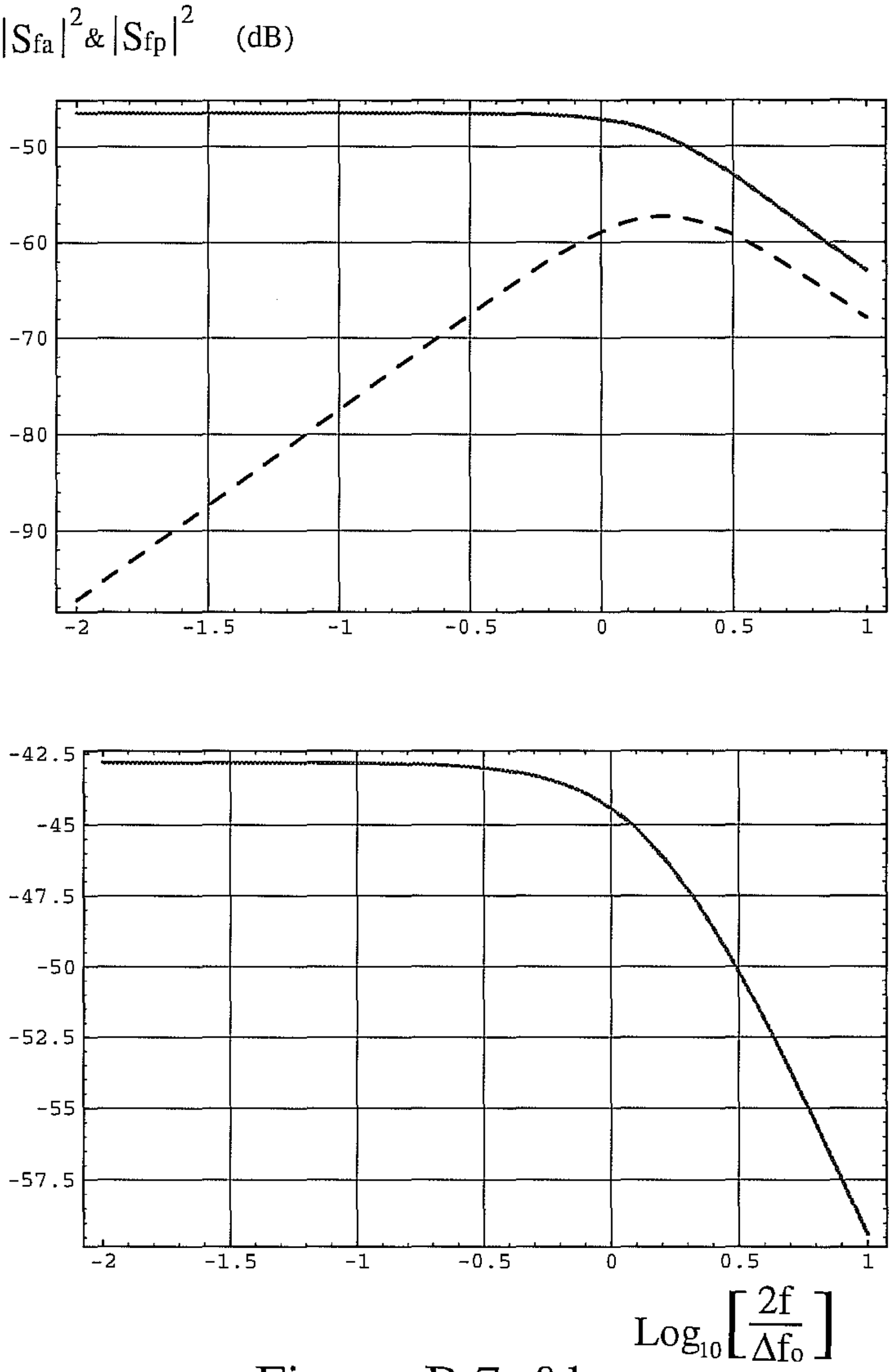

# Figure B.7a&b

Shows $|S_{fa}|^2$(bold line) and $|S_{fp}|^2$(dashed line) as a function of offset frequency from the carrier in terms of unloaded half bandwidths ($2f/\Delta f_o$). Figure B.4a above is with the driving oscillator incident below resonance at an pump offset frequency of $f_p = -(\beta_c+1)\Delta f_o/(2\mathrm{Sqrt}[3])$ Hz. Figure B.4b below is when the driving oscillator is on resonance. Here $|S_{fwd}| \sim 5\ 10^4$, $\beta_c = 0.5$ and $\omega_m/\Omega_o \sim 10^{-7}$.

From (Veitch, 1991) the spectral density of force noise $S_f(\omega)$ $(N/\sqrt{Hz})$, generated between the capacitor plates is;

$$S_f(\omega) = \left(\frac{C_o V_p^2}{2x_o}\right)\sqrt{S^{\dagger}_{am}(\omega)} \qquad \text{(B45).}$$

Here $V_p$ is the peak pump oscillator voltage across the capacitor and $S^{\dagger}_{am}(\omega)$ is the spectral density of a.m. noise inside the resonator in rads/Hz. The pump power inside the resonator is $p_o = V_p^2/(2\ R)$, where $R = Q_e/(\Omega_o C_o)$ with $P_{inc}$ related to the pump power inside the cavity by $p_o = |T(j\delta)|^2\ P_{inc}$. Combining these equations to eliminate $C_o V_p^2$ in (B39) and using $df/dx = f_o/(2x_o)$, it can easily be shown that;

$$S_f(\omega) = \left(\frac{P_{inc} Q_e |T(j\delta)|^2}{\pi\ f_o^2}\ \frac{df}{dx}\right)\sqrt{S^{\dagger}_{am}(\omega)} \qquad \text{(B46)}$$

Thus the complex reverse transductance components can be calculated to be;

$$S_{fp} = \left(\frac{\sqrt{P_{inc}}\ Q_e |T(j\delta)|}{\pi\ f_o^2}\ \frac{df}{dx}\right)\Gamma_{ap} \qquad \text{(B47),}$$

$$S_{fa} = \left(\frac{\sqrt{P_{inc}}\ Q_e |T(j\delta)|}{\pi\ f_o^2}\ \frac{df}{dx}\right)\Gamma_{aa} \qquad \text{(B48),}$$

which can be written in terms of the centre of resonance forward transductance as;

$$S_{fp} = \frac{\omega_m}{\beta_e \Omega_o}\ |S_{fwd}|\frac{T_{ap}}{\sqrt{1+4Q_e^2\delta^2}} \qquad \text{(B49),}$$

$$S_{fa} = \frac{\omega_m}{\beta_e \Omega_o}|S_{fwd}|\frac{T_{aa}}{\sqrt{1+4Q_e^2\delta^2}} \qquad \text{(B50).}$$

These scattering parameters are used in chapter 7 to calculate the back action noise components due to the incident pump oscillator's amplitude and phase noise. Figure B.7 shows the characteristics of $|S_{fa}|^2$ and $|S_{fp}|^2$ as a function of offset frequency from the carrier in terms of unloaded half bandwidths.

# APPENDIX C

# INVESTIGATION OF NOISE SIDEBAND BEHAVIOUR WHEN REFLECTING OFF A RESONANT CAVITY

## C.1 INTRODUCTION

In this appendix the reflection scattering parameters derived in appendix B are used to calculate the noise sideband components when reflecting of a resonant cavity. This model was developed so it could be applied to a gravitational wave antenna with a resonant parametric transducer. In chapter 7 specific use of the model is implemented when estimating the sensitivity of the gravitational wave detector at the University of Western Australia.

This theory can be applied to any measurement system which makes use of a high Q cavity in a phase bridge circuit as shown in figure C.1. First it is applied to the discriminator method of measuring oscillator noise. It is shown if the offset frequency of measurement is within the cavity bandwidth the theory gives the known form of oscillator noise sensitivity. Secondly the theory is applied to resonant cavity transducers.

## C.2 THEORY

The carrier side bands reflected off a resonant cavity can be expressed in terms of the incident side bands at a carrier offset of $\omega$ rads/sec, and the scattering parameters derived in appendix B as;

$$\begin{aligned} \mathbf{b}_{am}(\omega) &= S_{ap}\mathbf{a}_{pm}(\omega) + S_{aa}\mathbf{a}_{am}(\omega) \\ \mathbf{b}_{pm}(\omega) &= S_{pp}\mathbf{a}_{pm}(\omega) + S_{pa}\mathbf{a}_{am}(\omega) \end{aligned} \qquad \text{(C1)},$$

where $S_{pp}$, $S_{ap}$, $S_{aa}$ and $S_{pa}$ are given by (B32-33) and (B37-38) respectively.

From (C1) the reflected phase noise power may be expressed as;

$$P_{pm} = |S_{pp}|^2 S_{pm} P_{inc} + |S_{pa}|^2 S_{am} P_{inc} + 2\eta \left| S_{pp} S_{pa}^* \right| \sqrt{S_{pm} S_{am}}\, P_{inc} \qquad \text{(C2)},$$

where $\eta = S_{pm\ am} / \sqrt{S_{pm} S_{am}}$, is the correlation coefficient between the oscillator's phase and amplitude spectral densities $S_{pm}$ and $S_{am}$ respectively, and $S_{pm\ am}$ is the spectral density due to correlations between amplitude and phase fluctuations. When $\eta = 0$ phase and amplitude noise are uncorrelated, when $\eta = 1$ they are totally correlated.

Assuming the resonator is incorporated in a bridge circuit as in figure C.1, with the phase shifter set so the measurement is phase sensitive, then the in phase correlated phase noise

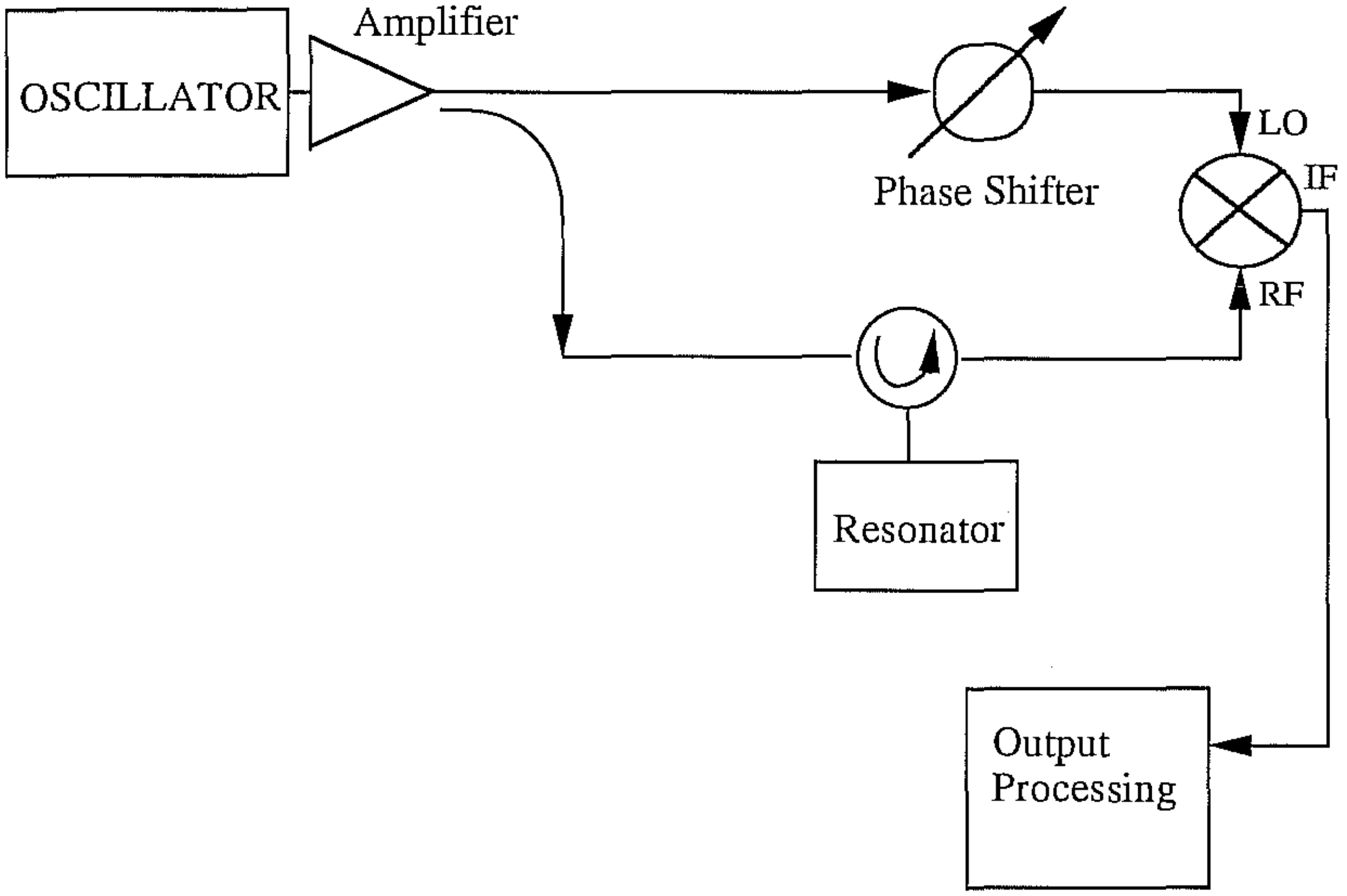

Figure C1

Bridge Circuit with a resonator in reflection in one arm
and a phase shifter in the other.

will cancel in the mixer and need not be considered. Breaking (C2) into these components it may be rewritten as (here $\theta_{pp}=\theta_{\Gamma 1}$ and $\theta_{pa}=\theta_{\Gamma 2}$ from appendix B);

$$P_{pm} = |S_{pp}|^2 Cos^2(\theta_{pp}) S_{pm} P_{inc} + |S_{pp}|^2 Sin^2(\theta_{pp}) S_{pm} P_{inc} + |S_{pa}|^2 (1-\eta) S_{am} P_{inc} + |S_{pa}|^2 \eta Cos^2(\theta_{pa}) S_{am} P_{inc} + |S_{pa}|^2 \eta Sin^2(\theta_{pa}) S_{am} P_{inc} + 2\eta \left|S_{pp} S^*_{pa}\right| Cos^2(\theta_{pp}-\theta_{pa}) \sqrt{S_{pm} S_{am}} P_{inc} + 2\eta \left|S_{pp} S^*_{pa}\right| Sin^2(\theta_{pp}-\theta_{pa}) \sqrt{S_{pm} S_{am}} P_{inc} \qquad (C3).$$

The first, fourth and sixth term in (C3) are correlated in phase components, and will cancel with the oscillator phase noise in the double balanced mixer. Eliminating these terms leaves;

$$P_{pm} = |S_{pp}|^2 Sin^2(\theta_{pp}) S_{pm} P_{inc} + |S_{pa}|^2 (1-\eta Cos^2(\theta_{pa})) S_{am} P_{inc} + 2\eta \left|S_{pp} S^*_{pa}\right| Sin^2(\theta_{pp}-\theta_{pa}) \sqrt{S_{pm} S_{am}} P_{inc} \qquad (C4).$$

In the following analysis it is assumed that $\eta$ -> 0. This may not always be true, however this may be justified for the ultra low noise SLOSC oscillators developed during this research project. The phase noise of this type of oscillator is given in figure 4.23, this spectrum depends on the feed back amplifier in the loop oscillator and the synthesiser mixed with the SLOSC oscillator. However the a.m. noise is dependent on the series a.m noise introduced by the amplifier in series with the oscillator. This fact is highlighted in figure 4.18 and 4.19 where the a.m. amplifier noise is seen above the a.m. noise of the composite SLOSC/synthesiser oscillator. Hence the main contributions towards the a.m. and p.m. noise should be uncorrelated, and $\eta$ -> 0. Thus (C4) becomes;

$$P_{pm} = |S_{pp}|^2_q S_{pm} P_{inc} + |S_{pa}|^2 S_{am} P_{inc} \qquad (C5),$$

where $|S_{pp}|^2_q = |S_{pp}|^2 Sin^2(\theta_{pp})$. The first term is the reflected quadrature phase noise due to dispersion, and the second term is the a.m. to p.m. conversion.

From (C1) the reflected amplitude noise may be expressed as;

$$P_{am} = |S_{ap}|^2 S_{pm} P_{inc} + |S_{aa}|^2 S_{am} P_{inc} + 2\eta \left|S_{ap} S^*_{aa}\right| \sqrt{S_{pm} S_{am}} P_{inc} \qquad (C6).$$

No amplitude noise will cancel in the mixer, therefore assuming that the amplitude and phase noise are uncorrelated leaves only the first two terms in (C6). The first term is the p.m. to a.m. conversion and the second term is the reflected a.m. noise.

## C.3 DISCRIMINATOR METHOD OF MEASURING OSCILLATOR PHASE NOISE

A discriminator phase noise measurement is similiar to the set up shown figure C.1, with the oscillator tuned to the resonant frequency of the cavity. For this case $|S_{pa}|^2 = 0$ and $P_{pm} = |S_{pp}|^2 \mathrm{Sin}^2(\theta_{pp}) S_{pm} P_{inc}$. When $\mathrm{Sin}(\theta_{pp}) << 1$ it is easy to show that;

$$P_{pm} = \left(\frac{2\beta_e(2f/\Delta f_o)}{(1+\beta_e)^2}\right)^2 S_{pm} P_{inc} \qquad \text{(C7)},$$

Figure C.2 compares the square of the term in brackets of (C7) with $|S_{pp}|_q^2 = |S_{pp}|^2 \mathrm{Sin}^2(\theta_{pp})$ for various values of coupling, as a function of offset frequency in terms of unloaded half bandwidths ($2f/\Delta f_o$). The approximation given by (C7) is invalid for offset frequencies larger than the half bandwidth of the cavity. Figure C.3 shows a graph of normalised sensitivity versus electric coupling, and clearly shows that this measurement is most sensitive when the electrical coupling to the resonator is unity.

## C.4 RESONANT CAVITY TRANSDUCERS

In this section it is assumed that a resonant cavity transducer (or parametric transducer) is sensitive to a displacement which causes a change in resonant frequency of the transducer. An example of this is the capacitance modulated transducer presented in appendix B. The reflected noise off a resonant parametric transducer is given by (C5) and (C6). Figure C.4 shows the normalised reflected phase noise $|S_{pp}|_q^2/|S_{pp}(\delta{=}0)|_q^2$ and the incident phase noise converted to reflected amplitude noise $|S_{ap}|^2/|S_{pp}(\delta{=}0)|_q^2$ as a function of pump oscillator offset frequency. The form of these conversions are exactly the same as the electro-mechanical coupling amplitude and phase components illustrated in figure 6.8. If the phase of the detection circuitry shown in figure C1 is set to the same phase given by (B24) that maximises the total electro mechanical coupling. Then as well as sensing the total electro-mechanical coupling, the circuit will also be sensitive to the incident oscillator phase noise and insensitive to the incident amplitude noise. The series noise added by the oscillator for this configuration will be;

$$N_{pm} = \left(|S_{pp}|_q^2 + |S_{ap}|^2\right) S_{pm} P_{inc} \qquad \text{(C8)}.$$

Hence the ratio of total electro-mechanical coupling (B18) to total reflected quadrature phase noise (C8) is constant as a function of pump oscillator offset frequency. Also the signal side band power has the same dependence on electrical coupling ($\beta_e$) to the resonator as the reflected noise power (compare (C7) and (B15)). This means that there is no optimum offset frequency or resonator coupling that maximises the signal to noise.

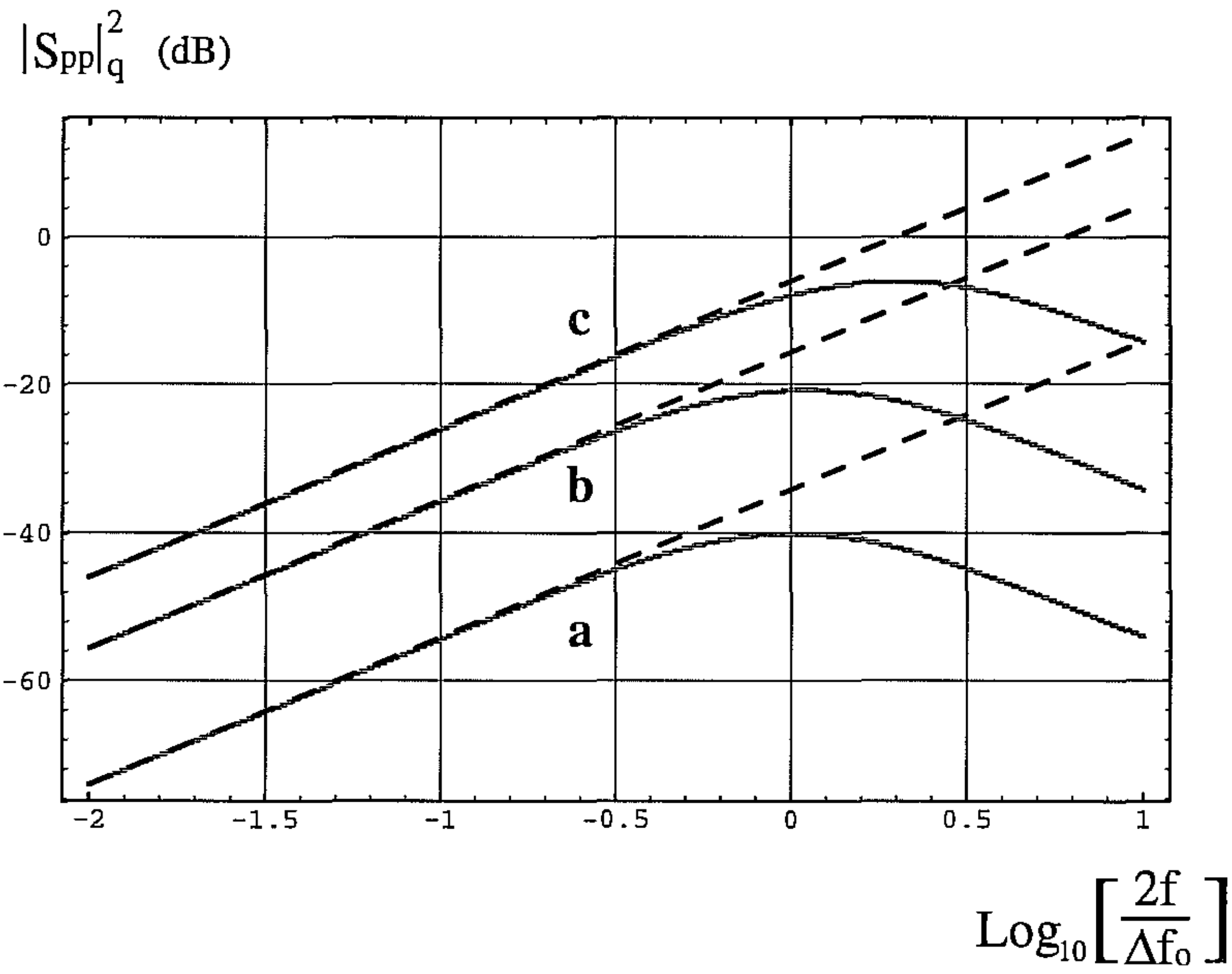

# Figure C.2

Compares $|S_{PP}|_q^2$ (bold lines) with the approximation inside the bandwidth of the discriminator $[2b_e(2f/\Delta f_0)/(1+b_e)^2]^2$ (dashed lines), as a function of offset frequency in terms of unloaded half bandwidths ($2f/\Delta f_0$). **a.** Coupling of $\beta_e = 0.01$. **b.** Coupling of $\beta_e = 0.1$. **c.** Coupling of $\beta_e = 1.0$.

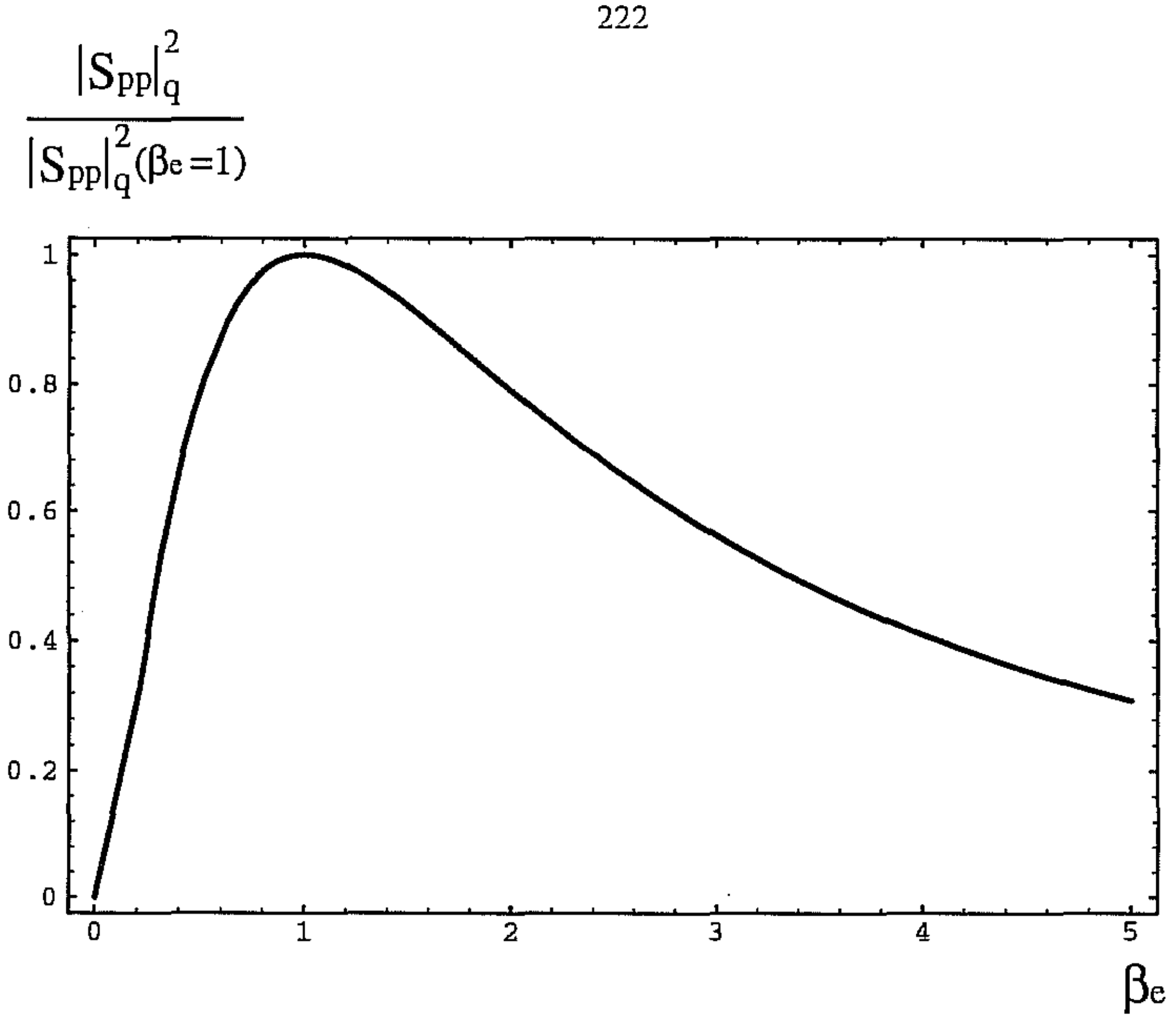

# Figure C.3

Normalised sensitivity versus electric coupling, which clearly shows that an oscillator noise measurement using a discriminator cavity in reflection is most sensitive when the electrical coupling to the cavity is unity.

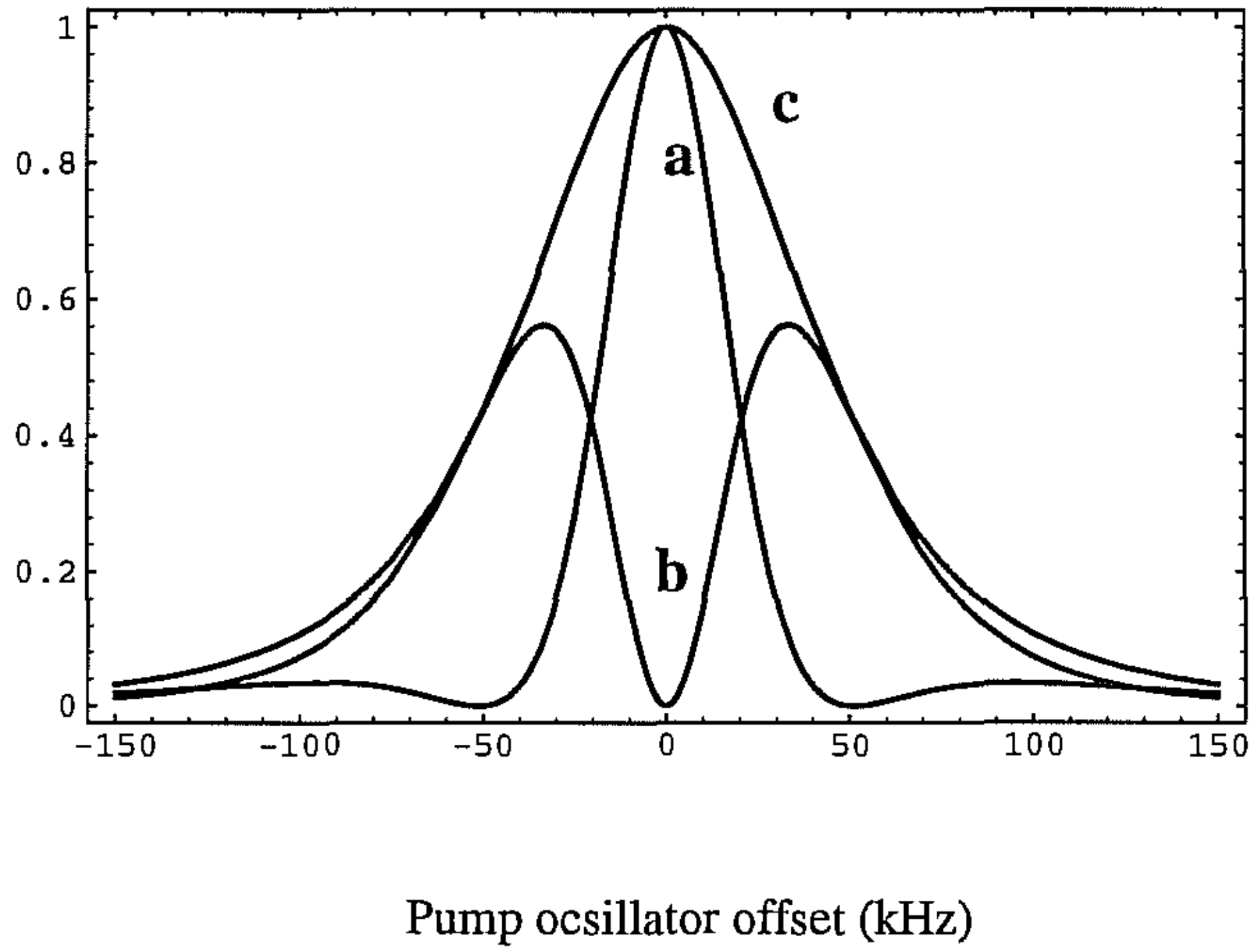

# Figure C.4

Assuming a half bandwith of 69.8 kHz for the resonator the following curves show:

**a.** Normalised reflected quadrature phase noise due to oscillator phase noise.
**b.** Normalised reflected amplitude noise due to oscillator phase noise.
**c.** Total normalised reflected side band noise not cancelled in the mixer, due to oscillator phase noise.

Comparing this plot to the electro-mechanical coupling plots in figure 6.8 reveals that the signal side band amplitude and phase power are the same form as the reflected phase to amplitude noise conversion and reflected quadrature phase noise respectively.

However in the presence of other noise sources (ie. amplifier noise) that are independent of offset frequency and resonator coupling, it is necessary to maximise the signal power. This will be achieved with $\beta_e = 1$ and $\delta = 0$. In chapter 6 and 7 it has been discussed that it should be optimum to operate at an offset given by $\delta$=-1/$\sqrt{3}$ (6.27a), to minimise seismic effects. At this offset the reflected phase noise and electro-mechanical coupling will both be reduced by a factor of 9/16.

# APPENDIX D

## PHASE NOISE CAUSED BY VIBRATIONS IN A MECHANICALLY TUNABLE RESONATOR

A mechanically modulated resonantor can be modelled as the LCR circuit shown in figure D.1. Here we assume vibrations modulate a gap displacement x(t) which will inturn modulate the equivalent circuit capacitance, and hence the resonant frequency. Here we assume;

$$x(t) = x_o\,(1+\alpha \mathrm{Sin}[\omega_m t]) \qquad \text{(D.1)},$$

where $\omega_m$ is the modulation angular frequency, $x_o$ is the steady state gap spacing and $\alpha << 1$. For a capacitor the capacitance is inversely proportional to the displacement, therefore one can write;

$$c(t) = C_o\,(1-\alpha \mathrm{Sin}[\omega_m t]) \qquad \text{(D.2)},$$

where $C_o$ is the steady state capacitance. The resonant frequency is inversely proportional to the square root of the capacitance, and since $\alpha << 1$ it can be written as;

$$f(t) = f_o\,(1-\alpha \mathrm{Sin}[\omega_m t]/2) \qquad \text{(D.2)},$$

where $f_o$ is the steady state resonant frequency. Thus using df/dx = df/dt dt/dx, from (D.1) and (D.2) it can be shown that $df/dx = f_o/(2x_o)$.

When a signal is transmitted through the resonant cavity, the phase modulation coefficient of a capacitance modulated transducer has been calculated to be (Veitch, 1991);

$$m_{pm}^2 = \frac{Q_L^2\alpha^2}{\left(1+\left(2Q_L f_m/f_o\right)^2\right)} \qquad \text{(D.3)},$$

where $Q_L$ is the loaded cavity Q factor. Substituting $S_{pm} = m_{pm}^2$ and $S_x = \alpha^2 x_o^2$ (D.3) becomes;

$$S_{pm} = \left(2Q_L(df/dx)/f_o\right)^2 \frac{1}{\left(1+\left(2Q_L f_m/f_o\right)^2\right)}\, S_x \qquad \text{(D.4)}.$$

This is equivalent to (4.5) used in chapter 4, given that $\Delta f = f_o/Q_L$. Equation (D.4) thus relates the phase noise spectral density in $rads^2/Hz$ to the displacement spectral density in $m^2/Hz$ exciting the tuning mechanism in the tunable resonator.

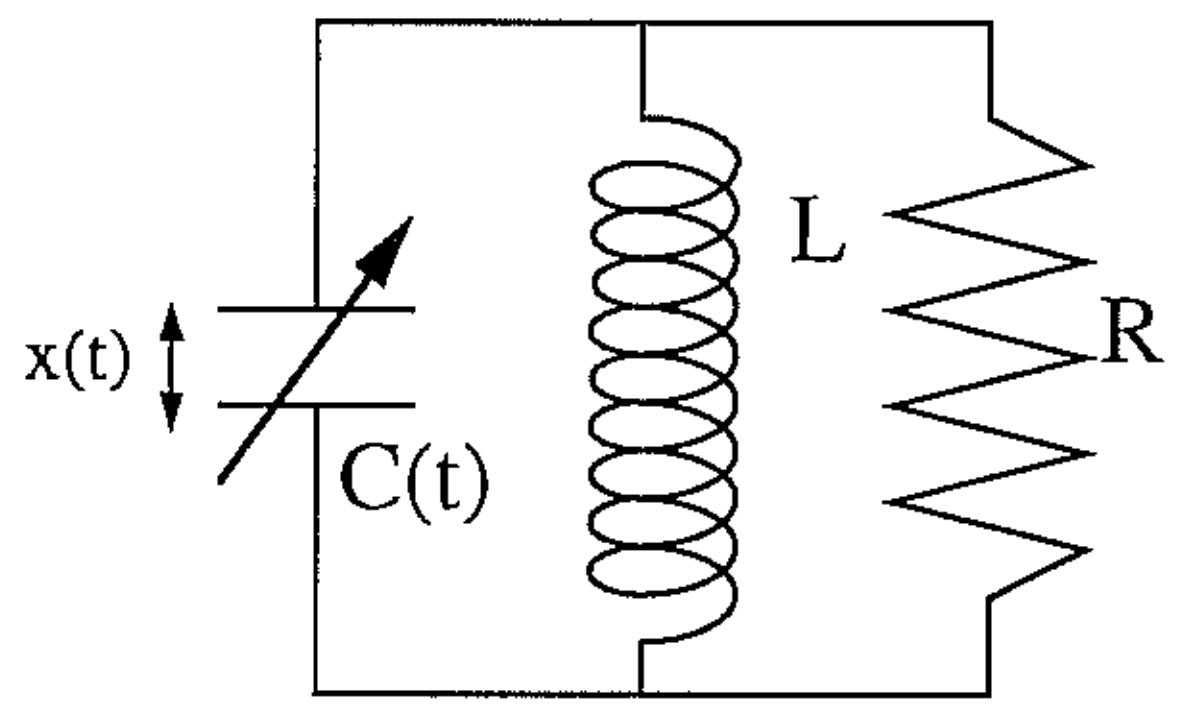

# Figure D.1

A tunable resonant cavity modelled as a capacitance modulated resonator due to a changing gap thickness.